\PassOptionsToPackage{}{xcolor}
\documentclass[PhD, colorlinks=true, linkcolor=black, citecolor=black, urlcolor=black]{dfaunictthesis}
\usepackage{lipsum}
\usepackage[Lenny]{fncychap}

\usepackage{amsthm}
\usepackage{amssymb}
\usepackage{amsmath}
\usepackage{amsfonts}
\usepackage{bbold}
\usepackage{comment}
\usepackage{siunitx}
\usepackage{graphicx}
\usepackage[autostyle]{csquotes}
\usepackage{emptypage}
\usepackage{subfig}
\usepackage{geometry}
\usepackage{tabularx}
\usepackage{booktabs}
\usepackage{siunitx}
\usepackage{multirow}

\usepackage[numbers,sort&compress]{natbib}
\newcommand{\medio}[1]{\langle{#1}\rangle}
\newcommand{\p}{\partial}

\newcommand{\tonde}[1]{\left({#1}\right)}
\newcommand{\quadre}[1]{\left[ {#1} \right]}

\newcommand{\invRe}{\text{Re}^{-1}}
\newcommand{\mfp}{ \lambda_{\text{mfp}}}
\newcommand{\vet}[1]{\mathbf{#1}}

\newcommand{\KnR}{ \text{Kn}^{-1}_R }
\newcommand{\IKn}{ \text{Kn}^{-1}}
\newcommand{\trento}{T\raisebox{-.5ex}{R}ENTo}
\newcommand{\kompost}{K{\o}MP{\o}ST} 

\begin{document}

\author{Vincenzo Nugara}
\title{Far-from-Equilibrium Attractors and Universality\\ in Ultra-Relativistic Heavy-Ion Collisions\\ within Relativistic Kinetic Theory}
\aayear{2024/2025}

\begin{supervisors}
   \supervisor{Chiar.mo}{Prof.}{V. Greco}
   \supervisor{Chiar.mo}{Prof.}{S. Plumari}
\end{supervisors}

\phdname{physics}

\maketitlepage

\tableofcontents

\chapter*{List of publications}

This work is largely based on the following publications:

\begin{description}
	\item[-] V.~Nugara, L.~Oliva, S.~Plumari and V.~Greco, ``Far-from-equilibrium attractors with full relativistic Boltzmann approach in boost-invariant and non-boost-invariant systems'',  \emph{Eur.Phys.J.C} 84 (2024) 8, 861
	[arXiv:2311.11921 [hep-ph]] \cite{Nugara:2023eku}.
	
	\item[-] V.~Nugara, V.~Greco and S.~Plumari, ``Far-from-equilibrium attractors with Full Relativistic Boltzmann approach in 3+1D: moments of distribution function and anisotropic flows $v_n$'',  \emph{Eur.Phys.J.C} 85 (2025) 3, 311 
	[arXiv:2409.12123 [hep-ph]] \cite{Nugara:2024net}.
	
	\item[-] V.~Nugara, N.~Borghini, V.~Greco and S.~Plumari, ``Knudsen number and universal behavior of collective flows'', [arXiv:2509.05495 [hep-ph]] \cite{nugara2025knudsennumberuniversalbehavior}.
	
	\item[-] G.~Parisi, V.~Nugara, S.~Plumari and V.~Greco, ‘‘Shear viscosity of a binary mixture for a relativistic fluid at high temperature'', [arXiv:2510.20704 [hep-ph]] \cite{Parisi:2025gwq}.
\end{description}

\chapter*{Abstract}

This PhD Thesis is devoted to the study of the emergence of attractors, universality and collectivity in ultra-relativistic collisions by means of relativistic kinetic theory. After an introduction about  Quantum Chromodynamics (QCD), Quark-Gluon Plasma (QGP) and the importance of heavy-ion collisions to investigate both, we give an overview about the two main models able to describe the hot QCD matter collective behaviour, namely kinetic theory and hydrodynamics. Afterwards, the Relativistic Boltzmann Transport (RBT) model, which has been employed to obtain most part of the results of this thesis, is carefully described, from the numerical and physical perspectives. The study of attractors and universality proceeds then by starting from a simple one-dimensional massless model, moving to increasingly more complex scenarios, involving the full 3+1D setup, non-conformal systems and realistic event-by-event fluctuations. Particular attention is paid to the physical scales which govern the system collectivity and their interplay. We show that a very good description of collective behaviour can be carried out by means of a few variables which characterise the systems under study.

\chapter*{Introduction}
\addcontentsline{toc}{chapter}{Introduction}
\markboth{INTRODUCTION}{INTRODUCTION}
The microscopic structure of matter has always been a fascinating field of research for humankind. Starting from the atomic theory of Democritus, through European philosophers during the Enlightenment, up to the modern Standard Model of particles and interactions, scholars have always asked what matter is made of. As said, the Standard Model offers today the most comprehensive description of the microscopic world, with the ability to predict observed quantities with astonishing precision, particularly in the context of the electroweak theory. There is however a section of the Standard Model which, despite having been so far extremely successful in its predictions, still resists complete comprehension, due to its inherent complexity and difficulty of calculation: the Quantum Chromodynamics (QCD). Albeit sharing the same mathematical structure of the electroweak interaction, namely that of a gauge theory, the non-abelianity of QCD prevents physicists from using the usual effective approaches to solve the equations and carry predictions out, and has forced them to find alternative routes in order to have a clear picture of what happens within a nucleus and, moreover, inside a nucleon. In this context one of the most effective methods has been to smash objects against each other. Nonetheless, these \emph{objects} are nuclei and \emph{smashing} means accelerating them basically at the velocity of light in huge rings, making these beams of ions collide an incredibly high number of times per second and building enormous detectors to observe the outcomes of such collisions. Surprisingly or not, this ended up to be a really successful way to probe the microscopic details of matter and made it possible to learn and discover a lot about QCD. In particular, one of the predicted and later observed phenomena was the signature of the transition beyond a certain extremely high temperature ($T\sim 10^{12}$ K) towards a new state of matter, in which quarks and gluons are no longer confined within hadrons, but can exist (even though not directly observed) as coloured states. Despite it is known that quarks and gluons, which represent the fundamental objects of the theory, usually live as colour singlets and despite increasing evidence has been collected proving that in these extreme regimes quarks and gluons experience deconfinement, no clear picture exists able to model how this process occurs. This is undoubtedly one of the most outstanding open issues in the context of the Standard Model. Even more excitingly, this deconfined state of matter is thought to have existed in the very early stages of the universe, when it was so hot and compressed that such temperatures were certainly reached and matter existed in a deconfined state. This stage lasted however only a few microseconds, when the temperatures were already low enough for the confinement to occur. It is certainly fascinating to think that heavy-ion collisions probably `free' quarks that have been confined in a hadron for basically 13 billions years.\\
The prediction of the existence of this state of matter, called Quark-Gluon Plasma (QGP), was however accompanied by the surprising discoveries that its progressive characterisation brought along. It was found to be the hottest (i.e. with the highest observed energy density), most perfect (i.e. with the smallest hypothesised value for the shear viscosity over entropy density), most magnetised and most vorticose medium ever observed. A huge quantity of features have been put under the focus of the scientific community: this work, in particular, is mostly focused on the analysis of the QGP collective behaviour. Studying QGP collectivity has raised several issues and its description pushed the boundary of the employed theoretical framework far beyond expectations: the surprising success of relativistic hydrodynamics led to a deep and profound study of this model and of its theoretical foundations, and the same could be said of relativistic kinetic theory, its possible formulations and applications. The phenomenological discoveries therefore led to a strong theoretical development, that was afterwards applicable also outside the initial realm of formulation.\\
More recent analyses and observations brought to attention the fact that small systems generated by proton-ion ($pA$), proton-proton ($pp$) and light-ion collisions do show features that could be interpreted as signatures of collectivity. This raised further questions concerning the nature of the produced quark matter and the dimension of the smallest possible QGP droplet. The small system realm, moreover, looks far away from the usual regimes in which hydrodynamics is applicable and, more generally, in which one expects a collective behaviour. On the one hand this seems to require a non-hydrodynamic theoretical framework to approach the problem, such as the aforementioned kinetic theory; on the other hand the unexpected success of fluid dynamical theory even in these strongly out-of-equilibrium contexts induced a strong inquire about its actual foundations. Phenomenologically, signatures of collectivity do not necessarily imply the presence of a QGP, but can be rather considered a necessary condition and encourage further studies: for instance, collectivity has been observed also in the context of atomic physics with manipulations of a few (order $10^1$) cold atoms.\\
A theoretical standard model has been roughly established for heavy-ion collisions, albeit there are still unsolved issues, mainly due to the necessity to switch from one framework to the another in different stages of the collisions. This may bring along the presence of discontinuities and, in principle, dependence on non-physical parameters such as the switching time itself, with the possible origination of unphysical effects. One could think about the necessity to switch between a conformal and a non-conformal equation of state in passing from early-time pre-equilibrium kinetic models (which, in 2D or 3D, are to date mainly conformal) to the non-conformal realistic EOS of the hydrodynamic codes. Even more interesting, however, is the lack of a similar standard model for small systems, where most of the approximations and assumptions which are reasonable for larger collision systems can no longer apply. This is certainly one of the most urgent challenges in the field, also in view of the likely new light-ion LHC runs (both in the collider and in fixed target (SMOG) setups) and of the incoming Electron-Ion Collider era.

\chapter[QCD, HIC \& QGP]{Quantum Chromodynamics, \\Heavy-Ion Collisions \& Quark Gluon Plasma}
\label{chap:qcd_qgp}

\section{Quantum Chromodynamics}

\subsection{A brief history}
The history of Quantum Chromodynamics has its roots in the discovery of the components of the nucleus: the proton (1911, Rutherford experiment) and the neutron (1932, Joliot-Curies and Chadwick, who was the first to draw the right conclusions from the experiments). It is in particular the discovery of the neutron which inspires the hypothesis of the existence of a nuclear force (Yukawa, 1934), since the electromagnetic interaction could not explain the stability of the nucleus. Around the mid-20th century, however, more and more massive particles appeared (pions, kaons, $\Lambda$'s, $\Sigma$'s...) suggesting that these particles are not elementary, as already happened with the elements of Mendeleev's table. In 1961 Gell-Mann and Ne'eman proposed a structure underlying the hadrons (\emph{The Eightfold Way}); in 1963, Zweig and Gell-Mann independently proposed a model able to reproduce the observed plethora of particles starting from a small number of constituents: quarks (following the winning nomenclature proposed by Gell-Mann); at the time three quarks were enough. Mesons were predicted to be bound states of quark-antiquark pairs; baryons to be made up of three quarks. These new unobserved particles were supposed to have fractional charge, different `flavours' and a new degree of freedom, called `colour', whose existence was hypothesised ad hoc to avoid the violation of Pauli principle and to explain why nobody had ever observed a free quark. Despite the initial understandable scepticism, experimental evidence seemed to confirm the hypothesis. Nonetheless, a rigorous theory was still missing. It was necessary to wait until 1973, when QCD was formulated (independently by the duo Gross-Wilczek and Politzer) in the modern language of a gauge theory and completed the Standard Model in its modern version. In the `90s the discovery of the top quark completed the picture. Quarks are then classified in three doublets: starting from the lightest and the only stable pair (up and down), then the charm and strange up to the heaviest (top and bottom). QCD is however blind to flavours, which matter only in the electroweak context, while it is sensitive to colours. The spin-1 massless bosons mediating this interaction are called gluons, due to the impossibility to isolate a single quark. This will remain the most intriguing problem of the theory.

\subsection{QCD Lagrangian}
Following the path of the electroweak interaction, QCD is formulated as an $SU(3)$ gauge theory: quarks carry three different charges and antiquarks the corresponding anticharges (red, anti-red, blue, anti-blue, green, anti-green) and eight different bosons are needed. This structure is beautifully depicted by the $SU(3)$ gauge group: quarks transform under the fundamental 3-representation of the colour SU (3) symmetry; antiquarks transform in the $\bar 3$-representation. The inner product of $3\otimes \bar 3$ is invariant under $SU(3)$ transformations. Under this perspective, colour singlets (i.e. colour-neutral objects) are represented by combinations of quark wave functions $q$ that remain invariant under $SU(3)$ transformations:
\begin{equation}
	q^i\bar q_i,\quad \varepsilon^{ijk}q_i q_j q_k, \quad \varepsilon_{ijk}\bar q^i \bar q^j \bar q^k,
\end{equation}
These correspond to mesons, baryons and antibaryons, respectively.
\\
The QCD Lagrangian takes the form:
\begin{equation}\label{eq:LQCD}
	\mathcal L_{QCD} = \bar q (i\gamma^\mu D_\mu - M) q - \frac 14 G^a_{\mu\nu} G_a^{\mu \nu},
\end{equation}
where $q$ is the quark wave function which has 4 components in the Dirac spinor space, 3 in colour space and $N_f$ (6 as far as we know) in flavour space; $\gamma^\mu$ are the Dirac matrices and identity operators in the other two spaces. $M$ is the $N_f\times N_f$ \emph{non-diagonal} mass matrix in the flavour space. $D_\mu$ is the covariant derivative defined as:
\begin{equation}
	D_\mu = \p_\mu + i g \frac{\lambda_a}{2} G^a_\mu,
\end{equation}
$g$ being the bare strong coupling constant, $\lambda_a/2=t_a$ the 8 Gell-Mann $3\times 3$ generators of the $SU(3)$ group, and $G^a_\mu$ are the 8 gluonic fields, each one with four Lorentz components. Lastly, the gluonic field tensor is
\begin{equation}
	G^a_{\mu \nu} = \p_\mu G^a_\nu - \p_\nu G^a_\mu - g f_{abc} G^b_\mu G^c_\nu,
\end{equation}
$f_{abc}$ being the structure constants of $SU(3)$.\\
The most striking feature of the QCD Lagrangian is the self-interacting term. Since QCD is a non-abelian theory, bosons not only mediate the interaction, but carry colour charge and participate in the interaction itself. The Feynman representation of the interaction vertex involves three different  possible vertices (unlike the unique vertex allowed in QED):
\begin{center}
    \centering
    \includegraphics[width=0.5\linewidth]{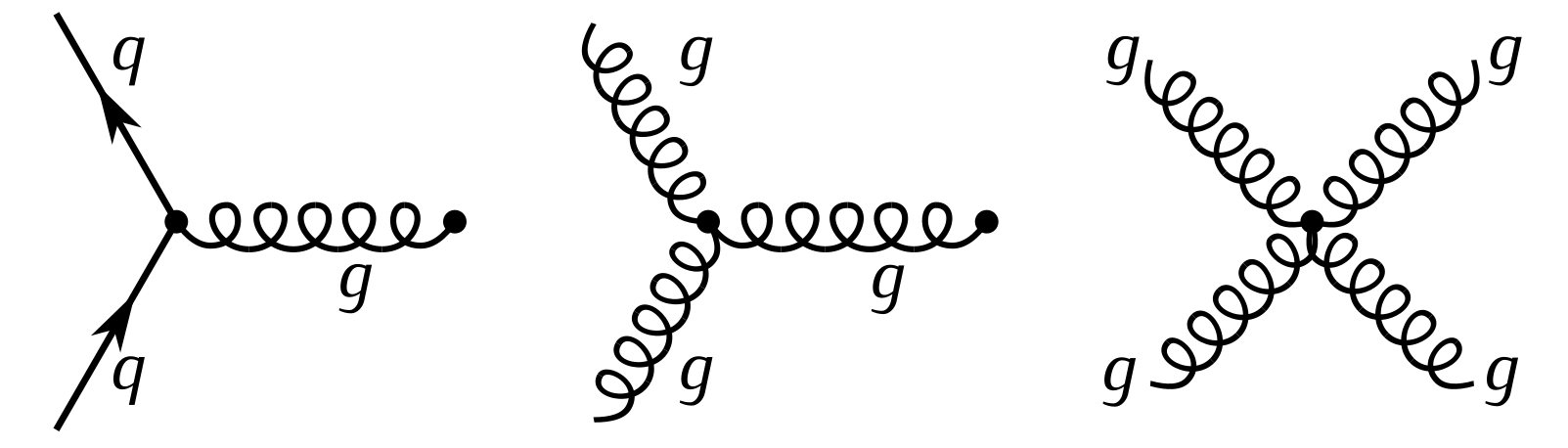}
\end{center}
The three- and four-gluon vertices render QCD far more challenging to study compared to the electroweak theory. \\

\subsubsection{QCD Lagrangian symmetries}
If one sets $m=0$ in Eq. (\ref{eq:LQCD}), the first term of the QCD Lagrangian splits into two independent terms. By defining $\gamma_5=\gamma_0\gamma_1\gamma_2\gamma_3\gamma_4$ and $q_{L,R} = \dfrac{1\mp \gamma_5}{2}\, q $, it is straightforward to show that
\begin{equation}\label{eq:cQCD}
	\mathcal L^c_{QCD} = i \,\bar q_L \gamma^\mu D_\mu q_L + i \,\bar q_R \gamma^\mu D_\mu q_R - \frac 14 G^a_{\mu\nu} G_a^{\mu \nu},
\end{equation}
where the superscript $c$ stands for ``chiral''.\newline
After separating left-handed and right-handed quark fields, it is possible to perform independent transformations which leave the Lagrangian density invariant, acting independently upon the two $q_L$ and $q_R$ fields. In an $N_f$-flavour theory with $\tau^a/2$ the generators of the $SU(N_f)$ group, the fields can be modified as follows:
\begin{gather}
	q_R\to e^{-i\alpha_R}e^{i\beta^R_a(x) \tau^a/2} q_R, \qquad q_L\to e^{-i\alpha_L}e^{i\beta^L_a(x) \tau^a/2} q_L;
\end{gather}
The invariance of the Lagrangian under such these transformations is the $U_R (1) \times SU_R (N_f) \times U_L(1)\times SU_L(N_f)$ symmetry, which leads, according to Noether's theorem, to the four conserved currents:
\begin{gather}
	J^\mu_{R/L} = \bar q_{R/L} \gamma^\mu q_{R/L};\\
	J^{\mu a}_{R/L} = \bar q_{R/L} \gamma^\mu t^a q_{R/L};
\end{gather}

It is possible to arrange them in a different fashion:
\begin{gather}
	V^\mu = J^\mu_R + J^\mu_L = \bar q \gamma^\mu q;\\
	A^\mu = J^\mu_R - J^\mu_L = \bar q \gamma^\mu \gamma^5 q;\\
	V^{\mu a} = J^{\mu a}_R + J^{\mu a}_L = \bar q \gamma^\mu t^a q;\\
	A^{\mu a} = J^{\mu a}_R - J^{\mu a}_L = \bar q \gamma^\mu \gamma^5 t^a q;
\end{gather}
Respectively, they account for:
\begin{itemize}
	\item Vector current conservation and consequently baryon number conservation. This remains exactly valid, even when quark masses are restored.
	\item Axial current conservation: holds in the classical massless QCD Lagrangian (if $m\ne 0$ the current divergence is proportional to $m$), but is broken by the quantisation of the theory (axial anomaly).
	\item $SU(N_f)$ charge conservation: it is exactly valid also in the quantised theory, but only if the quark masses are equal. Indeed, it is a quasi-symmetry for $u,d$ systems (isospin symmetry), and already significantly broken by the inclusion of the strange quark, whose mass is significantly larger than $m_u$ and $m_d$;
	\item Chiral symmetry: exactly valid only if quarks are massless. Nonetheless, it is spontaneously broken even in the massless case; the Goldstone bosons generated by this spontaneous symmetry breaking can be identified as the pions, whose small masses (with respect to the typical nucleon energy scales $\sim$1 GeV) are justified by the fact that the symmetry is already explicitly lightly broken by the small quark masses. The order parameter of this phase transition is the quark condensate $\langle \bar qq\rangle$: this may hint at a relation between the chiral phase transition and the confinement problem, even though to our knowledge it cannot be more than a suggestion. 
\end{itemize}

\subsection{Peculiarities of QCD}
Wilson, Gross and  Politzer proved that non-Abelian theories such as QCD are asymptotically free. This feature was suggested for the first time by deep-inelastic scattering findings; experimental results over the years perfectly agree with the theoretical predictions and show that, at increasing energy scales, the running coupling constant $\alpha_s(Q^2)$ sensitively decreases, indicating that quarks may behave as free at extremely high energy.\\
\begin{figure}
	\centering
	\includegraphics[width=0.7\linewidth]{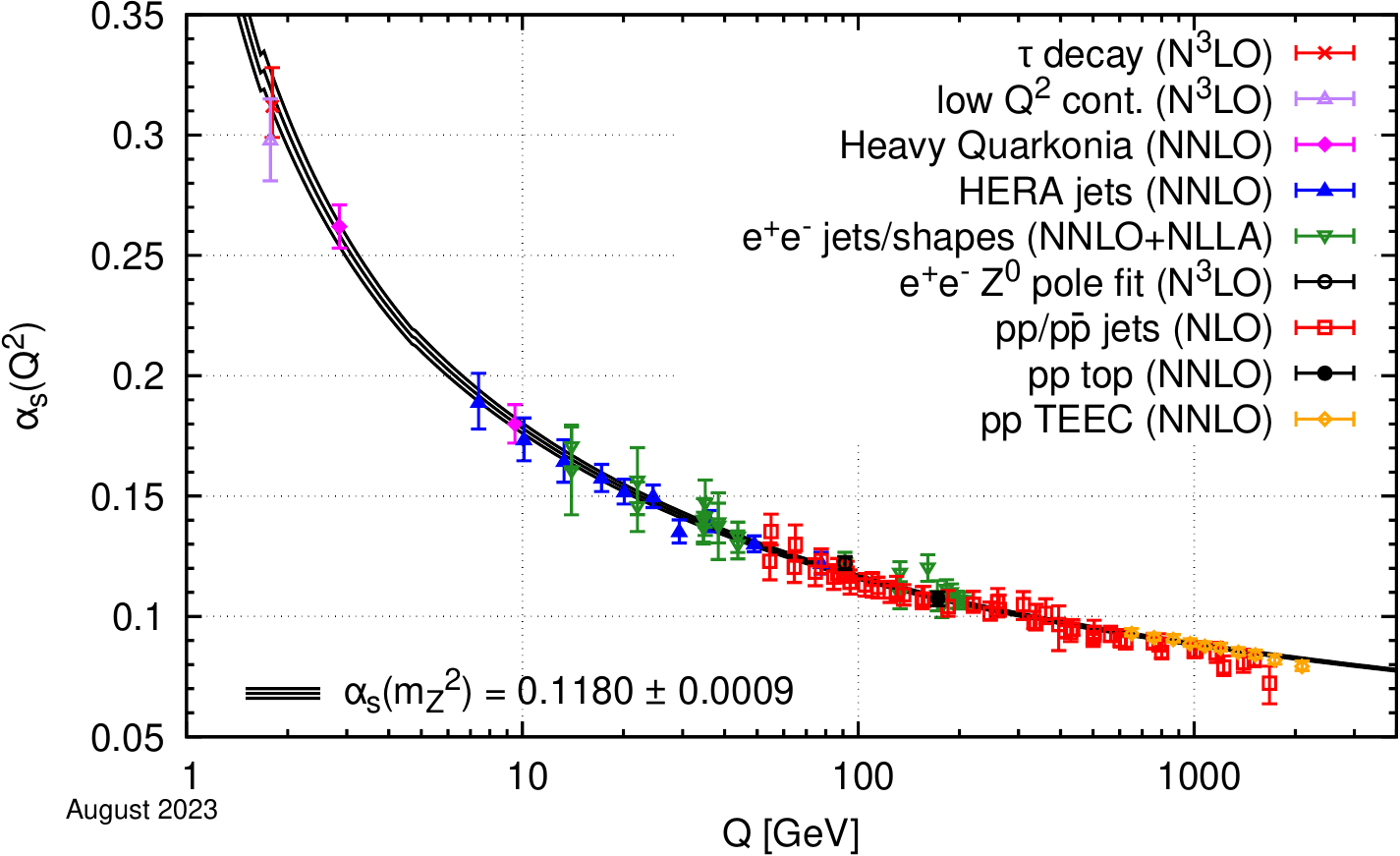}
	\caption{Experimental determinations on the running $\alpha_s(Q)$ compared to the predictions computed at five loops taking as an input the current PDG average, $\alpha_s(m^2_Z) = 0.1180\pm0.0009$ \cite{ParticleDataGroup:2024cfk}.}
	\label{fig:alphas}
\end{figure}
The running of the constant is, as in QED, a direct consequence of the renormalisation of the theory: as an outcome of the Callan-Symanzik equation at one loop, one finds:
\begin{equation}\label{eq:runningcc}
	\alpha_s(Q) = \frac{2\pi}{(11-2/3 N_f) \log(Q/\Lambda)},
\end{equation}
Here $\Lambda$ is the energy scale at which the coupling constant becomes large, and is experimentally found to be $\approx 200$ MeV.
As visible in Figure\,\ref{fig:alphas}, experimental data perfectly agree with the QCD predictions for the running of the coupling constant.\\
Despite continual advance in the understanding of the theory, a central unsolved problem remains: colour confinement. It has not been possible to theoretically prove the mechanism that confine quarks (and gluons) within hadrons and prevents their direct observation. Indeed, confinement occurs in a regime where QCD cannot be studied by standard theoretical methods. Perturbative quantum field theory, so successful in QED, turns out to be inapplicable at the QCD typical energy scale ($\Lambda_{QCD}\sim200$ MeV), since, as stated above, $\alpha_s\sim1$ and there is in principle no limit to the number of Feynman diagram which contribute to describe a process at such energies.\\
Confinement, and QCD more generally, are extremely interesting also because they are responsible for 99\% of the existing mass. As far as ordinary matter is concerned, the Higgs Boson mechanism contribution can account only for the electron and for the bare masses of $u$ and $d$ quarks, each of order MeV. In a hydrogen atom, whose mass is $\approx 1$ GeV, the masses of $2u+d+e$ sum to a few MeV: the remaining  is due exclusively to the interaction energy of quarks, i.e. their ``dressed'' masses. 

\subsubsection{How to study QCD}
\begin{figure}
	\centering
	\includegraphics[width=0.6\linewidth]{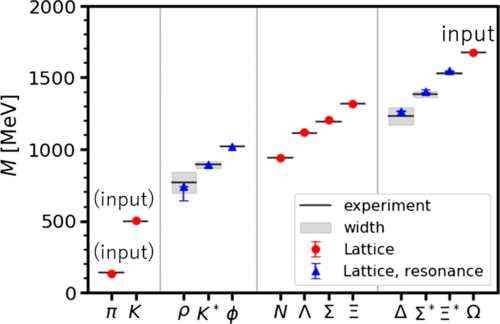}
	\caption{Summary of the hadron spectrum at the state-of-the-art point, where $m_\pi, m_K$ and $m_\Omega$ are used as inputs, while the others are accurately predicted by the theory. Figure from Ref.\cite{Aoyama:2024pts}}.
	\label{fig:hadronmasspredictions}
\end{figure}

As explained above, the difficulty of solving QCD following what has been done for electroweak theory has forced scientists to look for new methods.\\
The closest to the exact resolution is lattice QCD (lQCD). If one cannot approach the QCD Lagrangian analytically, the brute-force solution remains available. It is possible, at very high computational effort, to solve the theory on a discretised 3+1D lattice by computing a huge number of integrals via Monte Carlo methods, following the scheme introduced by Wilson \cite{Wilson:1974sk}: the continuum gauge theory is replaced by a discrete statistical mechanical system on a four-dimensional Euclidean lattice. The basic idea is to push the computational capabilities to more refined lattices and then to extrapolate the results in the continuum limit. Most recent lQCD calculations include $SU(3)$ gauge fields and dynamical $u$, $d$, and $s$ quarks with physical masses; they make use of improved lattice
actions and investigate the convergence as a function of the number of Euclidean time slices. Fully convergent lQCD results at zero baryon chemical potential have been achieved for the QCD equation of state and other thermodynamic properties. Another striking success of lQCD is the prediction of hadron mass values, as shown in Fig.\,\ref{fig:hadronmasspredictions}. \\
However, the method brings along some disadvantages: it can be rigorously applied only at vanishing baryon potential ($\mu_B=0$) due to the so-called sign problem: the fermion determinant in the path integral, which in the $\mu_B=0$ case is real and positive and serves as a probabilistic weight for the integral configurations in Monte Carlo simulations, turns complex and therefore renders importance sampling techniques ineffective.\\
In order to explore regions of the phase diagram which are unreachable by lQCD and to get qualitative insights into the QCD behaviour, a large number of effective theories have been developed, which can make use of different theoretical tools, such as AdS/CFT correspondence, the functional renormalisation group, chiral effective field theory, quasi-particle models etc., which aim at capturing the non-perturbative dynamics at finite $T$ and $\mu_B$. We are not going to delve into these models, but mainly concentrate on the method of inquiry followed in this thesis: the investigation of QCD matter at extreme energies, which are achievable by colliding ions against each other.

\section{Heavy-Ion Collisions (and beyond)}
\subsection{Why HICs}
The idea of investigating the microscopic nature of matter constituents by making them collide is the beginning of nuclear physics, with Rutherford's experiment. It was soon clear that in order to investigate a structure as small as the nucleus, whose size is $\sim 10$ fm, there would not have been many other possibilities. Deep inelastic scattering of electrons on protons played a major role in the development of QCD, for instance in proving the existence of quarks. Several experiments through the years succeeded in defining the current picture of the Standard Model: probably the most striking discovery in particle physics of the last years, i.e. the Higgs Boson finding, was due to the $pp$ collisions performed at LHC.\\
The first relativistic collisions (1-2 GeV per nucleon) between heavy ions were performed at the Bevatron at Berkeley. It paved the way to higher-energies collisions at AGS and then RHIC (Brookhaven, USA), and SPS and then LHC (CERN, Europe).
Today, the most powerful accelerators (RHIC at Brookhaven National Labs and LHC at CERN) can reach energies of $10^2-10^4$ GeV per nucleon, in events better known as ultra-Relativistic Heavy-Ion Collisions (uRHICs). RHIC has worked since 2000 at $\sqrt{s_{NN}}=7.7-200$ GeV, while LHC pushed energies at $\sqrt{s_{NN}}=2.76-5.5$ TeV. Both RHIC and LHC are ring colliders in which ion beams (usually Pb-Pb or Au-Au) are accelerated one against the other. \\
The main point of interest in heavy-ion collisions are the properties characterising the matter that is created at such energies: as illustrated more in detail in Sec.\,\ref{sec:qgp}, beyond the critical temperature $T_c \approx$ 150 MeV a new state of matter is produced, a deconfined medium of quarks and gluons whose collective behaviour will be the main focus of the present work. Studying and characterising the Quark Gluon Plasma (QGP) is a way of tackling the everlasting problem of confinement.\\

\begin{figure}[t]
	\centering
    \includegraphics[width=0.48\linewidth]{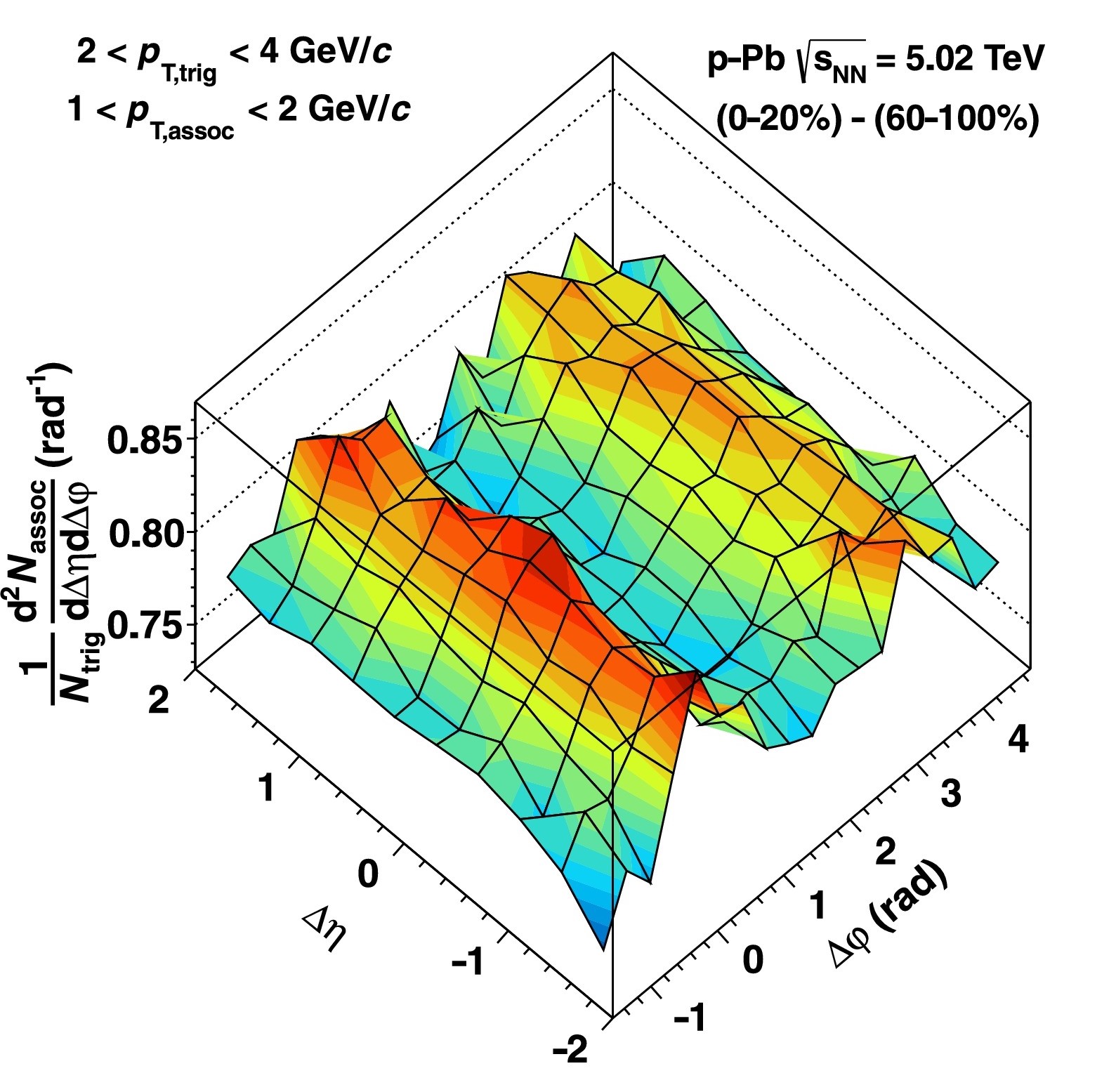}\includegraphics[width=0.48\linewidth]{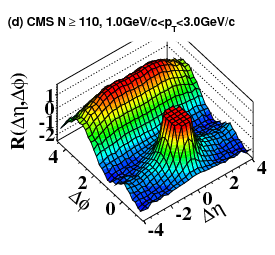}
	\caption{Signatures of collectivity in $pA$ (left panel, from \cite{CMS:2012qk}) and high multiplicity $pp$ (right panel. from \cite{ALICE:2012eyl}).}
	\label{fig:small systems_collectivity}
\end{figure}

It is worth noting, however, that the notion of ``heavy-ion'' is too restrictive: experiments have been performed with $pp$ or $pA$, or in general lighter ions (short runs for O-O and Ne-Ne were performed in July 2025), and more are planned at LHC in the next few years. Quite surprisingly,  signs of collectivity have been claimed to be found also in these smaller systems \cite{ALICE:2012eyl, CMS:2012qk}, even though the volume of the produced medium is expected to be 100--1000 times smaller than the one produced in $AA$ collisions and to live too shortly to reach equilibrium (Figure\,\ref{fig:small systems_collectivity}). It is not clear whether collective behaviour actually emerges in these collision systems or if, for instance, the observed effect is already present in the initial stage of the collisions: solving this puzzle is one of the most intriguing challenges in this field.\\

\subsection{Stages of a uRHIC}
\begin{figure}[t]
	\centering
	\includegraphics[width=0.58\linewidth]{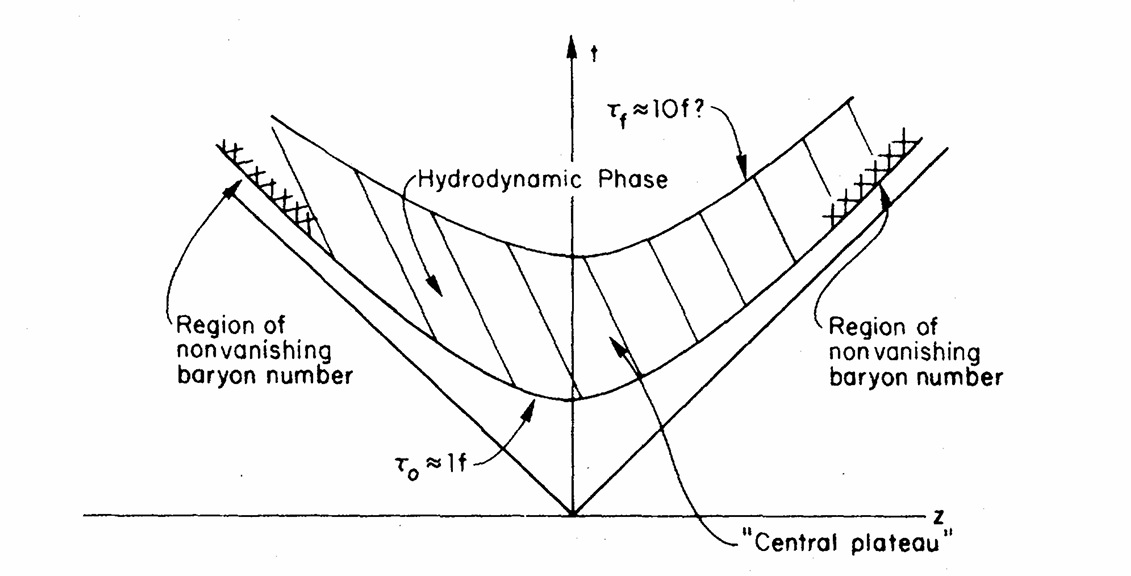}
	\includegraphics[width=0.36\linewidth]{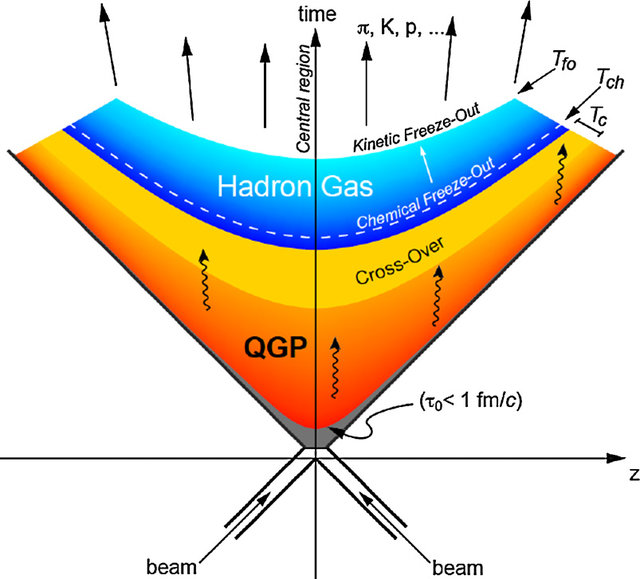}
	\caption{Schematic picture of an uRHIC as pictured by Bjorken in the '80s \cite{Bjorken:1982qr} in the left panel and as nowadays conceived in the right panel \cite{Braun-Munzinger:2018hat}.}
	\label{fig:urhics_scheme}
\end{figure}

We briefly report the standard model of uRHICs, even though more will be said in the following. Refer to Figure\,\ref{fig:urhics_scheme} for a schematic representation.

\begin{description}
	\item[Initial stage] ($t\sim0^+$) The ions are accelerated to reach velocities close to the velocity of light, therefore Lorentz contraction makes them appear as two disks (or ``pancakes'') with negligible transverse size. They almost pass  through each other, depositing in the central collision region a huge amount of energy, which however is only a small fraction of the energy the nuclei were carrying. At such energies, the created medium is almost completely baryon-free, and can be described by a purely gluonic plasma, the glasma. The initial geometry of the system, and its possible anisotropies, are already characterised at this stage.
	\item[Pre-equilibrium] ($t\sim0^+-1 \:\text{fm}$) This strongly interacting matter is initially highly out of equilibrium. It undergoes an extremely fast, nearly boost-invariant longitudinal expansion, responsible for the huge initial pressure anisotropy. This phase is expected to last for $\sim 1$ fm, which is the typical time scale within which it reaches at least partial thermal equilibrium. As explained more in detail in Chapter\,\ref{chap:kin_and_hydro}, this phase is commonly modelled by kinetic theory, which in principle is able to deal with highly non-equilibrated systems.
	\item[Hydro phase] ($t \sim 1- 10 \:\text{fm}$) When the QGP is close to thermal equilibrium, it can be suitably described as a collective medium and the macroscopic hydrodynamic picture has been successfully employed to this goal. In this quite long-lasting phase most observables are thought to be developed, for instance particle spectra, anisotropic flows and jet quenching.
	\item[Freeze-out(s)] ($t\gtrsim 10\:\text{fm}$) The cooling process proceeds until the temperature decreases below $T_c$, when the medium undergoes a smooth phase transition to confined matter, known as hadronisation. However, the system still exhibits collective behaviour until it reaches the chemical and kinetic freeze-out, which mark the freezing of particle species (no more inelastic collisions) and of particle momenta (no more collisions).
	\item[Hadron gas] The so-formed hadrons free stream towards the detectors. Huge detectors (especially ALICE, ATLAS and CMS, but recently also LHCb) are needed to collect as much information as possible: starting from these experimental data, the whole evolution has to be reconstructed.
\end{description}

It is interesting to wonder how this picture changes in the presence of small systems. Since the produced medium is much smaller, it cools down faster and lives for a shorter time. This can be understood also from trivial considerations about the outer layer of the system (at lower temperature, soon reaching the freeze-out) which is in small systems much more relevant than in HICs. It is thus difficult to figure a pre-equilibrium phase followed by a hydrodynamic evolution: the system likely expands and cools down well before reaching thermal equilibrium. This may suggest that no signatures of collectivity should be found in these collisions, which, as already stated, seem to be contradicted by experimental results.

\section{Quark-Gluon Plasma}
\label{sec:qgp}

\subsection{Predictions of QGP}

\begin{figure}
	\centering
	\includegraphics[width=0.85\linewidth]{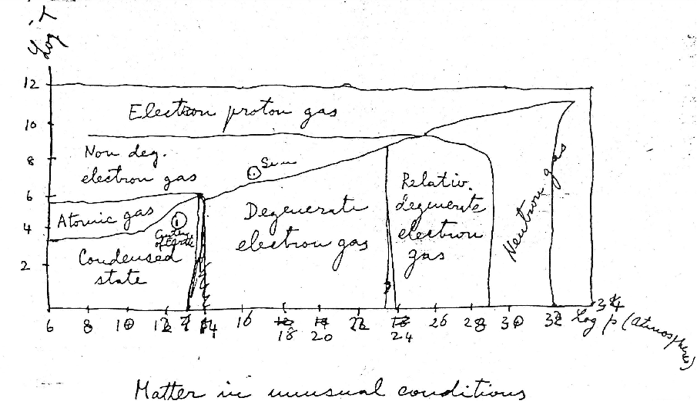}
	\caption{A hand-written phase diagram ($T$ vs $P$) by Enrico Fermi in which nothing was predicted beyond $10^{12}$ K $\approx 200$ MeV, since the inverse of this energy scale is approximately the size of a proton (1 fm), which was then considered an elementary particle. Notice that there is a typo in the $x$ axis label (it is dyne/cm$^2$) \cite{fermi1966notes}.}
	\label{fig:fermischeme}
\end{figure}

The first hint about the (non-)existence of something different from common nuclear matter at temperatures $T\sim 10^{12}$ K is due to Fermi, who sets this value as a sort of maximum for the temperature. It is not difficult to understand why: since nucleons were supposed to be elementary particles and their size is $r\sim 1$ fm, it would not have been possible to reach $T>1/r$.\\
A more refined argument was due to the Hagedorn model for the partition function of the density of hadronic states. Hagedorn established $T_0 \sim$ 160 MeV as an upper bound for the temperature, since above this value the density itself would diverge. He had observed that the number density of hadronic states $\rho(m)$ grows exponentially with mass, and therefore the partition function of an hadron gas reads:
\begin{equation}
	\log Z(T,V) \propto \int_{m_0}^\infty dm\, m^{3/2} \rho(m)\, e^{-m/T} \propto \int_{m_0}^\infty dm\, m^{\alpha + 3/2}\, e^{m (1/T_0 - 1/T)},
\end{equation}
Clearly the partition function diverges if $T\to T_0^-$.\\
In 1975 the first predictions about the would-be QGP appeared: Collins and Perry hypothesised that \emph{[…] matter at densities higher than nuclear consists of a quark soup. The quarks become free at sufficiently high density or temperature}, starting from considerations of asymptotic freedom  \cite{collins75}. It is worth noting that their prediction does not yet describe a QGP: it is true that deconfinement occurs, but still occurs far from the asymptotic freedom regime, since the created matter is extremely strongly interacting, as clear from its characterisation.\\
More interestingly, Cabibbo and Parisi interpreted the divergence in the Hagedorn partition function as a clear signal of a phase transition from hadronic to quark-gluon matter \cite{cabibbo1975}.\\
25 years later, CERN announced \emph{evidence for the existence of a new state of quark-gluon matter in which quarks […] are liberated to roam freely. […] in which quarks and gluons are no longer confined but free to move around over a volume} \cite{abbott2000}. It was one of the first pieces of evidence, and somewhat distant from recent achievements, but it started to be clear that predictions were right.

\subsection{QCD Phase Diagram}

\begin{figure}
	\centering
	\includegraphics[width=0.65\linewidth]{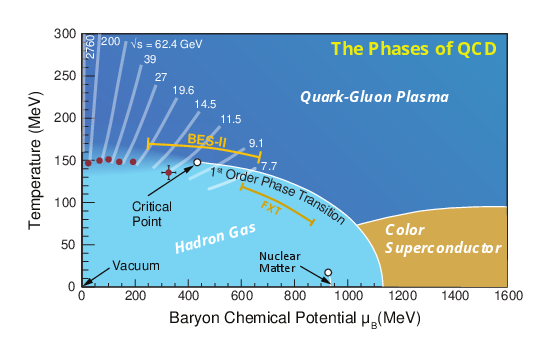}
	\caption{The conjectured QCD phase diagram according to current knowledge. Red circles signal the experimental findings for freeze-out points; white lines highlight the phase-diagram trajectories followed by hydrodynamically expanding matter in a collision whose energy is specified by the label. The range covered by the second phase of the Beam Energy Scan program is marked as ``BES-II'' (collider mode) and ``FXT'' (fixed target). The critical point position is obviously only conjectured \cite{Du:2024wjm}.}
	\label{fig:phasediagram}
\end{figure}

Today, the picture of the QCD phase diagram is much more involved. The region at which LHC and RHIC operate at top energies, i.e. the one on which this work is focused, is basically the one at $\mu_B\approx 0$, with temperatures of order $0.5$-$0.6$ GeV, well above $T_c$, which most recent lQCD estimates set at $148$ MeV. Moreover, the phase transition from nuclear matter to QGP at zero chemical potential is known to be a smooth crossover rather than a transition of definite order. This has been confirmed by lQCD calculations that, as stated above, could in principle be performed only at $\mu_B=0$.\\
The phase diagram of QCD (Figure\,\ref{fig:phasediagram}) is far richer, and still worthy of further investigation. Here we just mention the symmetric extreme case, at $T\sim 0$ and large chemical potential, which should describe for instance the neutron-star matter with a possible colour superconducting phase and the intermediate region, in which several effective theories predict the existence of a critical point, which should mark the onset of a first-order transition between ordinary nuclear matter and QGP. The position and the very existence of such a critical point remain one of the main questions of the field, and are currently under investigation from different perspectives, among which it is worth mentioning the RHIC Beam Energy Scan, in which the beam energies are changed to explore more intermediate regions of the phase diagram, the Compressed Baryonic Matter experiment at FAIR (a new facility under construction in Darmstadt) and extensions of lQCD at finite chemical potential.

\subsection{Characterisation of QGP}

Signatures of QGP formation have been found in uRHICs in abundance \cite{Niida:2021wut}. Most of them were already predicted and expected \cite{Harris:1996zx}, while some others arose along with the experimental findings \cite{Harris:2023tti}. Here only a selection of them is reported:
\begin{description}
	\item[Collective flow] Studying the anisotropic flows made it possible to characterise the QGP as the most perfect fluid ever found, due to the extremely small value of the shear viscosity over entropy density ratio $\eta/s$, with values between 0.05 and 0.2. Moreover, the $v_n(p_T)$ of mesons and baryons show an impressive scaling if normalised as $v_n/n_q (p_T/n_q)$  which provides evidence that the collective flow is generated at the quark level in a deconfined medium, according to the coalescence/recombination picture of hadronisation, which is the most successful at low $p_T$. Finally, radial flow and mean $p_T$ of hadrons are sensitive to the specific bulk viscosity $\zeta/s$ which Bayesian analysis constrains to small but non-zero values ($\approx 10^{-2}$).
	\item [Jet quenching and high-$p_T$ hadron suppression] Highly energetic gluons and quarks give rise to jets of hard particles, whose interaction with the medium via collisional and radiative processes leads to a quenching of the jet itself. The jet quenching is studied via the nuclear modification factor $R_{AA}$ which quantifies the yield suppression with respect to $pp$ collisions, in which no medium is expected to form. It is possible to observe a strong suppression for light high-$p_T$ hadrons, as expected since highly energetic partons strongly interact with the medium; the $R_{AA}$ is instead close to 1 for direct photons, which do not interact via the strong force and see the QGP as a transparent medium. By extracting an effective temperature from the photon spectrum, one finds  values ($T\sim 260-300$ MeV) that are well above the critical temperature of QGP and the temperature extracted from hadrons, proving that photons decouple from the hot medium well before the system freeze-out.\\
	\item [Suppression of quarkonia] Quarkonia, i.e. bound states of heavy quarks such as $c\bar c$ and $b\bar b$ are among the most important probes of the hot QCD medium, since they are mostly formed during the early stages of the collisions and poorly affected by the long time dynamics. Their formation is observed to be strongly suppressed in a deconfined medium, where colour screening makes the interaction between quark and anti-quark weaker, qualitatively similar to what happens for the Debye screening in QED, as predicted by first studies about QGP \cite{Matsui:1986dk}. Moreover, since different bound states of $c\bar c$ as well as $b\bar b$ have different binding energies, it has been possible  to observe their sequential melting happening for increasingly high $T$ (or equivalently energies) \cite{Digal:2001ue}. These observations have been carried out by computing the $R_{AA}$, i.e. the ratio between the yield of quarkonia in heavy-ion collisions and that in $pp$ collisions, in which a medium is not expected to be formed, similarly to what is done for jets. Interestingly, at very high energy, when the number of produced $c\bar c$ is large enough (up to 20 at top LHC energies), it is possible to observe a new increase of charmonium production, due to the regeneration process: $c\bar c$ bound states can be formed also at slightly later times by quark-antiquark pairs which are not necessarily produced together \cite{Thews:2000rj}.     
	\item [Enhancement of strange hadrons production] One of the strongest hints for the formation of QGP is the enhancement of strangeness in the apparent chemical equilibrium of the created hadrons: by comparing the yield of strange hadrons in $AA$ to that in elementary particle collisions, for instance $pp$, one sees that the production of strange hadrons is strongly suppressed in the latter. Since QCD perfectly conserves strangeness (only weak interactions can change the  flavour of a particle), this means that in a deconfined medium abundant gluon‐fusion processes generate $s$-$\bar s$ pairs on time scales much shorter than in hadronic gas, driving strangeness toward full chemical equilibrium and producing the enhanced strange‐hadron abundances observed.
	\item [Complete soft-hadron equilibration] In heavy-ion collisions the observation of the complete chemical equilibration of light hadrons exhibiting the same thermodynamic conditions has been considered a robust evidence that there exists a phase transition from a deconfined state to the hadronic gas at a temperature $T_c\approx 155$ MeV, which beautifully agrees with lQCD predictions. Figure\,\ref{fig:phasediagram} highlights how these freeze-out points perfectly lie on the predicted cross-over region.
    \item [Chiral symmetry restoration] As discussed above, chiral symmetry is spontaneously broken by ordinary matter. However, lattice QCD calculations predict the restoration of chiral symmetry above a critical temperature $T_c$, where the value for $\langle\bar q q\rangle$ is expected to vanish for massless quarks and to strongly decrease for the realistic massive ones. This may be observed experimentally via the degeneration of excitation modes which differ only by parity and whose masses are sensitively distinct for $T<T_c$, for instance the $\rho$ and $a_1$ mesons \cite{Kapusta:1993hq}. Despite the available data cannot exclude the symmetry restoration, the unequivocal detection of this effect is still beyond the so far reached precision and may be addressed by future experiments such as ALICE 3 \cite{ALICE:2022wwr}.
\end{description}

\section{Coordinates and units}
\label{sec:coords_and_units}
In studying heavy-ion collisions, the most suitable coordinates  to exploit the system symmetries in order to work with the minimum number of variables are the so-called Milne coordinates, which identify the beam direction as the preferred one and are particularly useful in the case of boost invariance. To pass from the familiar cylindrical coordinates $(t, r,\phi, z)$ to the Milne coordinates $(\tau, r, \phi, \eta_s)$, one has to use
$$ \tau = \sqrt{t^2 -z^2}, \qquad \eta_s = \tanh^{-1} \tonde{\frac{z}{t}}. $$
$\tau$ is the \emph{proper time}, i.e. the time as measured in the rest frame of the system, while $\eta_s$ is referred to as \emph{space-time rapidity}.\footnote{We will always use the $s$ subscript so to distinguish this symbol from the viscosity $\eta$.} Both these variables are properly defined only in the time-like region, where $t>z$, which is the physically interesting one, in which particles can be produced and travel. Ideally, massless particles could have $t=z$ (and therefore lie in the light-like region) if they travel exactly in the beam direction: however, this is an unphysical case, since these particles are actually spectators of the collision (they simply go on along their original path) or are exactly back-scattered; in any case they cannot be detected, since no detector could cover the 0 (or $\pi$) angle. The metric tensor of such a coordinate system is
\begin{equation}
	g_{\mu\nu} = \begin{pmatrix}
		1 & 0 & 0 & 0\\
		0 & -1 & 0 & 0 \\
		0 & 0 & -1 & 0 \\
		0 & 0 & 0 & -\tau^2 
	\end{pmatrix}.
\end{equation}
Notice that the metric is singular if $\tau=0$, reflecting what we have just said about the light-like limit.\newline
As far as the momentum coordinates are concerned, the \emph{rapidity} is defined as:
$$ Y=\tanh^{-1} \tonde{\frac{p_z}{E}}. $$
In the case of a one-dimensional motion
$$ \beta=\tanh(Y), \qquad \gamma=\cosh(Y), $$
where $\beta$ is the velocity and $\gamma$ the usual Lorentz factor. Rapidities in Minkowski space are additive, unlike velocities: it means that a particle with rapidity $Y_1$ in a given frame will have rapidity  $Y_1 + Y_2$ in a system moving with $Y_2$ with respect to the first one. If one defines also the transverse mass (or transverse energy):
$$ m_T=\sqrt{m^2 + p_x^2 + p_y^2}, $$
it is possible to write the four-momentum vector as
\begin{align*}
	p^\mu = &(E, p_x, p_y, p_z) = (m_T \cosh (Y), p_x, p_y, m_T\sinh(Y)) =\\=& (m_T \cosh (Y), p_T \cos (\phi), p_T \sin(\phi), m_T\sinh(Y)).
\end{align*}
$p_T, Y$ and $m_T$ are the most used variables in describing high-energy collisions: namely the $dN/dY$ particle spectrum is one of the most important observables in such experiments.\newline
For completeness, we report some useful identities:
\begin{gather}
	z=\tau \sinh(\eta_s);\\
	t = \tau \cosh(\eta_s).
\end{gather} 
In case of massless particles, we can exploit:
\begin{gather}
	\label{eq:w_tau_rel}
	p_z=p_T \sinh(Y);\\
	E=p_T \cosh(Y).
\end{gather}
We can also define the boost-invariant momentum space coordinates: the momentum associated to the proper time:
$$p^\tau= \sqrt{p_T^2 + w^2/\tau^2 + m^2},$$
and to the pseudorapidity $\eta_s$:
$$ w= p^{\eta_s} = tp_z - zE.$$
For a matter of convenience we will make use of $p_w=w/\tau$, so as to have momentum in conventional units.
In the massless case:
\begin{gather}
	p^\tau = p_T \cosh(Y-\eta_s);\\
	w/\tau = p_T \sinh (Y-\eta_s).
\end{gather}

As far as the units are concerned, we are going to use natural units $c=1$, $\hbar=1$, $k_B=1$. This means that space and time quantities are measured in fm, while energy, momentum and temperature will be given in GeV (or multiples), and simply $\hbar c=$ 0.197326 GeV $\times$ fm$=1$.

\chapter{Kinetic Theory \& Hydrodynamics}
\label{chap:kin_and_hydro}

The classical picture for modelling continuous media identifies two different regimes in which kinetic theory and hydrodynamics should be applied. If one introduces the Knudsen number (Kn) as the ratio between a microscopic scale, typically the mean free path $\mfp$, and a macroscopic scale, usually the length scale within which macroscopic quantities sensitively change such as the inverse of a gradient $1/\p$, the kinetic theory should model a strongly dilute medium (Kn $\gg 1$), while hydrodynamics should be applicable in the opposite limit (Kn $\ll 1$, ideally Kn $=0$) to an almost equilibrated system. In the context of HICs, the relativistic kinetic theory of transport seemed to provide the most suitable description of the pre-equilibrium phase ($0^+$--1 fm), while relativistic fluid dynamics was able to model the evolution of the quasi-thermalised QGP (1--10 fm) as an expanding almost-ideal fluid. The modern perspective is though much more involved. On one hand the kinetic theory can be interpreted as a mesoscopic model, which allows to map the distribution function $f$ without describing faithfully the microscopic processes; on the other hand hydrodynamics, especially in its viscous and then anisotropic formulation, has extended its regime of applicability even to far-from-equilibrium regimes. Moreover, it has been shown that in the (large) intermediate regime of small but non-vanishing specific shear viscosity $\eta/s$, which quantifies the degree of ``imperfection'' of the fluid, the two models give basically the same predictions and it is even possible to derive hydrodynamics from a microscopic perspective. In this chapter we present the usual formulation of kinetic theory (Sec.\,\ref{sec:kinetic_theory}) and fluid dynamics (Sec.\,\ref{sec:hydrodynamics}) and finally bridge the gap between the two (\ref{sec:microscopic_hydro})

\section{Relativistic kinetic theory of transport}
\label{sec:kinetic_theory}

Since its foundation, the basic idea of kinetic theory is to derive from a microscopic perspective the macroscopic characterisation of a system. Boltzmann was the first one to intuit this possibility, and was able to derive the fundamental thermodynamics laws starting from a microscopic approach and making use of statistics.

\subsection{One-body distribution function}
The statistical description of the system is made possible by the definition of the one-body distribution function:
\begin{equation}
	f(x,p) = \frac{dN}{d^3x\, d^3p/(2\pi)^3}
\end{equation}
which gives by definition the average number of particles within a certain infinitesimal region of the phase-space which is a neighbourhood of $(x,p)$ with volume $d^3x\,d^3 p/(2\pi)^3$. Physically, these differentials should be interpreted as $\Delta^3x$ and $\Delta^3p$ so that each volume is large enough to contain a huge number of particles and in the meanwhile very small with respect to the macroscopic scales of the system.\\
It is of great importance in view of the following to prove that the one-body distribution function $f(x,p)$ is a Lorentz scalar \cite{DeGroot:1980dk}. This can be intuitively understood by  the fact that it can be interpreted as a number or probability distribution: in both cases, the numerical value of the function cannot depend on the reference frame.\\
Consider a physical system that can be described by a certain number of ensembles. The distribution function for on-shell particles reads:
\begin{equation}
	f(x,p) = \left\langle \sum_i^{N} \delta^3 [x - x_i(t)] (2\pi)^3\delta^3 [p-p_i(t)] \right\rangle,
\end{equation}
where $N$ is the total number of particles and the angular brackets denote the average over  the ensemble. Firstly, let us define:
\begin{equation}
\label{eq:mathcalF_def}
	\mathcal F(x,p) = \frac{2\pi}{p^0} \delta (p^0 - \sqrt{p^2 + m^2}) f(x,p) = \left\langle  \sum_i^{N} \frac{(2\pi)^4}{p^0_i(t)} \delta^3 [x - x_i(t)] \delta^4 [p-p_i(t)] \right\rangle.
\end{equation}
If we introduce the identity $1=\int dt\, \delta(t)$:
\begin{equation}
	\mathcal F(x,p) =\left\langle \int dt \sum_i^{N} \delta(t-t_i) \frac{(2\pi)^4}{p^0_i(t)} \delta^3 [x - x_i(t)] \delta^4 [p-p_i(t)] \right\rangle
\end{equation}
Lorentz transformations ensure that $d\tau /dt = m/p^0$, since both time and energy are the time-component of a 4-vector. Therefore one finds, assuming $m$ is the same for all particles:
\begin{equation}
	\mathcal F(x,p) =\frac {(2\pi)^4}{m} \int d\tau \left\langle \sum_i^{N} \delta^4 [x - x_i(\tau)] \delta^4 [p-p_i(\tau)] \right\rangle
\end{equation}
This shows that $\mathcal F(x,p)$ is a Lorentz scalar, which proves that $f(x,p)$ is a Lorentz scalar as well by making use of the known relation:
\begin{equation}
	\theta(p^0) \delta(p^2 -m^2) = \frac{1}{2p^0} \delta(p^0 - \sqrt{p^2 + m^2}),
\end{equation}
where the first term is manifestly Lorentz invariant, while the second is the factor between  $\mathcal F(x,p)$ and $f(x,p)$ in Eq.\,\eqref{eq:mathcalF_def}.\\
Starting from the distribution function it is possible to derive some basics quantities:
\begin{description}
	\item[Density 4-flow] $$ N^\mu = \int \frac{d^3p}{(2\pi)^3p^0} p^\mu f(x,p); $$
	\item[Energy-momentum tensor] $$ T^{\mu\nu}  \int \frac{d^3p}{(2\pi)^3p^0} p^\mu p^\nu f(x,p) ;$$
	\item[Entropy 4-flow] $$ S^\mu = \int \frac{d^3p}{(2\pi)^3p^0} p^\mu f(x,p) [ 1- \log f(x,p) ] $$
\end{description}

These expressions bridge the gap between microscopic and macroscopic quantities. It is worth noting, however, that the distribution function contains much more information and allows for the definition and the extraction of several observables that will be introduced later.\\
One last remark: the formalism here shown can be straightforwardly extended to a mixture of different kinds of particles by simply adding a summation across the species.

\subsection{The relativistic Boltzmann equation}
It is clear that $f(x,p)$ is the unknown of the problem. Provided that the initial condition $f_0(x,p)$ is exactly known (despite in the context of HICs what is known is closer to the \emph{final} distribution function), an evolution equation is needed. Classically, the Boltzmann kinetic equation is derived with the assumptions of binary collisions and of a slowly-changing (in space and time) distribution function. One further hypothesis, i.e. that of molecular chaos, which requires the absence of correlations between particles before each collision, can be shown to be unnecessary. If one derives the equation from a Quantum Field Theory benchmark this hypothesis can be relaxed and the absence of correlations has to be assumed only at a given initial time \cite{DeGroot:1980dk}.\\
In principle, the kinetic equation can be interpreted as the result of two different contributions: the free-streaming and the collision term. Both make the distribution function evolve, despite the most physically interesting phenomena always lie in the collisions. As it will be clear later, this two-component structure is basically present in any resolution, analytical or numerical, of the Boltzmann Equation.

\subsubsection{Free-streaming term}

Consider a point $(x,p)$ in the phase-space. The number of particles, or, more precisely, of particle world lines, in a neighbourhood with volume $\Delta^3x \Delta^3p$ is:
\begin{equation}
	\Delta N (x,p) = \int_{\Delta^3x}\int_{\Delta^3 p} d^3x'\,\frac{d^3p'}{(2\pi)^3} f(x',p').
\end{equation}
Notice that such an interpretation is not so trivial if one generalises the equation to the quantum case, both for localisation issues in the phase space and for the interpretation of the distribution function itself. Consider now a space-like surface $\sigma$, $\Delta^3\sigma$ being a segment at $x$. The number of particle world lines with cross this segment are given by:
\begin{equation}
	\Delta N (x,p) = \int_{\Delta^3\sigma}\int_{\Delta^3 p} d^3\sigma_\mu'\,\frac{d^3p'}{(2\pi)^3p^0}p^\mu f(x',p').
\end{equation}
If there are no collisions between particles, considering a 4-volume $\Delta^4 x$, the net flow of particles through the surface surrounding it must vanish:
\begin{equation}
	\int_{\Delta^3\sigma}\int_{\Delta^3 p} d^3\sigma_\mu'\,\frac{d^3p'}{(2\pi)^3p^0}p^\mu f(x',p') =0
\end{equation}
By making use of Gauss' theorem, it becomes:
\begin{equation}
	\int_{\Delta^4x}\int_{\Delta^3 p} d^4 x'\,\frac{d^3p'}{(2\pi)^3p^0}p^\mu \p_\mu f(x',p') =0
\end{equation}
Due to the arbitrariness of the phase-space interval $\Delta^4x \Delta^3p$,\footnote{Notice that the disequilibrium between coordinate and momentum space arises simply from the fact that the time-space component of the 4-momentum is constrained by the on-shell condition.} this equation holds only if the integrand vanishes:
\begin{equation}\label{eq:boltzmann_free_streaming}
	p^\mu\p_\mu f(x,p)=0,
\end{equation}
which is the Boltzmann equation in the free-streaming limit.\\
Notice that the collision-less Boltzmann equation could in principle include also an external force term: particles change their momenta not due to collisions but by virtue of an external field. In this work we assume no external forces.

\subsubsection{Collision term}

As aforesaid, the physically interesting contributions to the kinetic equation arise from the collision term, i.e. the RHS of Eq.\,\eqref{eq:boltzmann_free_streaming}. It allows particles to exchange momenta; in particular, in the solution here proposed,  it includes elastic binary collisions. It is worth mentioning that more recent and complete formulations of the theory include inelastic collisions as well, that is $1\leftrightarrow2$ or $2\leftrightarrow3$ terms.\\
The modification of the particle number within a certain phase-space volume $\Delta^4x \Delta^3p$ can be written as:
$$ \Delta^4x \frac{\Delta^3 p}{(2\pi)^3p^0} C(x,p), $$
where the $(2\pi)^3p^0$ is introduced in the denominator so to make $C(x,p)$ a Lorentz-invariant quantity. Consider then an elastic collision between two particles which initially have momentum $p$ and $k$, and after the collision $p'$ and $k'$. The average number of collisions like that or, alternatively, the probability that such a collision occurs in a 4-volume $\Delta^4 x$, should be proportional to the number of particles with these initial momenta, namely $\Delta^3p\,f(x,p)$ and $\Delta^3k\,f(x,k)$, and to the final momentum-space volumes $\Delta^3p'$ and $\Delta^3k'$. Therefore, the number of particles that have an initial momentum $p$ and are lost from the $\Delta^3p$ interval after the collisions are:
\begin{align}
	\text{loss term}&=\Delta^4x \frac{\Delta^3 p}{(2\pi)^3p^0} C(x,p) =\\&= \frac 12 \Delta^4x \frac{\Delta^3 p}{(2\pi)^3(2\pi)^3p^0} \int \frac{d^3k}{(2\pi)^3k^0} \frac{d^3p'}{(2\pi)^3p^0} \frac{d^3k'}{(2\pi)^3k'^0} f(x,p) f(x,k) W(p,k| p'k').
\end{align}
The factor $1/2$ takes into account the impossibility to distinguish between a final state with momenta $(p',k')$ from one with $(k',p')$. The new invariant function $W(p,k| p',k')$ is the collision rate and contains the information about when and how collisions occur. Obviously, alongside the loss term, we must add a gain term: particles which after the collision assume a momentum within the $\Delta^3 p$ interval. With the same interpretation:
\begin{equation}
	\text{gain term}= \frac 12 \Delta^4x \frac{\Delta^3 p}{(2\pi)^3p^0} \int \frac{d^3k}{(2\pi)^3k^0} \frac{d^3p'}{(2\pi)^3p^0} \frac{d^3k'}{(2\pi)^3k'^0} f(x,p') f(x,k') W(p',k'| p,k).
\end{equation}
The net change is therefore given by:
\begin{align}
	C(x,p) = \frac 12 \int \frac{d^3k}{(2\pi)^3k^0} &\frac{d^3p'}{(2\pi)^3p^0} \frac{d^3k'}{(2\pi)^3k'^0}\times\\& \times[f(x,p') f(x,k') W(p',k'| p,k) - f(x,p) f(x,k) W(p,k| p',k') ].
\end{align}
Finally, if we put our initial expression inside the kinetic theory equation:
\begin{equation}
	\int_{\Delta^4x}\int_{\Delta^3 p} d^4 x'\,\frac{d^3p'}{p^0}p^\mu \p_\mu f(x',p') = \Delta^4x \frac{\Delta^3 p}{p^0} C(x,p).
\end{equation}
Again, for the arbitrariness of the phase-space interval, we get the full kinetic equation:
\begin{equation}
	p^\mu \p_\mu f(x,p) = C(x,p).
\end{equation}

In the following, we will see the precise shape of the collision term in the specific cases of relaxation time approximation and binary elastic collisions.

\subsection{Relaxation Time Approximation}
Apart from a few cases, an analytical solution of the kinetic theory equation with the full collision integral is beyond our mathematical capabilities. Therefore different approximations of the collision integral have been proposed, such as the Relaxation Time Approximation, Adiabatic Approximation, Small Angle Approximation... \cite{Anderson:1974nyl, Blaizot:2013lga, Brewer:2019oha} In every of these cases an assumption has to be made on the system and its evolution. Among all of them, we show shortly the RTA Boltzmann Equation. We will present in the next paragraph the toy model of 0+1D RTA equation, and immediately after we will summarise the main features of 2+1D RTA Boltzmann equation.

In the RTA framework, the collision term is simply replaced by:
\begin{equation}
	C[f] \to -(u \cdot p) \frac{f -f_{eq}}{\tau_{eq}}.
\end{equation}
where $f_{eq}$ is the equilibrium distribution and $\tau_{eq}$ the relaxation time. One should give a hint for both the equilibrium distribution and the relaxation time by starting from physical considerations: changing $f_{eq}$, the equation is modified and therefore also the solution $f(x,p)$. The most common method is to choose the J\"uttner distribution $f_{eq}\propto \exp(-p\cdot u/T)$, assuming that the system relaxes to equilibrium. This approximation allows to bypass the involved definition of the collision integral and could be useful in case the microscopic processes involving the particles are not known, as there is no need to specify an expression for the collision rate. The counterpart is, however, a proper definition of the relaxation time $\tau_{eq}$, i.e. the time within which the distribution function (approximately) reduces to the equilibrium one: the microscopic information about the system, concerning the interaction and therefore the velocity of thermalisation, is hidden within $\tau_{eq}$. One possible definition wants the relaxation time to be equal to the mean free time between two collisions:
\begin{equation}
	\tau_{eq} = \frac{1}{n\sigma_{tr} v},
\end{equation}
where $\sigma_{tr}$ is the transport cross section, $n$ the mean particle density and $v$ the mean particle velocity. This is still involving microscopic quantities which in general are not known. In the conformal case, the most used ansatz for the relaxation time is to put it in relation with the temperature and the specific shear viscosity:
\begin{equation}
	\tau_{eq} = C \frac{\eta/s}{T},
\end{equation}
with $C$ being a constant factor often fixed $\approx 5$. The advantage in using this definition is the possibility to fix the transport coefficient $\eta/s$, instead of, for instance, the cross section, which allows a direct comparison with hydrodynamics and does not need the knowledge of the microscopic details of the system. This ansatz is used in both the formulations we show in the next paragraphs.\\
It is worth mentioning that recently several non-conformal formulations of the RTA Boltzmann equation have been proposed \cite{Jaiswal:2014isa}, some of them with a consistent relaxation-time \cite{Alalawi:2022pmg} which takes into account the presence of a non-zero bulk viscosity. These formulations are, up to now, limited to simplified 0+1D scenarios.

\subsubsection{0+1D}
\label{subsubsec:rta_0+1D}
In this section we will write down and solve the RTA Boltzmann equation for a one-dimensional, conformal, boost-invariant expanding system \cite{Strickland:2019hff, Florkowski:2013lya, Florkowski:2014sfa}. These symmetries allow to eliminate many functional dependencies from $f(x,p)$, which will be perfectly defined if only $T=T(\tau)$ is known. This is a widespread approach, as the resolution of the equation is straightforward and it is known that in the early stages of the collision the expansion is almost boost-invariant  \cite{Bjorken:1982qr}. By making use of the boost-invariant coordinates $(\tau, x, y, \eta_s)$, the distribution function will depend only on proper time, since there is no transverse dimension and the boost-invariance can be seen as an $\eta_s$-independence ansatz. The equilibrium distribution function is the J\"uttner, which in this coordinates appear as:
\begin{equation}
	f_{eq}(w,p_T; \tau) = \exp \left( -\sqrt{(w/\tau)^2 + p_T^2}/T(\tau) \right).
\end{equation}
Here $w = \tau \,p_T \sinh (Y-\eta_s)$, as defined in Sec.\,\ref{sec:coords_and_units}.
The initial condition $f_0=f(\tau_0)$ in principle can be chosen as an arbitrary function. We specify it as the Romatschke-Strickland distribution function:
\begin{equation}
	f_0 (w,p_T) = \exp \left( -\sqrt{(w/\tau)^2 (1+\xi_0) + p_T^2}/T_0 \right),
\end{equation}
with $\xi_0$ being a parameter quantifying the asymmetry of the momentum distribution of the system and therefore of the pressure anisotropy $P_L/P_T$.\\
The RTA Boltzmann equation assumes the simple form:
\begin{equation}
	\frac{\p f(\tau; w, p_T )}{\p \tau} = \frac{f_{eq} - f(\tau; w, p_T)}{\tau_{eq}};
\end{equation}
whose solution is:
\begin{equation}\label{eq:rta_0+1d}
	f(w, p_T; \tau) = D(\tau, \tau_0) f_0(w, p_T) + \int_{\tau_0}^{\tau} \frac{d\tau'}{\tau_{eq}(\tau')} D(\tau, \tau') f_{eq}(\tau', w, p_T).
\end{equation}
where \( D(\tau_2, \tau_1) \) is defined as:
\begin{equation}
	D(\tau_2, \tau_1) = \exp \left[ - \int_{\tau_1}^{\tau_2} \frac{d\tau''}{\tau_{eq}(\tau'')} \right].
\end{equation}
The common choice in order to solve Eq.\,\eqref{eq:rta_0+1d} by integrating both sides under $\int d^3p\, p$, so that the LHS appears as the time-dependent energy density $e(\tau)$. Notice that the RHS only depends on the temperature of the system, which appears in the equilibrium $f_{eq}$ and in the relaxation time: the two addends account respectively for the free-streaming evolution and for the interaction contribution. To link the temperature with the distribution function one imposes the Landau matching condition  $e=3T_{\text{\text{eff}}}^4/\pi^2$, which can be considered as the definition of an effective temperature assuming that the energy density is always expressed by the formula valid at equilibrium. By doing like that the RTA equation reduces to an integro-differential equation for the temperature that can be numerically solved by iteration on a discrete lattice in $\tau$. For instance one may insert the ideal hydro solution (see below Par.\,\ref{subsec:ideal_hydro}) $T(\tau)=T_0(\tau_0/\tau)^{1/3}$ and then iterate until convergence is reached. The equation is:
\begin{equation}
	T^4(\tau) = D(\tau, \tau_0) T_0^4 \frac{\mathcal{H}^{20}\left(\dfrac{\alpha_0 \tau_0}{\tau}\right)} {\mathcal{H}^{20}(\alpha_0)} + \int_{\tau_0}^{\tau} \frac{d\tau'}{2 \tau_{\text{eq}}(\tau')} D(\tau, \tau')	T^4(\tau') 	\mathcal{H}^{20}\left(\dfrac{\tau'}{\tau}\right).
\end{equation}
where the auxiliary functions $\mathcal H^{nm} (y)$ are defined as
\begin{equation}\label{eq:H_nm_strick}  
	\mathcal H^{nm} (y) = \frac{2 y^{2m+1}}{2m+1} {}_2F_1\tonde{ \frac 12 +m , \frac{1-n}{2}; \frac 32 + m; 1-y^2 }, 
\end{equation}
with ${}_2F_1(a,b; c; z)$ being the ordinary hypergeometric function.\\
Notice that in this way we have imposed only the system energy conservation. This in principle does not mean that also the number of particles is conserved: by doing such a choice for the collision integral we are going beyond the binary collision hypothesis that was originally advanced by Boltzmann.\\
If one wants to fix also the number of particles a new parameter has to be introduced: the fugacity $\Gamma$.\footnote{It is worthwhile to highlight that the hot QCD matter does not conserve the number of particles, even though it probably also deviates from the ideal $\Gamma(\tau)=1$ case. We focus here on the conserving-particle systems in view of Chapter 3.} Therefore the previously defined quantities have to be modified according to:
\begin{gather}
	f_{eq}(w,p_T; \tau) = \Gamma(\tau) \exp ( -\sqrt{(w/\tau)^2 + p_T^2}/T(\tau) );\\
	f_0 (w,p_T) = \Gamma_0 \exp( -\sqrt{(w/\tau)^2 (1+\xi_0) + p_T^2}/T_0 ).
\end{gather}
The effective temperature and fugacity are fixed by two independent Landau matching conditions, respectively for energy and particle density:
\begin{equation}
	e=\Gamma_{\text{eff}}\frac{3T^4_{\text{eff}}}{\pi^2}, \qquad n=\Gamma_{\text{eff}}\frac{T^3_{\text{eff}}}{\pi^2},
\end{equation}
which can be considered as definitions for the effective temperature and fugacity of the system.\\
Starting again from Eq.\,\eqref{eq:rta_0+1d}, by imposing the two conditions one gets two coupled integro-differential equations that can be solved with the same method proposed above:

\begin{gather}
	     \Gamma(\tau) T^4(\tau) = D(\tau,\tau_0) \Gamma_0 T_0^4 \frac{\mathcal H^{20}(\alpha_0 \tau_0 /\tau)}{\mathcal H^{20}(\alpha_0)} + \int_{\tau_0}^\tau \frac{d\tau'}{2\tau_{eq}(\tau')} D(\tau,\tau') \Gamma(\tau') T^4(\tau') \mathcal H^{20}\tonde{{\tau'}/{\tau}},\\
	     \Gamma(\tau)T^3(\tau')= \frac{1}{\tau} \Big[ D(\tau,\tau_0) \Gamma_0 T_0^3 \tau_0 + \int_{\tau_0}^\tau \frac{d\tau'}{\tau_{eq}(\tau')} D(\tau,\tau') \Gamma(\tau') T^3(\tau') \tau'\Big].
\end{gather}

\subsubsection{2+1D RTA and opacity parameter}
\label{subsubsec:rta_3+1D}
It is possible to solve the RTA Boltzmann equation in 2+1D, which takes into account the expansion in the transverse plane and time evolution, i.e. still imposing the boost-invariance hypothesis $f(x,p) = f(\tau, \mathbf x_\perp; p^\tau, p^\eta, \mathbf p_\perp)$ \cite{Kurkela:2018qeb, Kurkela:2018ygx,  Kurkela:2019kip, Kurkela:2019set, Kurkela:2020wwb, Ambrus:2021fej, Ambrus:2022koq, Ambrus:2022qya, Ambrus:2024hks, Ambrus:2024eqa}. The definition of $p^\tau$ and $p^\eta$ are in Section\,\ref{sec:coords_and_units}; for simplicity, we also define $v_z= \tau p^\eta /p^\tau$ and $v^\mu=p^\mu/p^\tau$. It can be shown that the RTA Boltzmann equation reduces to:
\begin{equation}
	\left( \partial_\tau + \mathbf{v}_\perp \cdot \nabla_\perp 
	- \frac{v_z (1 - v_z^2)}{\tau} \partial_{v_z} 
	- \frac{v_z^2 p^\tau}{\tau} \partial_{p^\tau} \right) f 
	= -\frac{T[f] v^\mu u_\mu [f]}{5\eta/s} (f - f_{\text{eq}}[f]).
\end{equation}

The equation is solved in terms of the $p^\tau$-integrated distribution:

\begin{equation}
\label{eq:ptau_integrated_f}
	\mathcal F(\tau, \mathbf x_\perp; \phi_p, v_z) = \frac{\nu_{\text{\text{eff}}}\pi R^2 \tau}{(2\pi)^3} \left( \frac{ dE_\perp^{(0)} }{d\eta_s} \right)^{-1} \int_{0}^{\infty} dp^\tau (p^\tau)^3 f(\tau, \mathbf x_\perp; p^\tau, \phi_p, v_z).
\end{equation}
Here $\nu_{\text{eff}}$ is the effective number of degrees of freedom of the system, $R$ the root-mean-square radius of the initial energy distribution and ${ dE_\perp^{(0)} }/{d\eta_s}$ the initial transverse energy, so that:
\begin{gather}
    e(\tau_0)= \frac{\pi^2\nu_{\text{eff}}}{30}T_0^4\\
    R^2 = \frac{\int d^2\mathbf x_\perp\, x_\perp^2\, e(\tau_0)}{{ dE_\perp^{(0)} }/{d\eta_s}} = \frac{\int d^2\mathbf x_\perp\, x_\perp^2\, e(\tau_0)}{\int d^2\mathbf x_\perp\, e(\tau_0)}
\end{gather}
Notice that since the system is boost-invariant by construction, $dE/d\eta_s$ does not depend on $\eta_s$.

The prefactors in Eq.\,\eqref{eq:ptau_integrated_f} are chosen on purpose so that the Boltzmann Equation assumes the dimensionless fashion:

\begin{equation}\label{eq:rta_2+1D}
	\left( \tilde{\partial}_{\tilde{\tau}} 
	+ \vec{v}_{\perp} \cdot \tilde{\partial}_{\vec{x}_{\perp}} 
	- \frac{v_z (1 - v_z^2)}{\tilde{\tau}} \partial_{v_z} 
	+ \frac{4 v_z^2 - 1}{\tilde{\tau}} \right) F 
	= - \frac{\hat{\gamma} v^\mu u_\mu [F] }{\tilde{\tau}^{1/4}}
	 \tilde{T} [F] 
	\left( F - \mathcal{F}_{\text{eq}} [F] \right),
\end{equation}
where dimensionless quantities are identified by the tilde and defined as:
\begin{gather}
	\tilde \tau \equiv \tau/R; \qquad \tilde {\mathbf x}_\perp \equiv {\mathbf x}_\perp/R\\
	\tilde e \equiv \frac{\tau \pi R^2}{dE^{(0)}_\perp / d\eta_s} e; \qquad \tilde T \equiv \tilde e ^{1/4}.
\end{gather}

With all these definitions, every dimensionfull quantity is now present only within the dimensionless opacity parameter $\hat \gamma$, defined as \cite{Ambrus:2021fej}:

\begin{equation}\label{eq:opacity}
\hat \gamma \equiv \frac{1}{5\eta/s} R^{3/4} \left( \frac{1}{\pi R^2 \pi^2\nu_{\text{\text{eff}}}/30 } \frac{dE^{(0)}_T }{d\eta} \right)^{1/4}\propto \frac{R^{3/4} (e_0\tau_0)^{1/4}}{\eta/s}.
\end{equation}
In the last term we highlighted the dependences of the $\hat \gamma$ on basic physical quantities that characterise the system.
The opacity quantifies the degree of interactiveness of the system under analysis: one could think at the opacity parameter as introduced in astrophysics to describe how much the stellar atmospheres are transparent to the photons produced in the inner part of the star.\\
It is possible to have a microscopic interpretation of the opacity as done in Ref. \cite{Kurkela:2019kip}. In the specific case of an initial Gaussian profile at $\tau=R$ and $r=0$, assuming the conformal relaxation time $\tau_R=(5\,\eta/s)/T$:
\begin{equation}
	\frac{R}{\tau_R(\tau=R)} = \frac{R}{5\eta/s} T(\tau=R) \propto \frac{\hat \gamma}{ (e_0\tau_0)^{1/4} }  (e(\tau=R)R)^{1/4}.
\end{equation}
One can estimate the ratio
\begin{equation}
	\frac{e(\tau=R) R}{e(\tau=\tau_0) \tau_0},
\end{equation}
using a very simple 0+1D toy model. Starting from a distribution with vanishing $P_L$, i.e. with identity of momentum and space-time rapidity $Y=\eta_s$, the system can be assumed to free-stream (constant $e(\tau)\tau$) until $\tau_R$, when it at least partially isotropises. Then, in the limit of large opacity, it can be modelled as an ideal fluid (constant $e(\tau)\tau^{4/3}$)  until $\tau=R$, when the transverse expansion sets off. Therefore, one can solve:
\begin{equation}
	\tau_R (\tau=\tau_R) \sim \frac{5\eta/s}{e(\tau_R)^{1/4}} \sim  \frac{5\eta/s\,\tau_R^{1/4} }{(e_0\tau_0)^{1/4}} \implies \tau_R(\tau=\tau_R)= \frac{(5\eta/s)^{4/3} }{ (e_0\tau_0)^{1/3} } = \hat \gamma^{4/3} R.
\end{equation}
Following the same toy model and the previous equation:
\begin{equation}
	\frac{e(\tau=R) R}{e(\tau=\tau_0) \tau_0} \approx \frac{e(\tau=R) R}{e(\tau=\tau_R) \tau_R} \approx \tonde{ \frac{\tau_R}{R} }^{1/3} \approx \hat \gamma^{4/9}
\end{equation}
And one eventually gets, in the limit $\hat \gamma \gg 1$ 
\begin{equation}
	\frac{R}{\tau_R(\tau=R)} \sim \hat \gamma^{8/9},
\end{equation}
which means that by fixing this quantity one fixes the ratio between the transverse size of the system and the relaxation time (or equivalently the mean free path) computed at $\tau=R$. This reminds of the previously introduced Knudsen number: the proximity between these ideas will be fully investigated in Chapter 6. However, also at this point it is easy to realise that a large opacity implies a small Knudsen number and therefore a strongly interacting system which can be suitably described by hydrodynamics, whilst in the opposite limit the mean free path is comparable with the system size and therefore the hydro regime is quite far away.\\
The numerical solution of Eq.\,\eqref{eq:rta_2+1D} is much more cumbersome than the 0+1D case, being it a partial differential equation. One possible approach is to solve it by linearising it at first order in opacity and perturbations \cite{ Kurkela:2018ygx, Kurkela:2018qeb}, using the free streaming solution as the background upon which the perturbations are propagated. There have been developed, however, solutions beyond the first-order corrections, that make use of the spherical harmonic moments expansion as in \cite{Kamata:2020mka} or the Relativistic Lattice Boltzmann (RLB) code in which the PDE is solved via a finite-difference algorithm \cite{Romatschke:2011hm, Ambrus:2016aub, Succi:2018book, Gabbana:2019ydb, Bazzanini:2020pxb}. We remark however that the solution of this equation is beyond the goals of this thesis, and we will consider the RTA outcomes and above all the opacity definition as a reference for our results in Chapter\,\ref{chap:RBT}.

\section{Relativistic hydrodynamics}
\label{sec:hydrodynamics}
Fluid dynamics is the effective field theory that describes the macroscopic dynamics of fluids \cite{rischke2021}. A more detailed analysis of this definition will be able to picture the main characteristic of the theory. \\
Fluid dynamics studies the dependence of a set of macroscopic functions on the space-time coordinates. These functions usually involve $e(x,t), P(x,t), T(x,t) \dots$, to which a value has to be assigned for every space-time point. It is therefore different from a `one-body' theory, in which a particle, or a fluid-element, is followed in its evolution and thus the main unknown is its trajectory $x=x(t)$. In this sense, also the kinetic theory can be considered a field theory, in which the \emph{phase-space} dependent function is the distribution function $f(x,p)$.
There is however, a major difference between the two. Seeking the macroscopic description of a system may not require the knowledge of its microscopic details: if the system has a huge number of degrees of freedom it would be computationally impossible. However, mesoscopic models such as the kinetic theory manage to deal with the microscopic physics by means of the distribution function and of an evolution equation, such as the Boltzmann Equation, which allows to bridge the gap between the micro- and macro-scopic regimes. The goal of fluid dynamics, in principle, is to infer a macroscopic description of the system without knowing which is the underlying physics. If one is interested in describing the long-distance and long-time behaviour, one simply chooses a set of a few macroscopic variables which vary slowly (hopefully continuously) and then look for a set of equations, starting from physical considerations such as energy and momentum conservation or thermodynamics laws. The main point is then to understand when a physical system allows such a description. If the microscopic degrees of freedom do not average to some convenient macroscopic variable, and are still relevant in its global behaviour, an effective field theory will certainly fail. This occurs when both microscopic and macroscopic scales are relevant in order to achieve a faithful picture of the system. On the contrary, if the microscopic and macroscopic scales are well-separated, an effective theory can be considered a good candidate to model the system. This condition recalls what has been said about the Knudsen Number in the previous section. Lastly, we are still missing the definition of a \emph{fluid}. Rigorously speaking, a fluid is a continuous system in which the microscopic and macroscopic characteristic lengths are far apart. One can consider an infinitesimal volume (a \emph{fluid element}) which is small enough relatively to the macroscopic scales (for instance, the scale along which a macroscopic function changes sensitively), but in the meanwhile is large enough to be considered as a thermodynamic system which \emph{must be close to thermodynamic equilibrium}. This requires its dimensions to be quite larger than the microscopic length scales. In every-day physics we have direct experience of what is a fluid. For instance it is possible to study the motion of water neglecting at all the fact that it is not continuous but made up of a huge amount of discrete bodies (molecules), by taking a fluid element of volume 1 mm$^3$.\\
\begin{figure}
	\centering
	\includegraphics[width=0.7\linewidth]{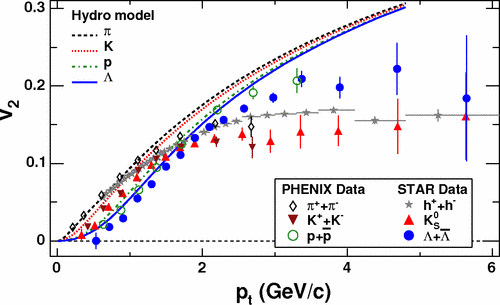}
	\caption{ One of the first good predictions of elliptic flow for $v_2$ vs. $p_T$ for light particles from minimum bias collisions. Data point are from STAR experiment, while the curves are ideal hydro results \cite{STAR:2004jwm}.}
	\label{fig:first_v2_vs_pt_hydro}
\end{figure}

Despite all these constrains may look rigid, a large amount of systems can actually be studied by fluid dynamics. In a medium that is too small or too rapidly changing (microscopic space or time scales are relevant for its evolution), some of the hypotheses listed above are likely to be violated. It was quite surprising therefore, at the beginning of the century, to find out that the first experimental data available at RHIC concerning the elliptic flow of the hot QCD matter, could be very well described (Fig.\,\ref{fig:first_v2_vs_pt_hydro}) by ideal relativistic hydrodynamics. The medium created in the heavy-ion collisions, such as Au-Au at RHIC, however, is expected to thermalise in a short time scale ($\sim$1 fm) and to live for $\sim$ 10 fm, until particlisation sets off. Moreover, the interaction between quarks and gluons should have a very short characteristic length ($1/g^4T$), which is small if compared with the macroscopic length of the system, e.g. the Au radius is $\approx$ 5-6 fm. These considerations may justify the success of hydrodynamics for such systems. Further refinements of the experimental data, also from LHC runs, however, required to move to dissipative hydrodynamics with an extremely small viscosity, namely the smallest ever observed.\\
\begin{figure}
	\centering
	\includegraphics[width=\linewidth]{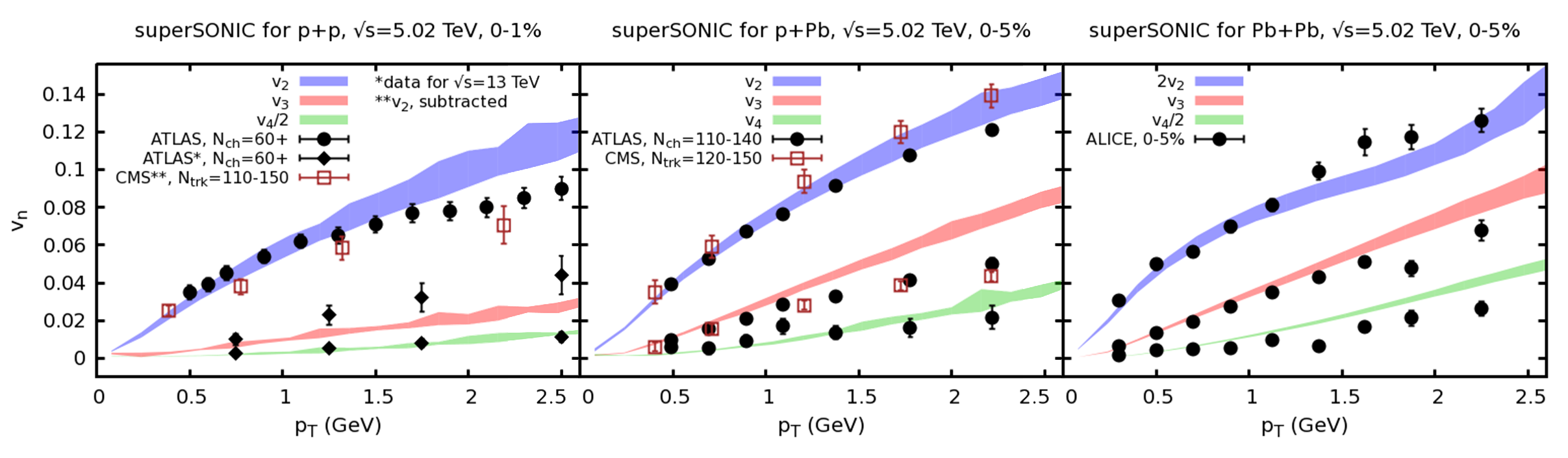}
	\caption{Elliptic ($v_2$), triangular ($v_3$) and quadrupolar ($v_4$) flow coefficients from superSONIC simulations compared to experimental data for $pp$ (left panel), $p$-Pb (center panel) and Pb-Pb (right panel) collisions. Hydro parameters are fixed once for all simulations as $\eta/s=0.08$ and $\zeta/s=0.01$ \cite{Weller:2017tsr}.}
	\label{fig:onefluidtorule}
\end{figure}

The most recent surprise in this field, however, is the successful application of hydrodynamics also in much smaller systems, such as proton - nucleus ($pA$) or proton - proton ($pp$) collisions (see Fig.\,\ref{fig:onefluidtorule}). The presence of finite anisotropic flows has been interpreted as the hint for the possible presence of collective behaviour, and dissipative fluid dynamics proved to be a good tool to describe these outcomes. Nevertheless, it is far from clear whether these signals, such as the `ridge' observed in the differential spectra, are due to the initial state anisotropies, a collective (but non-fluid-like) evolution or a fully hydro behaviour. This scepticism is due to the fact that for such collisions the system size is much smaller (the volume is approximately $10^{-2}-10^{-3}$ smaller than an $AA$ collision) and therefore should not come even close to local equilibration, also because the medium lives for much shorter before the hadronisation. The success of hydrodynamics is nonetheless out of doubt and has been giving a huge incentive to the development of the subject. Nowadays we are supplied with various extensions of hydrodynamics that could be applied in several cases of interests, along with deep theoretical investigations on the true regimes of applicability of hydrodynamics, which can be likely pushed beyond the conventional limits. Further insight about the so-called \emph{hydrodynamisation} problem will be given in the next chapters.

This section is devoted to illustrate the main features of the hydrodynamic theory according to the usual macroscopic approach, while the next one aims to bridge the gap between the hydrodynamics and the kinetic theory regime according to a more modern formulation of the hydrodynamic equations. In the following, most fluid dynamics equations will not be derived since the calculations are usually cumbersome and are not the main focus of this work, but references will be provided: if no indication is present, the reference textbook is the one by Denicol and Rischke \cite{rischke2021}.

\subsection{Ideal relativistic hydrodynamics}
\label{subsec:ideal_hydro}

A fluid is said to be ideal if it fulfils everywhere the local thermodynamic equilibrium condition: it is possible to assign point by point a value to quantities such as temperature $T(x)$, chemical potential $\mu(x)$, energy density $\varepsilon(x)$, flow vector  $\vet u(x)$, etc; $x$ is the position 4-vector $x^\mu=(t,\vet x)$. Temperature, chemical potential and velocity are known as \emph{primary fluid dynamic variables}.\\
As anticipated above, to obtain the equations of motion one must start from general physics laws: here we consider the conserved quantities of the system. For an ideal fluid no dissipation effects are present and, moreover, the system is assumed isolated so that there is no net changing in the particle number.\footnote{It is important to specify \emph{net} since at relativistic energies in principle a particle-antiparticle pair can be created, but with no change for the \emph{net} particle number, i.e. the difference between particles and antiparticles.} In the fluid local rest frame, where $\vet u(x)=0$ (notice that such a frame has to be defined locally in space-time and changes point by point, due to the fact that velocity itself is a function of the position), the system is in \emph{static thermodynamic equilibrium} for hypothesis, and its energy-momentum tensor takes the form:
\begin{equation}
	T^{\mu\nu}_{LRF} = \begin{pmatrix}
		\varepsilon & 0 & 0 & 0 \\
		0 & P & 0 & 0\\
		0 & 0 & P & 0 \\
		0 & 0 & 0 & P \\ 
	\end{pmatrix}.
\end{equation}
There is no energy flow ($T^{i0}_{LRF}=0$) and the stress tensor $T^{ij}$ takes the easy diagonal form in which the force per surface element is isotropic and can be identified with the thermodynamic pressure.\\
Moreover, in this frame no particle nor entropy flows could appear, therefore the two 4-vectors are simply:
\begin{gather}
	N^\mu_{LRF} = (n,0,0,0) = n u^\mu;\\
	S^\mu_{LRF} = (s,0,0,0) = s u^\mu;
\end{gather}
where $n$ and $s$ are respectively the particle and entropy density in the LRF. Notice that the flow vector $u^\mu = \gamma (1 , \vec v)$ is a relativistic 4-velocity and  is therefore normalised as $u^\mu u_\mu=1$, so that it only has three independent components.\\
Having all these definitions clear, the equations of motions are a straightforward consequence of the continuity equations:
\begin{gather}
	\p_\mu N_0^\mu=0;\\
	\p_\mu T_0^{\mu \nu} =0;
\end{gather}
which are valid only in the case of ideal fluid (the subscript $0$ stands for `ideal'). By projecting these equations into the subspaces parallel and orthogonal to $u^\mu$, one gets:
\begin{gather}
	\dot\varepsilon + (\varepsilon + P) \theta=0;\\
	(\varepsilon + P) \dot u^\mu - \nabla^\mu P=0;\\
	\dot n + n \theta=0.
\end{gather}
with $ \theta=\nabla_\mu u^\mu$ being the expansion scalar.
As one can immediately see, there are five equations for six variables ($\varepsilon, P, n $ and the three independent components of $u^\mu$). In order to close the system of equations a further one is necessary, which must bring information about the fluid itself. So far, indeed, no hypothesis has been advanced upon it, beside the fact of being `ideal'. Such an equation is known as \emph{equation of state}, and takes the general form
$$ F(\varepsilon, P, n)=0 .$$
In the case of a conformal system the equation of state is simply: $\varepsilon=3P$.

\subsection{Dissipative relativistic hydrodynamics}

As anticipated, though, a more refined picture of a realistic system needs some of the previous restrictive assumptions to be removed. The conservation laws and the Lorentz invariance must be preserved, but the local equilibrium hypothesis cannot always be assumed: in general it is not true that each fluid element is in thermodynamic equilibrium with its surroundings, since most of the times it exchanges heat with them; furthermore, fluid elements are in relative motion with respect to each other, yielding to energy dissipation. The dissipation prevents the energy-momentum tensor to be diagonal in the LRF and the stress subtensor to be isotropic. These effects can be taken into account by the introduction of the dissipative currents $n^\mu$ and $\tau^{\mu\nu}$:
\begin{gather}
	N^\mu = n_0 u^\mu + n^\mu;\\
	T^{\mu\nu} = \varepsilon_0 u^\mu u^\nu - \Delta^{\mu\nu} P_0 + \tau^{\mu\nu},
\end{gather}
where the $0$ subscript denotes the equilibrium quantities and $\Delta^{\mu\nu}$ is the projection operator onto the 3-space orthogonal to $u^\mu$:
\begin{equation}
    \Delta^{\mu\nu}=g^{\mu\nu}- u^\mu u^\nu.
\end{equation}
In the ideal case, macroscopic thermodynamic functions such as energy, pressure or temperature have been unambiguously defined starting from the diagonal elements of $T^{\mu\nu}$ in the LRF. This can no longer be the case for a dissipative fluid: the thermodynamic variables can now be defined \emph{as if the system were in local thermodynamic equilibrium}. In doing so a certain arbitrariness is necessarily introduced, since no univocal definition is present. Following the most common choice in the literature, we opt for the \emph{Landau matching conditions}:
\begin{gather}\label{eq:Landmatch}
	\varepsilon= u_\mu u_\nu T^{\mu\nu};\\
	n = u_\mu N^\mu.
\end{gather}
All the other functions can be derived \emph{as if} the system were actually at equilibrium. The thermodynamic entropy density can be computed by making use of the equation of state one would use in the ideal case. Once energy, particle and entropy densities have been fixed, we can define temperature, chemical potential and pressure by exploiting the thermodynamic relations:
\begin{gather}
	\frac 1T = \beta_0 = \frac{\p s}{\p \varepsilon} \Big|_n, \qquad \frac{\mu}{T} = \frac{\p s}{\p n}\Big|_\varepsilon;\\
	P_0 = -\varepsilon + T_0 s_0 + \mu_0 n.
\end{gather}
These equations are true as physical relations \emph{only in the ideal case}: now they are nothing more than definitions. It is worthwhile to underline that rigorously speaking all these quantities have lost their original physical meaning: for instance the `temperature' defined this way for a dissipative fluid has to be considered nothing more than an effective temperature.

\subsubsection{Definition of the LRF}\label{par:LRF}
There is however one last delicate point, regarding the definition of the 4-velocity field $u^\mu(x)$. For an ideal fluid, the LRF was implicitly defined as the frame where there is no flow of energy and net particle number; the definition of the velocity field was a natural consequence. In the dissipative case there is no unique choice. Under a physical perspective, there are two natural choices \cite{Landau1987fluid, Eckart:1940zz}:\footnote{There is a third commonly-used frame, which is useful especially in a quantum-mechanical perspective. It is known as thermometric or J\"uttner frame. For further reading, cf.\,\cite{van2013}. We also remark that the \emph{naturalness} of these choices is mainly a matter of convention, and many other definitions of the LRF have been given, some of them with the aim to solve intrinsic issues present for in the Landau or Eckart definition.}
\begin{description}
	\item[The Landau frame,] in which velocity is defined by the energy flow:
	\begin{equation}
		T^{\mu\nu} u_\nu= \varepsilon u^\mu,
	\end{equation}
	which means, trivially, that the 4-velocity vector is nothing more than the eigenvector of $T^{\mu\nu}$ associated with its maximum eigenvalue. Notice that this definition is perfectly coherent with the Landau matching condition in Eq. (\ref{eq:Landmatch}), since $u_\mu u^\mu =1$;
	\item[The Eckart frame,] in which velocity is defined by the particle flow:
	\begin{equation}
		N^\mu = n u^\mu.
	\end{equation}
\end{description}
We will always work in the Landau frame in the following.

\subsubsection{Tensor decomposition and equations of motion}
Dealing now with dissipative currents, we are now forced to decompose them in order to write down the equations of motion for a dissipative fluid.
\begin{gather}
	N^\mu = N^\mu_0 + n^\mu = n_0 u^\mu + n^\mu;\\
	T^{\mu\nu} = T^{\mu\nu}_0 + \tau^{\mu\nu} = \varepsilon_0 u^\mu u^\nu - \Delta^{\mu\nu} P_0 + \tau^{\mu\nu}.
\end{gather}
Some constraints on $\tau^{\mu\nu}$ can be immediately found: it must be symmetric in order to preserve the conservation of angular momentum (it can be easily proved that the complete energy-momentum tensor must be symmetric to ensure it), and, applying the Landau matching conditions:
$$ u_\mu \tau^{\mu\nu} u_\nu =0. $$
Its most general decomposition is therefore:
\begin{equation}
	\tau^{\mu\nu} = -\Pi \Delta^{\mu\nu} + 2 u^{(\mu} h^{\nu)} + \pi^{\mu\nu},
\end{equation}
where in general $f^{(\mu\nu)} = (f^{\mu\nu} + f^{\nu\mu})/2$. We have defined:
\begin{equation}
	\Pi = -\frac 13 \Delta_{\alpha\beta} \tau^{\alpha\beta},\qquad h^\mu=\Delta^\mu_\alpha u_\beta  \tau^{\alpha\beta},\qquad \pi^{\mu\nu} = \Delta^{\mu\nu}_{\alpha\beta} \tau^{\alpha\beta}.
\end{equation}
The rank-4 projector $\Delta^{\mu\nu}_{\alpha\beta}$ is:
$$ \Delta^{\mu\nu}_{\alpha\beta} =\frac12 (\Delta^\mu_\alpha \Delta^\nu_\beta + \Delta^\mu_\beta \Delta^\nu_\alpha) - \frac{1}{3} \Delta^{\mu\nu} \Delta_{\alpha\beta}. $$ 
All the introduced quantities have a precise physical meaning:
\begin{itemize}
	\item The \emph{particle diffusion} $n^\mu$;
	\item The \emph{bulk viscous pressure} $\Pi$;
	\item The \emph{energy diffusion 4-current} $h^\mu$;
	\item The \emph{shear stress tensor} $\pi^{\mu\nu}$. Note that $\pi^{\mu}_{\mu}=0$.
\end{itemize}
It is straightforward to prove that in the Landau frame $h^\mu=0$, while $n^\mu\ne 0$; on the contrary, in the Eckart frame the particle diffusion vanishes while the energy diffusion does not.
By projecting again the conservation laws into the subspaces parallel and orthogonal to $u^\mu$ we find the most general equations of motion for a dissipative system:
\begin{gather}
	u_\alpha \p_\beta T^{\alpha\beta} = \dot \varepsilon + (\varepsilon + P_0 + \Pi) \theta - \pi^{\alpha\beta}\sigma_{\alpha\beta} =0;\\
	\Delta^\mu_\alpha \p_\beta T^{\alpha\beta} = (\varepsilon + P_0 + \Pi) \dot u^\mu - \nabla^\mu (P_0 + \Pi) + \Delta^\mu_\alpha \p_\beta \pi^{\alpha\beta}=0;\\
	\p_\mu N^{\mu} = \dot n + n\theta + \p_\mu n^\mu =0.
\end{gather}
Here \emph{shear tensor} $\sigma^{\mu\nu}$ is defined as:
\begin{equation}
	\sigma^{\mu\nu} = \p^{\langle \mu} u^{\nu\rangle} = \frac 12 (\nabla^\mu u^\nu - \nabla^\nu u^\mu) - \frac 13 \Delta^{\mu\nu} \nabla_\lambda u^\lambda.
\end{equation} 
These equations, however, are only a trivial consequence of the conservation laws and of the decomposition of the tensor: they are exact and most general, but there is little physics inside them. We again have only 5 equations but now for 14 unknowns functions. We need further nine dynamical or constitutive relations in order to close the system. The different choices in closing this system lead to the different formulations of dissipative hydrodynamics.

\subsubsection{Navier-Stokes equations}

The most straightforward attempt may be to follow what is done in non-relativistic hydrodynamics, that is to use some constitutive equations to relate the dissipative currents and $\theta, \sigma^{\mu\nu}$ and $\nabla^\mu\alpha_0$, respectively the already defined expansion scalar and shear tensor and what will be shown to be the 4-gradient of the chemical potential. This choice is justified by the fact that it ensures the entropy production to be positive. Indeed, one can write the second law of thermodynamics as:
\begin{equation*}
   \p_\mu S^\mu_0 = \dfrac{u_\nu}{T}\p_\mu T^{\mu\nu}_0 - \alpha_0 \p_\mu N^\mu_0. 
\end{equation*}
Since the total currents have to be conserved, it is not the case for the equilibrium components of the currents. This means that the equation becomes:
\begin{equation}
	\p_\mu S^\mu_0 = \alpha_0 \p_\mu n^\mu + \frac 1T (-\Pi \theta + \pi^{\mu\nu} \sigma_{\mu\nu});
\end{equation}
by rewriting the first term in the RHS one can isolate in the LHS the divergence of an \emph{off-equilibrium} entropy:
\begin{equation}
	\p_\mu (S^\mu_0 - \alpha_0 n^\mu ) = - n^\mu \nabla_\mu \alpha_0  + \frac 1T (-\Pi \theta + \pi^{\mu\nu} \sigma_{\mu\nu})>0.
\end{equation}
Therefore, to ensure the positivity of the 4-divergence of the entropy flow the simplest choice is:
\begin{equation}\label{eq:navier_stokes_const}
	\Pi=-\zeta \theta;\quad n^\mu = \kappa \nabla^\mu \alpha_0; \quad \pi^{\mu\nu} = 2\eta \sigma^{\mu\nu}.
\end{equation}
The proportionality constant introduced are known as transport coefficients and are interpreted as the bulk viscosity ($\zeta$), the particle diffusion ($\kappa$) and the shear viscosity ($\eta$). Constraining them to be positive, $\p_\mu S^\mu \le 0$ always.\\
This procedure, firstly proposed by Landau and Lifshitz with this definition of the LRF, leads to the relativistic Navier-Stokes equations:
\begin{align}
	\dot{\varepsilon} &= - (\varepsilon + P_0 - \zeta \theta) \theta + 2 \eta \sigma_{\alpha \beta} \sigma^{\alpha \beta}, \\
	(\varepsilon + P_0 - \zeta \theta) \dot u^\mu &= \nabla^\mu P_0 - \nabla^\mu (\zeta \theta) - 2 \Delta^\mu_\alpha \partial_\beta (\eta \sigma^{\alpha \beta}), \\
	\dot{n} &= - n \theta - \partial_\mu (\kappa \nabla^\mu \alpha_0).
\end{align}
Specifying the equation of state and knowing all the transport coefficients would close the system of equations and provide a first-order relativistic theory of fluid dynamics.
Unfortunately, Navier-Stokes equations implicitly require instantaneous signal propagation within the fluids which are not an issue in the Newtonian dynamics (non-relativistic hydrodynamics) but become catastrophic in the relativistic one. Indeed, the constitutive equations\,\eqref{eq:navier_stokes_const} imply that a perturbation in the primary variables instantaneously generate a dissipative current. Furthermore, this causality violation is strictly connected (actually it is an \emph{if and only if} relation) with the stability of the equations themselves, which become parabolic (instead of comfortably hyperbolic): first-order Navier-Stokes relativistic equations are not stable and violate causality \cite{Hiscock:1985zz, Denicol:book}. Moreover, this instability is not so manifest at a first sight, but shows up only if we try to Lorentz-transform the equations into a moving frame. The problem is quite involved and research is still in progress in this field, since a difference choice of the LRF could solve these issues \cite{Bemfica:2019knx}. If one wants to work in the Landau or Eckart frame, it is necessary to go up to second order.

\subsubsection{Israel and Stewart approach}
The first attempt to move to a second-order dissipative relativistic hydrodynamics was performed by Israel and Stewart \cite{Israel:1976tn, Stewart:1977}. Their starting point was again the second law of thermodynamics; but since the instantaneous propagation of information from the primary variables to the dissipative currents could be considered as the origin of the ill-behaving of the equations, they promoted the dissipative currents to dynamical variables, which relax to the first-order definition within certain characteristic relaxation times:
\begin{gather}\label{eq:relaxation_of_diss_curents}
	\tau_\Pi \dot \Pi + \Pi = -\zeta \theta + \dots ,\\
	\tau_n \Delta^\mu_\alpha \dot n^\alpha + n^\mu = \kappa \nabla^\mu \alpha + \dots,\\
	\tau_\pi \Delta^{\mu\nu}_{\alpha\beta} \dot \pi^{\alpha\beta} + \pi^{\mu\nu} = 2\eta \sigma^{\mu\nu}+\dots.
\end{gather}
One sees trivially that if the relaxation times go to 0 it is possible to recover the Navier-Stokes definitions. Let us show why this ansatz is successful in solving some of the issues of Naiver-Stokes hydrodynamics. First of all, since these quantities are now dynamical,  the equation for the entropy production looks a bit different:
\begin{equation}
	\p_\mu S^\mu = -\frac 1T \Pi \theta + \frac 1T \pi^{\mu\nu}\sigma_{\mu\nu} - n^\mu\nabla_\mu \alpha_0 + \p_\mu Q^\mu,
\end{equation}
in which $Q^\mu$ has to contain all the possible second-order terms:
\begin{equation}
	Q^\mu \equiv -\frac{1}{2} u^\mu \left( \delta_0 \Pi^2 - \delta_1 n_\alpha n^\alpha + \delta_2 \pi_{\alpha \beta} \pi^{\alpha \beta} \right) - \gamma_0 \Pi n^\mu - \gamma_1 \pi_{\nu}^\mu n^\nu.
\end{equation}

By imposing the positiveness of the entropy production, further relations can be found that constrain all the introduced coefficients. This means that the expression for the entropy production must be:
\begin{equation}
	\partial_\mu S^\mu = \beta_0 \bar\omega_\Pi \Pi^2 - \bar\omega_n n^\mu n_\mu + \beta_0 \bar\omega_\pi \pi_{\mu\nu}\pi^{\mu\nu},
\end{equation} 
with ensuring the positiveness of $\bar \omega_\pi, \bar\omega_\Pi, \bar\omega_n\ge0$. By combining this equations with what found for $S^\mu$ and $Q^\mu$ the relaxation equations for the dissipative currents are found, which all are of the shape of Eq.\,\eqref{eq:relaxation_of_diss_curents}, with all the coefficients constrained. The derivation is quite cumbersome and out of the goals of this work; what is worth to underline is that up to this point nothing has been said on the microscopic physics of the medium: conservation laws and general thermodynamics laws have been sufficient to derive these hydrodynamics equations. This approach will be totally flipped in the next section, in which the path to derive the same equations from the kinetic theory will be shown.

\section{Microscopic foundations of hydrodynamics}
\label{sec:microscopic_hydro}

Similarly to what was done by Boltzmann in giving microscopic foundations to thermodynamics laws, several approaches have been developed which derive the relativistic hydrodynamic equations starting from kinetic theory. The underlying idea is always to obtain fluid dynamics as an effective theory, by integrating on \emph{some} microscopic degrees of freedom of the `more fundamental' theory. It is important to stress the fact that ideal and even dissipative hydrodynamics can be seen as completed disentangled from kinetic theory, as illustrated in the previous section, and that, strictly theoretically speaking, there are still some caveats in its microscopic derivation: these issues are, however, out of the focus of the present work and we will give for granted the well-posedness of this problem.\\
It is easy to understand that the choice about \emph{which} degrees of freedom have to be integrated on is not trivial. This explains the wide flourishing of different hydrodynamic theories, that bring pretty much to the same equations with the most of the difference lying in the determination of the transport coefficients. Here the derivation of the equations will be mostly skipped and only a few of these methods are rapidly illustrated. 

\subsection{Chapman-Enskog Theory}
The first attempt to derive fluid dynamics from the Boltzmann equation was to follow the Chapman-Enskog originally non-relativistic approach \cite{ChapmanCowling:1974}, that Israel first adapted to the relativistic context. The basic idea is to start form the primary fluid dynamical variables ($T, \mu$ and three independent components of $u^\mu$) and to add their gradients. Therefore the local distribution function is expanded to higher terms which are then arranged in powers of the Knudsen number:
\begin{equation}\label{eq:f_exp_chapman}
	f = f^{(0)} + \text{Kn}\, f^{(1)} + \text{Kn}^2 \, f^{(2)}+ \dots.
\end{equation}
Trivially, the smallest the Knudsen number, the most accurate the truncation at lower order is, and thus the most `hydrodynamic' the system is. The macroscopic length scale $L$ is determined by the space gradient of the distribution function in the LRF, i.e. $\nabla^\mu \sim L^{-1}$, and the same can be done with the time scale and the covariant derivative $D\sim  \tau^{-1}$. With a redefinition of the derivatives in order to have dimensionless variables ($\widetilde \nabla_\mu = L\nabla_\mu$ and $\widetilde D=\bar \tau D$), is possible to write the Boltzmann equation in a dimensionless fashion:
\begin{equation}\label{eq:boltzmann_eq_chapman}
	\text{Kn}_t \widetilde D f + \frac{\text{Kn}}{E} p^\mu \widetilde\nabla_\mu f = \frac{\lambda}{E} C[f].
\end{equation}

Then Eq.\,\eqref{eq:f_exp_chapman} is inserted in Eq.\,\eqref{eq:boltzmann_eq_chapman} and the solution is found using perturbation theory. As anticipated, we are not going to develop the equations, but just to highlight the most important characteristics of the theory. Namely, one recovers ideal and Navier-Stokes hydro going to zero- and first-order in Kn; moving to second order the so-called Burnett equations are obtained \cite{Burnett:1935}. However, Chapman-Enskog expansion is an asymptotic series \cite{Grad:1963, Denicol:2016bjh} and leads to unstable equations of motion, therefore cannot be used systematically for realistic simulations.

\subsection{14-moments approximation}
This was the first proposed method making us of an expansion in terms of moments. The approach, due again to Israel and Stewart \cite{Israel:1979wp}, is conceptually different to the Chapman-Enskog method: there is no expansion in powers of a small quantity, such as the Kn, but `some' moments of the one-particle distribution function are taken into account up to a certain truncation order. The basic idea is to recover the macroscopic spirit of hydrodynamics and therefore to have equations in terms of $N^\mu$ and $T^{\mu\nu}$. Indeed, these currents carry exactly 14 degrees of freedom: 4 from the particle current and 10 independent components from the energy-momentum tensor.\\
Israel-Stewart ansatz for the out-of-equilibrium distribution function of a classical gas is:
\begin{equation}
	f(p) = \exp\tonde{ \mu/T - u_\mu p^\mu/T + \epsilon + p^\mu\epsilon_\mu + p^\mu p^\nu \epsilon_{\mu\nu}+ \dots }=\exp\tonde{ \mu/T - u_\mu p^\mu/T + \delta\epsilon_p }
\end{equation}
in which all the terms included in $\delta\epsilon_p$ are an expansion around the equilibrium distribution $f\propto \exp[(\mu - u\cdot p)/T]$ and all the $\epsilon$ coefficients have to be found. Therefore, one can expand the distribution function for small momenta as:
\begin{equation}
	f(p) = f_0(p) + f_0(p) \delta\epsilon_p + O(\delta\epsilon_p^2).
\end{equation}
By using the usual definitions of the currents:
\begin{gather}
	N^\mu = \int dP p^\mu (1 + \epsilon + \epsilon_\nu p^\nu + \epsilon_{\nu\rho} p^{\nu\rho} + \dots ) f_0(p);\\
	T^{\mu\nu} = \int dP p^\mu p^\nu (1 + \epsilon + \epsilon_\rho p^\rho + \epsilon_{\rho\sigma} p^{\rho\sigma} + \dots ) f_0(p).
\end{gather}
The idea followed by Israel and Stewart was to stop this expansion at second order, neglecting all the $\dots$ we have introduced. Therefore, one is left with (again) 14 degrees of freedom: $\epsilon, \epsilon_\mu, \epsilon_{\mu\nu}$, since the tensor can be chosen to be symmetric and traceless (the trace can be included in the scalar $\epsilon$). Then these 14 unknowns can be matched to the 14 independent components of the currents $N^\mu$ and $T^{\mu\nu}$, so that one closes the system of equations and is able to fully specify the distribution function just on the basis of the currents themselves. Once these moments are computed, the equations of motion are obtained starting from the second moment of the Boltzmann Equation (i.e. the Boltzmann Equation integrated on $\int dp^\mu \int dp^\nu$). In doing this there is a new source of arbitrariness in the theory, since whatever moment of the Boltzmann Equation would provide a set of equations for the moments. These equations would share the same structure, but the values for the transport coefficients would be different. This arbitrariness is one of the drawbacks of Israel-Stewart theory; however, provided that certain relations between the transport coefficients are respected, the theory hence developed is stable and causal and can give a faithful description of a relativistic hydrodynamical medium.

\subsection{DNMR hydrodynamics}
One of the most successful implementation of a microscopic derivation of hydrodynamics is due to Denicol-Niemi-Molnar-Rirschke (DNMR) \cite{Denicol:2012cn}. The basic idea is to eliminate the ambiguity inherently present in the Israel-Stewart theory: the set of moments chosen for deriving the hydrodynamic equations is that of an irreducible set of moments, which appears to be an orthogonal and complete basis and does not allow for an arbitrary nor \emph{ad hoc} truncation of the set of equations. This is obtained, in a sense, recovering the original idea by Chapman-Enskog of a power counting in Knudsen number and adding also the Reynolds number power counting, which allow to integrate exactly on the microscopic degrees of freedom up to a certain order. If we choose to stop at second order in Knudsen and inverse Reynolds numbers, we end up with just 14 dynamic variables (analogously to the 14 moments), but the transport coefficients carry information about all the moments of the distribution function. This method can be used to easily improve the equations going to higher orders (23-, 32- and 41-moment equations have been computed) and allows also to perform the expansion around a reference state which is not the usual local thermal equilibrium, but, for instance, can include momentum anisotropy.\\
We have already addressed the issue of the Knudsen number, while we miss the definition of the Reynolds number. As already stated above, the basic hypotheses of fluid dynamic require both the separation of micro- and macroscopic scales (Kn$\ll 1$) and the nearly-thermalisation of the medium. In other words, it is needed that the system is always close to local thermal equilibrium ($\delta f \ll f_{eq}$). This deviation can be quantified in terms of the ratio between dissipative and equilibrium quantities. In particular:
\begin{equation}
	\text{Re}_\Pi^{-1} \equiv \frac{|\Pi|}{P}, \qquad \text{Re}_n^{-1} \equiv \frac{|n^\mu n_\mu|}{n_0}, \qquad \text{Re}_\pi^{-1} \equiv \frac{|\pi^{\mu\nu}\pi_{\mu\nu}|}{P},
\end{equation}
which represent respectively the bulk, the diffusion and the shear inverse Reynolds number. In principle, there is no direct relation between Kn and Re numbers: on one hand, the system could be close to equilibrium but extremely dilute or weakly-interacting; on the other hand, the system can have large Reynolds numbers and be in a fluid-dynamical collision regime, despite this configuration would be transient, having the strong coupling the consequence to relax the system close to equilibrium.\\

We are not going into the details of DNMR derivation, but illustrate just the general method. We start by expanding the distribution function with respect to the equilibrium one:
\begin{equation}
	f(p) = f_{eq}(p) (1 + \phi (p)).
\end{equation}
By comparing it with the Israel-Stewart formalism, one sees that according to the latter $\phi(p) = \epsilon + \epsilon^\mu p_\mu + \epsilon^{\mu\nu} p_\mu p_\nu$. What is argued by DNMR is that this expansion is not suitable, since the tensors making up the basis ($1, p^\mu, p^\mu p^\nu, p^\mu p^\nu p^\sigma, \dots$) are not irreducible with respect to Lorentz transformations, and since the expansion is truncated, when one looks for an expression for the expansion coefficients $\epsilon^{\alpha...\beta}$, it cannot be the complete one.\\
Their proposal is to expand $\phi(p)$ in terms of a complete and orthogonal basis of irreducible tensors:
\begin{equation}
	1, p^{\langle \mu \rangle}, p^{\langle \mu} p^{\nu \rangle}, p^{\langle \mu} p^\nu p^{\sigma \rangle}, \dots,
\end{equation}
where $A^{\langle \mu_1 \dots \mu_n\rangle} = \Delta^{\mu_1\dots \mu_n}_{\nu_1\dots\nu_n} A^{\nu_1\dots \nu_n}$; details about the $\Delta$ projectors can be found in \cite{DeGroot:1980dk, Anderson:1974}.\\
One can also define the irreducible moments of the distribution function deviation $\delta f_p$:
\begin{equation}
	\rho_n^{\mu_1\dots \mu_l} \equiv \int dP E_p^n p^{\langle \mu_1}\dots p^{\mu_l\rangle} \delta f_p,
\end{equation}
where $E_p^n=(u^\mu p_\mu)^n$, and due to the Landau matching one has:
\begin{equation}
	\rho_1 = \rho_2 = \rho_1^\mu=0,
\end{equation}
which basically means that the energy and particle density and the four flow $u^\mu$ are that of an equilibrated system, which is what we imposed. These moments can be shown to obey the evolution equation:
\begin{equation}
\dot\rho_r^{\langle \mu_1\dots\mu_l \rangle}= \Delta^{\mu_1\dots \mu_l}_{\nu_1\dots\nu_l} \frac{d}{d\tau}  \int dP E_p^{r} p^{\langle \mu_1}\dots p^{\mu_l \rangle} \delta f(p).
\end{equation}
Eventually, one can replace in this equation the expression for the distribution function deviation got by the Boltzmann Equation:
\begin{equation}
	\delta  f_p = - \dot f_{0p} - \frac{1}{E_p} \tonde{ p^\nu \nabla_\nu f_{0p} + p^\nu \nabla_\nu \delta f_p - C[f] }.
\end{equation}
This allows to construct an infinite tower of equations for all the moments that are completely exact. However, in order to construct a reliable second-order hydrodynamical theory, it is sufficient to stop at order two. In the following we will show the complete equations in the specific case of Bjorken flow.

\subsubsection{Conformal Bjorken flow 0+1D}
The simplest case in which the equations can be computed is that of a boost-invariant 1D system. In Milne coordinates $(\tau,\vec x_\perp, \eta_s)$, all the quantities will depend just on the proper time $\tau$, and therefore the equations reduce to easily solvable Ordinary Differential Equations (ODEs). In Paragraph\,\ref{subsubsec:rta_0+1D} the same scenario was addressed in the context of the RTA Boltzmann Equation.\\
By restricting to the case of a conformal system (particle mass $m=0$), the bulk viscous pressure is $\Pi=0$, and we are left with a system of two coupled evolution equations for $e$ and $\pi$ \cite{Denicol:2010xn, Denicol:2012cn}:
\begin{subequations} \label{eq:vHydro}
	\begin{gather}
		\label{eq:vHydro_eps}
		\partial_\tau e = -\frac{1}{\tau} (e + P -\pi),\\
		\partial_\tau \pi = - \frac{\pi}{\tau_\pi} + \frac 43 \frac{\eta}{\tau_\pi \tau} - \lambda \frac{\pi}{\tau}, \label{eq:vHydro_pi}
	\end{gather}
\end{subequations}
where the shear viscous pressure is $\pi=\sqrt{2\pi^{\mu\nu}\pi_{\mu\nu}/3}$ and the shear relaxation time is $\tau_\pi=\tau_{eq}=5(\eta/s)/T$.
The transport coefficient $\lambda$ according to DNMR theory in 14-moment approximation is computed as 124/63. In order to quantify the deviation from equilibrium one can also compute the $\delta f$ in this specific case:
\begin{equation}
	\delta f_{\text{DNMR}} = \frac{3}{16} \frac{f_{eq} (T)}{T^2}[(p\cdot u)^2 - 3 (p\cdot z)^2] \frac{\pi}{e}.
	\label{eq:df_IS}
\end{equation}
Notice that in the Chapman-Enskog approach \cite{Romatschke:2011qp} $\delta f_{\text{CE}}^{(1)}/f_{eq} \sim p/T$, while $\delta f_{\text{DNMR}}/f_{eq} \sim (p/T)^2$; the former agrees much better with the outcomes of RBT transport theory in the case of isotropic cross section \cite{Plumari:2015sia}, whose spectra are nicely fit by $\delta f/f_{eq} \sim p_T ^ {0.98}$.

These equations are solved by making use of a 4th-order Runge-Kutta method to get the results shown in Chapter\,\ref{chap:attractors_1D}.

\subsubsection{0+1D anisotropic hydrodynamics}
\label{subsubsec:0+1D_ahydro}
As said before, DNMR easily allows to expand the full distribution function around a reference state different from the $f_{eq}$, which is always assumed to be a Boltzmann-J\"uttner. For instance, one can use the Romatschke-Strickland distribution function:
\begin{equation}
	f(p, \tau) = \gamma (\tau) \exp \tonde{ - \frac{\sqrt{ p_T^2 + (1+\xi(\tau))(p\cdot z)^2 }}{\Lambda (\tau)} }.
\end{equation}   
In this case we also take into account the number conservation of particles in the system, by including a fugacity in the description of the system evolution. This is certainly not the case for gluons and more generally partons in the QGP, but will be useful in the following to have a proper comparison with RBT model. Using the same formalism explained above, it is possible to derive  a system of three coupled ODEs \cite{Alqahtani:2017mhy} which describe the evolution of the medium in terms of three macroscopic functions: $\gamma$, that is related to the effective fugacity, $\Lambda$, that in linked to the effective temperature, and $\xi$, that is connected to the pressure anisotropy $P_L/P_T$. Even though this nomenclature is widely spread, one should more precisely talk about $P_L/P$, which is related to the inverse Reynolds number. This latter quantity depends on the $\xi$ parameter via:

\begin{equation}\label{eq:pi_over_e_xi}
	\frac{\pi}{e}(\xi)= \frac 13 \tonde{1 - \frac{\mathcal R(\xi)}{\mathcal R_L(\xi)}},
\end{equation}
where
\begin{gather}
	\label{eq:R(xi)}
	\mathcal R(\xi)=\frac 12 \quadre{ \frac{1}{1+\xi} + \frac{\arctan \sqrt{\xi}}{\sqrt{\xi}} },\\
	\mathcal R_L(\xi) = \frac{3}{\xi} \quadre{ \frac{(\xi+1) \mathcal R(\xi) -1 }{\xi +1} }.
\end{gather}

The zeroth- and the first-moment equations account for the number and energy-momentum conservation.
The second-moment equation is the difference between the $zz$ projection and one third of the sum of the $xx$, $yy$ and $zz$ projections of the equation:
\begin{equation}\label{eq:2nd_moment}
	\partial_\lambda I^{\lambda \mu\nu} = \frac{1}{\tau_{eq}} (u_\lambda I_{eq}^{\lambda \mu \nu} - u_\lambda I^{\lambda\mu\nu})
\end{equation}
where
\begin{equation}
	I^{\lambda\mu\nu} = \int dP \, p^\lambda p^\mu p^\nu f.
\end{equation}
By imposing the energy-momentum and particle conservations and rewriting Eq.\,\eqref{eq:2nd_moment} in terms of $\Lambda$, $\gamma$ and $\xi$ one gets the three coupled ODEs:   
\begin{equation}
	\label{eq:aHydro}
	\begin{split}
		& \partial_\tau \log \gamma + 3 \partial_\tau \log \Lambda - \frac 12 \frac{\partial_\tau \xi}{1+\xi} + \frac{1}{\tau} =0,\\
		& \partial_\tau \log \gamma + 4 \partial_\tau \log \Lambda + \frac{\mathcal R'(\xi)}{\mathcal R(\xi)} \partial_\tau \xi = \frac{1}{\tau} \quadre{ \frac{1}{\xi(1+\xi) \mathcal(\xi)} - \frac{1}{\xi} -1 },\\
		& \partial_\tau \xi - \frac{2(1+\xi)}{\tau} + \frac{\xi (1+\xi)^2 \mathcal R^2(\xi)}{\tau_{eq}}=0,
	\end{split}
\end{equation}
where the relaxation time is given by the usual $5\eta/s /T$ and the effective temperature $T$ and fugacity $\Gamma$ are related to $\Lambda$ and $\gamma$ via the function $\mathcal R(\xi)$ defined in Eq.\,\eqref{eq:R(xi)}:
\begin{gather}
	T= \mathcal R(\xi) \sqrt{1 + \xi} \Lambda,\\
	\Gamma = \frac{\gamma}{(1+\xi)^2 \mathcal R^3(\xi)}.
\end{gather}
Notice that, if $\xi=0$, $\Lambda$ and $\gamma$ reduce respectively to $T$ and $\Gamma$.

\chapter{Relativistic Boltzmann Transport Code}
\label{chap:RBT}

The Relativistic Boltzmann Transport Code has been developed for the last few years, parallelly to the research performed. Here we illustrate its main features up to this point. The code exploits the same numerical methods (test particle methods and stochastic collisions) as in \cite{Xu:2004mz} and in \cite{Ferini:2008he, Plumari:2012ep, Scardina:2012mik, Scardina:2014gxa, Plumari:2015sia}. However, this implementation is completely new, with a more flexible structure in order to reach better statistics and performance and to host new features and extensions.\\
The aim is to solve the Boltzmann kinetic equation with the full collision integral in the case of elastic $2\leftrightarrow2$ collisions for a gas of classical particles, by evolving the full distribution function $f(x,p)$ which is sampled with a large number of test particles (see Section \ref{sec:test_particles}). This can be done in several scenarios, depending on the goal of the research: a box with periodic boundary conditions, a 1D (non-)boost-invariant medium or a 3D expanding fireball.

\section{The distribution function and the test particle method}\label{sec:test_particles}
As the distribution function is sampled by a set of point-like particles, it can be seen as a sum of delta distributions:
\begin{equation}\label{eq:f_discr}
	f(x,p) = \sum_{i=1}^N  \delta^3(\vet x - \vet x_i(t) ) (2\pi)^3\delta^3 (\vet p - \vet p_i(t) ),
\end{equation}
where $\vet x_i(t), \vet p_i(t)$ are the phase-space coordinates of the $i$-th particle and $N$ is the total number of particles. From the definition of distribution function, 
\begin{equation}\label{eq:f_norm}
	\int d^3 \vet x \int \frac{d^3 \vet p}{(2\pi)^3 p^0} p^0 f(x,p) = N.
\end{equation}
All the physically relevant quantities, such as the components of the energy-momentum tensor or the particle density 4-vector, are determined from the distribution function itself.\\
As in every numerical simulation, one of the main issues is the convergence of the results with the increasing number of performed events: hence, to achieve a meaningful outcome, we must average over a sufficient number of numerical events. Nonetheless, this may not be enough. Since the aim of the computation is to follow the evolution of the distribution function, in order to sample it a large number of particles is often necessary, especially if one deals with an expanding system, or with a strongly non-homogeneous medium; moreover, thermodynamic relations are massively used in solving the Boltzmann equation, and we now they are strictly valid only in the $N\to \infty$ limit. Nevertheless, the number of physical particles cannot be chosen arbitrarily, since it is constrained by the initial conditions $f_0(x,p)$. To overcome such a complication the number of fictitious particles is artificially incremented, by simulating $N_{\text{test}}$ particles per each \emph{physical} particle: the simulation is performed over $N_\text{tot} = N \times N_{\text{test}}$ particles, which ensure proper statistics. This numerical artefact can be shown to be equivalent to the simulation of $N_{\text{test}}$ physical ensembles. \\
To allow this modification, the expression of the distribution function has to be modified: Eq. (\ref{eq:f_norm}) still has to hold, since $N$ is a physical observable and the distribution function has to give back physical quantities; therefore Eq. (\ref{eq:f_discr}) becomes:
\begin{equation}\label{eq:f_discr_ntest}
	f(x,p) = A \sum_{i=1}^{N_{\text{tot}}}  \delta^3(\vet x - \vet x_i(t) ) (2\pi)^3\delta^3 (\vet p - \vet p_i(t) ),
\end{equation}
where $A$ is a proper normalisation factor that is fixed by Eq. (\ref{eq:f_norm}):
\begin{equation}
	A \int d^3 \vet x \int \frac{d^3 \vet p}{(2\pi)^3 p^0} p^0 f(x,p) = A N_{\text{tot}} = N \implies A=\frac{1}{N_\text{test}}.
\end{equation}
For instance, in a volume $V$ the energy-momentum tensor elements for an homogeneous system is computed as:
\begin{equation}\label{eq:emtensor}
	T^{\mu\nu} (x) = \int \frac{d^3 \vet p}{(2\pi)^3 p^0} p^\mu p^\nu f(x,p) = \frac{1}{V N_{test}}\sum_{i=1}^{N_{tot}} \frac{p_i^\mu p_i^\nu}{p^0_i};
\end{equation}
where we exploit $\int d^3\vet x\, \delta^3(\vet x - \vet x_i(t)) =1 $ to write $\delta^3(\vet x - \vet x_i(t)) = 1/V, \forall i$.
Considering for instance the energy density, one gets:
\begin{equation}
	e = T^{00} =  \frac{1}{V N_\text{test}} \sum_{i=1}^{N_\text{tot}} {p^0_i},
\end{equation}
which is simply the sum of the single particles' energies divided by the volume and the $N_\text{test}$ factor: the latter division is an average over the different ensembles we have introduced with the test particles artifice.

\section{The code design}
In this section the main features of the code are illustrated, without delving into technical details.\\
The two essential components, which are initialised at the beginning of the simulation and are almost entirely responsible for the occupied RAM, are
\begin{description}
	\item[The test particle array.] It is a structure array of length $N_\text{tot}$, which can however be changed if particles are created or lost. Each test particle is a structure that contains all the information concerning the test particle itself: phase-space coordinates, $\eta_s$, $Y$, mass, flavour, further indices or information. Every particle  is identified by an id number, corresponding to the array index.
	\item[The cell matrix.] Coordinate space is discretised in a certain number of cells. It is possible to choose between Cartesian or Milne coordinates: in both cases, cells are parallelepipeds of volume $\Delta x \Delta y \Delta z $. In the case of Milne coordinates, cell dimensions are fixed in $\eta_s$, and are therefore expanding in $z$, changing their volumes at every time step. Each cell keeps memory of the number of particles contained therein and of their id numbers. Cells are necessary in the collision phase, since only particles within the same cell are allowed to collide; the propagation routine, instead, ignore the discretisation of space: particles can occupy whatever position, and are assigned to a given cell after their propagation is completed. At the end of every collision routine, all the information within each cell is reset. Finally, the cells are useful also to extract local information of the distribution function. In the following, when local quantities are shown, they are always computed within a single cell. 
\end{description}

One can choose between three different timescales:
\begin{description}
	\item[Linear] Fixed $\Delta t$; used preferably when there is no expansion of the medium (e.g. static box).
	\item[Logarithmic] Given the initial and final time $t_0\ne0$ and $t_{fin}$ and the number of time steps $N_t$, the time interval is $\Delta (\log t)= \dfrac{\log t_{fin} - \log t_0 }{N_t}$ and the $i$-th time step is $t_i=t_0 \, e^{i \Delta (\log t) }$.
	\item[Linearly expanding] Given the initial and final time $t_0\ne0$ and $t_{fin}$,
	$$ t_i = t_{i-1} [ 1 + a \Delta(\tanh\eta)], $$
	in which $a$ is a constant fixed $<1$ and $\Delta(\tanh\eta)$ gives the $z$ width of a cell. This adapting time scale was first introduced in \cite{Xu:2004mz} and is the most suitable to model an expanding medium.
\end{description}

In principle, the numerical solution will converge to the exact one in the limit $\Delta t\to 0$, $\Delta^3 \vet x\to 0$. Under a computational perspective, we discretise in a more refined way until convergence has been reached.\\
This discretisation, however, cannot be arbitrary: in order to preserve the collision locality the condition to be imposed wants $\Delta x< \lambda_\text{mfp} = 1/n \sigma$, where $\lambda_\text{mfp}$ is the mean free path. If it were not the case, indeed, particles inside the same cell, which are allowed to collide, may be far apart a distance $L>\lambda_\text{mfp}$, with the consequent loss of locality. For the very same reason, we ask $\Delta t < \lambda_\text{mfp}/v$, $v$ being the average velocity of the particles.\\

\begin{figure}
	\centering
	\includegraphics[width=0.3\linewidth]{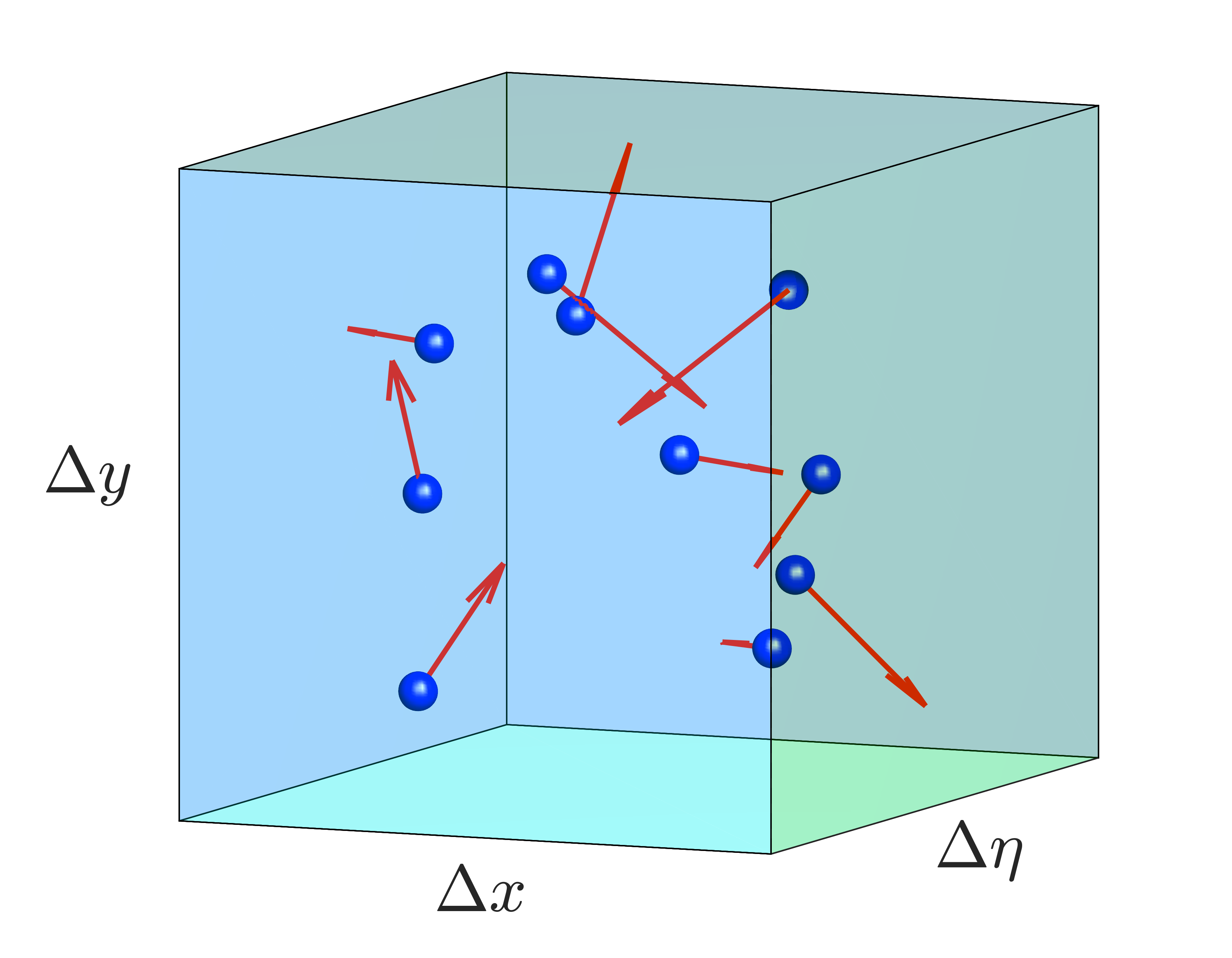}
	\includegraphics[width=0.3\linewidth]{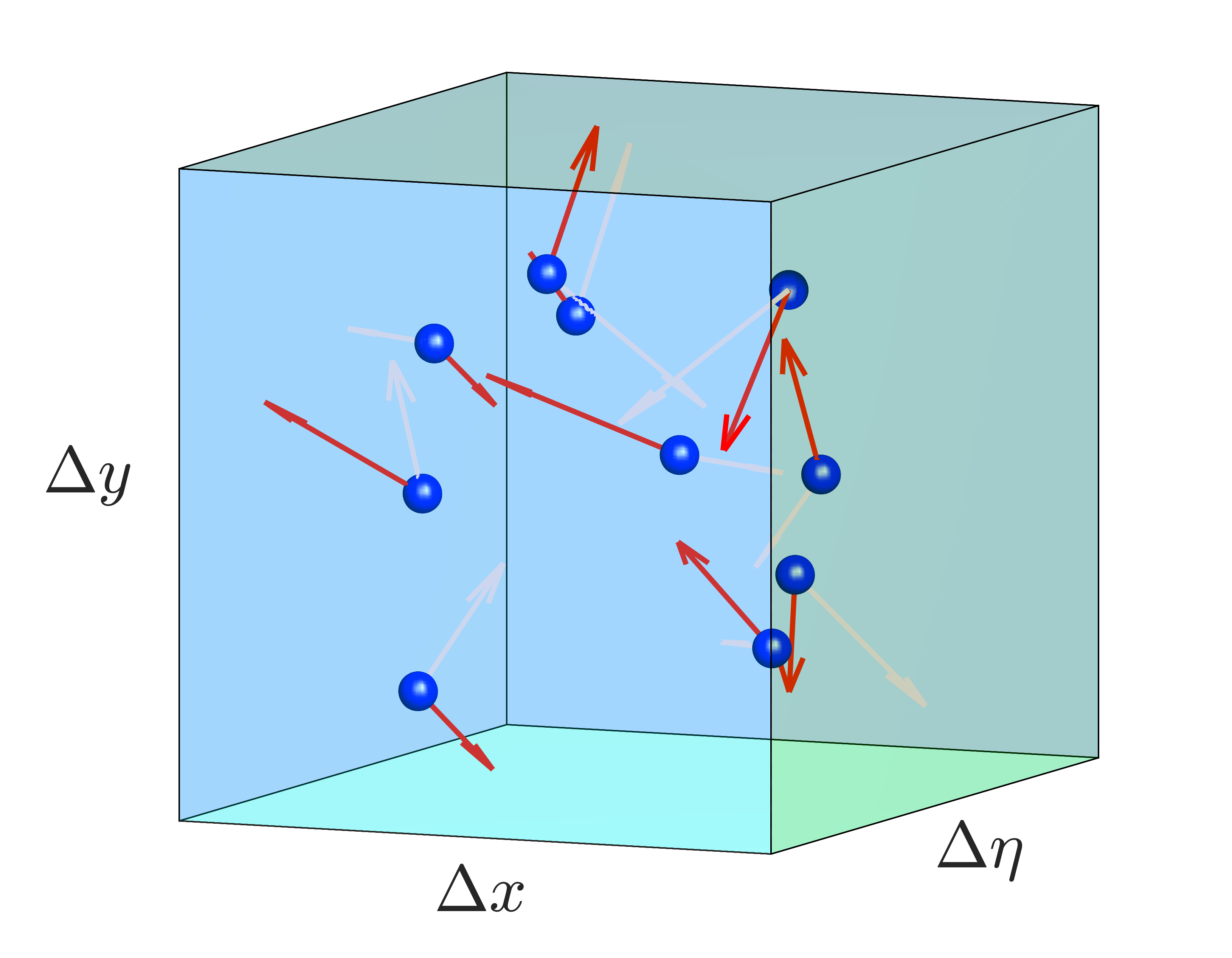}
	\includegraphics[width=0.3\linewidth]{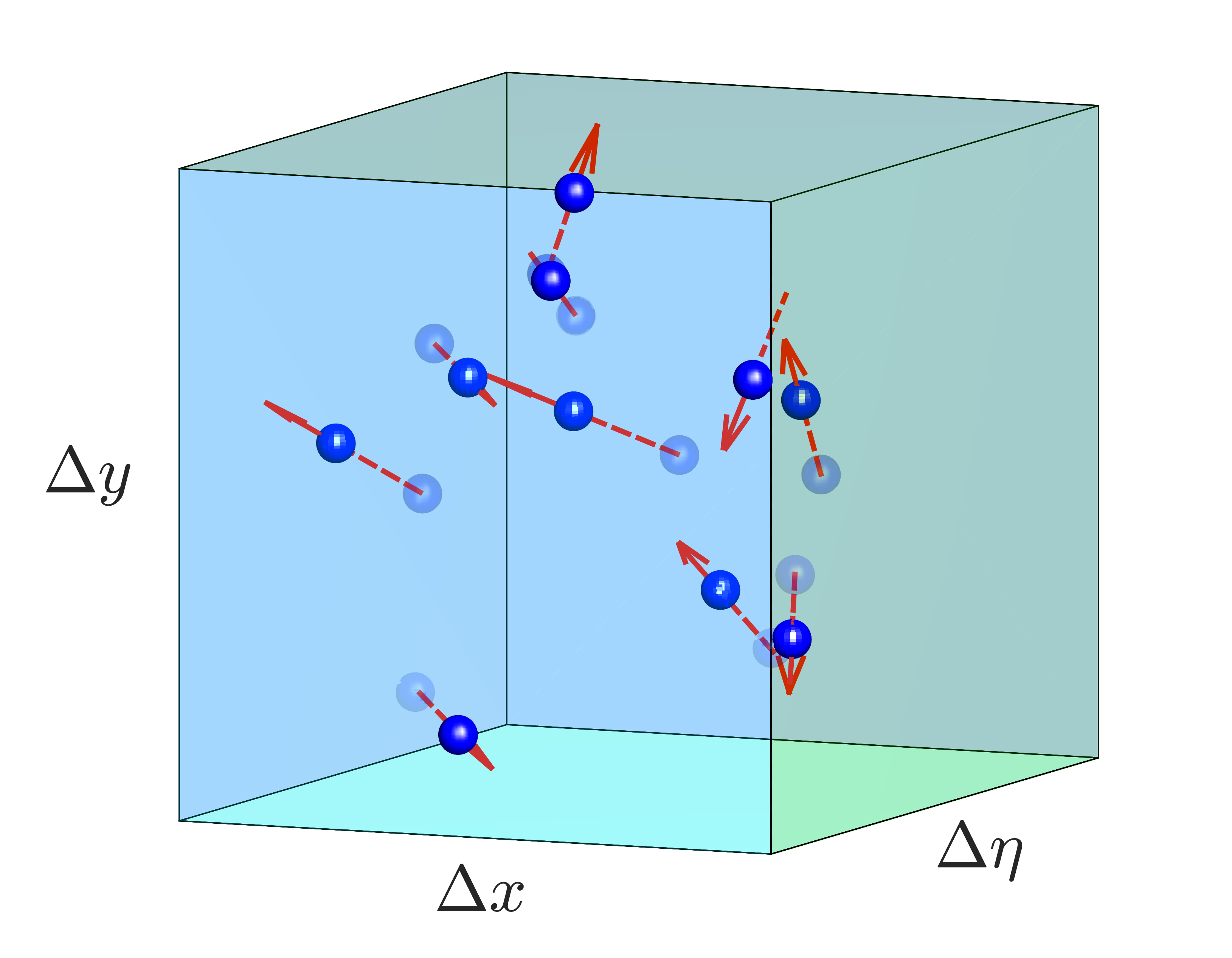}
	\caption{The three blocks of the code algorithm in a specific cell. On the left panel, point-like test particles are initialised with a given position and momentum. In the central panel, particles have changed their momenta due to the implementation of the collision integral. In the right panel, particles are propagated for a time interval $\Delta t$ according to the new momenta.}
	\label{fig:cellpre}
\end{figure}

The code is basically made up of three main blocks:
\begin{description}
	\item[Initialisation] Once the geometry of the system and the initial conditions are specified, the cell matrix and the test particles array are initialised. While the former process is straightforward and is not commented further, the latter is largely illustrated in\,\ref{sec:initialisation}. The cell matrix is generated only once for each execution of the program; the initialisation of the particles is obviously repeated at the beginning of every event. After the initialisation is complete, a loop on the time steps begins.
	\item[Collision] The stochastic implementation of the collision integral is performed: only particles within the same cell can collide according to the given differential cross section. 
	\item[Propagation] Particles are propagated from $t$ to $t+\Delta t$ according to the Hamilton equations of motion.
\end{description}

\section{Initialisation}\label{sec:initialisation}
In order to solve the Boltzmann Equation an initial condition for the distribution function $f_0(x,p)$ has to be specified: this means that for every test particle initial phase-space coordinates, mass, flavour have to be known. If the simulation is running in a box, the best option is to assign them at a given initial time $t_0$, while in case of a 1D or 3D expansion it is more meaningful to specify them at a fixed initial proper time $\tau_0$; in this latter case the code assigns a `birth time' to each particle $t_0 = \tau_0 \cosh \eta_s$, that depends on the space-time rapidity. The particle is `activated' at the first time step $t^*>t_0$ and propagated freely for the (short) time interval $t^* - t_0$. The assignment of the coordinates can be performed in three ways, depending on the chosen configuration.

\subsection{Smooth initial conditions}

\begin{figure}
	\centering
	\includegraphics[width=1\linewidth]{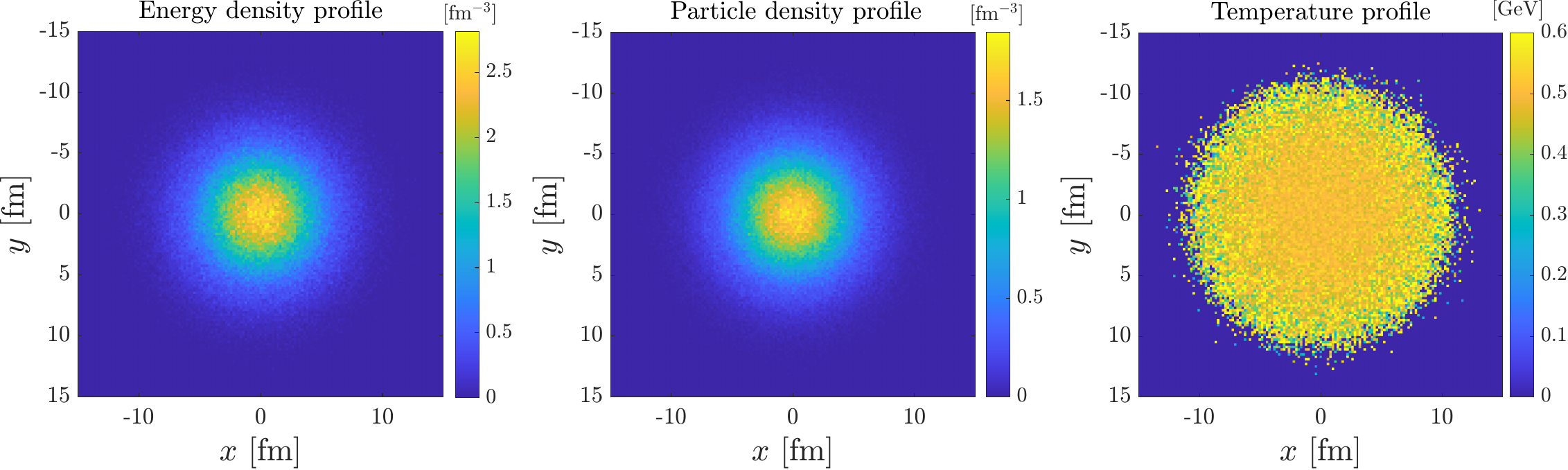}
	\caption{Energy density, particle density and temperature profiles for an initial symmetric gaussian in the transverse plane with constant temperature $T_0=0.5$ GeV.}
	\label{fig:profilesgaussian}
\end{figure}

In order to perform academic investigations on an equilibrated medium or to start a simulation with a simple initial condition, the analytical expression of the distribution function can be specified in coordinate and momentum space. We list just a few possibilities for the three typical configurations:

\begin{description}
	\item [Box] Simulation of an infinite medium in a box with periodic boundary conditions starting with a distribution function uniform in coordinate space which follows a Boltzmann-J\"uttner law in momentum space; if one wants to study the thermalisation, the momentum distribution can be strongly out-of-equilibrium, e.g. a delta function in the modulus of $p$.
	\begin{equation}
		f(x,p) = e^{-E/T}; \qquad f(x,p) = \frac{2\pi^2N}{V \bar p^2} \delta(p - \bar p).
	\end{equation}
	\item [1D expansion] The distribution function is uniform in the transverse plane with area $A_\perp$ and its profile in the $\eta_s$ direction can be specified. If one wants to simulate a boost-invariant medium, it is enough to initialise a flat distribution in $\eta_s \in [-\eta_{max}, \eta_{max}]$, with $\eta_{max}$ being large enough to prevent information from propagating from the border to the midrapidity region. Otherwise, an arbitrary profile in $\eta_s$ can be specified to study a non-boost-invariant system. The momentum distribution can be a Boltzmann-J\"uttner or, for instance, a Romatschke-Strickland:
	\begin{equation}
		f(x,p) =  \theta(|\eta_s| - \eta_{max} )\exp \tonde{ - \frac{\sqrt{ p_T^2 + (1+\xi_0)(p\cdot z)^2 }}{\Lambda_0} },
	\end{equation}
    where $z^\mu$ is the  4-vector orthogonal to $u^\mu$ and defines the longitudinal direction.
	\item [3D expansion] The distribution function has a certain profile on the transverse plane, such as a sharp circle, a Gaussian or a Wood-Saxon; a profile in $\eta_s$ has to be specified as well and is usually chosen to be a flat distribution. As far as the momentum distribution, one can choose to put $\eta_s = Y$, that is the identity between space-time and momentum rapidity, ensuring an initial vanishing $P_L$. In the case of a gaussian profile in the transverse plane (see Figure\,\ref{fig:profilesgaussian}):
		\begin{equation}
			 f_0(x,p) =\gamma_0 \exp(-x_\perp^2/R^2)\,\theta(|\eta_s|- 2.5)\delta(Y-\eta_s)\exp{ \left(- \sqrt{p_x^2 + p_y^2 + m^2}/\Lambda_0 \right)}.
		\end{equation}
\end{description}

All these configurations are obtained with Monte Carlo algorithms in which particles coordinates are generated randomly in order to follow the given distribution in momentum or coordinate space. When possible, an inverse-CDF approach is used; otherwise a standard rejection algorithm is chosen.

\subsection{Monte Carlo Glauber fluctuations}
\label{subsec:mc-glauber}

\begin{figure}
	\centering
	\includegraphics[width=1\linewidth]{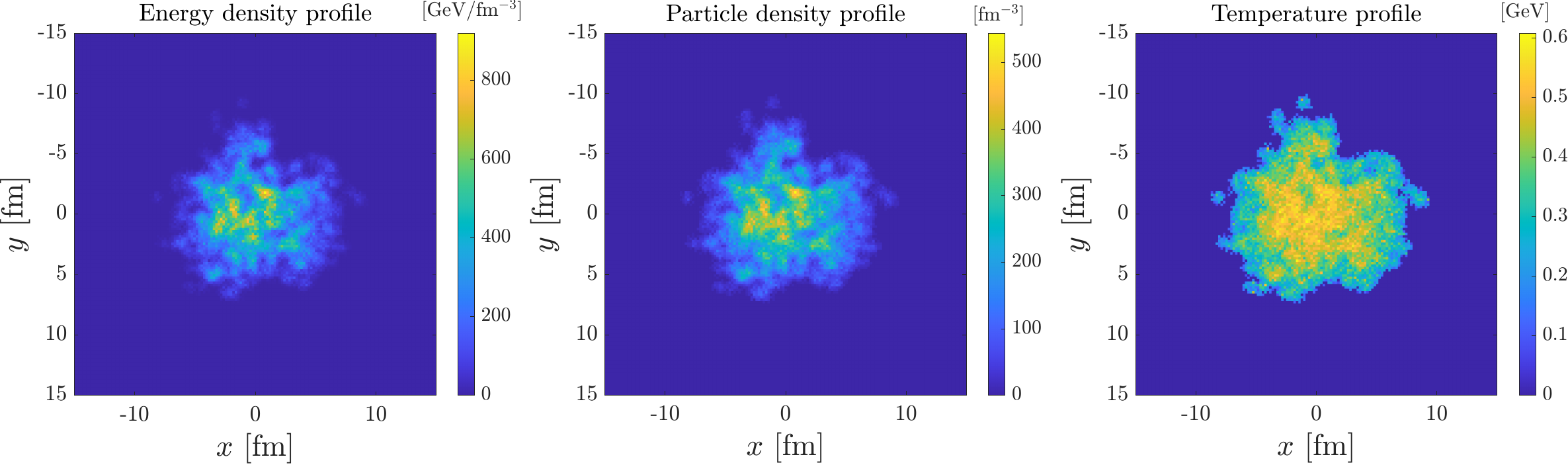}
	\caption{Initial profile generated with the Monte Carlo Glauber algorithm mimicking a PbPb collision @2.76 TeV with impact parameter $b=6$ fm.}
	\label{fig:profilemcg}
\end{figure}

To simulate more realistic conditions a Monte Carlo Glauber algorithm is implemented in the code. The idea of the Monte Carlo Glauber approach is to sample each of the two colliding nuclei as a set of $N_n$ nucleons, whose positions are chosen randomly according to a given distribution \cite{Bialas1976, Miller2007, Alver2008}. Afterwards, the collision between the two nuclei is simulated by generating a random impact parameter $b$ between $b_{min}$ and $b_{max}$ and then measuring the transverse distance $d_T$ between each nucleon pair in order to implement a geometric collision algorithm. Indeed, if $d_T < \sqrt{\sigma_{NN}/\pi}$, $\sigma_{NN}$ being the nucleon-nucleon cross section which depends mostly on the collision energy, the two nucleons are active in the collision and are enumerated among the participants $N_\text{part}$. Every participant nucleus is modelled as a peak in an energy/particle/entropy density profile; in principle the shape of this peak is arbitrary but it usually chosen as a gaussian; the global profile is obtained by summing all the transverse profiles of the participating nucleons. The extra parameter to be fixed is an overall normalisation factor in the initial profile, which is usually fixed by comparing results with the final charged multiplicity $dN_{ch}/d\eta_s$ or the deposited energy $dE/d\eta_s$ of the event with available experimental data. It means that one ends up with this distribution:
\begin{equation}\label{eq:mcg_nucleon}
	\rho(\vet x_\perp) = \rho_N\sum_{i=1}^{N_\text{part}} e^{ - (\vet x_i-\vet x_\perp)^2/2\sigma_n },
\end{equation}
in which $\sigma_n$ is the width of the gaussian.\\
Obviously such an approach models just the transverse coordinate profile of the distribution. In order to have a complete specification of the initial distribution function a profile in $\eta_s$ (usually chosen as flat in a given interval) and a momentum distribution have to be given.\\
We give some details on the implementation built in the code:
\begin{itemize}
	\item Nucleons can be distributed according to a Wood-Saxon or to a deformed Wood-Saxon \cite{Schenke:2020mbo} distribution:
	\begin{align*}
		\text{standard Wood-Saxon:}&\quad \rho (\vet x_\perp, z) =  {\rho_0}\quadre{\exp\tonde{ \dfrac{\sqrt{x_\perp^2 + z^2} -R_{WS}}{\delta_{WS}} } +1 }^{-1}\\
		\text{deformed Wood-Saxon:}&\quad \rho(r,\theta) = {\rho_0}\quadre{\exp\tonde{ \dfrac{r -R'(\theta)}{\delta_{WS}} } +1}^{-1},\\
		R'(\theta) &= R_{WS} [ 1+ \beta_2 Y_2^0 (\theta) + \beta_4 Y_4^0(\theta) ].
	\end{align*}
	Notice that these are distributions in the 3D space: afterwards, the nucleons are `flattened' in a 2D disk.\\
	The standard parameters are the radius $R_{WS}$ and the skin depth $\delta_{WS}$, while in the deformed case the two further parameters $\beta_2$ and $\beta_4$ are introduced in order to weight the two spherical harmonics. A set of these parameters is given in Table 1 of \cite{Schenke:2020mbo} for large nuclei.
	\item A minimal distance $d_{min}$ between nucleons within the same nucleus can be imposed, in order to account for the finite size of the nucleons. When generating the coordinates according to the (deformed) Wood-Saxon distribution, a check is performed on the distance $d$ of the newly generated nucleon with respect to the already generated ones. If for one or more cases $d<d_{min}$ the angular coordinates $(\theta,\phi)$ are resampled until the condition on $d_{min}$ is fulfilled. If one simply rejects all the nucleons violating this condition the distribution will eventually be tilted, showing slightly higher values at large $r$. However, to prevent the generating algorithm to stuck, after a user-defined number of failed attempts, the generated nucleon coordinates can be rejected.
	\item A nucleon substructure with a finite (preferably $n_q=3$) number of constituent quarks can be implemented. In this case the $n_q$ quarks are distributed according to the Eq.\,\eqref{eq:mcg_nucleon}. Each quark position is the centre of another gaussian with a newly specified $\sigma_q$.
	\item After the nucleon position generation, a recentering is performed in order to have the most correct estimation of the centre-of-mass of each nucleus, which is used to measure the impact parameter. The recentering is also performed at the end of the particle generation in order to have the centre of mass of the system to coincide with the origin of the coordinate space.
\end{itemize}
Eventually, one ends up with a function:
\begin{gather}
	 \rho(\vet x_\perp) = \rho_N \sum_{i=1}^{N_\text{sources}} \exp\tonde{ (\vet x_\perp-\vet x_i)^2 / \sigma };\\
\end{gather}
$\rho_N$ being the overall normalisation factor, $N_\text{sources}$ is the number of participant nucleons or quarks, depending on the chosen setup; accordingly $\sigma$ can be $\sigma_n$ for a nucleon or $\sigma_q$ for a quark. This function is interpreted as an energy/particle/entropy density and used as the coordinate space distribution according to which particles are generated with a Monte Carlo rejection algorithm. Moreover, chemical equilibrium is assumed, therefore the momentum distribution follows a temperature $T(x)$ which is determined locally by the value of $\rho(x)$.\\
Fig.\,\ref{fig:profilemcg} shows a typical initial profile obtained with the following parameters: 208 nucleons per nucleus (Pb), $\sigma_{NN}=6.4$ fm$^2$, $d_{min}=0.8$ fm, $R_{WS}=6.38$ fm, $\delta_{WS}=0.535$ fm, $\sigma_n=0.8$ fm, $\sigma_q=0.12$ fm.

\subsection{Arbitrary initial conditions}
Finally, an arbitrary initial profile can be read from an external file, which can be interpreted as energy, particle or entropy density or temperature. The specified profile can be 2D, in which case a boost-invariant $\eta_s$ distribution is assumed, or fully 3D. The grid is read as a discretised function $\rho(x_i)$; at each grid cell a value for the particle density and the temperature is computed, respectively to allow for the coordinate space and momentum space initialisation of particles. This grid does not have to coincide with the space grid of the cell matrix.\\
This function is very suitable to attach the transport code to other models or to event generators. For instance, it has been used to couple the Relativistic Boltzmann Transport approach to \kompost\, \cite{Kurkela:2018wud,Kurkela:2018vqr}, MUSIC \cite{Schenke:2010nt, Schenke:2010rr, Paquet:2015lta} and \trento\, \cite{Moreland:2014oya} event generator. A few initial profiles generated with \trento\,are reported in Chapter 6 in Fig.\ref{fig:profileoo60-80}.

\section{Collisions}
\label{sec:collision}

The collision function includes the core of the RBT code, which is the implementation of the full collision integral $C[f]$ in the case of elastic collisions. In the future, we plan to extend it to involve also inelastic collisions; we illustrate in this section how the collision integral is solved up to this point.

\subsection{The stochastic method}
\label{subsec:collision}
The stochastic method was firstly introduced in \cite{xu2004}: the aim is to solve the Boltzmann Equation in a consistent Lorentz-invariant fashion, going beyond the limitations of the geometrical model. The basic idea is to derive the collision probability between two particles from the collision integral of the Boltzmann equation. If two particles are in a spatial volume $\Delta^3x$ and have momentum respectively in the range $(\vet p_1, \vet p_1 + \Delta^3 \vet p_1)$ and $(\vet p_2, \vet p_2 + \Delta^3 \vet p_2)$, their collision rate must be:
\begin{align*}
	\frac{\Delta N^{22}_\text{coll}}{\Delta t} &= \Delta^3 x \frac{\Delta^3 p_1}{2 (2\pi)^3 E_1}\frac{\Delta^3 p_2}{2 (2\pi)^3 E_2} f(\vet x_1,\vet p_1) f(\vet x_2,\vet p_2)\\
	&\times \int \frac{d^3 \vet p'_1}{2 (2\pi)^3 E'_1}\frac{d^3 \vet p'_2}{2 (2\pi)^3 E'_2} |\mathcal M_{12\to 1'2'}|^2 (2\pi)^4 \delta^4(p_i + p_2 - p_1' - p_2').
\end{align*}
Here $|\mathcal M_{12\to 1'2'}|^2= \pi\,\mathfrak{s}^2 d\sigma/d\mathfrak{t}$ with {$\mathfrak{s}$} and  $\mathfrak{t}$ being the Mandelstam variables.\\
Using the discretisation of the distribution function:
\begin{equation}
	f(\vet x_i, \vet p_i) = \frac{\Delta N_i}{\Delta^3 x_i \Delta^3 p_i/(2\pi)^3}
\end{equation}
and the expression for the total cross section \cite{DeGroot:1980dk}:
\begin{equation}
	\sigma = \int \frac{d^3p_1'}{2 (2\pi)^3 E_1'} \frac{d^3p_2'}{2 (2\pi)^3 E_2'} \frac{2}{\sqrt{s(s-4m^2)}}  |\mathcal M_{12\to 1'2'}|^2, 
\end{equation}
the expression for the probability reads:
\begin{equation}\label{eq:p22_1}
	P_{22} = \frac{\Delta N^{22}_\text{coll}}{\Delta N_1 \Delta N_2 }= v_{rel} \sigma_{22}\frac{\Delta t}{\Delta^3 x},
\end{equation}
where 
\begin{equation}
	v_{rel}= \sqrt{\frac{(p_1 \cdot p_2)^2 - m_1^2 m_2^2}{e_1 e_2}}
\end{equation}
is the relative velocity between the two particles, with $e_i$ being the $i$-th particle energy.\\
For each time-step and each cell, a loop on all the possible particle pairs is performed. Once the probability $P_{22}$ is computed,  a real random number $r$ in the $[0,1]$ interval is extracted and the collision takes place only if $P_{22}>r$. Notice that such a probability is a Lorentz-invariant quantity, because such are, separately, $v_{rel}, \sigma_{22}$ and $\Delta t/\Delta^3 \vet x$, thus it can be plainly computed in the lab frame. If the collision does occur, though, the new momenta are to be computed in the two particles' centre-of-mass (COM) frame, which in the relativistic case is defined as the frame in which $\vet p_1 + \vet p_2 =0$. Here, the $\theta$ and $\phi$ angles are generated according to the differential cross section for one of the particles and the modulus of the momentum is fixed by the energy conservation. Once the new momenta are assigned in COM frame, they are Lorentz-transformed back to the lab frame. The results shown are mostly obtained with isotropic cross section; in Par.\,\ref{sec:shear_viscosity_mixture}, however, we will show results in the case of an anisotropic differential cross-section.\\
The test particle method allows to have a certain statistic within each cell even when the number of physical particles is too low, so to avoid large fluctuations. In order to have nonetheless the physical collision rate, the cross section has to be rescaled according to $\sigma\to \sigma/N_\text{test}$. Hence, Eq. (\ref{eq:p22_1}) turns into:
\begin{equation}\label{eq:p22_nt}
	P_{22} = v_{rel} \frac{\sigma_{22}}{N_\text{test}} \frac{\Delta t }{\Delta^3 \vet x}.
\end{equation}

A further simplification is possible:  to reduce the computational complexity of the code, one can avoid to compute, for each cell, the probability of $N_\text{cell}(N_\text{cell} -1)/2$ collision candidate pairs, $N_\text{cell}$ being the number of particles within the cell. This makes the computational effort of the code scale with $N^2$, and strongly discourages from  increasing the number of test particles. A numerically advantageous solution consists in choosing a certain number $N_{22}$ of particle pairs which are candidate to collide, compute only for these the probability $P_{22}$ and, in case, calculate their new momenta. The $N_{22}$ number can be chosen arbitrarily provided that the collision probability is amplified in the following way:
\begin{equation}\label{eq:p22_st}
	P_{22} \to P'_{22} = P_{22} \frac{N_\text{cell}(N_\text{cell}-1)/2}{N_{22}} = \frac{N_\text{cell}(N_\text{cell}-1)/2}{N_{22}} \frac{\sigma_{22}}{N_\text{test}} \frac{\Delta t }{\Delta^3 \vet x}.
\end{equation}
The results obtained with this implementation of the collision integral perfectly agree with those got by the standard $\sim N_\text{cell}^2$ number of candidate pairs.\\
Eq. (\ref{eq:p22_st}) is the final expression for the collision probability, the one we have implemented into the transport code. In particular, we have set $N_{22} = N_\text{cell}$, so that the computational complexity grows just linearly with $N$, instead of quadratically.\\
Notice that, since the probability has to be by definition smaller than 1, in choosing the space time discretisation, the cross section and the test particle number, there is a constraint we need to obey to: since $v_{rel}<1$ by construction,
\begin{equation}
	P_{22}<1 \implies \frac{\sigma_{22}}{N_\text{test}} \frac{N_\text{cell}-1}{2} \frac{\Delta t }{\Delta^3 \vet x} <1.
\end{equation}
This means that space cannot be discretised arbitrarily, since $\Delta^3 \vet x$ appears at the denominator. Moreover, the quantity $N_\text{cell}/(N_\text{test}\Delta^3\vet x)$ is basically the physical particle density of the cell and therefore can not be tuned by hand; being the cross section also fixed one can reduce $\Delta t$ to avoid unitarity violation. In case $P_{22}>1$, though, the code {does not stop}: it would be pointless to cancel an entire simulation if a few cells do not fulfil this unitarity condition, thus a warning is sent to the user, with the space-time coordinates of the unitarity violation.\\

\subsubsection{Cross section determination}
The code allows for two different scenarios:
\begin{description}
	\item [Fixed cross section] A single (differential) cross section or multiple cross sections (in case of a multi-species mixture of particles) are fixed. In this case the $\sigma_{22}$ is known once a candidate pair of test particles is extracted. This is mostly suitable to study the well-known system of hard spheres.
	\item [Fixed $\eta/s (T)$] In order to give a more realistic description of the hot QCD matter, it is far more convenient to fix globally the shear viscosity over entropy density ratio $\eta/s(T)$, which appears to be the transport coefficient governing the dissipative shear component in the hydrodynamic equations. Most works have been performed with a fixed $\eta/s$, but a realistic ansatz for this quantity requires a strong temperature dependence in proximity of the critical temperature and milder dependence at large $T$. The method firstly proposed in \cite{Ferini:2008he} is to fix globally $\eta/s$ by computing locally in space-time the cross section $\sigma_{22}$. It is therefore necessary to find an expression which relates the cross section and $\eta/s$.
\end{description}

\begin{figure}
	\centering
	\includegraphics[width=0.6\linewidth]{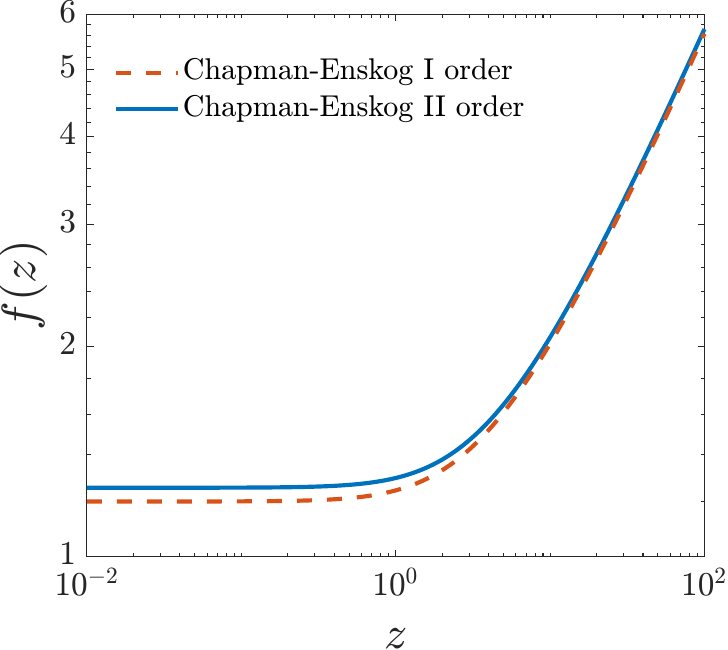}
	\caption{Determination of the $f(z) = \sigma \eta /T$ function in terms of $z=m/T$ according to the first-order and second-order Chapman-Enskog formalism.}
	\label{fig:fz_chapmanenskog}
\end{figure}

The Chapman-Enskog expansion up to second order provides a satisfying framework to compute the $\eta/s$ in a medium at thermal equilibrium with fixed cross section. As shown afterwards in this chapter, the Chapman-Enskog approach agrees very well with the Green-Kubo results obtained from the RBT code itself in a wide range of scenarios.\\
In case of a single-component gas and isotropic cross section, one finds
\begin{equation}\label{eq:CE_sigma}
	\sigma = f(z) \frac{T}{(\eta/s) \cdot s},
\end{equation}
where $f(z)$ is a certain function of $z=m/T$ and depends on the chosen approximation order; in Fig.\,\ref{fig:fz_chapmanenskog} the first- and second-order results are shown. For the 2\textsuperscript{nd} order formula used in the code, cf. \cite{wiranata2012}. In the conformal limit $f(z)\to 1.2558$, which differs from the 16-th order result 1.267 by $<1\%$. Equation\,\eqref{eq:CE_sigma} is used in the RBT code to compute locally the cross section in order to fix globally $\eta/s$. This implicitly requires, however, the system to be close to local equilibrium so that the Chapman-Enskog expression can hold; moreover, one has to compute locally the temperature $T$ and entropy density $s$. In order to achieve this, we choose to use the Landau frame for the LRF and to use the Landau matching conditions. This means that for each cell we compute the primary fluid dynamical variables starting from the energy-momentum tensor $T^{\mu\nu}$ and the density current $N^\mu$:
\begin{equation}
	T^{\mu}_\nu u^\nu = e_0 u^\mu,\quad n_0 = N^\mu u_\mu,\quad T=g^{-1}(e/n), \quad \Gamma=\frac{n_0}{n_{eq}}
\end{equation}
The first equation implies the resolution of the eigenvalue problem for the energy-momentum tensor: the largest (actually the only positive) eigenvalue is the energy density $e_0$, and the associated eigenvector is interpreted as the cell four-flow (or equivalently the fluid-element velocity). The eigenvalue problem is solved making use of GNU Scientific Library (GSL). The obtained $u^\mu$ is then used to compute also the particle density $n_0$. Starting from $e_0$ and $n_0$ one computes the temperature by inverting the function $g$:
\begin{equation}
	\frac{e_0}{n_0} = g(T) = \tonde{3 T + m \frac{K_1(m/T)}{K_2(m/T)}} \overset{m\to 0 }{\rightarrow} 3T.
\end{equation}
The local fugacity is computed by dividing the number density $n_0$ with the density at chemical equilibrium $n_{eq}$:
\begin{equation}
	n_{eq} (T) =  \frac{g}{2\pi^2} z^2\, T^3\, K_2(z),
\end{equation}
where $g$ is the number of degrees of freedom of the medium.\\
Finally, the entropy density has to be calculated. Since the number of particles is conserved, the system is out of chemical equilibrium: starting from the definition $s = \int dP f \log(1+f)$ and assuming a locally thermalised medium, one gets :
\begin{equation}
	s= \frac{e_0}{T} + n_0(1 - \log\Gamma).
\end{equation}

All these quantities are not only essential for the resolution of the collision integral, but are massively used in the computation of the thermodynamic and kinetic quantities, such as Knudsen and Reynolds numbers, in order to probe locally the distribution function. 

It should be emphasised that the collisions occurring between particles are not physical collisions. For instance, in the same time interval one particle can `collide' (i.e. change its momenta) multiple times without changing its position. The numerical method employed is indeed a way to map the evolution of the distribution function according to a given collision integral, as one can understand by looking at its derivation, without giving a physical interpretation to the microscopic processes involved. This goes with the fact that also the test particles are not physical particles, but just a tool to sample the distribution function.

\subsubsection{Freeze-out}
There is the possibility of implementing a freeze-out in the case of an expanding medium. In each cell and for every time step, once the temperature or the energy density is below a certain threshold, the cell is frozen: no collisions are performed and the particles contained therein are frozen too, which means that they will not propagate nor take part in collisions for the rest of the simulation. 

\section{Propagation}
\begin{figure}
	\centering
	\includegraphics[width=0.48\linewidth]{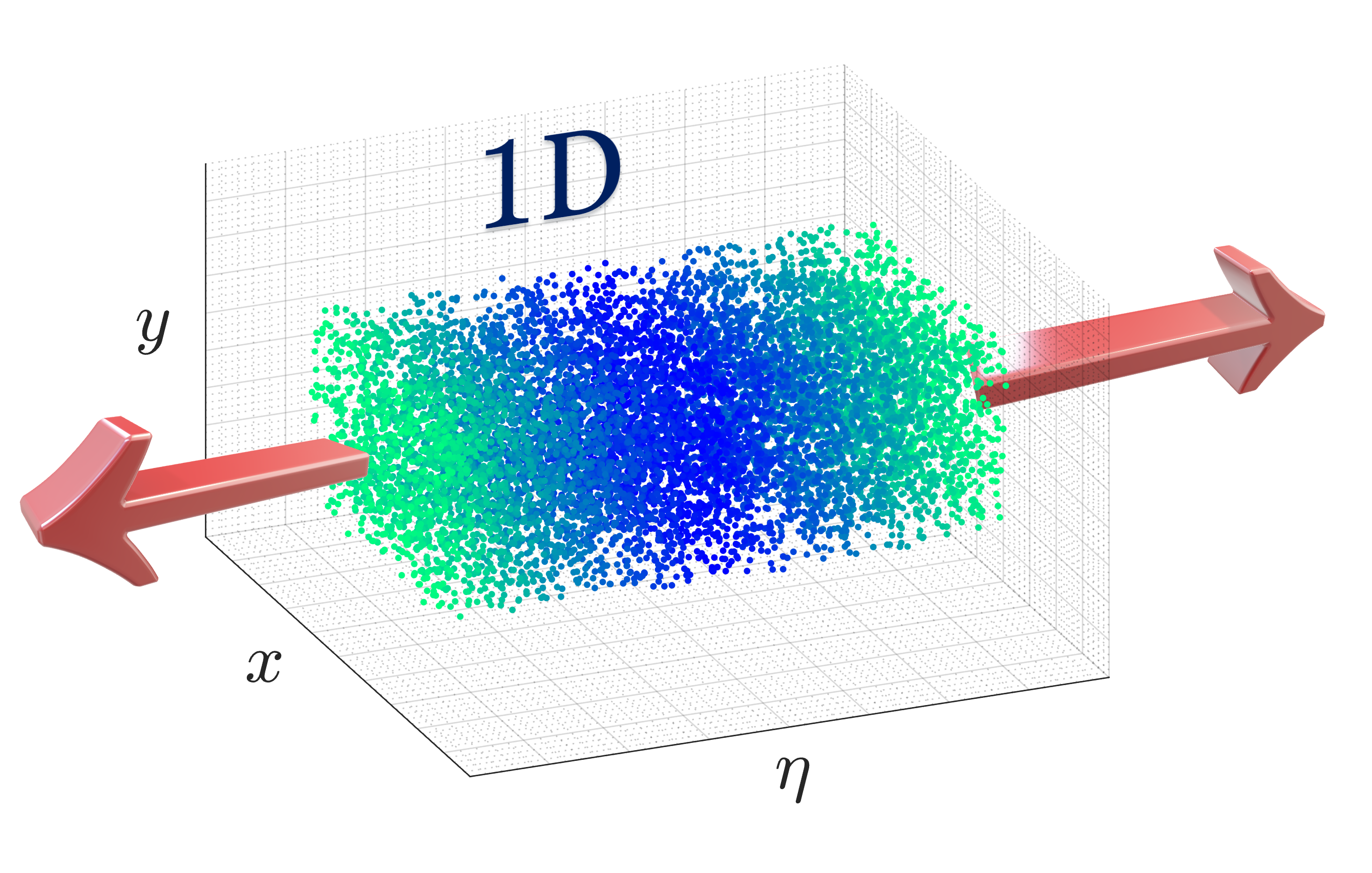}\hspace{10pt}\includegraphics[width=0.48\linewidth]{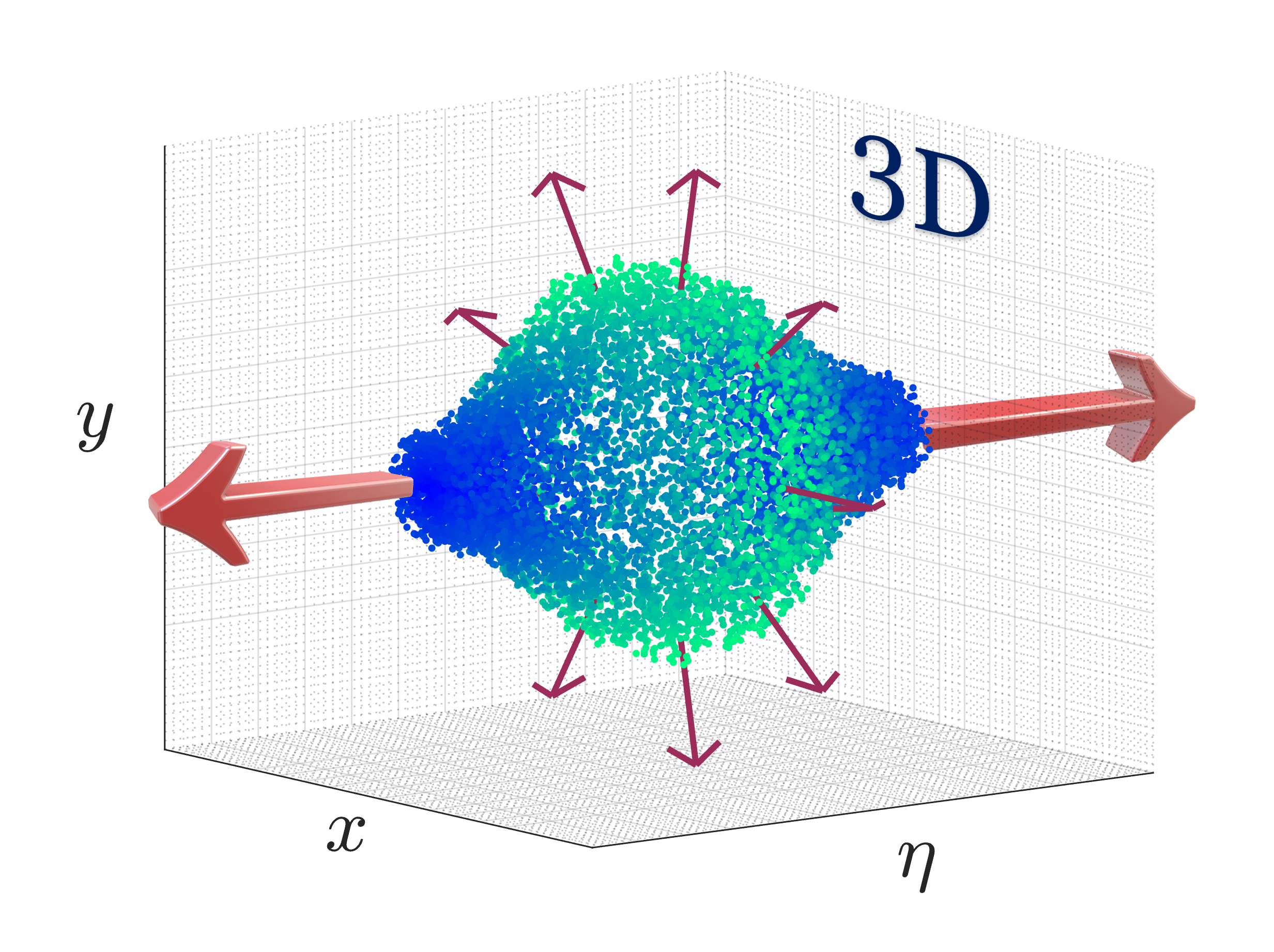}
	\caption{Two screenshots at fixed $t$ of a few representative test particles of a simulation in the two cases of 1D expansion (left panel) and 3d expansion (right panel). The arrows highlight the direction of the expansion, which is only longitudinal for the 1D case, and also transverse for a 3D simulation.  }
	\label{fig:1dexpansion}
\end{figure}
Between one collision and another, particles propagate for a given $\Delta t$. In particular, the code solves for each particle $i$ the system of relativistic equations:
\begin{equation}\label{eq:rel_newton}
	\begin{cases}
		\dfrac {d{\vet x}_i (t)}{dt} = \dfrac{\vet p_i(t)}{p_i^0(t)};\\
		\dfrac {d{\vet p}_i (t)}{dt} = \vet F (\vet x_i(t), \vet p_i(t)).
	\end{cases}
\end{equation}
The force vector $\vet F(x,t)$ accounts for possible external fields acting on the particles; for instance, if there is an electromagnetic field, $\vet F$ is the Lorentz force. It could also describe mean-field interactions. Up to now, all the relevant results have been obtained with $\vet F=0$, i.e. with free propagation of particles between collisions; however, the inclusion of such external fields could be of great interest in the uRHICs field, in which the electric and magnetic fields reach extremely high values. \\
Depending on the computation performed, the algorithm to solve the equations can be chosen among the trivial first-order Runge-Kutta (Euler method), which is suitable for the free-streaming case, the Heun method or the 4-order Runge-Kutta.\\
It should be also highlighted that not all particles are propagated: some of them could be not generated yet, since their birth-time is less than the running time, others (if a freeze-out is chosen) will be `dead' and do not propagate any more.

The propagation function includes also the implementation of the boundary conditions. The code is designed to accommodate arbitrary conditions at the borders, despite in our simulations only two different choices were needed:
\begin{description}
	\item [Periodic boundary conditions.] If a particle exits the domain borders, it re-enters from the opposite point with the same momentum. The periodic boundary conditions can be implemented in the three spatial directions $(x,y,z)$ so to have a box which actually simulates an infinite homogenous medium; implementing these conditions only in the transverse plane $(x,y)$, instead, allows to model a 1D expanding system, as in Bjorken flow. This latter configuration is used to get the results in Chapter\,\ref{chap:attractors_1D}.
	\item[Dynamic boundary conditions.] If a particle overcomes the boundaries, the code is stopped (or, alternatively, a warning is sent to the user). In this case the numerical domain has to be designed in order to contain the whole evolution of the particles: by knowing the initial and final time of the simulation and considering $v=1$ as the limiting velocity, this estimation is straightforward. Sometimes when we work with Milne coordinates, we could be contented also with losing some particles which have too large rapidity: these ones will not be detected by the experimental devices, which can cover only a finite interval in $\eta_s$. The limit is ideally $|\eta_s|\to \infty$, which corresponds to the direction of the beams. This is the setting used in Chapters\,\ref{chap:attractors_3D} and\,\ref{chap:knudsen}.
\end{description}

\section{Performance}

The CPU time required for a simulation scales approximately linearly with the number of time steps $N_t$ and the total number of test particles $N_\text{tot}$. In contrast, the spatial grid has a negligible impact on the running time, due to the fact that the current implementation of the collision integral scales with the number of particles in each cell, and therefore with $N_\text{tot}$. A rough estimate of the running time can be computed by means of
$$\text{CPU time} \approx 20' \times \tonde{\frac{N_\text{tot}}{1M}} \times \tonde{\frac{N_t}{1000}}.$$
A realistic simulation with $N_\text{tot}=15$M and $N_t=700$ takes approximately 3.5h CPU-hours. \\
The Resident Set Size (RSS) for the process is predominantly occupied by the test particle array and by the cell matrix, and it slightly increases (up to $\sim 10\%$) during the evolution if a dynamical fireball is simulated and additional memory is allocated to accommodate test particles in previously empty cells. A realistic simulation involving 15M test particles and 4.4M cells requires approximately 1.7 GB of RSS.

\section{Numerical Tests}
In this section some of the first results obtained with the code are shown. Here, the configuration in which the code is employed is that of a box with periodic boundary conditions in order to simulate an infinite medium. The first paragraphs show a few of the consistency checks performed where results can be compared to exact analytical solutions computed for a thermalised medium. Finally, the last paragraph includes original results of calculations of shear viscosity with the Green-Kubo formalism.\newline

\subsection{Collision rate}
\label{subsec:coll_rate}

\begin{figure}[t]
	\centering
	\includegraphics[width=.7\linewidth]{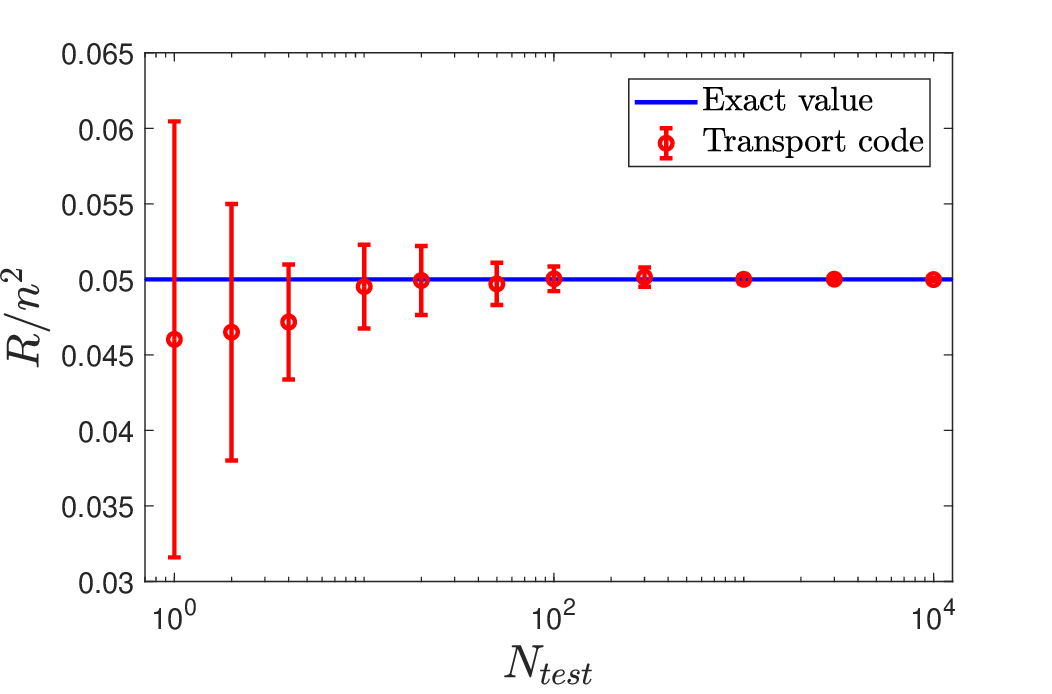}
	\caption{Convergence of $R/n^2$ with increasing $N_{test}$.}
	\label{fig:collrateconeps}
\end{figure}

One of the first checks which can easily be performed concerns the collision rate in the case of an homogeneous system with Boltzmann momentum distribution and constant isotropic cross section. The collision rate is defined as the number of collisions occurring in a given volume in a given time interval:
\begin{equation}
	R= \frac{N_\text{coll}}{\Delta t V};
\end{equation} 
in the code one has to further divide by $N_\text{test}$ in order to have the physical meaningful quantity:
\begin{equation}
	R \to R= \frac{N_\text{coll}}{N_\text{test}\Delta t V}.
\end{equation} 
This number strongly characterises the system: it is related to the mean free path between particles and affects, in case we are not at equilibrium, the time scale within which the system thermalises. It is possible to derive an analytical formula that connects the collision rate with the ratio $m/T$ and with the cross section $\sigma$.
The general formula for two colliding particles $a$ and $b$, with an energy-dependent cross section, reads \cite{koch1986}: 
\begin{equation}\label{eq:coll_rate}
	R= \frac 12 n^2 \medio{\sigma v} = \frac 12 n^2 \frac{\int_{\sqrt{s_0}}^\infty d\sqrt{s}\, \sigma(\sqrt{s})\, [s-(m_a+m_b)^2][s-(m_a - m_b)^2] K_1(\beta \sqrt{s}) }{m_a^2\, m_b^2\, K_2(\beta m_a)\, K_2(\beta m_b)}
\end{equation}
where $s=(p_1 +p_2)^2$ is the Mandelstan variable. In the massless case for an energy-independent cross section, the formula reduces to:
\begin{equation}
	R= \frac 12 n^2 \sigma.
\end{equation}

\begin{figure}[t]
	\centering
	\includegraphics[width=0.7\linewidth]{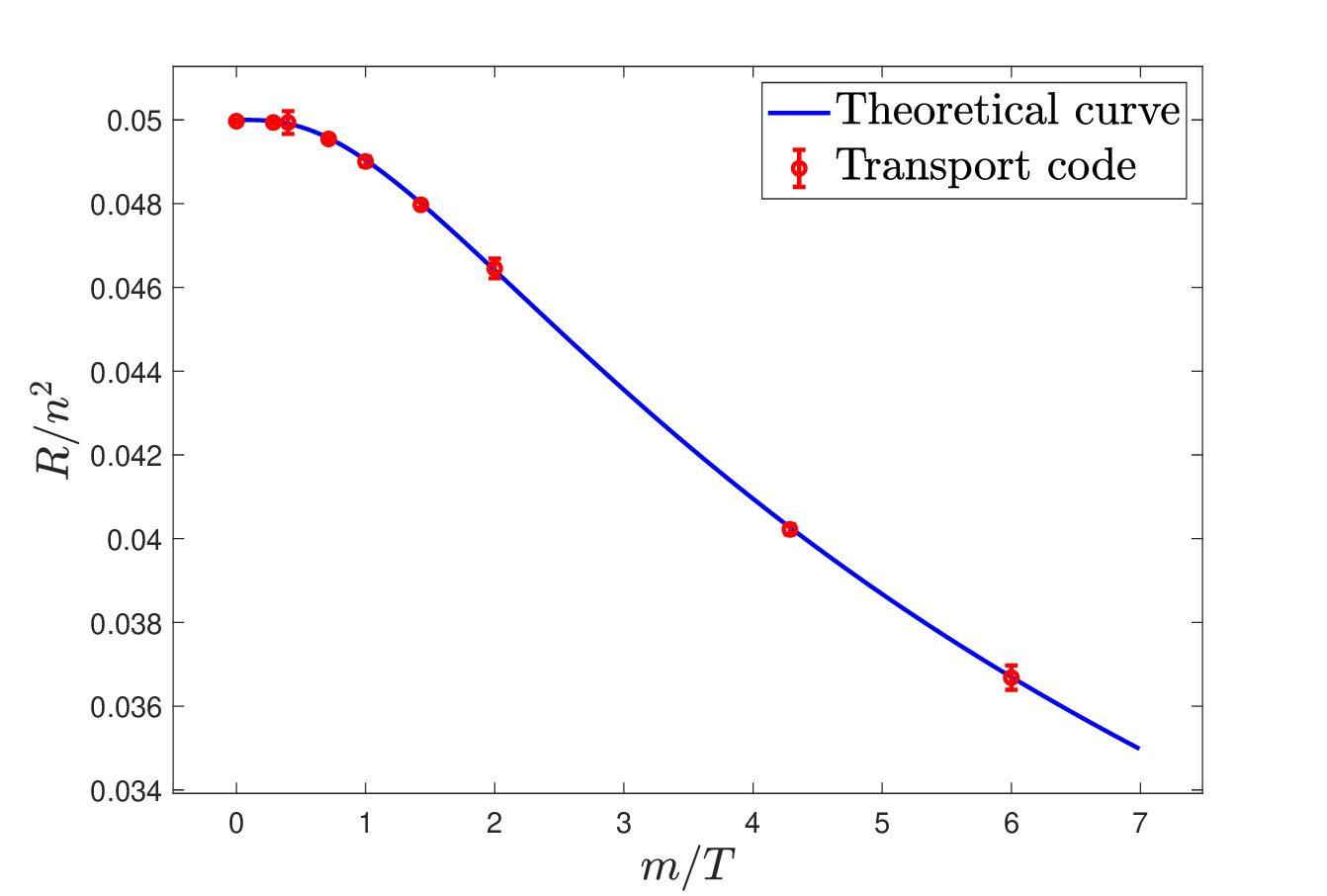}
	\caption{Expected and computed collision rate in unit of $n^2$ as a function of $z=m/T$. We plot $R/n^2$ instead of $R$ since the former quantity is dependent only on $z$ and not on $T$. Larger error bars are due to larger mass values.}
	\label{fig:collrate}
\end{figure}

The simulation is performed in a cubic volume $V=(\SI{3}{fm})^3$, made up of cubic cells $\Delta v=(\SI{0.2}{fm})^3$; $t_\text{fin}=10$ fm and $N_t=100$. The number of real particles is constraint by the medium temperature by imposing chemical equilibrium; the cross section is fixed $\sigma=0.1\, \text{fm}^2$. Given, for instance, an initial temperature $T=0.4$ GeV, the mean free path is $\lambda = 1/n \sigma \simeq \SI{12}{fm}/g $, which is considerably larger than $\Delta x = 0.2$ fm.
As far as the number of test particles is concerned, in Figure\,\ref{fig:collrateconeps} it is possible to observe how the result converges by increasing the number of test particles: already with some hundreds of particles the exact result is approximately obtained. Nonetheless, it should be outlined that the present static configuration needs much less particles then an evolving one, such as those used in the following Chapters. In Figure\,\ref{fig:collrate}, instead, some of the simulated collision rates are compared with the exact formula (\ref{eq:coll_rate}).

\subsection{Relaxation towards the equilibrium state}
\begin{figure}[t]
	\centering
	\includegraphics[width=.75\linewidth]{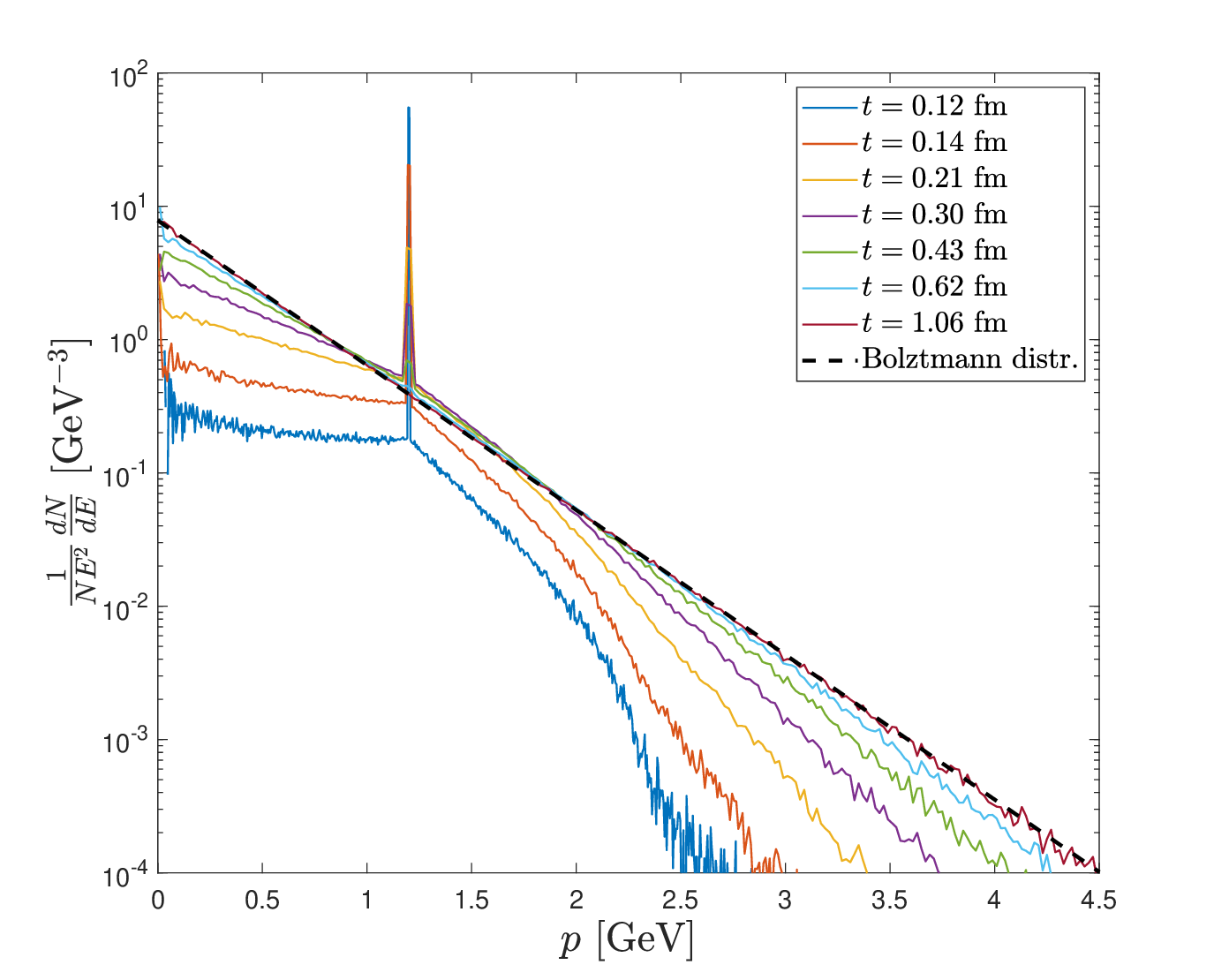}
	\caption{Evolution of the system towards equilibrium. The simulation is performed with $N=365, N_\text{tot}\sim 10^6$, $V=27$ fm$^3$ and $\sigma=10$ mb. Time scale is logarithmic with $t_0= .1$ fm and $t_f = 1.5$ fm.}
	\label{fig:thermalizationmassless}
\end{figure}
As anticipated above, the collision rate leads the relaxation of the system towards equilibrium. One can study this process by distributing particles homogeneously in space and giving all of them the same momentum modulus with random direction. Within a certain time interval the system is observed to converge to equilibrium: looking at the different curves in Figure\,\ref{fig:thermalizationmassless}, one can sees how the distribution function converges to the Boltzmann one:
\begin{gather}
	\frac{dN}{d^3\vet p} = \mathcal{N} e^{-E/T};\\
	\frac{dN}{d^3\vet p} = \frac{dN}{p^2\, dp \,d\Omega} = \frac{dN}{E^2 \,dE\, d\Omega} = \mathcal{N} e^{-p/T},
\end{gather}
where the second equality holds for a system of massless particles, while $\mathcal N$ is the proper normalisation constant. In order to have a certain equilibrium temperature, say 0.4 GeV, we make use of:	$ E/N = 3T$ and therefore assign each particle an initial momentum modulus $p=1.2$ GeV.

\subsection{Equation of state}

\begin{figure}[t]
    {\includegraphics[width=.48\textwidth]{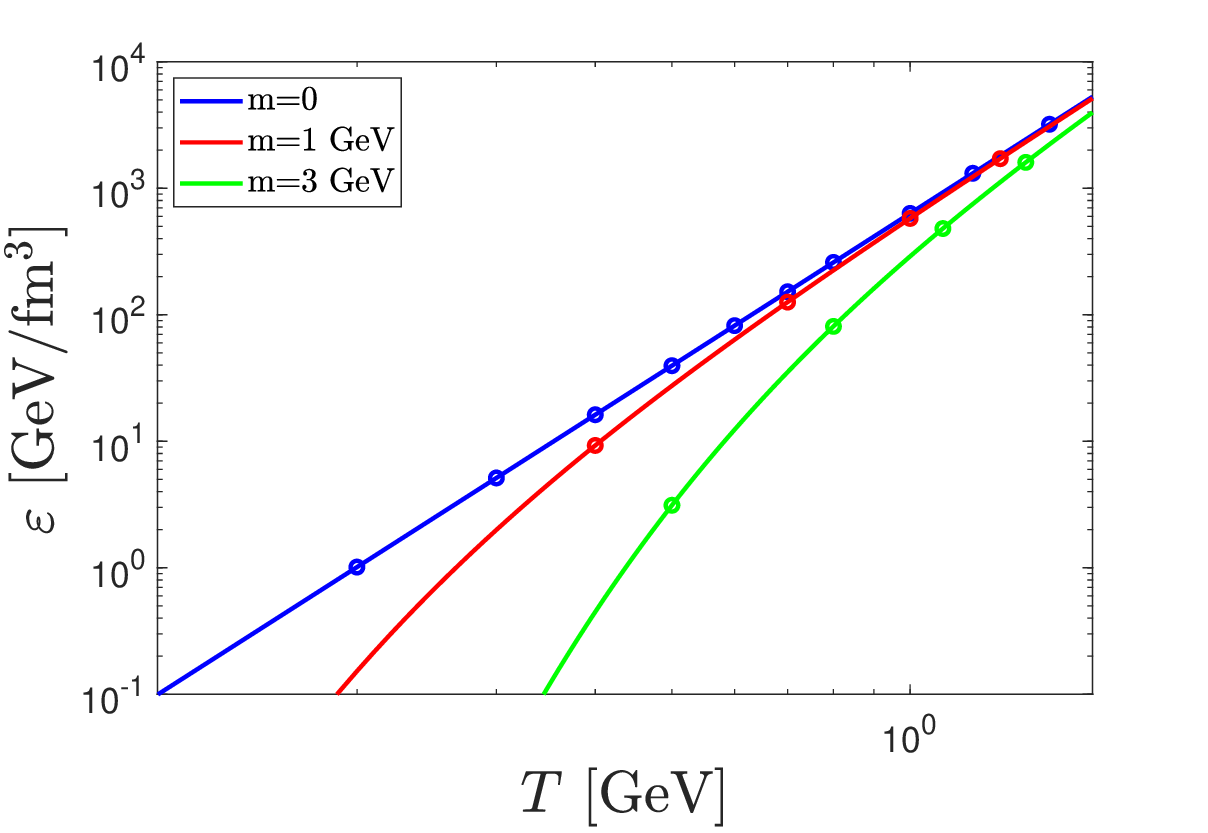}}
	{\includegraphics[width=.48\textwidth]{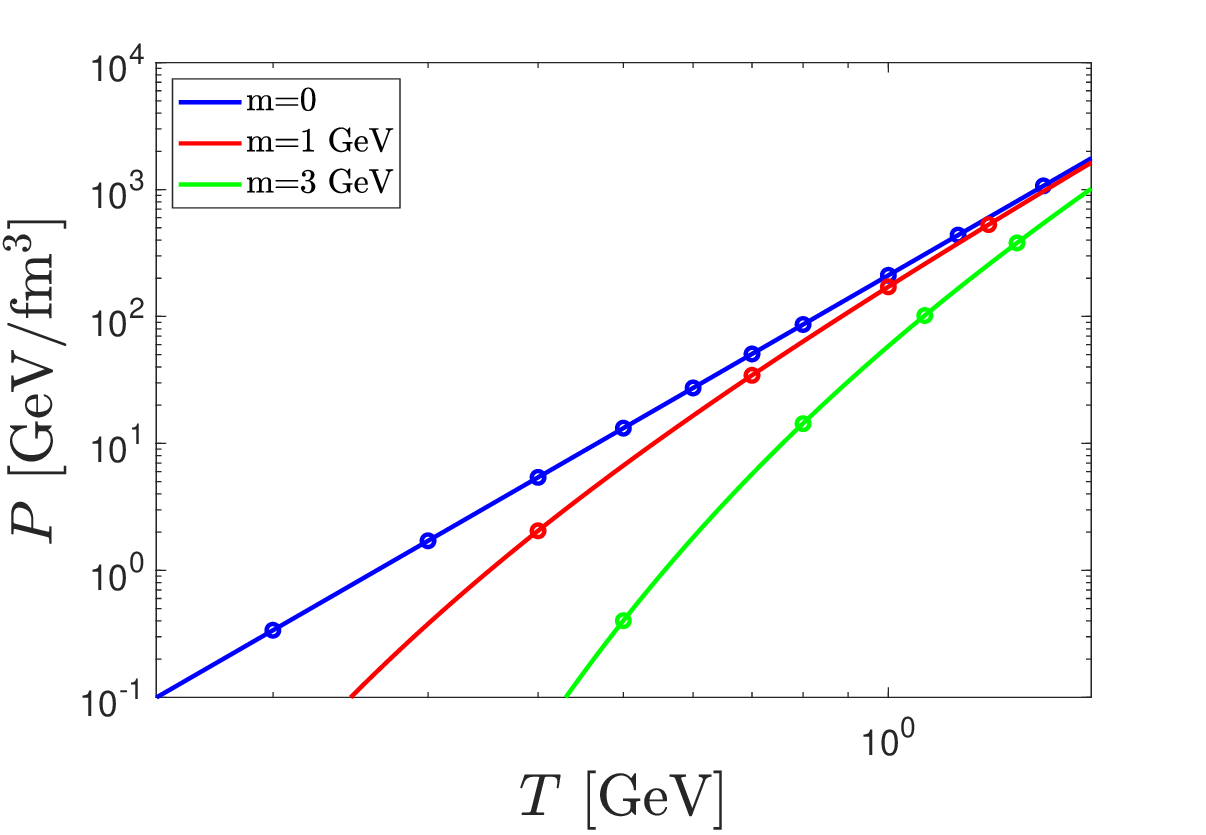}}
	\caption{Energy density (left panel) and pressure (right panel) dependence on temperature for different mass values. The box setting is always the same, the number of test particles is chosen to achieve convergence, guaranteed if $N_\text{tot}\sim 10^5$. The solid lines are the theoretical curves, the points are the code results.}
	\label{fig:en_pre_t}
\end{figure}

Going backwards to the setup in Par.\,\ref{subsec:coll_rate}, we can make further consistency checks, exploiting the fact that the system is at global thermodynamic equilibrium. Therefore, a few thermodynamic relations can be checked involving, for instance, energy and pressure. We change the values of the particle mass and of the temperature and verify if the Boltzmann thermodynamic relations hold:
\begin{subequations}\label{eqs:term_rel}
	\begin{align}
		\label{eq:density}
		n &= N^0 = \medio{p^0} = \frac{g}{2\pi^2} z^2\, T^3\, K_2(z),\\
		\label{eq:en_density}
		e &= T^{00} = \medio{p_0^2} = \frac{g}{2\pi^2} z^2\, T^4\, \quadre{ 3 K_2(z) + z K_1(z)},\\
		P& = \frac 13 (T^{11} + T^{22} + T^{33}) = \medio{p_x^2 + p_y^2 + p_z^2} = \frac{g}{2\pi^2} z^2\, T^4\, K_2(z),\\
		s &= S^0 = \frac{e + p}{T},
	\end{align}
\end{subequations}
In Figure\,\ref{fig:en_pre_t} the comparison between the expected and the computed data shows perfect agreement.

\section{Shear viscosity of a mixture}
\label{sec:shear_viscosity_mixture}

\subsection{The Green-Kubo method}
\label{subsec:green_kubo}
In Chapter\,\ref{chap:kin_and_hydro} the transport coefficients have been introduced in the context of dissipative hydrodynamics. The derivation of this quantities can be carried out in several ways: the Chapman-Enskog and the DNMR approaches, for instance, provide relations between the transport coefficients and the microscopic details of the system, such as temperature and cross section. As shown in introducing non-ideal hydrodynamics, transport coefficients are related to the relaxation to equilibrium of dissipative currents ($\pi^{\mu\nu}, \Pi, n^\mu,\dots$). This relation is not trivial but has a strong physical fundament: since dissipation of fluctuations has the same physical origin as the relaxation towards equilibrium, both dissipation and relaxation time are determined by the same transport coefficients. This connection is exploited by the Green-Kubo method \cite{Green:1954ubq,Kubo:1957wcy}, which makes use of the linear response theory to compute the transport coefficients. In particular, the focus of this section is on the shear viscosity $\eta$ and thus on the relaxation towards equilibrium of the shear viscous tensor $\pi^{\mu\nu}$.\\
In this framework, the expression of the shear viscosity according to the Green-Kubo formula \cite{zubarev1996statistical} reads:
\begin{equation}
	\eta^{ij}=\beta \,V\lim_{T_{\text{max}}\to +\infty}\int_0^{T_{\text{max}}}d t  \langle \pi^{ij}(t)\pi^{ij}(0)\rangle,
	\label{4.1}
\end{equation}
where $\beta$ is the inverse temperature and $\pi^{ij}$ is the $ij$ matrix element of the shear component of the energy momentum tensor. Here $\langle ... \rangle$ denote the following convolution procedure:
\begin{equation}
	\left\langle \pi^{ij}(t)\right.\left.\pi^{ij}(0) \right\rangle = 
	\lim_{\mathcal{T}_{\text{max}} \to +\infty} \frac{1}{\mathcal{T}_{\text{max}}} \int_0^{\mathcal{T}_{\text{max}}} dt' \, \pi^{ij}(t + t')\pi^{ij}(t'). \label{eq:pi_pi_time_correlator}
\end{equation}
In principle, the shear viscosity $\eta^{ij}$ is a traceless symmetric tensor (so is $\pi^{xy}$), due to the fact that the system  may react in a different way if the same shear stress is applied in different directions. For our purpose (and for the common use of the shear viscosity in hydrodynamics and kinetic theory), this response can be considered isotropic and thus the shear viscosity is simply a scalar $\eta$. This means that one can choose whatever off-diagonal components of the shear stress tensor; we choose $\pi^{xy}$.

The shear stress correlations $\langle \pi^{xy}(t)\pi^{xy}(0)\rangle$ can be easily computed by making use of the RBT code, with the static configuration of a box with periodic boundary conditions, similarly to what has been done in \cite{Plumari:2012ep}. 
The $xy$ shear component of the energy-momentum tensor is given by:
\begin{equation}
	\pi^{xy}(\mathbf{x},t)=T^{xy}(\mathbf{x},t)=\int \frac{d^3 \mathbf{p}}{(2\pi)^3}\frac{p^xp^y}{E}f(\mathbf{x},\mathbf{p},t);
	\label{eq:pi_xy}
\end{equation}
since we are already in the LRF of the medium, the $xy$ component of the shear stress tensor is given by the corresponding component of the energy-momentum tensor itself. Moreover, there is no spatial inhomogeneity (e.g. external fields or dynamical expansion), thus one can consider simply the volume-averaged shear tensor $\pi^{xy}(t)$. Its numerical evaluation is given by:
\begin{equation}
	\pi^{xy}(t)=\frac{1}{N_{\text{test}}}\frac{1}{V}\sum_{i=1}^{N_{\text{tot}}}\frac{p_i^xp_i^y}{E_i},
	\label{eq:discretized_pi_xy}
\end{equation}
where $i$ runs over all the test particles. The shear viscosity expression can be derived by the discretisation of Eq.\,\eqref{4.1}:
\begin{equation}
	\eta=\beta\,V \Delta t\sum_{j=0}^{N_{T_{\text{max}}}-1}\left\langle\pi^{xy}(j\Delta t) \pi^{xy}(0)\right\rangle,
	\label{4.1_discretized}
\end{equation}
where $N_{T_{\text{max}}}=T_{\text{max}}/\Delta t$, being $T_{\text{max}}$ the maximum time chosen in the simulation. The convolution procedure $\langle ...\rangle$ in Eq.\,\eqref{eq:pi_pi_time_correlator} is performed as
\begin{equation}
	\left\langle \pi^{xy}(j\Delta t)\pi^{xy}(0) \right\rangle=\nonumber\frac{1}{N_{\mathcal{T}_{\text{max}}}} \sum_{k=0}^{N_{\mathcal{T}_{\text{max}}}-1} \pi^{xy}(j\Delta t + k\Delta t)\pi^{xy}(k\Delta t)\label{eq:pi_pi_time_correlator_discretized}
\end{equation}
where $N_{\mathcal{T}_{\text{max}}}=\mathcal{T}_{\text{max}}/\Delta t$ ($\Delta t$ is kept fixed for both Eq.\,\eqref{4.1_discretized} and Eq.\,\eqref{eq:pi_pi_time_correlator_discretized}). In order to have a good accuracy, $\mathcal{T}_{\text{max}}$ has to be chosen significantly larger than $T_{\text{max}}$: for each event we fix $\mathcal{T}_{\text{max}}/T_{\text{max}}=10$, both being much larger than the typical relaxation time of the system, mimicking the analytical limit $\mathcal{T}_{\text{max}},\,T_{\text{max}}\to \infty$. The specific values of each quantity, however, are empirically fixed in order to ensure convergence of the result.\\
In particular, every Green-Kubo value for $\eta$ has been computed as an average over 45 numerical events, each running with a total number of 1--1.5M test particles. The values of the grid parameters are $\Delta x=\Delta y=\Delta z=0.4 $ fm, with a total box volume of $V=(5.2$ fm)$^3$; the timestep is fixed as $\Delta t=0.0625$ fm, whereas the total time $T_{\text{max}}$ is varied for each configuration, since it is massively affected by the chosen temperature and cross section: indeed, $T_{\text{max}}$ has to be fixed large enough so that the exponential behaviour of each correlator is appropriately dumped ($T_{max}\gg\tau_{eq}\propto\beta^3/\sigma$), but not too large in order to exclude the noise which arises from the statistical fluctuations of each event and that have a major impact when the correlator approaches its vanishing value at late times.\\

\begin{figure}
	\centering
	\includegraphics[width=.7\columnwidth]{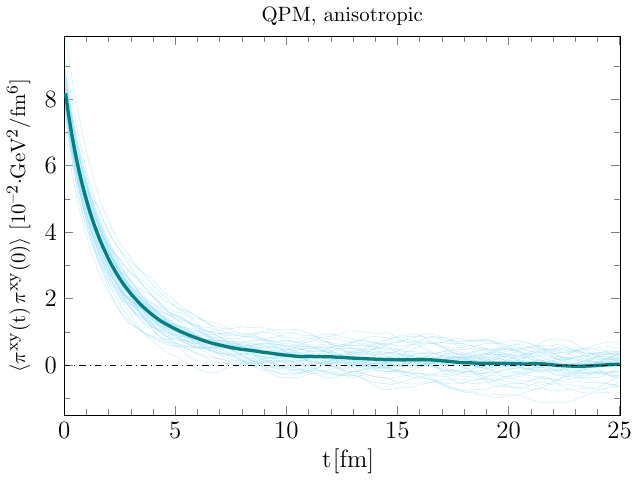}
	\caption{The behaviour of $\langle \pi^{xy}(t) \pi^{xy}(0)\rangle $ with respect to $t$ at a temperature $T=0.5$ GeV in the case 7. In cyan the results for 45 different numerical events, in dark blue their average, which exhibits a clear exponential behaviour.}
	\label{average_correlator_qpm_T_0.5}
\end{figure}

The calculation of the shear viscosity has been carried out for different scenarios:
\begin{enumerate}
	\item One-component gas with constant isotropic cross section, in the massive and massless case. 
	\item Two-component gas with constant isotropic cross section, different for the various species, in the massless case.
	\item Light quarks with equal Quasi-Particle-Model (QPM) inspired mass and gluons interacting with a single constant isotropic cross section $d\sigma/d\Omega=1/4\pi$ fm$^2$. The expression used for the masses is:
	$$ m_g^2 = \frac 16 \tonde{N_c + \frac 12 N_f}g^2 T^2,\quad m_q^2 = \frac{N_C^2 -1}{8N_C} g^2T^2, $$
	where $N_f$ is the number of flavours, $N_C$ the number of colours and the coupling constant $g(T)$ is determined by fitting lattice QCD data for the equation of state and interaction measure.
	\item Light quarks and gluons interacting with 5 different constant isotropic cross sections for the different processes:
	\begin{gather}
		\frac{d\sigma}{d\Omega}(qq' \rightarrow qq') = \frac{0.1}{4\pi} \, \text{fm}^2, \quad 
		\frac{d\sigma}{d\Omega}(q\bar{q} \rightarrow q\bar{q}) = \frac{1}{4\pi} \, \text{fm}^2,\\
		\frac{d\sigma}{d\Omega}(qq \rightarrow qq) = \frac{10}{4\pi} \, \text{fm}^2, \quad 
		\frac{d\sigma}{d\Omega}(qg \rightarrow qg) = \frac{3}{4\pi} \, \text{fm}^2,\\
		\frac{d\sigma}{d\Omega}(gg \rightarrow gg) = \frac{3}{4\pi} \, \text{fm}^2.
	\end{gather}
	 
	\item Single component (gluon) gas interacting with a perturbative QCD (pQCD) inspired differential cross section:
	$$ \frac{d\sigma^{gg\to gg}}{dt} = - 9\pi \alpha_s^2 \frac{1}{(t - m^2_D(T))^2}, $$
	where $m_D(T)=g(T)T$ is the Debye screening mass \cite{Philipsen2001}. 
	\item Single component (gluon) gas interacting according to the full tree-level pQCD differential cross section.
	\item A mixture of light quarks (and antiquarks) and gluons interacting according to the full tree-level pQCD interaction matrices, with differential cross sections depending on the species involved in the collision and with temperature dependent masses taken from the QPM.
\end{enumerate}

\begin{figure}
	\centering
	\includegraphics[width=1\linewidth]{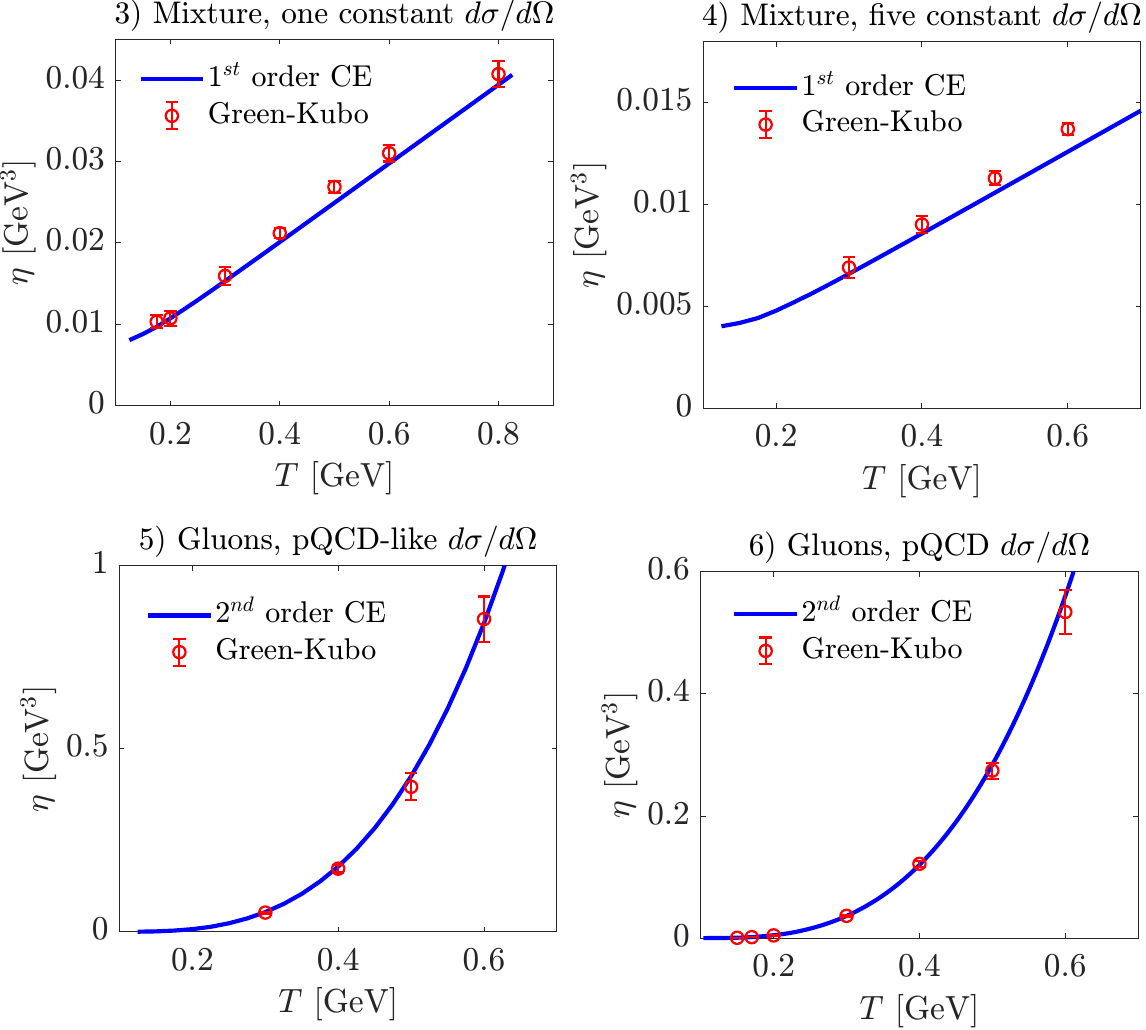}
	\caption{Comparison between the Green-Kubo and the Chapman-Enskog in the different scenarios illustated in the text.}
	\label{fig:gk_vs_ce}
\end{figure}

For the details concerning the matrix elements of the two latter cases see \cite{Parisi:2025gwq}. In the cases involving a mixture of species, the relative abundance is fixed by the number of degrees of freedom and by the respective masses. Since the first two cases had already been addressed in the literature \cite{Plumari:2012ep}, here we show the results obtained in the remaining ones. In \cite{Parisi:2023tesi} and \cite{Parisi:2025gwq}, the results obtained with the Green-Kubo formalism are compared with the first- or second-order Chapman-Enskog outcomes developed therein. As shown in Figure\,\ref{fig:gk_vs_ce} for the cases 3-6, the agreement is extremely good: the regular underestimation by the Chapman-Enskog method which appears in the top panels is due to the first order expansion. For instance, in the simple case of a one-component massless gas with a constant isotropic cross section, the first order Chapman-Enskog gives $1.2$, which is $\approx 5\%$ smaller than the 16-th order results 1.267. Finally, in Figure\,\ref{fig:gk_vs_ce} the data for the shear viscosity over entropy density ratio $\eta/s$, which is the coefficient governing the shear relaxation time and therefore the hydrodynamical equations, is shown for the most realistic case 7 \cite{Parisi:2025gwq}. The Green-Kubo results perfectly agree with the first-order Chapman-Enskog calculations; for reference, some lQCD points are present. For this specific case, in Figure\,\ref{average_correlator_qpm_T_0.5} the result of the 45 different calculations is shown for $\langle \pi^{xy}(t) \pi^{xy}(0)\rangle $ at a temperature of $T=0.5$ GeV. Along with the results for each event, which carry a significant amount of fluctuations, their ensemble average (dark blue) is shown, which highlights the typical exponential decay of the correlator.

\begin{figure}
	\centering
	\includegraphics[width=.7\columnwidth]{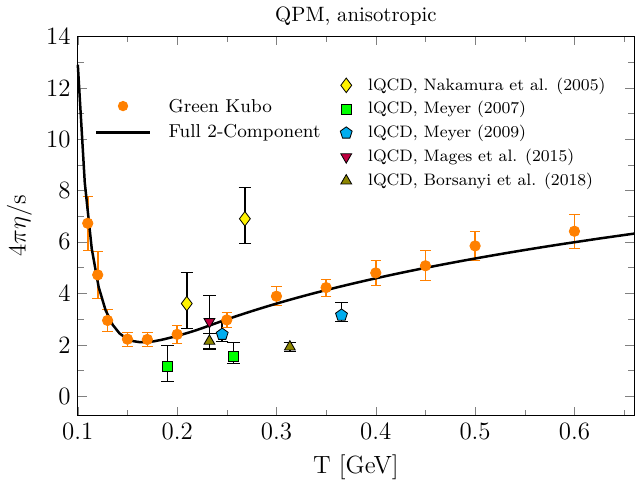}
	\caption{Shear viscosity over entropy density ratio for a mixture with pQCD cross sections as obtained via the Green--Kubo and Chapman--Enskog methods and compared with available lQCD data.}
	\label{fig:shear_viscosity_medie_green_kubo	}
\end{figure}

\chapter{Attractors in 0+1D and 1+1D systems}
\label{chap:attractors_1D}

In Chapter\,\ref{chap:kin_and_hydro} the success of hydrodynamics, especially in its viscous and anisotropic formulations, in modelling the evolution of the hot QCD matter has been discussed: even when the dynamics is certainly far from equilibrium, hydrodynamic predictions surprisingly agree with experimental measurements on hadron transverse momentum spectra and anisotropic flows not only in uRHICs but also in smaller systems, such as $pp$ and $pA$ collisions in the case of high-multiplicity events, raising questions about the possible formation of a small droplet of QGP \cite{Yan:2013laa, Romatschke:2015gxa, Habich:2015rtj, Weller:2017tsr, Shen:2016zpp, Mantysaari:2017cni, Romatschke:2017acs, Grosse-Oetringhaus:2024bwr}. This discovery has raised much interest in the scientific community concerning the theoretical foundations of hydrodynamics, which have been deeply investigated recently, as well as the importance of understanding the thermalisation mechanism, the role of the system size and the peculiar time scales to comprehend the nature of the observed collective behaviour. Among these new advancements, one will be mainly addressed in this work: the emergence of the attractor behaviour. The far-from-equilibrium evolution of macroscopic quantities, irrespective of the initial conditions, exhibit universal scaling, denoted as non-equilibrium attractors, which capture the long-term collective behaviour of the fluid system. This universality has been found not only in the context of relativistic hydrodynamics, but also in kinetic theory, classical Yang-Mills equations as well as in AdS/CFT calculations.  (hydrodynamics \cite{Heller:2015dha, Strickland:2017kux,  Chattopadhyay:2019jqj, Jaiswal:2019cju,Blaizot:2019scw, Alalawi:2020zbx, Heller:2020anv}, Effective Kinetic Theory (EKT) \cite{Kurkela:2018vqr, Kurkela:2019set, Almaalol:2020rnu}, Relaxation Time Approximation (RTA) kinetic theory \cite{Behtash:2017wqg, Blaizot:2017ucy,  Strickland:2018ayk, Heller:2018qvh, Kamata:2020mka, Frasca:2024ege}, small-angle kinetic theory \cite{Tanji:2017suk, Brewer:2022vkq}, BAMPS \cite{Ambrus:2021sjg}, RBT \cite{Nugara:2023eku, Nugara:2024net}, Yang-Mills equations \cite{Berges:2013eia,Berges:2013fga}, AdS-CFT \cite{Kurkela:2019set, Heller:2011ju}).\\
This lack of memory on the initial conditions can be noteworthy at least under two different perspectives. On one hand it can be related with this success of hydrodynamics out of its rigorous applicability regime; the system undergoes the \emph{hydrodynamisation} process, which can be understood with the fast decay of the non-hydrodynamic modes, while a few long-living hydrodynamic modes survive and characterise the late-time behaviour of the system. The appearance of the attractor is therefore related with the physical scales which characterise the system evolution, as shown in the following. On the other hand, by looking at which observables keep or not memory about the initial conditions, one can select those carrying information about the initial state of the collision, which is obviously out of direct experimental reach.\\
These studies have been widely performed in the context of boost-invariant 0+1D systems with fixed $\eta/s$ in the aforementioned theoretical frameworks; recent research in the realm of effective kinetic theory has extended these studies to the more realistic scenario of 3+1D simulations \cite{Ambrus:2021sjg, Ambrus:2022koq, Boguslavski:2023jvg}. In this work a 0+1D investigation is carried out in the context of RBT, with a comparison with RTA Boltzmann kinetic theory, viscous and anisotropic hydrodynamics, and then several extensions are introduced: for the first time attractors are studied for non-boost-invariant systems and with a more realistic $T$-dependent $\eta/s$; later-on emergence of universality is studied in 3+1D so to analyse the impact of the transverse dynamics on the observed universality and finally the appearance of attractors is also connected to the scaling in anisotropic flows between different collision systems. Following what had been done in the literature, the quest for attractors is primarily focused on sets of moments of the distribution function $f(x,p)$, while in the final Chapter it also addresses the anisotropic flows $v_n$.

\section{RBT configuration and initial conditions}

With the intent of studying the emergence of attractors in 1D systems, the RBT Code is used in its 1D configuration (see Chapter\,\ref{chap:attractors_1D}): periodic boundary conditions are implemented in the transverse plane (a square of $5.2\times5.2$ fm$^2$), over which the medium is fully homogeneous. The longitudinal distribution in $\eta_s$ is flat in a given interval $[-\eta_{s,max}, \eta_{s, max}]$: $\eta_{s,max}$ is chosen large enough to prevent information to travel from the boundaries to the midrapidity region, while it is given a finite value when non-boost-invariant simulations are performed. The grid parameters are fixed as $\Delta x = \Delta y = 0.4 $ fm and $\Delta \eta_s = 0.4$. We employ about 1500-3000 time-steps, depending on the final time $t_\text{fin}$. Finally, we are able to simulate systems with up to $3\cdot 10^8$ total particles, achieving quite good statistics. Results are extracted by considering a selection in $\eta_s$, whose thickness is equal to $\Delta\eta_s = 0.4$, unless otherwise specified. The setup described above has been verified to guarantee convergence and stability of the RBT method.\\

The initial condition for the $f(p)$, which has a pure momentum-dependence due to the coordinate space homogeneity, is the Romatschke-Strickland distribution function \cite{Romatschke:2003ms}, which models a spheroidally deformed thermal initial condition for an ideal gas of massless particles:
\begin{equation}\label{eq:frs}
	f_0 (p) = \gamma_0 \exp \tonde{ - \frac{\sqrt{ p_T^2 + (1+\xi_0)(p\cdot z)^2 }}{\Lambda_0} },
\end{equation}
where the parameters $\Lambda_0$ and $\gamma_0$ are computed to match the initial energy and particle density to the Boltzmann equilibrium values, while $\xi_0$ quantifies the system momentum anisotropy (see Section\,\ref{subsubsec:0+1D_ahydro}). It is also useful for the discussion to introduce the elliptical anisotropy parameter $\alpha_0 = (1+\xi_0)^{-1/2}$. The isotropic Boltzmann distribution is recovered when $\xi_0=0$; in this case the parameter $\Lambda_0$ reduces to the initial temperature $T_0$ and $\gamma_0$ to the standard fugacity $\Gamma_0$.  For the results shown in this section, we fix $T_0=0.5$ GeV and $\Gamma_0=1$; different initial conditions are implemented by changing the anisotropy parameter $\xi_0$, in particular $\xi_0=[-0.5,\,0,\,10]$ which correspond, respectively, to a prolate, spherical and oblate distribution in momentum space. {We will also consider the limit $\xi_0\to \infty$, which means to initialise a system with $\eta_s = Y$ and thus initial $P_L=0$.}\\
We compare the results obtained with the RBT approach with other models, which have been already described in Chapter\,\ref{chap:kin_and_hydro}: the solution of the Boltzmann equation in Relaxation Time Approximation, second-order dissipative viscous hydrodynamics and anisotropic hydrodynamics. The same initial conditions have to be specified: in the case of RTA equation the Romatschke-Strickland distribution is given as $f_0$; in the case of DNMR equations the energy density is used to fix the initial temperature, while the anisotropy parameter $\xi_0$ fixes the $\pi/e$ ratio according to Eq.\,\eqref{eq:pi_over_e_xi}; in anisotropic number-conserving hydrodynamics the initial $\gamma_0$, $\Lambda_0$ and $\xi_0$ parameters fully determine the initial conditions.

As anticipated, we look for attractors by analysing the moments of the distribution function in the various models. Following Ref. \cite{Strickland:2018ayk}, we define the momentum moments of the distribution function for on-shell particles as
\begin{equation}\label{eq:momentum_moments}
\mathcal{M}^{{nm}}(\tau)=\int dP \, (p\cdot u)^n \, (p\cdot z) ^{2m} \, f(\tau,p),
\end{equation}
where $dP=d^3p/[(2\pi)^3 p^0]$ is the momentum integration measure.
The full (infinite) set of moments encode all the information of the distribution function: in principle the larger the number of known moments, the better the full distribution function is specified.\\
It is useful to define the normalised moments $\overline M^{{nm}}=\mathcal{M}^{{nm}}/\mathcal{M}^{{nm}}_{eq}$ \cite{Strickland:2019hff}, where the moments are rescaled by their corresponding equilibrium values $\mathcal{M}^{nm}_{eq}$. Notice that in the limit of isotropic and thermal equilibrium $\overline M^{{nm}}\to 1$.
In the case of massless Boltzmann distribution with particle number conservation the equilibrium moments are given by {\cite{Strickland:2019hff}:
\begin{equation}\label{eq:MeqBoltz}
    \mathcal{M}^{nm}_{eq} (\tau) = \frac{(n+2m+1)!\,\Gamma(\tau)\, T^{n+2m+2}(\tau)}{2\pi^2 (2m+1)},
\end{equation}
which are exactly the moments in Eq.\,\eqref{eq:momentum_moments} computed when $f_{eq} (\tau,p) = \Gamma(\tau) \exp (-p\cdot u/T(\tau))$, with $T$ and  $\Gamma$ being the effective time-dependent temperature and fugacity.\\
In the context of the RBT code the evaluation of such moments is straightforward: one has just to insert the expression for the distribution function as a sum of delta functions to get the full expression. In particular, in the context of Bjorken flow, these moments reduce to:
\begin{align}\label{eq:momentum_moments_BJ}
	\mathcal{M}^{{nm}}(\tau)&=\int dP \, (p\cdot u)^n \, (p\cdot z) ^{2m} \, f(\tau,p)\\&=\int \frac{d^2 p_T dp_w}{(2\pi)^3 p_\tau} p_\tau^n p_w^{2m} \, f(\tau,p)
\end{align}
where the integration measure in the momentum space $dP$ in this case becomes $\dfrac{d^2 p_T dp_w}{(2\pi)^3 p_\tau}$, being
\begin{subequations}\label{eq:strick_moments}
	\begin{gather}
		p_w=p_z\cosh(\eta_s) - p_0 \sinh(\eta_s),\\
		p_\tau=p_0\cosh(\eta_s) - p_z \sinh(\eta_s),
	\end{gather}
\end{subequations}
the momenta associated with the proper time $\tau$ and the space-time rapidity $\eta_s$, in which $p_z$ is the longitudinal momentum and $p_0$ for an on-shell massless particle corresponds to the energy.
The system flow velocity $u^{\mu}$ and the orthogonal versor $z^{\mu}$ in the case of the Bjorken flow reduce to \cite{Bjorken:1982qr}
\begin{subequations}\label{eq:umu_bjorken}
	\begin{gather}
		u^\mu = (\cosh \eta_s, 0,0,\sinh\eta_s), \\ z^\mu=(\sinh\eta_s, 0,0, \cosh\eta_s).
	\end{gather}
\end{subequations}
Some of the previous moments have a clear physical interpretation: $\mathcal{M}^{10}$, $\mathcal{M}^{20}$ and $\mathcal{M}^{01}$ correspond, respectively, to the number density $n$, the energy density $e$ and the longitudinal pressure $P_L$. For a conformal system $e=2P_T+P_L$; therefore, we can compute also the transverse pressure from $\mathcal{M}^{20}$ and $\mathcal{M}^{01}$.
Notice that we will relax the assumption of Bjorken flow in Sec.\,\ref{subsec:no_boostinv}; in that case the expression in Eq.\,\eqref{eq:momentum_moments_BJ} is no more valid and we will use the general definition Eq.\,\eqref{eq:momentum_moments}.\\

In order to have a proper comparison with the other models, a definition of the momentum moments in these different frameworks is necessary:
\begin{description}
	\item[RTA Boltzmann]
	\begin{multline}
		M^{nm}(\tau) = \frac{(n+2m+1)!}{(2\pi)^2}\times\\ \times
		\bigg [
		D(\tau,\tau_0) \alpha_0^{n+2m-2} T_0^{n+2m+2}\Gamma_0 \frac{ \mathcal H^{nm} (\alpha\tau_0/\tau) }{ [\mathcal H^{20}(\alpha_0)/2]^{n+2m-1} } +\\
		+ \int_{\tau_0}^\tau \frac{d\tau'}{\tau_{eq}(\tau')} D(\tau',\tau') \Gamma(\tau') T^{n+2m+2}(\tau') \mathcal H^{nm} \tonde{ \frac{\tau'}{\tau} } \bigg ].
	\end{multline}	
	where the $D(\tau_1,\tau_2)$ and $\mathcal H^{nm} $ functions were introduced in Par.\,\ref{subsubsec:rta_0+1D}.\\
	\item[DNMR] In the framework of hydrodynamics there is no distribution function to be integrated in order to get its moments. It is nonetheless possible to exploit the ansatz that in the DNMR theory is proposed for $\delta f = f - f_{eq}$ to define an equivalent set of moments. Exploiting the expansion in Eq.\,\eqref{eq:df_IS}, the normalised moments read as:
	\begin{equation}\label{eq:vhydro_moments}
		\overline M^{nm}_{\text{DNMR}} = 1- \frac{3m(n+2m+2)(n+2m+3)}{4(2m+3)} \frac{\pi}{\varepsilon}.
	\end{equation}
	However, it is easy to understand that only a few of them are meaningful in the context of viscous hydrodynamics.
	\item[Anisotropic hydrodynamics] The same can be done with anisotropic hydrodynamics:
	\begin{equation}\label{eq:ahydro_moments}
		\overline M^{nm}_{\text{aHydro}} (\tau) = (2m+1) (2\alpha)^{n+2m-2} \frac{\mathcal H^{nm} (\alpha) }{ [\mathcal H^{20} (\alpha)]^{n+2m-1} },
	\end{equation}
	
\end{description}

\section{Isotropisation and thermalisation}\label{sec:iso_therm}

\begin{figure*}[t]
    \centering
    \includegraphics[width=\textwidth]{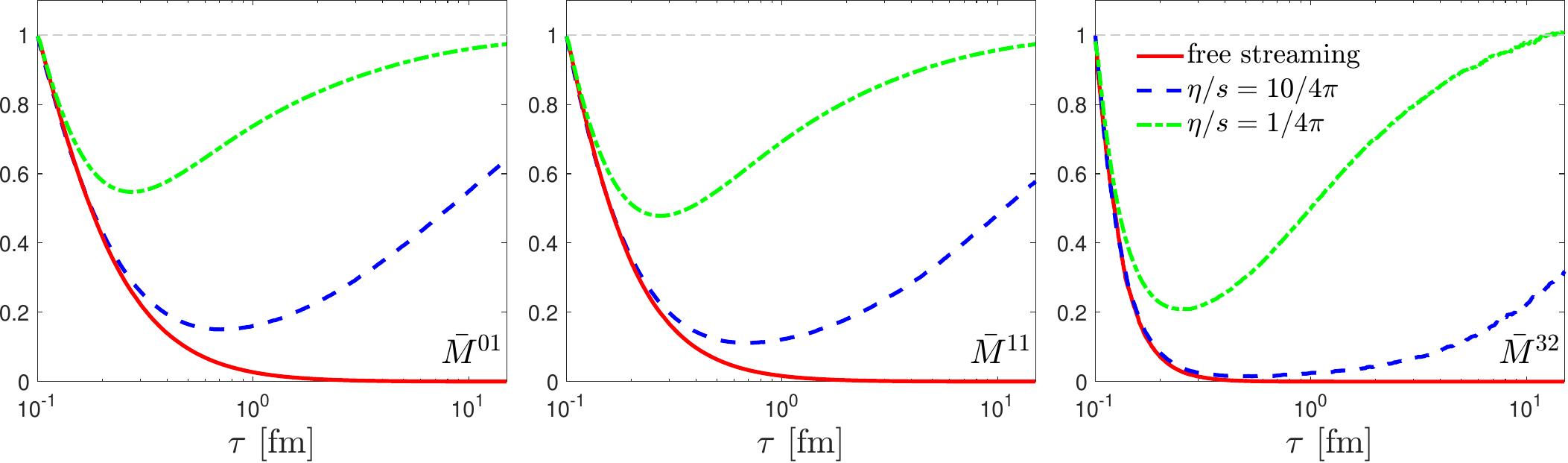}
    \caption{Normalised moments as a function of proper time $\tau$ at midrapidity obtained with the RBT approach initialised with $\tau_0= 0.1$ fm, $T_0 = 0.5$ GeV and $\xi_0=0$. Different curves correspond to different values of $\eta/s$: $1/4\pi$ (dot-dashed green), $10/4\pi$ (dashed blue), free streaming $\eta/s \to \infty$ (solid red).}
    \label{fig:freestream}
\end{figure*}

In Fig.\,\ref{fig:freestream}, we show the results of the RBT approach for some of the normalised momentum moments of the distribution function as a function of the proper time at midrapidity $|\eta_s|<0.2$. The calculation starts at initial proper time $\tau_0= 0.1$ fm, with initial temperature $T_0 = 0.5$ GeV and without initial anisotropy ($\xi=0$, $\alpha=1$). We consider two different values of the specific shear viscosity: $4\pi\eta/s=1$ and $4\pi\eta/s=10$. All moments reach the isotropic and thermal limit for large times. In particular, in the case of smaller $\eta/s$ the equilibrium is reached earlier with respect to the case with larger specific viscosity: for all the moments presented $\overline M^{{nm}}\to 1$ at a time $\tau\sim2$ fm for $4\pi\eta/s=1$, while that limit is reached at $\tau>20$ fm for $4\pi\eta/s=10$. This is not surprising, since the scattering rate is inversely proportional to the specific viscosity.
Momentum moments with different $m,n$ probe the behaviour of the distribution function in different regions in $p$ and $p_z$ of the phase-space.
Due to the chosen initial condition ($P_L/P_T=1$), the moments start from $\overline M^{{nm}}=1$, deviate from the equilibrium limit and then approach it again, with different time scales depending on the specific viscosity of the system, as mentioned before. However, the amount of deviation from the equilibrium during the evolution depends on the powers of $p_z$ and $p$. Moving from the left to right in the plot, we probe higher powers of $p_z$ and $p$. We observe that for higher order $m,n$ there is a larger deviation of the normalised moments from the equilibrium limit.
This is easily understood since high-energy particles contributing more to the high order moments are expected to thermalise later, as they need a larger number of scatterings in order to equilibrate with the surrounding medium.
In Fig.\,\ref{fig:freestream} the curves obtained with finite specific viscosity are compared with the case of a free-streaming medium, that corresponds to $\eta/s\to\infty$. In the RBT code this limit is achieved imposing that the scattering cross section is zero.
We observe that, even though the free-streaming lines start from the equilibrium limit $\overline M^{{nm}}=1$, they quickly deviate from it due to the longitudinal expansion and, since there are no microscopic mechanisms that tend to recover the local equilibrium, the moments move away indefinitely from the equilibrium limit.
It is however interesting to notice that the curves at finite viscosity follow the free-steaming case at initial time and then deviate from it approaching again the equilibrium limit. This is due to the competition between two different effects: the longitudinal expansion and the collisions between particles.
The momentum exchange due to collisions tends to isotropise the system, producing the effect of reducing the strong anisotropy along the longitudinal direction and drives the normalised moments toward the isotropic limit $\overline M^{{nm}}\to 1$: at initial times, the longitudinal expansion is so violent that its effect dominates over the collisions. 
Therefore, the appearance of a minimum in the curves $\overline M^{{nm}}$ with $n>0$ and the subsequent increase toward the isotropic limit is connected to the time-scale in which collisions start to dominate over the expansion. 

\begin{figure*}[t]
    \centering
    \includegraphics[width=\textwidth]{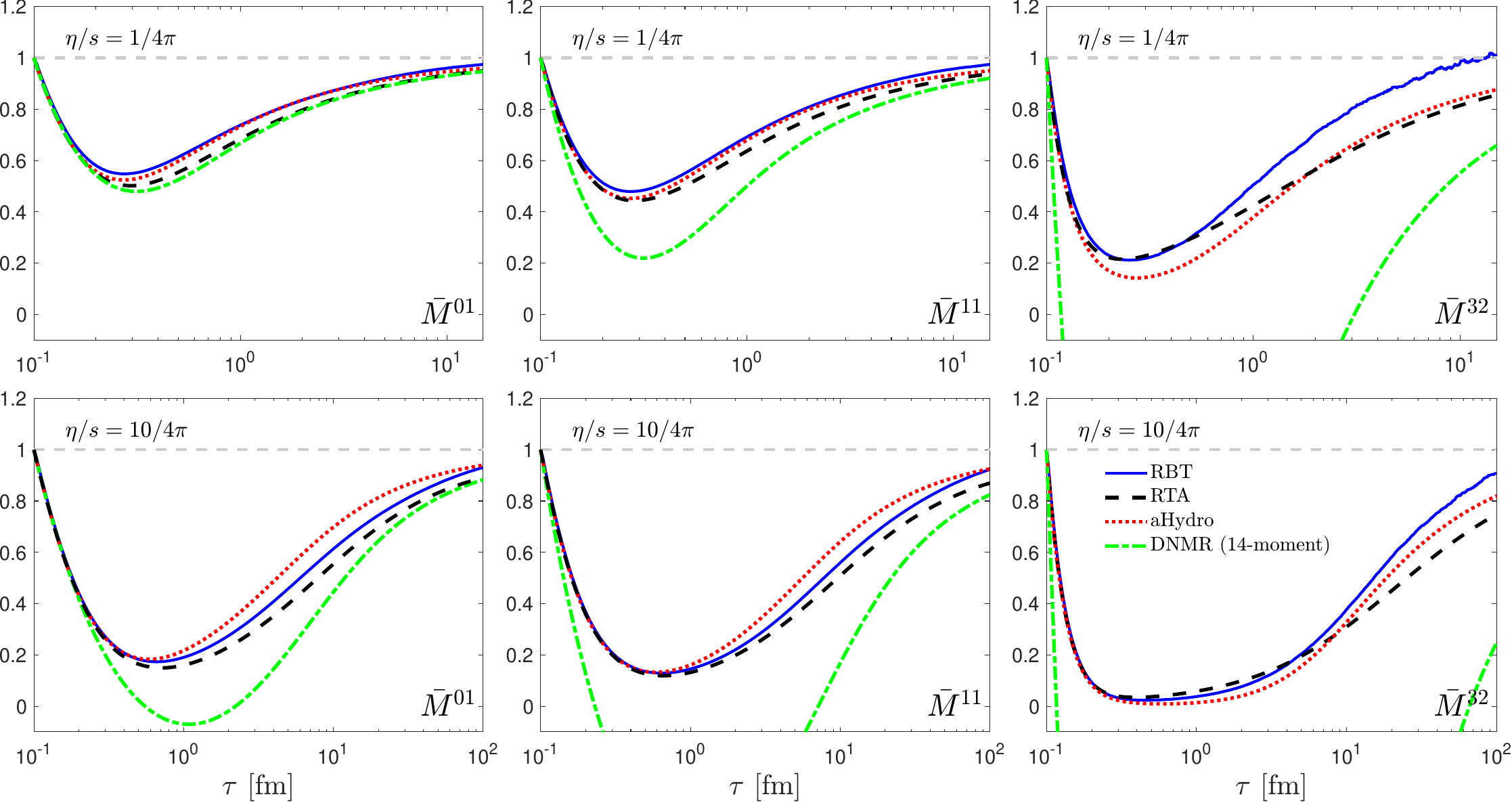}
    \caption{Normalised moments as a function of proper time $\tau$ at midrapidity, for $\tau_0= 0.1$ fm, $T_0 = 0.5$ GeV and $\xi_0=0$. Different curves correspond to different models: RBT (solid blue), Boltzmann RTA (dashed black), anisotropic hydrodynamics (dotted red) and DNMR with 14-moment approximation (dot-dashed green). The upper panels correspond to the case with $\eta/s=1/4\pi$, while the lower ones to $\eta/s=10/4\pi$.}
    \label{fig:Mmn_approaches}
\end{figure*}

In Fig.\,\ref{fig:Mmn_approaches} we show, for the same initial conditions in Fig.\,\ref{fig:freestream} and the two values of specific viscosity $4\pi\eta/s=1$ and $4\pi\eta/s=10$, the normalised moments $\overline M^{01},\overline M^{31}, \overline M^{32}$ obtained with the RBT approach in comparison to the results of different approaches.
The solid blue curve is the result of the simulation obtained with RBT, while the dashed black, the dotted red and the dot-dashed green lines are the results, respectively, of the RTA Boltzmann equation, anisotropic hydrodynamics and DNMR viscous hydrodynamics in the 14-moment approximation.
The agreement between full RBT and RTA depends on both the viscosity and the order of the moments. Indeed, for larger viscosity we get a better agreement between the two approaches for all the three moments, but for smaller viscosity the discrepancy increases, especially for $\overline M^{32}$.
A similar trend is present for the difference between aHydro and the full Boltzmann approach. However, it is interesting to notice that, especially for higher order moments, the equilibration is faster for RBT than for ahydro and RTA. As far as DNMR is concerned, it is not surprising that the agreement is better for lower order moments and breaks down for higher ones, since DNMR is constructed by taking just the first order moments of the Boltzmann distribution function, and therefore turns even negative for higher order ones. It would be interesting, however, to compare these results with more recent formulations of DNMR, in which one goes beyond the 14-moment approximation \cite{Denicol:2012cn, Wagner:2023joq}.\\

\begin{figure*}[th!]
	\centering
	\includegraphics[width=\textwidth]{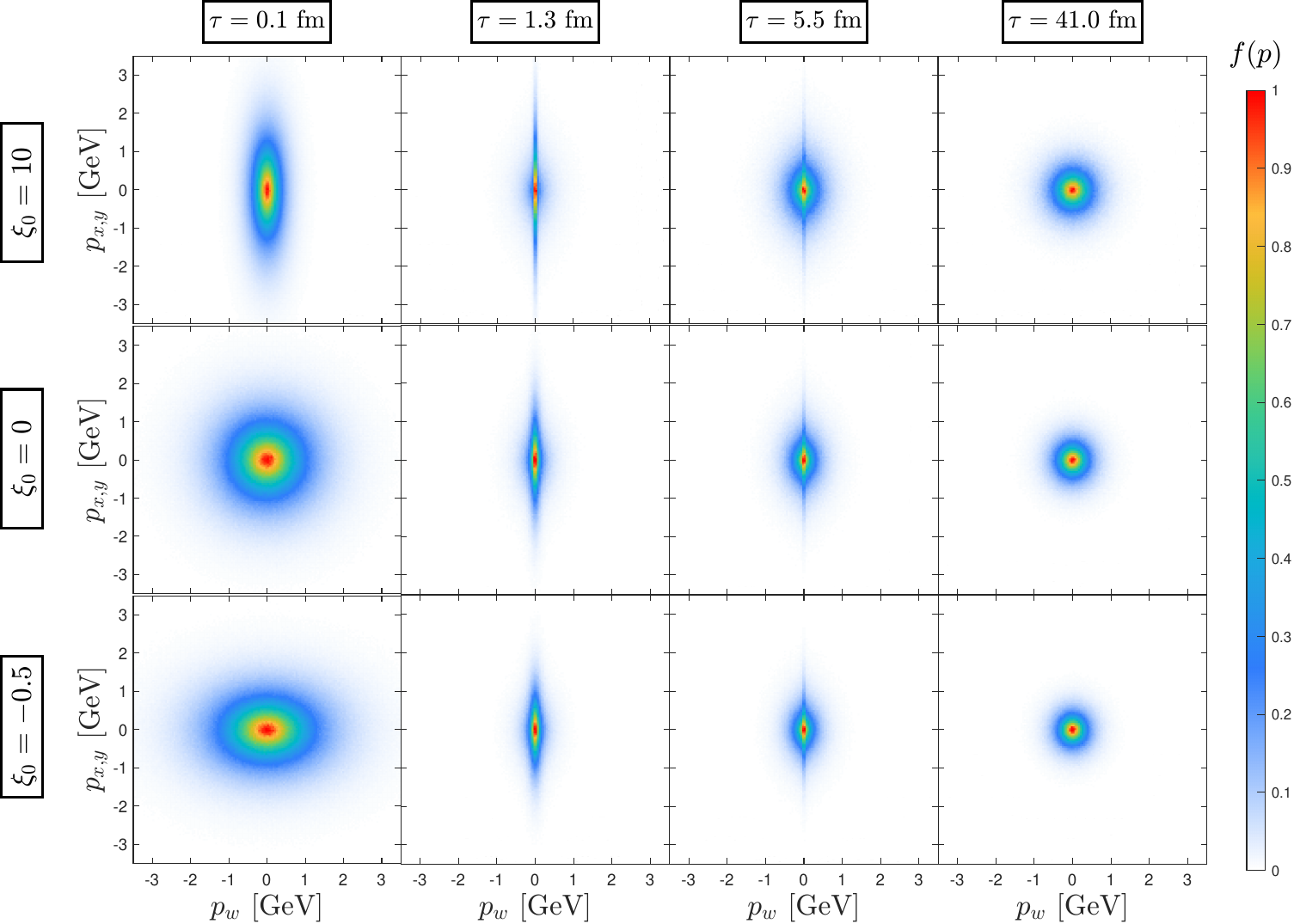}
	\caption{Contour plots of the normalised distribution function in the ($p_{x,y},p_w$) space for RBT computed at midrapidity. Different columns refer to different $\tau$ values; different rows to different initial anisotropies. We fix $T_0 = 0.5 $ GeV, $\tau_0= 0.1$ fm and $\eta/s = 10/4\pi$. }
	\label{fig:distr_contour}
\end{figure*}

In order to make clear the process of isotropisation of the system, we show in Fig.\,\ref{fig:distr_contour} the time evolution of the profile of the distribution function on the $p_w$-$p_T$ plane for {$\eta/s=10/4\pi$} and three different values of the initial anisotropy parameter: $\xi_0=[-0.5,\,0,\,10]$. We observe that, although the shape in momentum space is very different at the initial time of the simulation $\tau_0=0.1$ fm, after about {1.3 fm} the distributions become very similar and, in all three cases, elongated more along the $p_T$ than along the $p_w$ axis. This convergence of the anisotropic shape of the distribution function is driven by both the strong initial longitudinal expansion and the collisions. After this convergence the profiles continue to evolve in a similar way toward the isotropic limit.
The profile of the distribution function has been studied in Ref. \cite{Strickland:2018ayk}, pointing out that the distribution function contains two components: an anisotropic part which becomes more and more squeezed in longitudinal momentum as time evolves; a more isotropic piece which dominates at late times.
A similar distinction is clearly visible also in our simulations, especially by looking at the third column ($\tau = 5.5$ fm).

\begin{figure}[th!]
	\centering
	\includegraphics[width=.48\columnwidth]{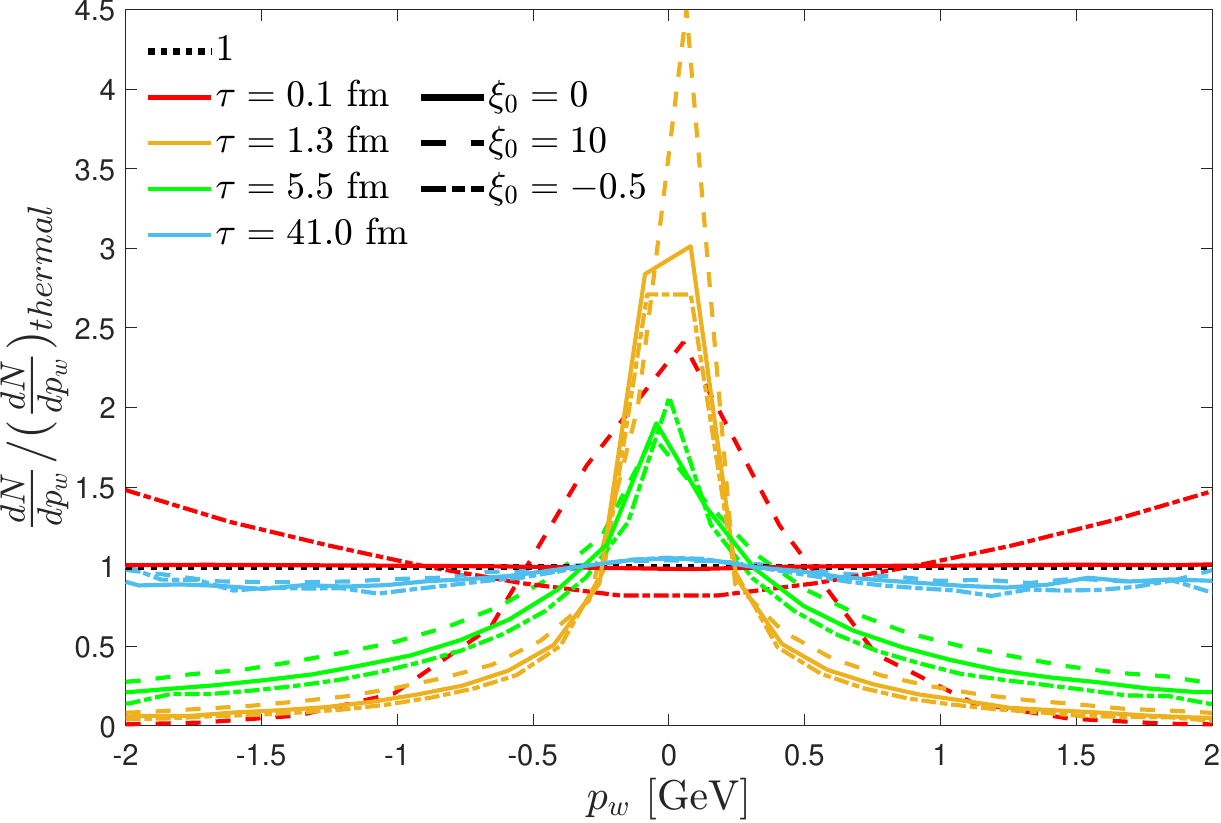}\hspace{10pt}\includegraphics[width=.48\columnwidth]{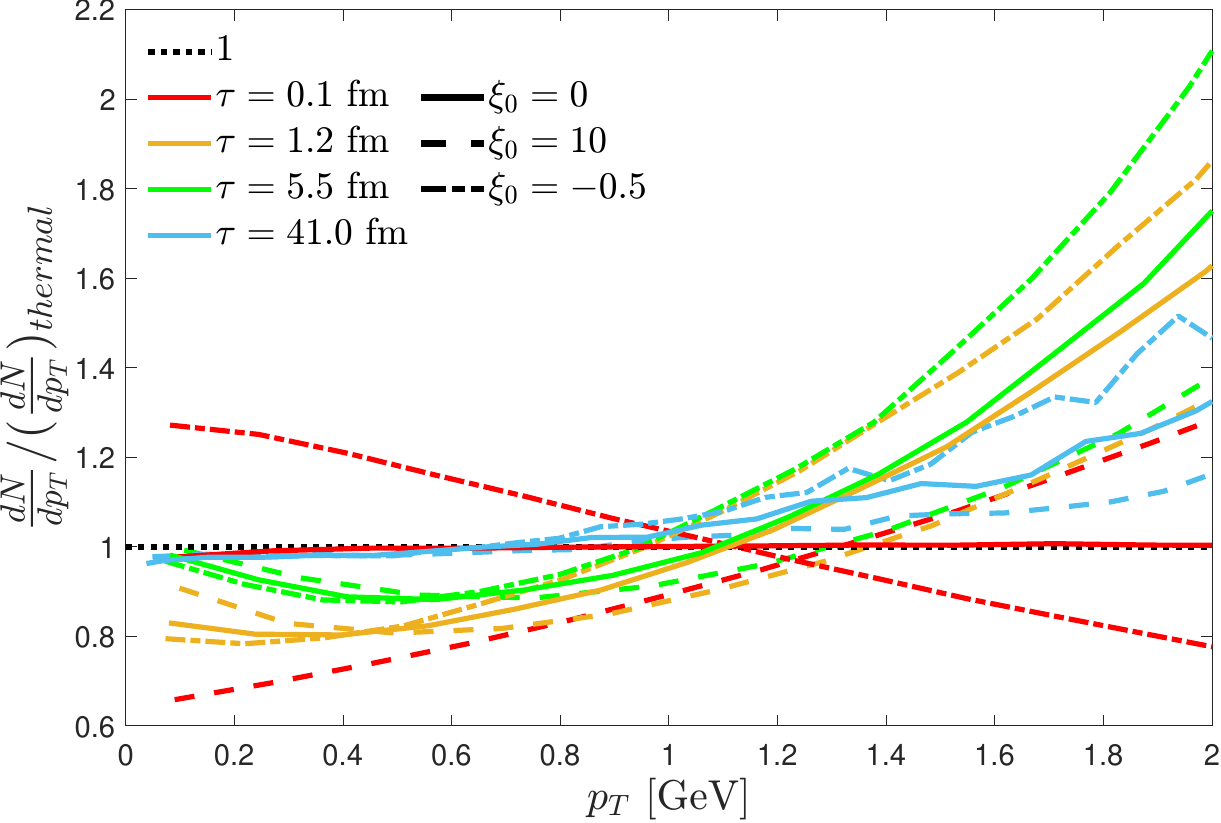}
	\caption{Left panel: ratio between the computed $dN/dp_w$ and the thermal one corresponding to the effective temperature $T$. Right panel: ratio between the computed $dN/dp_T$ and the thermal one. In both panels, different colors correspond to the same proper times shown in Fig.\,\ref{fig:distr_contour} and different styles to the same $\xi_0=[0,\,10,\,0.5]$ respectively for solid, dashed and dot-dashed lines. The initial conditions are those of Fig.\,\ref{fig:distr_contour}. }
	\label{fig:ratio_spectrum}
\end{figure}

While Fig.\,\ref{fig:distr_contour} gives us information about the isotropisation of the system, in order to investigate its thermalisation, we show in Fig.\,\ref{fig:ratio_spectrum} the time evolution of the ratio of the phase-space distribution function over the thermal Boltzmann distribution corresponding to the effective temperature of the system at the considered time. In particular, the left panel presents the distribution in longitudinal momentum {$dN/dp_w$} integrated over $p_T$ and the right panel is the distribution in transverse momentum {$dN/dp_T$} integrated over $p_w$, both computed at midrapidity {$|\eta_s|<0.75$}.
From the top plot, we see that at initial time $\tau_0=0.1$ fm the lines corresponding to different initial anisotropies are very different between each other. In particular, $\xi_0=0$ corresponds initially to a thermal distribution by construction. We observe that after 1.0 fm due to the collisional dynamics for $\xi_0=0$ the spectrum becomes more populated at high longitudinal momentum with respect to the thermal distribution, showing large deviations from 1. The other two curves $\xi_0=-0.5$ and $\xi_0=10$ are from the beginning far from equilibrium. As time evolves, the three simulations get closer to each other, suggesting the universal behaviour in the longitudinal momentum component of the phase-space distribution function, and show a similar trend in approaching the equilibrium limit 1.
From the bottom panel, we observe a similar behaviour in approaching the thermal distribution, as observed in $dN/dp_w$. The collisional dynamics affects the transverse momentum spectrum (both thermal and non-thermal) by moving particles from high to low $p_T$ and the rate of this shift in $p_T$ depends on the initial anisotropy of the distribution. From the plot we notice that in the case $\xi_0=10$ the shift of particles from high to low $p_T$ happens faster than for the other two cases; moreover, the approach toward equilibrium is slower for particles with high $p_T$.
	
\section{Forward attractors}
\label{subsec:forward_attractors_1D}

\begin{figure*}[t]
	\centering
	\includegraphics[width=\textwidth]{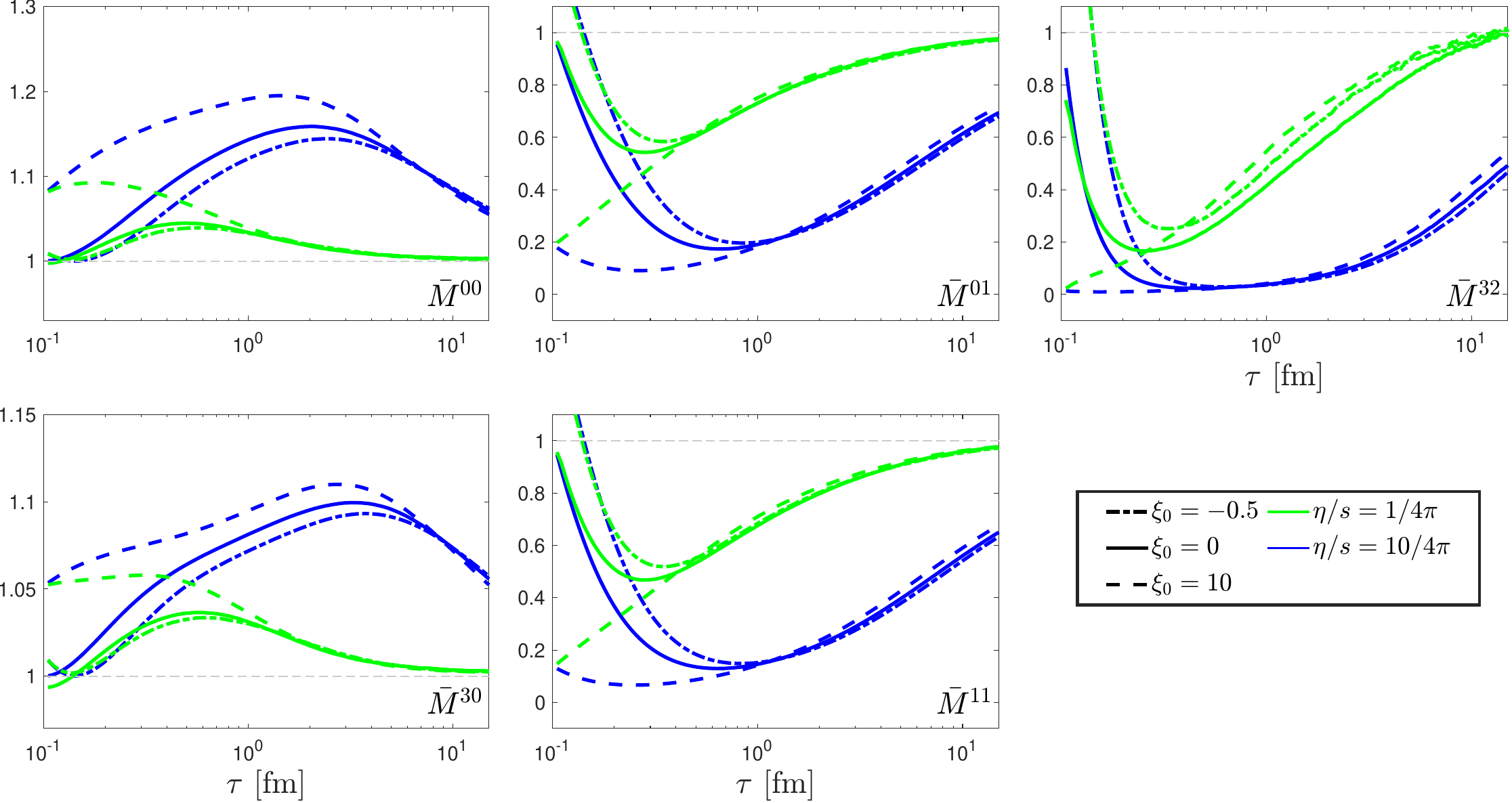}
	\caption{Normalised moments as a function of $\tau$ within the RBT approach. Different line styles correspond to four different values of initial anisotropy $\xi_0 = [-0.5,\,0,\,10,\, +\infty]$, highlighting the forward attractor behaviour. The two colours correspond to two different values of $\eta/s$.}
	\label{fig:forward_tau}
\end{figure*}

As seen in the previous paragraph, the full distribution function exhibits universality even when it is far from the complete equilibration: despite being known that the late-time shape has to be the same for every possible initial condition, it is non-trivial that the convergence towards a unique behaviour appears well before the thermalisation and isotropisation processes are complete. To look more in detail into this aspect, we analyse the appearance of the attractors in the momentum moments of the distribution function. In Figure\,\ref{fig:forward_tau} we plot the evolution of a set of moments for systems with the same  $\tau_0, T_0$, two values of $\eta/s$ and different initial anisotropy parameter $\xi_0$. Curves with the same $\eta/s$ (same colour) share the initial energy density and the interaction rate, and therefore are expected to thermalise in a similar way. In fact, one clearly sees that starting from a particular $\tau$, which depends both on the interaction measure $\eta/s$ and on the momentum order $m,n$, the equilibration process happens \emph{exactly in the same way}: the system loses memory of the initial conditions and exhibits an attractor, called \emph{forward attractor}. The unique curve is clearly visible for $n=1$ and $n=0$, while it is slightly less defined for $n,m=3,2$. This feature is easily explained by considering that the higher the order of the moments, the larger the momentum of the particles which strongly determine them, since higher powers of $p_w$ and $p_T$ appear in the integral; one could expect that for high $p$ particles the thermalisation is slower, since more collisions are necessary to equilibrate with the surrounding bulk, and thus they remain sensitive to the initial conditions for a longer time. One may also observe that, for $n=0$ moments, the proper time at which the attractor is reached is far larger than for the other cases ($\overline M^{10}$ and $\overline M^{20}$ are identically 1 due to the matching conditions). This peculiar behaviour can be accounted for by the two-component shape of the distribution function highlighted in Figure\,\ref{fig:distr_contour}. As outlined in \cite{Strickland:2018ayk} this is due to the fact that these moments are strongly sensitive to the region of the phase space where $p_w \sim 0$, which is exactly that populated by the squeezed component of the distribution function. As a consequence, these moments are forced to reach the attractor and to thermalise later.} This trend can be compared to what is shown in Fig.\,\ref{fig:ratio_spectrum}, where the universality at 1.2 fm is far more evident in particles with small $p_T$ with respect to the small $p_w$ ones: the peak at small $p_w$ in $dN/dp_w/ (dN/dp_w)_{thermal}$ is large ($\sim 2$) at $t=5.5$ fm when the correspondent ratio at small $p_T$ is $\sim 0.8$. \\
Finally, from Fig.\,\ref{fig:forward_tau} we observe that, in the case of smaller specific viscosity, the attractor is reached when the normalised moments are closer to 1 with respect to the case with larger $\eta/s$. For example $\overline{M}^{01}=\overline{P}_L=0.6$ for $\eta/s=1/4\pi$ and $\overline{M}^{01}=\overline{P}_L=0.23$ for $\eta/s=10/4\pi$.
This is due to the fact that for larger specific viscosity the effect of the strong longitudinal expansion dominates for longer time, so that the system has reached a higher degree of anisotropy when it start deviating from the free-streaming trend (see the minima in Fig.\,\ref{fig:freestream}). It is clear that the convergence of the normalised moments of systems with different anisotropies is reached earlier for smaller specific viscosity, because a smaller $\eta/s$ corresponds to a larger scattering rate, leading to a quicker loss of memory of the initial condition details. However, it is interesting to notice that despite there is a factor 10 between the two different specific viscosities $\eta/s$, there is a much smaller factor between the times at which attractors are reached. This suggests that the mechanism which brings the system to the attractor is not due only to the collisions, but there is also a strong contribution by the initial quasi-free longitudinal expansion. Indeed, by observing the location of the minimum and following the considerations done in the previous paragraph, one easily realise that it is exactly the initial expansion which leads the system towards the attractor and thus make it forget about its initial conditions, namely its initial anisotropy. This is particularly interesting also in view of more realistic studies: it is known that in uRHICs the first stage can be quite well described by the one-dimensional abrupt Bjorken flow, and therefore such an attractor may appear in the very beginning of the medium evolution and cancel physically interesting details of the initial energy deposition \cite{Giacalone:2019ldn}.

\section{Relaxation time}

As stated in the introduction, the emerging of a universal behaviour is always a matter of scales: systems with very different initial conditions can resemble each other once the short-living modes have decayed and the only surviving modes are characterised by a few scales which fully determine the remaining evolution. In particular, in the case of a one dimensional Bjorken expanding medium, there is only one natural time (or equivalently length) scale emerging: the average time between two successive collisions for each particle, or equivalently the mean free path. In this section we will mainly talk of \emph{average collision time per particle}, while when moving in 3+1D we will prefer \emph{mean free path}.\\
In the framework of the RBT this natural time scale can be computed as:
\begin{equation}\label{eq:taucoll}
	\tau_\text{coll}= \frac 12 \tonde{ \frac{1}{N} \frac{\Delta N_{\text{coll}} }{\Delta t} }^{-1},
\end{equation}
where $\Delta N_{\text{coll}}$ is the number of collisions occurring in a certain time interval $\Delta t$ in a given volume with $N$ test particles, and {the factor $1/2$ accounts for double counting}. The quantity $\tau_\text{coll}$ determines how quickly the system approaches equilibrium: the system equilibration, within a kinetic approach, is connected to the particle collisions, which have the role to isotropise and thermalise the system, inducing also the loss of information about the initial conditions.\\
It is natural to compare this time scale to the one characterising the models we are comparing to, which is the relaxation time $\tau_{eq}=5(\eta/s) /T$. In the context of RTA and in hydrodynamics (in the equations used the only dissipative quantity is $\pi$ and $\tau_\pi=\tau_{eq}$) the relaxation time is needed for the formulation of the equations: it governs the whole evolution of the medium and obviously its convergence towards equilibrium. 
More precisely, what is relevant for the system dynamics is defined as transport relaxation time, that for a system of massless particles interacting through an isotropic cross-section, is \cite{Plumari:2012ep}: 
\begin{equation}
	\tau_{tr} = \frac{1}{n\, \sigma_{tr}} = \dfrac{1}{\frac{2}{3}n\, \sigma_{tot}} = \frac 32 \tau_{coll},
\end{equation}
where we exploited for isotropic differential cross section:
\begin{equation}
    \sigma_{tr} = \int d\Omega\frac{d\sigma}{d\Omega}\sin^2\theta = \int_0^{2\pi}d\phi\int_{-1}^{1}d(\cos\theta) \frac{\sigma_{tot}}{4\pi}(1-\cos^2\theta)=\frac{2}{3}\sigma_{tot}
\end{equation}
Hence one can naturally define the RBT relaxation time:
\begin{equation}\label{eq:tau_RBT}
	\tau_{eq}^{\text{RBT}} = \tau_{tr} = \frac 32 \tau_{coll}.
\end{equation}
For a system in chemical equilibrium the relation between entropy density and particle density is given by $s=4 n$. From the first-order Chapman-Enskog expansion $\eta = 1.2\, T/\sigma$, therefore one finds $\tau_{eq}^{RBT}=\tau_{tr}=\tau_{eq}^{RTA}$. In the case of RBT, as explained in Chapter\,\ref{chap:RBT}, we have to take into account the fugacity $\Gamma$ of the system, due to the particle number conservation. Therefore, the relation between entropy density and particle density becomes $s = n\, (4-\log \Gamma)$. This means that, in order to have a consistent comparison, $\tau_{eq}^{RTA}$ must be corrected by a factor $(1 - \log \Gamma /4)$. In principle, this definition also gives a constraint on the fugacity values to ensure the positiveness of entropy: $\Gamma<e^4\sim 55$. However, this is far from the usual range of fugacity, which, since the system is particle-conserving, goes mainly to values smaller than 1.
\begin{figure}[t]
	\centering
	\includegraphics[width=.48\columnwidth]{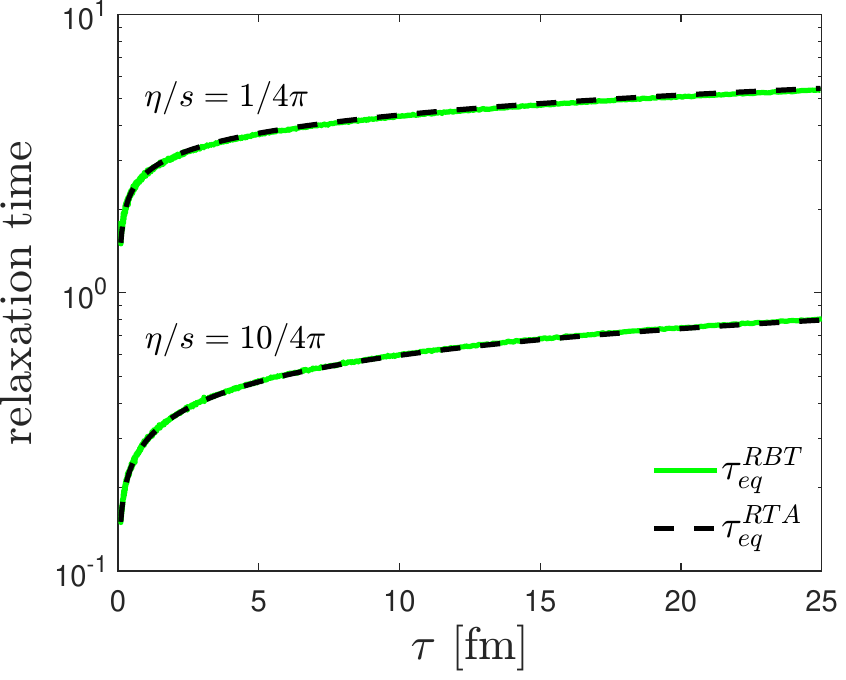}\includegraphics[width=.48\columnwidth]{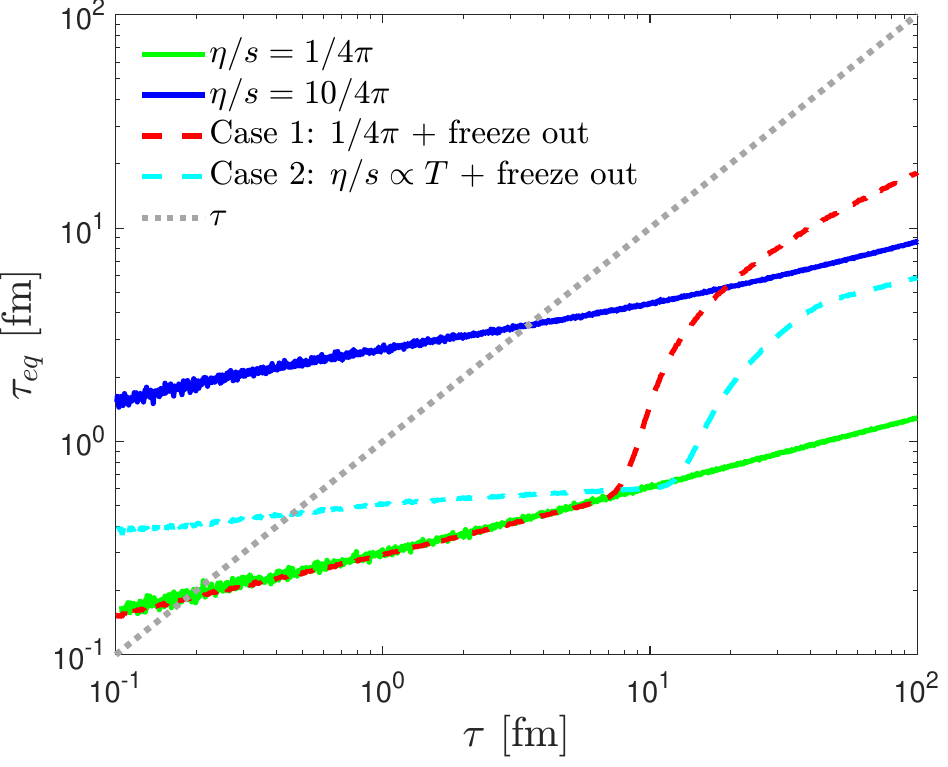}
	\caption{Left panel: Comparison for two different values of specific viscosity between the relaxation time $\tau_{eq}^{\text{RBT}}$ (solid green line) according to Eq.\,\eqref{eq:tau_RBT} and $\tau_{eq}^{\text{RTA}}=5(\eta/s)/T$ (dashed black line) where $T(\tau)$ is the one calculated in the simulation. The initial conditions are the same used in Fig.\,\ref{fig:freestream}. Right panel: Comparison among relaxation times  $\tau_{eq}^{\text{RBT}}$ in the RBT approach for different values of $\eta/s$: $1/4\pi$ (green line), $10/4\pi$ (blue line), and, referring to Sec.\,\ref{subsec:Tdep_visco}, Case 1 (red dashed line) and Case 2 (cyan dashed line).}
	\label{fig:tau_eq_RBT}
\end{figure}
In the left panel of Fig.\,\ref{fig:tau_eq_RBT}, we show the comparison of the $\tau_{eq}$ adopted in RTA and hydrodynamics with the one used in our approach, suitably modified to remove the effect of fugacity. There is a full agreement during the whole time evolution for the range of specific viscosity explored, as evident from the lines with $4\pi\eta/s=1$ and $4\pi\eta/s=10$ shown in the plot. In the right panel of the same figure, we plot $\tau_{eq}^{RBT}$ with respect to $\tau$ for $4\pi\eta/s=\{1,10\}$ and for $\eta/s(T)$ as parametrized in Sec.\,\ref{subsec:Tdep_visco}, as well as the bisector of the $\tau$-$\tau_{eq}$ plane. By comparing this plot and Fig.\,\ref{fig:freestream} one can see that when the collisions start to dominate ($\tau\lesssim \tau_{eq}$), the curves pass by the minima. This is quite patent by looking at the $\eta/s=1/4\pi$ case, while the case with $\eta/s = 10/4\pi$ is slightly more difficult to be read, since the position of the minimum and even the extension of the free-streaming-dominated region strongly depend on the moment order. 

\section{Pull-back attractors}
Once the main time scale of the system is identified, it is legitimate to wonder whether it can play a role in the emerging of attractors: this is widely done in literature, where  the system evolution, for instance the moments of the distribution function, is usually plotted with respect to the scaled time $\tau/\tau_{eq}$ \cite{Jankowski:2023fdz, Kurkela:2018vqr, Giacalone:2019ldn}. Also in the framework of the RBT it is possible to highlight the universal behaviour of attractors rescaling the proper time by $\tau_{eq}^{RBT}$, which is the timescale naturally emerging in our kinetic approach.\\
\begin{figure*}[t]
	\centering
	\includegraphics[width=\textwidth]{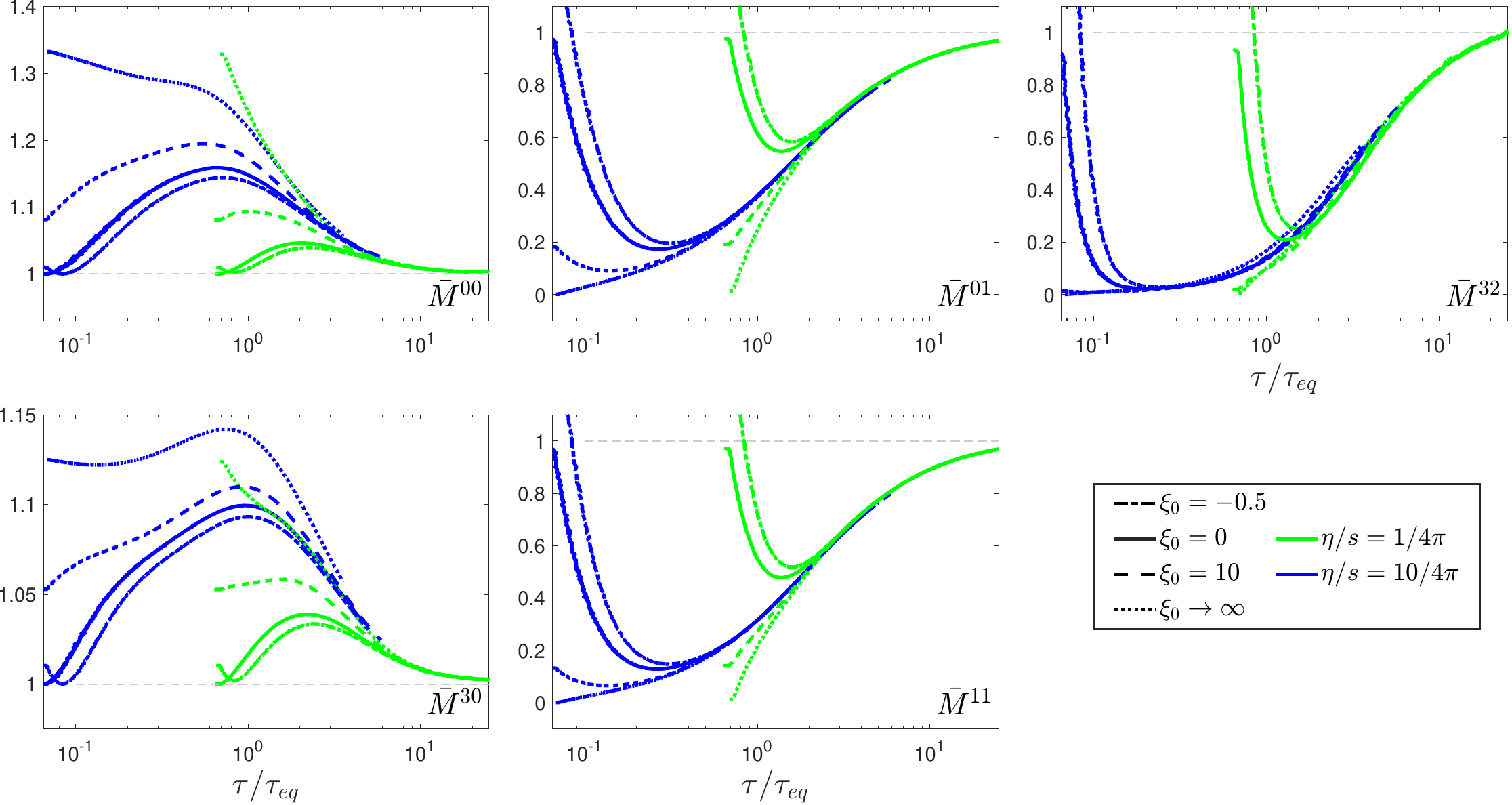}
	\caption{Normalised moments as a function of $\tau/\tau_{eq}$ within the RBT approach. Different line styles correspond to four different values of initial anisotropy $\xi_0 = [-0.5,\,0,\,10,\, +\infty]$, highlighting the forward attractor behaviour. The two colors correspond to two different values of $\eta/s$. We fix $T_0= 0.5$ GeV and $\tau_0 = 0.1$ fm.}
	\label{fig:forward}
\end{figure*}
To begin with, in Figure\,\ref{fig:forward} we show the same curves of Figure\,\ref{fig:forward_tau} plotted with respect to the scaled proper time. The curves obviously begin from a different $(\tau/\tau_{eq})_0$, but now a unique curve for the attractor is present. This convergence is usually referred to as \emph{pull-back attractor}, since in principle it is possible to go to smaller and smaller values of $(\tau/\tau_{eq})_0$ and reconstruct the `full' attractor curve. One should keep in mind that the origin of this scaling is different than the forward attractor introduced in the previous paragraph: here the major  role is played by the relaxation time which serves a renormalisation time scale. If the effect of this scale is removed via the aforementioned rescaling, \emph{every possible system} follow, after a short transient, the same attractor curve.

\begin{figure*}[t]
	\centering
	\includegraphics[width=1\linewidth]{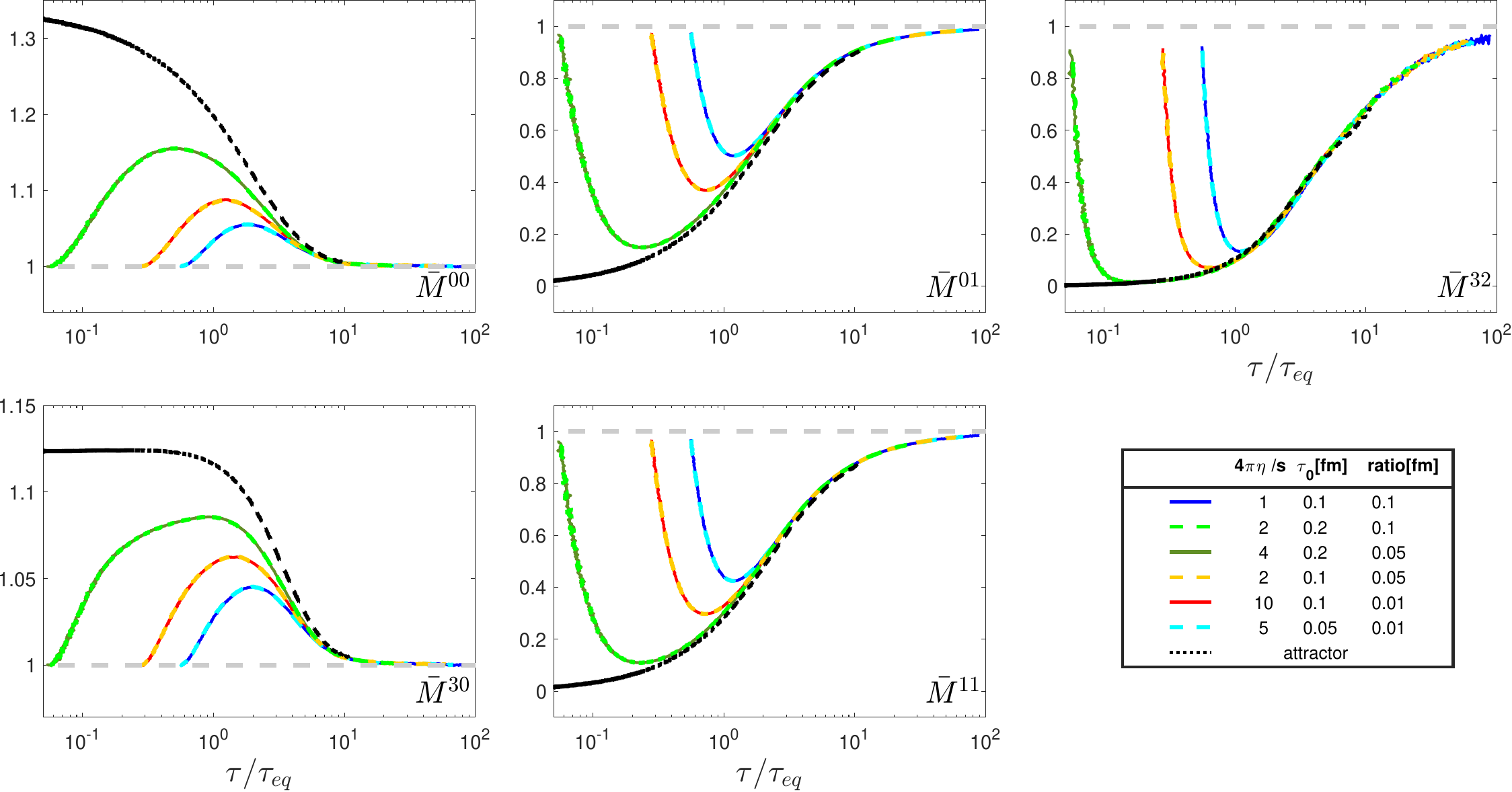}
	\caption{Normalised moments as a function of $\tau/\tau_{eq}$ within the RBT approach. Different curves correspond to three different values of the ratio $\tau_0/(4\pi\eta/s) = [0.1,\,0.05,\,0.01]$ fm, highlighting the pull-back attractor behaviour. The black dotted line is the attractor curve in RBT as described in the text. The initial temperature $T_0 = 0.5$ GeV is fixed.}
	\label{fig:pullback}
\end{figure*}

To investigate further this point, in Fig.\,\ref{fig:pullback} we show the momentum moments of the distribution function as a function of the scaled time $\tau/\tau_{eq}$ obtained with the RBT approach with the same initial temperature $T_0=0.5$ GeV and initial zero anisotropy ($\xi_0=0$). The different curves are obtained by changing the value of $\tau_0$ and $\eta/s$. We notice, however, that simulations with different $\tau_0$ and $\eta/s$ but with the same ratio $\tau_0/(4\pi\eta/s)$ give curves which lie on top of each other, as visible in the figure by looking for example at the {solid light green} and {dashed dark green} lines with have both $\tau_0/(4\pi\eta/s)=0.01$ (and similarly for the other two cases). We kept $T_0$ fixed for the sake of simplicity: one can check that it is only the ratio $(\tau/\tau_{eq})_0$ which matters when plotting curves as functions of the scaled time. By removing the effect due to the initial anisotropy, one can understand more clearly that the role of the $\tau_{eq}$ is that of renormalising different systems; $\tau_0$, instead, is the only scale which characterises the initial extension of the medium, since there are no dimensions beyond the proper time.
This scaling in viscous and anisotropic hydrodynamics  is clearly visible when the system of ODEs is written as a single equation for $\varphi = \varphi(\pi/e) = 2/3 + \pi/(3e)$ as a function of $w = \tau/\tau_{eq}$ \cite{Strickland:2017kux}. Namely, in the case of DNMR theory, one gets the equation:
\begin{equation}
	w\phi \phi' + \tonde{w - \frac{34}{7}}\phi - \frac{48}{63} - \frac{2w}{3}=0
\end{equation}
A similar equation is obtained in the context of anisotropic hydro.
Eventually, the normalised moments in Eqs.\,(\ref{eq:vhydro_moments}, \ref{eq:ahydro_moments}) depend only on the function $(\pi/e) (w)$, which is dependent only on the initial value $w_0 = \tau_0 T_0 /(\eta/s)$. Thus, if one changes two or more of these parameters without changing $w_0$, this does not affect the solutions $\overline M^{nm}(w)$, since the same ODE is solved with the same initial condition $\phi(w_0)=\phi_0$. The same can be proved to be valid in the context of the RTA Boltzmann Equation. \\

Referring now to the discussion about the forward attractors, we notice that the larger the specific viscosity, the smaller the scaled time needed to reach the attractor curve: for $\eta/s=1/4\pi$ ({green lines}) the simulations converge at $\tau_\text{attr} \approx 1.5\,\tau_{eq}$, whereas for $\eta/s =10/4\pi$ ({blue lines}) the attractor is reached at about $\tau_\text{attr} \approx 0.2\, \tau_{eq}$. \\

\begin{figure*}[t]
	\centering
	\includegraphics[width=\textwidth]{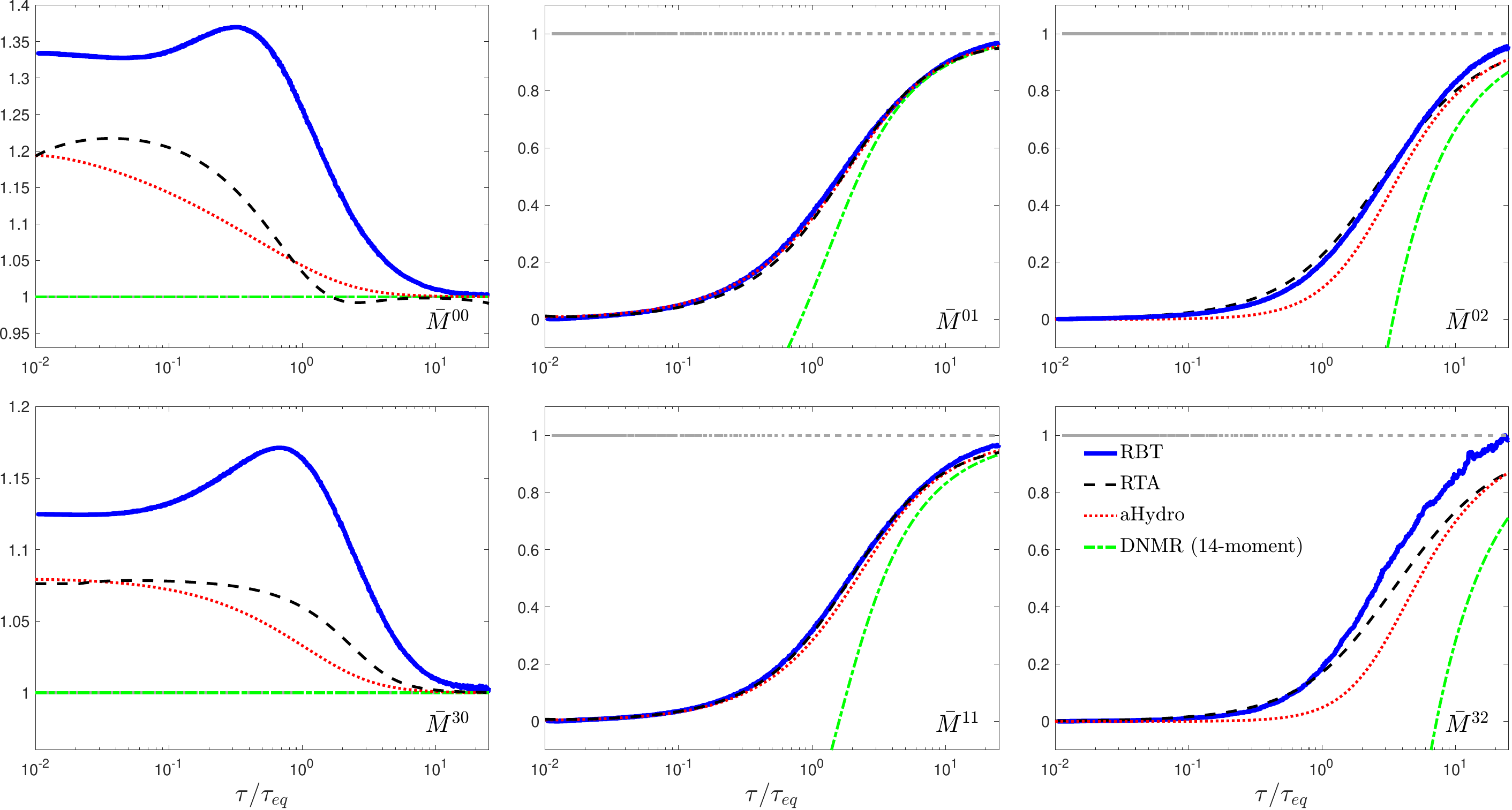}
	\caption{Attractor curves as function of the scaled time $\tau/\tau_{eq}$ for different models as described in the text: RBT (solid blue), RTA (dashed black), aHydro (dotted red) and DNMR with the 14-moment approximation (dot-dashed green).}
	\label{fig:attractor_models}
\end{figure*}

In Fig.\,\ref{fig:attractor_models} we show the comparison of the attractor curves obtained with the RBT approach ({solid blue line}) with those found in RTA ({dashed black line}), aHydro ({dotted red line}) and DNMR theory ({dot-dashed green line}). In our simulation the prescription used to obtain the attractor curve is the one used in several approaches \cite{Blaizot:2017ucy, Romatschke:2017vte,Heller:2015dha}, which is equivalent to initialise the distribution function with an initial zero longitudinal pressure and exploit the meaning of pull-back attractor by taking the limit $(\tau/\tau_{eq})_0\to 0$. In our RBT approach the initial vanishing $P_L=0$ is obtained by imposing the condition $Y=\eta_s$ (momentum rapidity equal to space-time rapidity), which implies $p_w=0$ for every test particle and, consequently, $\mathcal M^{nm}=0$ with $m>0$, including $M^{01}=P_L=0$. 
In the RTA case, the attractor curve is obtained by taking as the initial condition $\xi_0 \to \infty$ and the initial proper time $\tau_0\to 0$ \cite{Strickland:2018ayk}. These two constraints are equivalent to those we impose: the condition $\xi\to \infty$ requires $p_w=0$ in the distribution function and therefore $P_L=0$ as an initial condition; concerning the second one, in our prescription we require $(\tau/\tau_{eq})_0\to 0$, as we have already mentioned that the latter is the dynamic-governing quantity. 
{In the hydrodynamic calculations \cite{Strickland:2017kux} the attractor curve is obtained by taking the limit $w\to 0$ for the previously defined function $\varphi(w)$ and imposing $\varphi$ to be finite. In aHydro, the condition is fulfilled when $\varphi=3/4$, which implies $P_L=0$ \cite{Strickland:2017kux}}.\\
We observe that the full Boltzmann, RTA and aHydro approaches give similar results for the moments with one power of $p_w$; for higher order moments we see a deviation between {RBT} and RTA which gets larger with the increasing order, up to 10-15\% for $\overline M^{32}$  for $\tau/\tau_{eq}>10$. The discrepancy is still more important when comparing {RBT} and aHydro for $\tau/\tau_{eq}<10$, that is also a region of disagreement between aHydro and RTA. The discrepancy is quite significant for higher order moments where however the RBT approach is by construction more appropriate than hydrodynamics. For instance in RBT $\overline M^{32}$ reaches the near equilibrium value of 0.8 at $\tau/\tau_{eq}\sim 5$, while in aHydro this occurs only at $\tau/\tau_{eq} \sim 10$. 
For completeness we added also the curves corresponding to calculations with DNMR theory.
We see that all the normalised moments in the DNMR theory, as well as in the other viscous hydrodynamics models, become negative in a certain range, therefore losing their physical meaning. This is due to the fact that in the initial stage the system is dominated by the free-streaming which drives it in a regime where the near-equilibrium assumption of DNMR is no longer valid {\cite{Strickland:2018ayk}}. {Furthermore, the $\overline M^{n0}$ in DNMR are identically 1 for construction. As far as these modes are concerned, it is quite relevant that they show the strongest disagreement between the different frameworks. As outlined before, these moments explore the $p_w\sim0$ region of the phase space (the squeezed component of Fig.\,\ref{fig:distr_contour}) which is heavily sensitive to the free streaming expansion. Differently than $\overline{M}^{n0}$ with $n>0$, these moments thermalise later in RBT than in RTA or aHydro and even converge to quite different attractor curves. This suggests that, although the isotropic component of the distribution function is described by these approaches similarly than in RBT, the same cannot be said about the squeezed one, which is the most far-from-equilibrium component of the evolving distribution function.} 
{In order to account for this two-component picture an extension of aHydro has been introduced in \cite{Alalawi:2020zbx}, in which the distribution function ansatz explicitly has a free-streaming and an equilibrating component, as we report in Fig.\,\ref{fig:distr_contour}. This of course allows a better agreement with RTA especially for $\overline M^{n0}$.}\\

\section{Realistic temperature-dependent
 \texorpdfstring{$\eta/s$}{eta/s}}\label{subsec:Tdep_visco}

The studies on attractors in the literature have been performed assuming a constant specific viscosity $\eta/s$. However, lattice QCD calculations and phenomenological models show that $\eta/s$ should depend on the temperature and is expected to exhibit a minimum close to the critical temperature \cite{Csernai:2006zz, Plumari:2013bga, Yang:2022ixy}. In this section, we extend the previous analysis performed at fixed $\eta/s$, by studying the role of its temperature-dependence, which cannot be ignored in a more realistic description of hot QCD matter.

\begin{figure}[t]
    \centering
    \includegraphics[width=.6\columnwidth]{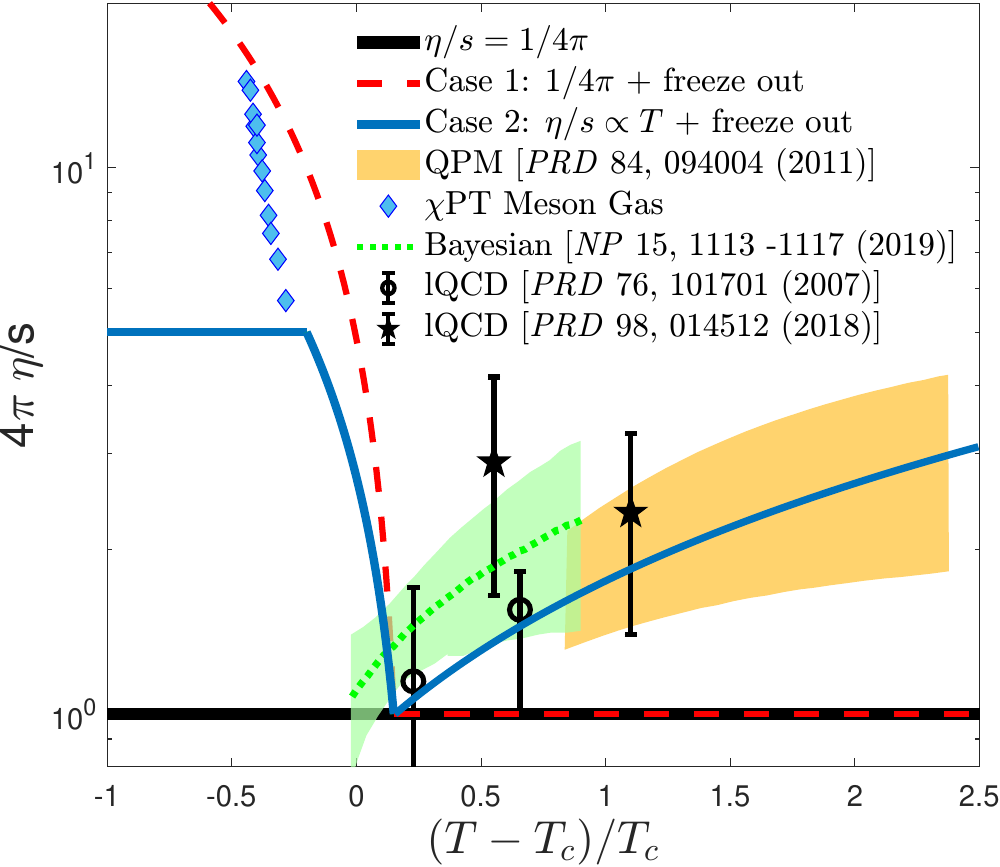}
    \caption{Different temperature-dependent $\eta/s$ parametrisations. Constant $\eta/s=1/4\pi$ (solid black), Case 1 (red dashed) and Case 2 (blue solid) are the ones used in our simulations. Case 1 corresponds to constant $\eta/s=1/4\pi$ for higher temperature and linearly rising at lower temperature, to simulate an almost sudden freeze-out; Case 2 is a more realistic parametrization with a minimum close to $T_C$.}
    \label{fig:etas_temp}
\end{figure}
\begin{figure*}[th!]
    \centering
    \includegraphics[width=\textwidth]{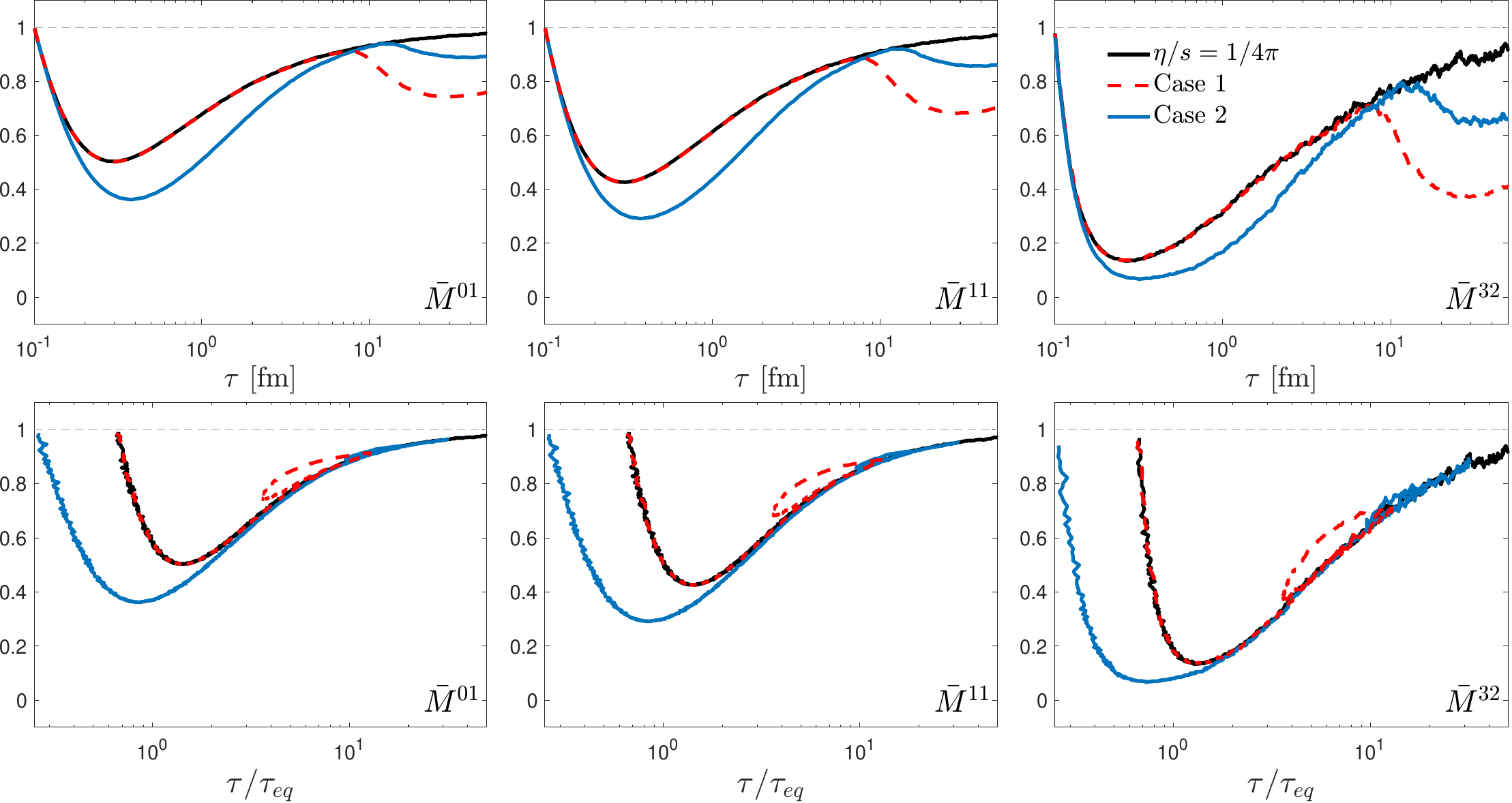}
    \caption{Upper panels: normalised moments as function of $\tau$ at midrapidity and for three different $\eta/s(T)$; colors are the same of Fig.\,\ref{fig:etas_temp}. Lower panels: normalised moments as function of the scaled time $\tau/\tau_{eq}$. Initial conditions are $T_0 = 0.3$ GeV, $\xi_0=0$. }
    \label{fig:moments_etas_temp}
\end{figure*}

We consider various parametrisation of $\eta/s (T)$, as shown in Fig.\,\ref{fig:etas_temp} in comparison to estimates from lattice QCD \cite{Meyer:2007ic, Borsanyi:2018srz} calculations, chiral perturbation theory ($\chi$PT) \cite{Chen:2007xe}, hadron-resonance gas models and a general Bayesian estimation \cite{Bernhard:2019bmu}. We have considered an extreme case (Case 1) where the specific shear viscosity is constant for $T>T_c$ and changes abruptly at the critical temperature, increasing indefinitely towards smaller temperatures, and a more realistic case (Case 2) with a minimum close to the critical temperature as suggested by combining information from lattice QCD and estimations for the hadronic phase.

In the upper panels of Fig.\,\ref{fig:moments_etas_temp} we show the time dependence of the moments of the distribution function for the different parametrisations of $\eta/s(T)$.
While for fixed specific viscosity the moments increase smoothly with increasing time after the minimum, as seen in detail in the previous sections, this behaviour changes when the $\eta/s$ increases at lower temperature. Indeed, at $\tau \approx 2-3$ fm the moments start to have a non-monotonic trend, especially for the Case 1 of $\eta/s(T)$.
This can be understood considering that in the kinetic approach, a change in $\eta/s$ corresponds to a change in the scattering cross section; since collisions drive the system towards the equilibration, a lowering in the scattering cross section allows a departure from the monotonic increase of the moments towards the equilibrium values after the initial free streaming dominance.\\

In the lower panels of Fig.\,\ref{fig:moments_etas_temp} the moments are presented as a function of scaled time $\tau/\tau_{eq}$.
In this case we recover the universal scaling at large $\tau/\tau_{eq}$, so that the curves for temperature-dependent $\eta/s$ lie on top of those with fixed specific viscosity after reaching the attractor.
Nevertheless, at intermediate values of the scaled time, we notice an interesting departure from the universality. At $\tau/\tau_{eq}\sim 2$--$3$ a small loop is present in the moments calculated with $\eta/s$ that increases at lower temperatures right below the critical value. This region in scaled time corresponds to the temperature region in which $\eta/s$ increases so that $\tau/\tau_{eq}$ is no more a monotonic function of the proper time. {Since, as we know from RTA, $\tau_{eq}\propto (\eta/s) /T$, if $\eta/s$ is an increasing function of the proper time while $T$ decreases, there could be some intervals in which $\tau/\tau_{eq} (\tau)$ is no more monotonic, but instead shows a maximum and a minimum. This can be read out from Fig.\,\ref{fig:tau_eq_RBT} noticing that the difference between $\tau_{eq}$ and $\tau$ increases and decreases again with time. Specifically for the cases shown in Figure\,\ref{fig:moments_etas_temp}, one can verify by looking at Fig.\,\ref{fig:tau_eq_RBT} that the non-monotonic behaviour, i.e. the loop in the plots, begin exactly when $\eta/s (T)$ starts increasing. This of course depends on the specific dependence of $T(\tau)$ and $\eta/s (T)$: the steeper the $\eta/s (T)$ curve, the more pronounced the loop. Physically, what happens is that an increase of the specific shear viscosity makes the system recede again from equilibrium, which reflects in a temporary decreasing of the normalised moments and results in a minimum. More quantitively, the rise of $\eta/s(T)$ in the hadronic phase ($T<T_C$) implies a very fast increase from $\tau_{eq} \simeq  0.5$ fm at $\tau < 10$ fm up to an order of magnitude, $\tau_{eq} \simeq  5$ fm,  at $\tau \simeq 15$ fm (Fig.\,\ref{fig:tau_eq_RBT}).
Therefore, as a function of temperature one can notice that $\eta/s(T)$ induces a receding from full thermalisation at $\tau \sim \, 10-20\rm \,fm$, which are comparable to those where the freeze-out hypersurface in $AA$ collisions lies. Indeed, in the upper panels of Fig.\,\ref{fig:etas_temp} one can see that for times typical of QGP lifetime the system can remain significantly out of equilibrium, especially in case of a quite rapid increase of $\eta/s(T)$ for $T<T_c$: the deviation is about 10-15\% for Case 1 and 30\% for Case 2.  }
These findings support the idea that an increase of the specific viscosity in the low-temperature region, as expected from lattice QCD and phenomenological calculations, determines a partial breaking of the universal attractor behaviour. Obviously, only a realistic 3+1D simulation can evaluate if such a second receding from equilibrium is actually expected to occur: at such time scales, indeed, also the transverse expansion is playing a major role, and thus there are different competing mechanisms determining the medium evolution. With this goal in mind, the Boltzmann approach can be particularly useful, since the large $\eta/s$ regime is in principle far from the hydrodynamic realm of applicability, since for instance the Knudsen number cannot be assumed to be a small quantity.\\
Finally, the impact of a temperature-dependent $\eta/s$ is expected to be particular relevant in small systems, such as $pA$ and $pp$, in which the space-time regions with a temperature close (or even lower) to the critical temperature are widely extended and play a crucial role starting from much smaller time scales, typically of the order of the transverse size of the initial profile.

\section{Attractors with breaking of boost invariance}
\label{subsec:no_boostinv}

The evolution equations of hydrodynamic and RTA used both in our work (see Chap.\,\ref{chap:kin_and_hydro}) and in previous studies on attractors are boost-invariant.
The RBT approach, however, is not inherently boost-invariant, hence we have the advantage to easily explore the impact of the breaking of boost invariance. Firstly, it is not granted that a boost-invariant initial condition leads through the dynamical evolution to boost-invariant final phase space distribution functions and thermodynamic quantities, such as temperature ($T$), density ($n$) and energy density ($e$).

 \begin{figure}[t!]
     \centering
     \includegraphics[width=.5\columnwidth]{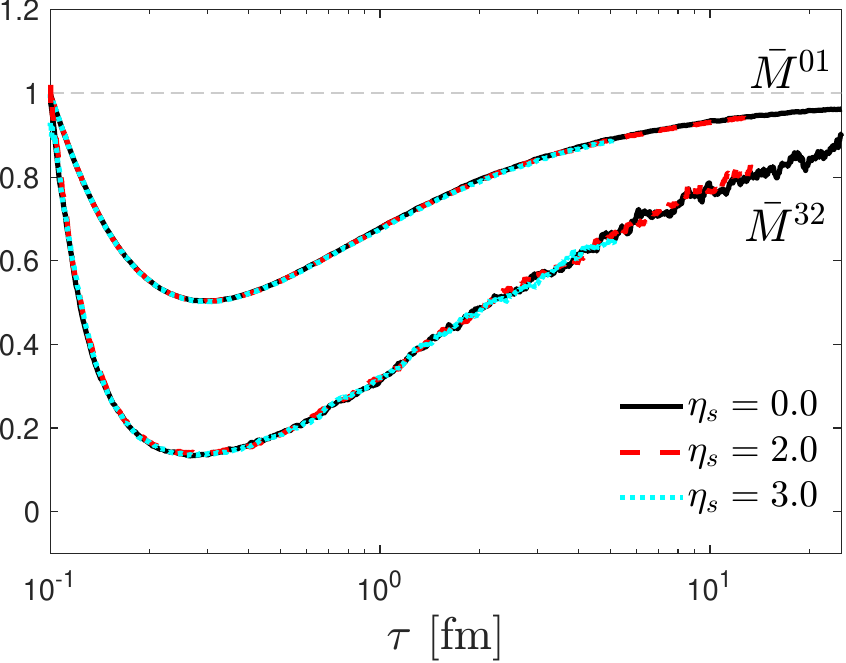}
     \caption{Normalised moments at different rapidities in the case of boost-invariant conditions. Different line styles correspond to: $\eta_s = 0.0$ (black solid), 2.0 (red dashed), 3.0 (cyan dotted). Normalised moments computed at different rapidities perfectly overlap.}
     \label{fig:checking_boost}
 \end{figure}
In the following, we present the results on the attractor behaviour obtained, first in the  `artificial' boost-invariant case used in the previous sections and then in a case where the boost invariance is explicitly broken during the dynamics. In the following, all the results are obtained within a slice of width $\Delta \eta_s=0.4$, centered around the reported value.

\begin{figure}
	\centering
	\includegraphics[width=0.7\linewidth]{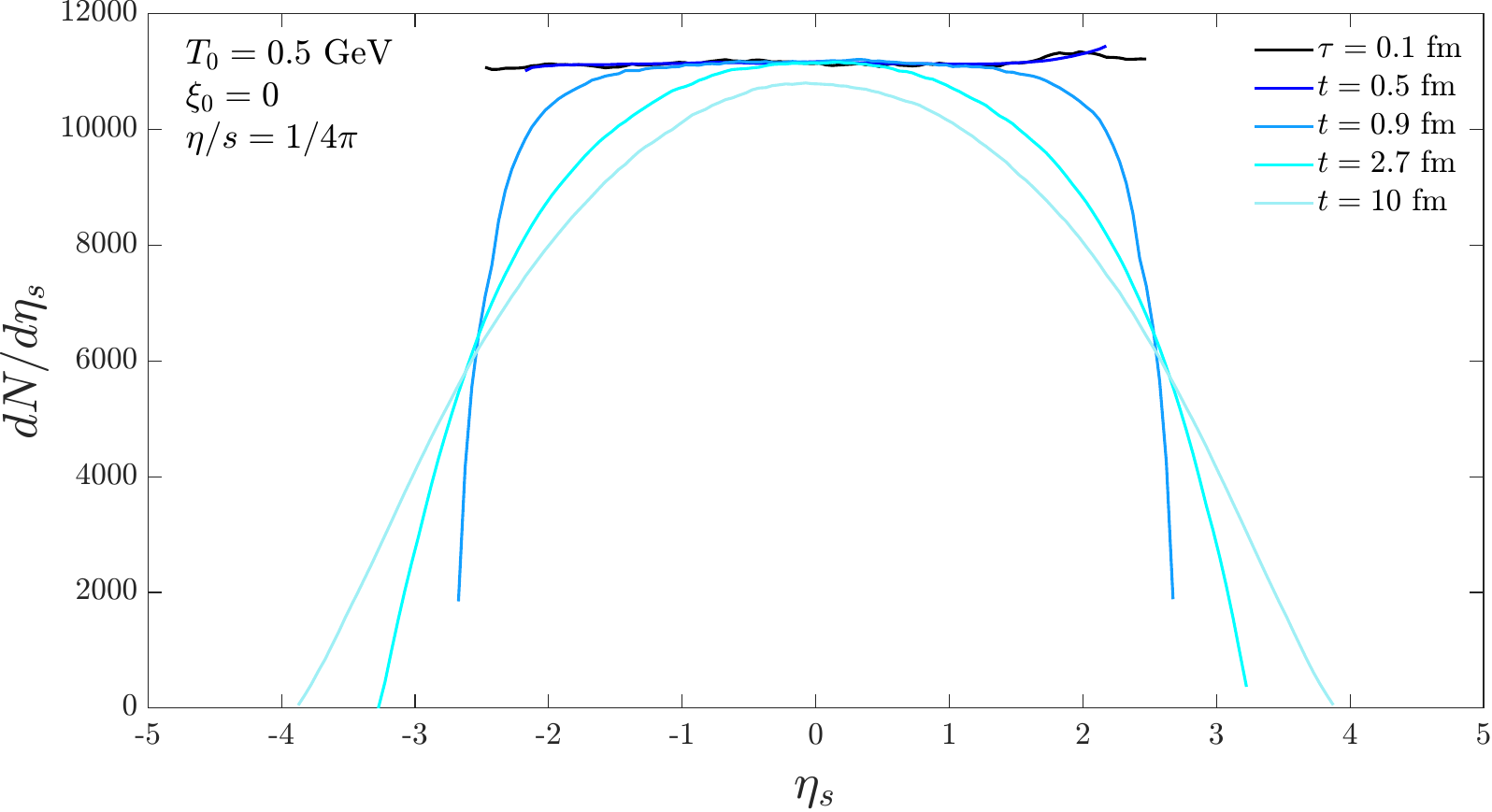}
	\caption{Longitudinal distribution $dN/d\eta_s$ for test particles with a finite initial distribution in $\eta_s$}
	\label{fig:non_boost_distribution}
\end{figure}

 \begin{figure*}[th!]
     \centering 
     \includegraphics[width=.9\textwidth]{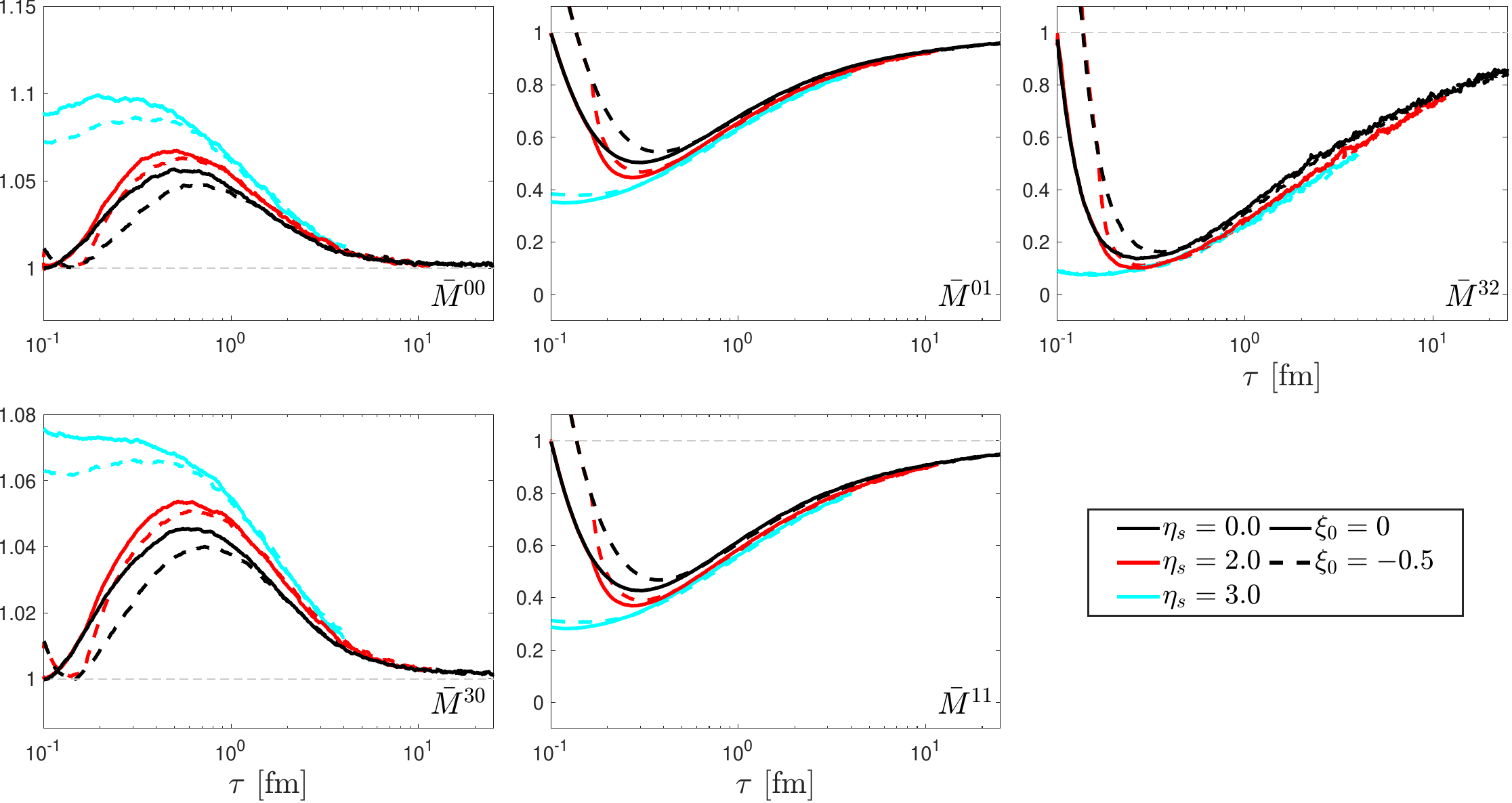}
      \caption{Normalised moments as a function of $\tau$ within the RBT approach. Different styles correspond to two different values of initial anisotropy $\xi_0 = [-0.5,\,0]$, highlighting the forward attractor behaviour. The three colors correspond to different rapidities: $\eta_s = 0.0$ (black), 2.0 (red), 3.0 (cyan). We fix $T_0= 0.5$ GeV and $\tau_0 = 0.1$ fm.}
     \label{fig:forwardrap_non-Bj_forward}
\end{figure*}
In the RBT approach the boost-invariant dynamics is achieved simulating a system with a large extension in the longitudinal direction, i.e. in $\eta_s$, with respect to the region which we are interested in, so that the information from the boundaries does not have the time to propagate to the considered region. 

In Fig.\,\ref{fig:checking_boost} we show two moments of the distribution function as a function of proper time for different values of the space-time rapidity $\eta_s$.
The initial longitudinal extension of the system is $|\eta_s|<8.5$ and the observed rapidities are $\eta_s=[0,\,2,\,3]$. 
All curves lie one on top of each other, indicating that the attractor behaviour seen at midrapidity is maintained when looking to the system at non-central rapidity. We have checked that the same scaling is observed for the other momentum moments not shown in the Figure and for different values of initial anisotropies and specific viscosity.
This proves that this implementation of the code is with excellent approximation boost-invariant. Indeed, we have checked that $T$, $n$ and $e$ do not differ for the considered values of $\eta_s$ during the whole time evolution. By computing the fluid four-velocity $u^\mu$ as the eigenvector of the energy-momentum tensor, we find that, for each cell, it fulfils the relation $u^\mu =(\cosh\eta_s ,0,0,\sinh\eta_s)$, 
which, according to Eq.\,(\ref{eq:umu_bjorken}), is exactly the boost-invariant condition. It means that the Lorentz factor $\gamma=\cosh\eta_s$ and $\gamma \beta_z = \sinh(\eta_s)$.

\begin{figure*}[ht]
	\centering 
	\includegraphics[width=\textwidth]{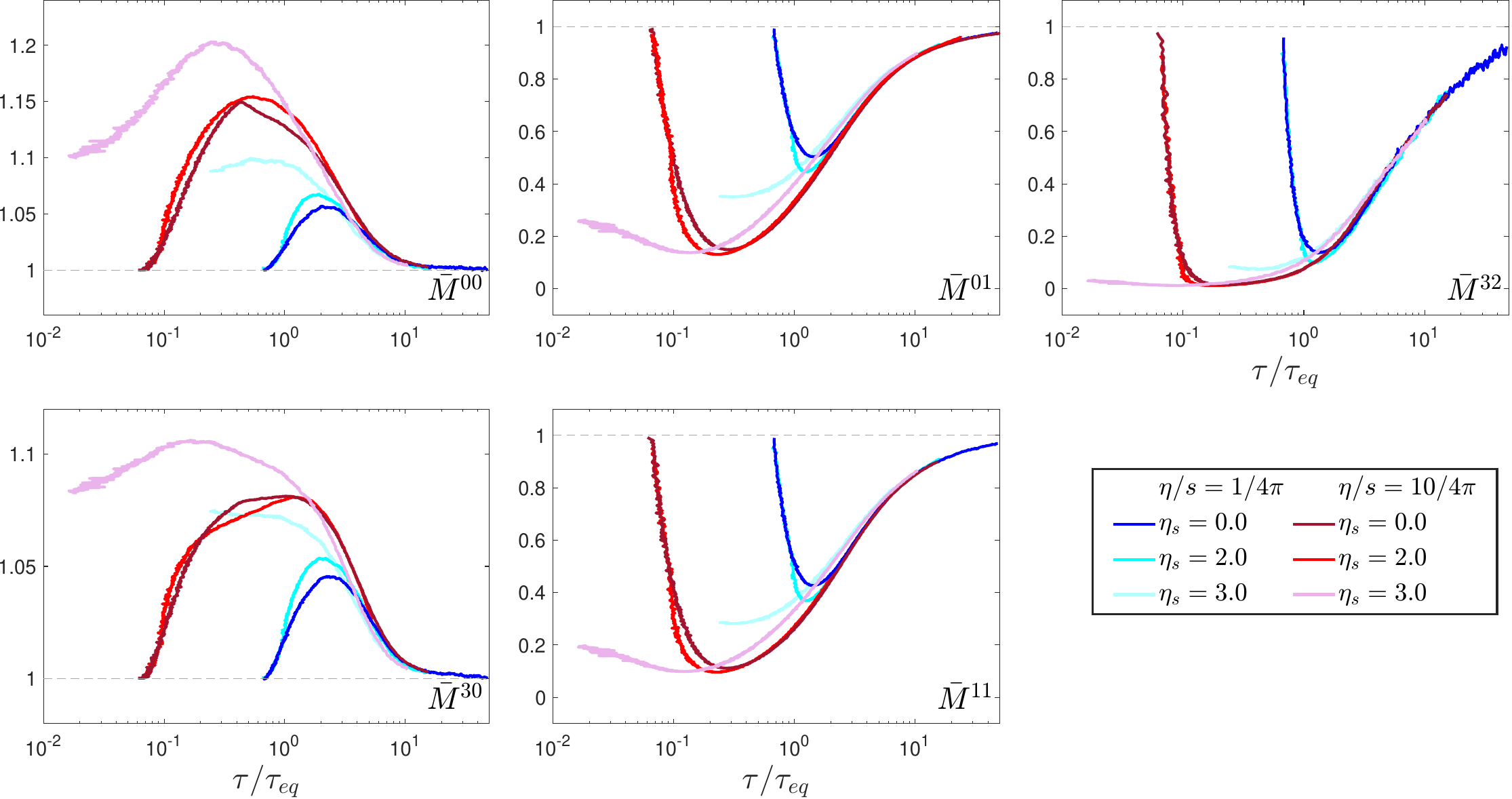}
	 \caption{Normalised moments as a function of $\tau/\tau_{eq}$ within the RBT approach. Different colour scales correspond to two different values of $\eta/s$ highlighting the pull-back attractor behaviour. The three shades correspond to different rapidities: $\eta_s = 0.0,\, 2.0,\, 3.0$. We fix $T_0= 0.5$ GeV and $\tau_0 = 0.1$ fm.}
	 \label{fig:forwardrap_non-Bj_pullback}
\end{figure*}

We present now the results obtained with simulations in which the initial longitudinal extension of the system is limited ($|\eta_s|<2.5$), so that the information from the external surface of the system reaches the $\eta_s$ region under consideration. In Figure\,\ref{fig:non_boost_distribution} we show the evolution of the longitudinal distribution $dN/d\eta_s$ exhibiting the strong breaking of the boost-invariance. Moreover, this boost-invariance breaking implies that $\gamma\neq\cosh\eta_s$. This proves that the fluid is no more in Bjorken flow, and therefore $u^\mu$ and $z^\mu$ are no more related with $\cosh\eta_s$ and $\sinh\eta_s$. This means that we cannot use anymore the expressions in Eq.\,\eqref{eq:strick_moments}, but compute directly $p^w=p\cdot z$ and $p^\tau = p\cdot u$ and use these quantities in the calculation of the momentum moments in Eq.\,\eqref{eq:momentum_moments}.
In Fig.\,\ref{fig:forwardrap_non-Bj_forward} we show some of the moments at different values of the space-time rapidity as a function of the proper time $\tau$.
The various curves correspond to $\eta/s=1/(4 \pi)$ and different values of the initial anisotropy $\xi_0$; we clearly see that the so-called forward attractor behaviour emerges regardless of the rapidity considered.
In Fig.\,\ref{fig:forwardrap_non-Bj_pullback} the simulations have been performed with $\xi_0=0$ and for different values of the specific viscosity; even in this case the universal scaling with respect to the scaled time $\tau/\tau_{eq}$ is reached and the pull-back attractor does not depend on the space-time rapidity. It is remarkable that the loss of information driving the system towards the attractor is effective also for $\eta_s=3.0$, where at initial time there are no particles due to the initialisation limited to $|\eta_s|<2.5$. Once such space-time region is populated due to the momentum transferred between particles by means of collisions, the corresponding distribution function moment approaches the attractor. This seems to suggest that, despite boost-invariance is broken, the two competing mechanisms of strong longitudinal expansion and successive onset of the collisions are still sufficient for a fundamental characterisation of the system, even far from midrapidity and in $\eta_s$ regions where the initial distribution is vanishing.

\chapter{Attractors in 3+1D systems}
\label{chap:attractors_3D}
In this Chapter the analysis carried out in 0+1D and 1+1D in the previous Chapter is extended to a more realistic 3+1D simulation where the role of the transverse dynamics for different system sizes and $\eta/s$ is studied, with particular attention to the appearance of the attractors in moments of the distribution function. Indeed, the impact of the Bjorken attractor on realistic collision systems can be fully understood only by studying a full 3+1D expanding fireball, since one can expect an impact due to the onset of the transverse expansion, which is absent in the 0+1D and 1+1D cases and introduces a new relevant time scale. If the longitudinal dynamic, indeed, is led by the almost-free initial expansion which is afterwards overridden by the collisions and all this information is encoded in $\tau_{eq}$, the transverse motion is dominated by the transverse size of the system. We will show that the interplay between these two scales fully determine the evolution of the medium and can be expressed in term of the well-known Knudsen number Kn. Being in 3+1D, the distribution function can be expanded not only in terms of the momentum moments introduced in the previous Chapter, but also in the basis of the collective flows $v_n$. The latter, however, will be addressed in the next Chapter.

\section{Initial conditions}
The RBT code is used in its most general form: the fireball expands in a 3D coordinate space: the numerical domain is defined larger than the initial distribution function so to contain the full evolution of the fireball.\\ 
The initial distribution is uniform in space-time rapidity $\eta_s$ in the range $[-2.5,2.5]$; the initial density profile in the transverse plane is given by a gaussian $f(x_\perp)\propto \exp(-x_\perp^2/R^2)$, where $\mathbf{x}_\perp = (x,y)$ and the transverse radius $R$ determines the initial extension of the fireball. In momentum space we give the same distribution as the previous Chapter, i.e. the Romatschke--Strickland. Thus globally one has:
\begin{equation}
	f_0(x,p) =\gamma_0 \exp(-x_\perp^2/R^2)\,\theta(|\eta_s|- 2.5)\exp{ \left(- \sqrt{p_x^2 + p_y^2+p_w^2(1+\xi_0)}/\Lambda_0 \right)}
\end{equation}
We fix $\Lambda_0$ and $\gamma_0$ thorough this Chapter so to have $T_0=0.5$ GeV and $\Gamma_0=1$. The $\xi_0$ parameter determines the longitudinal anisotropy in momentum space, as in the previous Chapter. In some specific case  we choose an initial distribution with $Y=\eta_s$. This corresponds to the limiting case of Eq.\,(\ref{eq:modifiedRS}) with  $\xi_0\to+\infty$ (or equivalently $p_w=0$); Eqs.\,\eqref{eq:Lambdacondition} reduce to $\Lambda_0 = 3/2\, T_0$ and $\gamma_0=27/8\, \Gamma_0$.\\
\begin{figure}
	\centering
	\includegraphics[width=.7\linewidth]{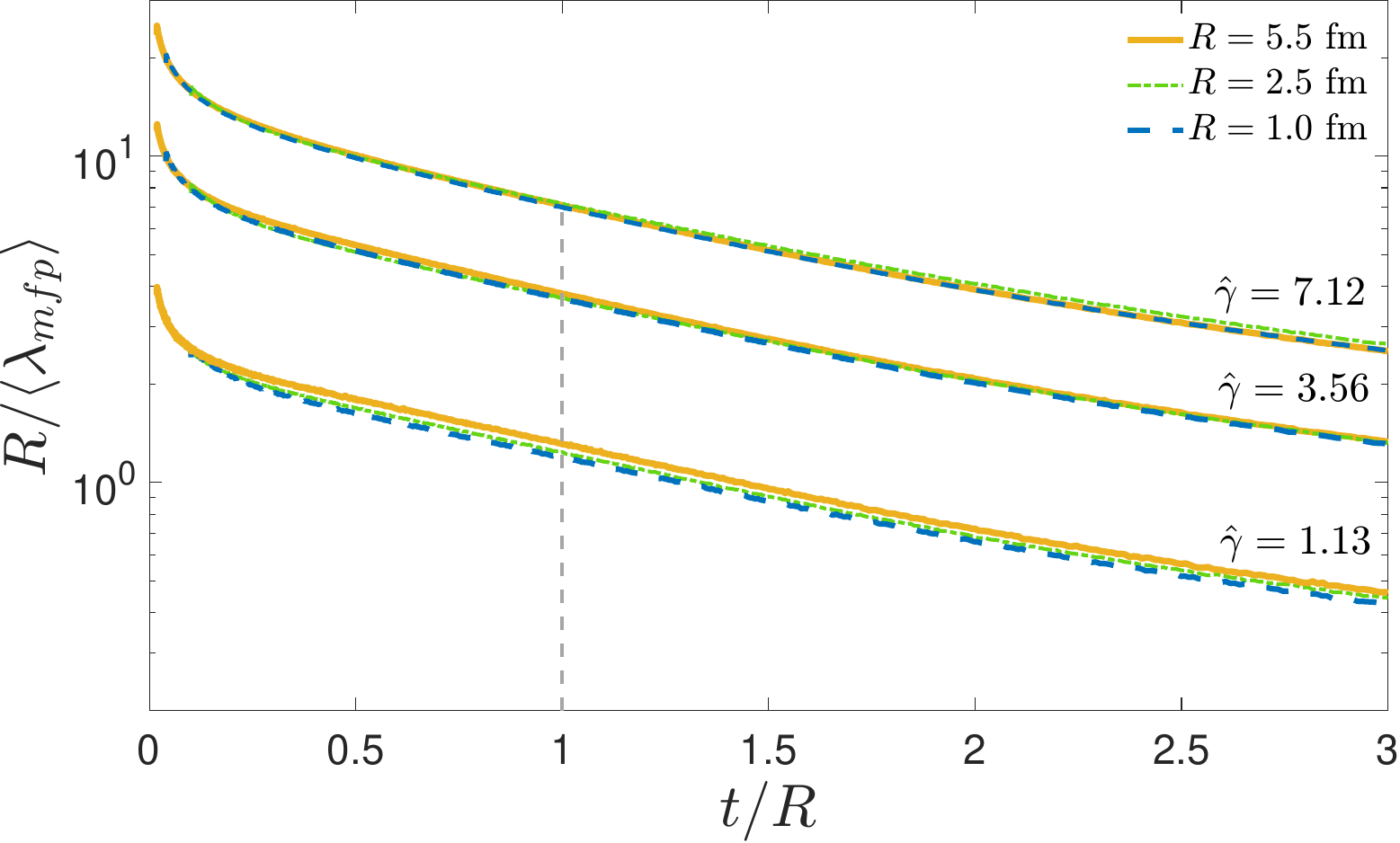}
	\caption{Ratio between transverse size $R$ and {average} mean free path $\langle\lambda_{\text{mfp}}\rangle$ as a function of scaled time $t/R$. Overlapped curves share the same opacity $\hat \gamma$. Corresponding $\eta/s$ values can be found in Table\,\ref{tb:opacity_values}.}
	\label{fig:gamma_def}
\end{figure}

\begin{table}[ht]
	\centering
	\begin{tabular}{ccc}
		\toprule
		{$\quad\hat \gamma\quad$} & {$\quad R\, \text{[fm]}\quad$} & {$\quad 4\pi\eta/s\quad $} \\
		\midrule
		\multirow{3}*{1.13} & 1.0 & 3.18  \\
		& 2.5 & 6.33  \\
		& 5.5 & 11.4  \\
		\midrule
		\multirow{3}*{3.56} & 1.0 & 1.00  \\
		& 2.5 & 2.00  \\
		& 5.5 & 3.61  \\   
		\midrule
		\multirow{3}*{7.12} & 1.0 & 0.503   \\
		& 2.5 & 1.00   \\ 
		& 5.5 & 1.81  \\
		\midrule
		\multirow{3}*{12.8} & 1.0 & 0.278    \\  
		& 2.5 & 0.554   \\ 
		& 5.5 & 1.00  \\
		\bottomrule
	\end{tabular}
	\caption{Values of the opacity $\hat \gamma$ for different radii and shear viscosity over entropy density ratio. Initial temperature and proper time are fixed: $T_0=0.5$ GeV and $\tau_0=0.1$ fm.}
	\label{tb:opacity_values}
\end{table}

\section{Opacity and Knudsen number}
As stated in the introduction of this Chapter, the role of multiple time/length - scales have to be taken into account to model an expanding 3D medium, and a very good candidate parameter is the Knudsen number Kn. However, to begin with, we follow what has been widely done in the literature in the context of RTA and ITA Boltzmann Equation (see Section\,\ref{subsubsec:rta_3+1D}), where a unique dimensionless parameter emerges from the resolution of the equation. It is called opacity and can be expressed in terms of the quantities relevant for our work as:
\begin{equation}\label{eq:our_opacity}
	\hat \gamma = \frac{1}{5\eta/s} \left( \frac{R}{\pi a} \frac{dE^0_\perp}{d\eta_s} \right)^{1/4} = \frac{ T_0}{5\,\eta/s} R^{3/4} \tau_0^{1/4},
\end{equation}
where $a=\varepsilon/T^4$, $R$ is the initial root mean square of the distribution function and ${dE^0_\perp}/{d\eta_s}$ the initial energy at midrapidity.
As outlined in \cite{Kurkela:2018qeb} and showed in Chapter 3 this parameter can be put in relation with the Knudsen number value computed at $r=0$ and $t=R$, i.e. when the transverse expansion starts to dominate, as will be clear in the following. It is noteworthy to repeat that, in the case with smooth initial conditions, simulations with different parameters but same $\hat \gamma$ are indistinguishable in the context of RTA or ITA; in our approach, instead, we are able to identify an equivalence class of different physical scenarios for each value of opacity $\hat \gamma$.\\

As far as the Knudsen number Kn is concerned, it is commonly defined as the ratio between a microscopic length (i.e. the mean free path $\mfp$) and a macroscopic scale. There is a certain degree of arbitrariness in choosing this macroscopic scale: following what is done in previous works \cite{Kurkela:2019kip,Ambrus:2021fej} we adopt the root-mean-square radius of the initial density distribution $R$; analogously, one could have used the root-mean-square of the initial energy density. One may also argue whether a certain role is played by the time-dependence of this transverse radius, since the system is expanding; however, we verified that our conclusions are not qualitatively affected by replacing the initial $R$ with the time-dependent $R(t)$. Therefore we define the local Inverse Knudsen number as:
\begin{equation}
	\IKn(x) = \frac{R}{\mfp(x)}.
\end{equation}
Since we work on a discretised grid, $\mfp(x)\to\mfp^j$, $j$ being the cell index. We define the local mean free path as:
\begin{align*}
	\lambda_{\text{mfp}}^j &=  \frac{1}{N_\text{cell}^j} \sum_i^{N^j_\text{cell}}  \lambda_{\text{mfp}}^i = \frac{1}{N_\text{cell}^j} \sum_i^{N^j_\text{cell}} v_i\,t_i \approx \\
	&\approx\frac{t_\text{coll}}{ N^j_\text{cell} } \sum_i^{N^j_\text{cell}} v_i = {t_\text{coll}}\, \langle v_j\rangle =   \frac {\Delta\tau N_\text{cell}^j }{ 2 \,N_\text{coll}^j}  \langle v_j\rangle
\end{align*}
where $N_\text{cell}^j$ is the number of test particles inside the $j$-th cell, $N_\text{coll}^j$ is the number of collisions occurring in the given time-step, $\langle v_j\rangle$
is the average velocity and $t_i$ is the time occurring between two successive collisions for each particle. Notice that we assume $t_i=t_\text{coll}$ for every test particle according to the hypothesis of homogeneity within each cell, with $t_\text{coll}$  being the average collision time per particle. In the massless case $v_i=1$, and therefore $\lambda_{\text{mfp}}=t_\text{coll}$ exactly.

In order to have a unique quantity to describe the whole system, one has to compute a global $\IKn$. We can define an average weighted by the number of particles: 
\begin{align*}
	\langle\text{Kn}^{-1}\rangle &= 
	\frac{1}{\sum_j N_\text{cell}^j} \sum_{j} \frac{R}{\mfp^j} N^j_\text{cell} =\\
	& =\frac{1}{\sum_j N_\text{cell}^j} \sum_{j} \frac{2RN_\text{coll}^j}{\langle v_j\rangle\, \Delta\tau\, N_\text{cell}^j} N_\text{cell}^j =\\
	&=\frac{1}{\sum_j N_\text{cell}^j} \frac{2R}{\Delta \tau} \sum_j \frac{N^j_\text{coll}}{\langle v_j \rangle},
\end{align*}
where the summation runs over the cells at midrapidity.

{The role of the Knudsen number is strictly related to the regime of applicability of hydrodynamics. Indeed, as reported in Chapter\,\ref{chap:kin_and_hydro}, fluid dynamics is rigorously applicable if the microscopic ($\mfp$) and macroscopic ($R$) scales of the system are well separated, that is $\IKn\gg 1$. This means that we can understand, by looking at $\IKn(t)$, whether or not the evolution of the system could be modeled by fluid dynamics in its different stages. Similar considerations have been done about the opacity, since depending on the $\hat \gamma$ value one can categorise a certain system as fluid- or particle-like. Being a unique parameter, however, opacity does not allow to study different space-time regions of the expanding medium.} \\
The connection between these two quantities is confirmed by computing the Knudsen number for simulations sharing the same opacity: one finds that, if the initial geometry is approximately fixed, the different systems show a {similar} evolution of $\IKn$ in terms of the scaled time $t/R$ at every time of the simulation, {with deviation $<5\%$,} as shown in Figure\,\ref{fig:gamma_def},
where we plot this quantity for three different values of $\hat \gamma = [1.13, 3.56, 7.18]$, each one for three different transverse radii $R=[1.0, 2.5, 5.5]$ fm. 
It should be outlined, however, that if two systems share the same $\IKn$ at a certain $t/R$, they will share it throughout the whole scaled-time evolution. Thus, the Knudsen Number appears as the key scaling parameter of the theory, according to which  different collision systems can be clustered in universality classes. each universality class can be labelled by the value $\KnR\equiv \IKn(t=R)$. It should be emphasised that, $\KnR \approx \hat \gamma$ in each of the three cases. This suggests that the opacity does not encode all the information in the RBT approach, which is totally non-surprising, since it arises from a different method of resolution of the Boltzmann equation.  We anticipate here that the difference between the two parameters emerges exclusively in the context of the collective flows (Chapter\,\ref{chap:knudsen}), which are particularly sensitive to the details of the transverse dynamics; the momentum moments, instead, are poorly affected if one moves from one framework to the other, and therefore will be presented only once.\\
We refer to Sec.\,\ref{sec:fluctuations} and in particular to Table\,\ref{tab:event_by_event} to associate  physical systems to universality classes, ranging from OO to PbPb at LHC and RHIC energies.

\section{Attractor for the inverse Reynolds number}
\label{sec:reynolds}
\begin{figure}
	\centering
	\includegraphics[width=.8\textwidth]{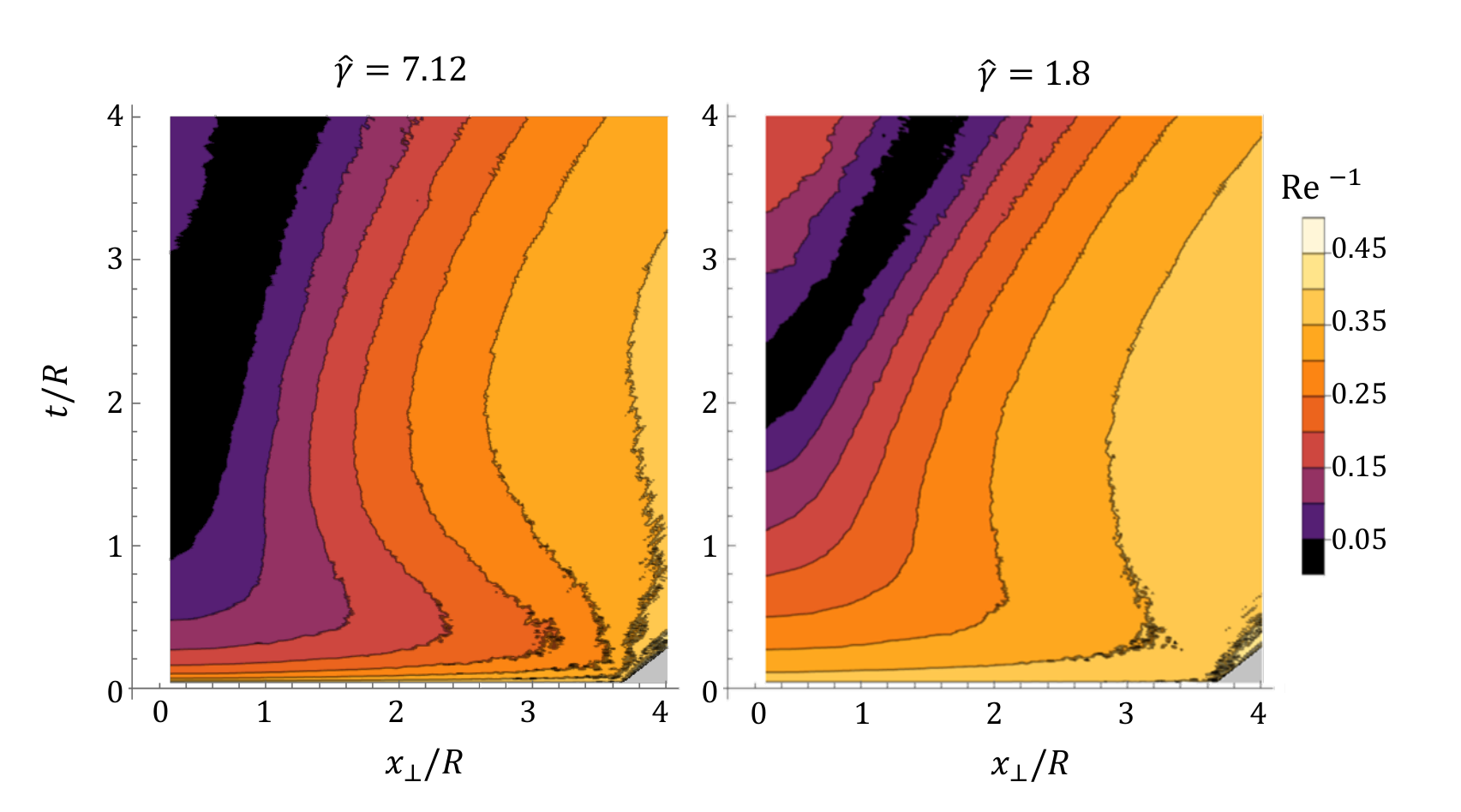}
	\caption{Contour plots of $\text{Re}^{-1}$ with respect to $t/R$ and $x_{\perp}/R$ at midrapidity $|\eta_s| < 0.125$ for two different values of opacity: $\hat\gamma = 7.12$ (left panel) and $\hat \gamma= 1.8$ (right panel). Both systems have $R=2.5$ fm; $\eta/s$ is respectively $1/4\pi$ and $4/4\pi$.}
	\label{fig:Q0_contour}
\end{figure}

\begin{figure}
	\centering
	\includegraphics[width=.9\linewidth]{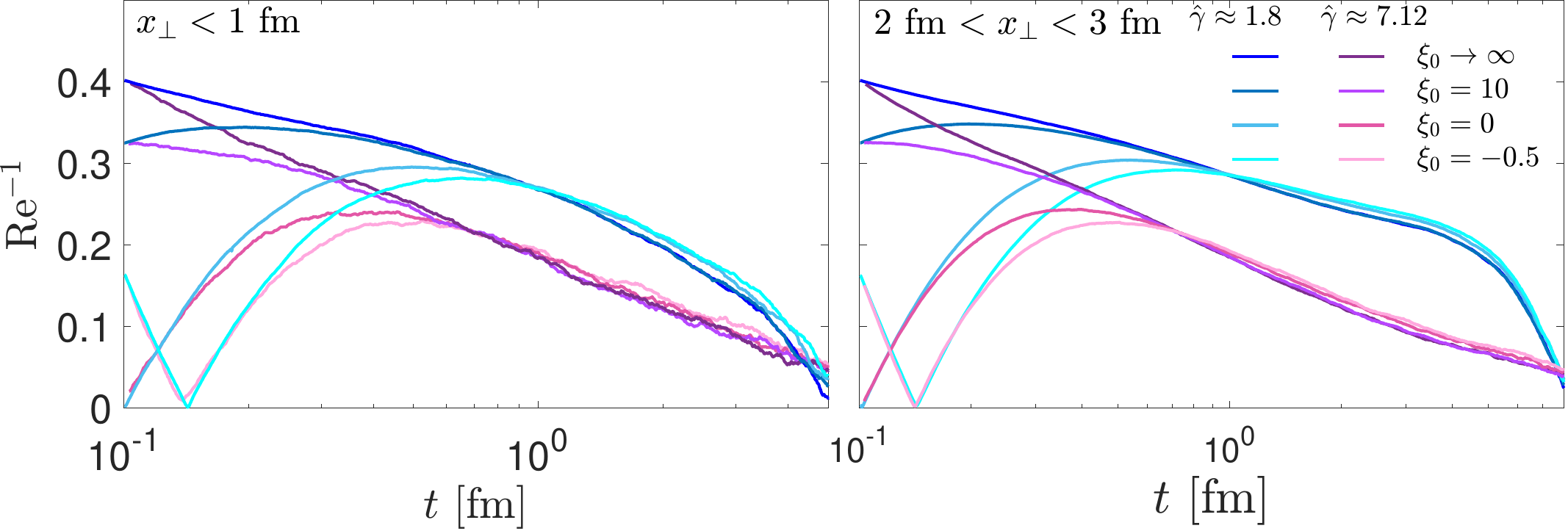}
	\caption{The inverse Reynolds number as a function of $t$ for $\hat\gamma = [1.8, 7.12]$, respectively for $R=2.5$ fm ($\eta/s=4/4\pi$) and $R=5.5$  fm ($\eta/s=1.8/4\pi$) and for the two anuli $x_\perp<$ 1 fm (left panel) and 2 fm $<x_\perp<$ 3 fm (right panel); different shades correspond to different initial anisotropies.}
	\label{fig:reynolds_attractors}
\end{figure}

We have previously introduced (Chapter 3) the inverse Reynolds numbers \cite{Denicol:book}, which quantify the deviation of a medium from the thermodynamic equilibrium. In our context, the only non-zero quantity among them is the shear inverse Reynolds number defined as (we omit the subscript $\pi$ for simplicity):
\begin{equation}
	\text{Re}^{-1} = \dfrac{\sqrt{\pi^{\mu\nu}\pi_{\mu\nu}}}{e}.
\end{equation}
We recall the expression for the shear stress tensor $\pi_{\mu\nu}$ in the conformal case:
\begin{equation}
	\pi^{\mu\nu} = T^{\mu\nu} - T^{\mu\nu}_{id}.
\end{equation}
The ideal energy-momentum tensor is
\begin{equation}
	T^{\mu\nu}_{id} = \text{diag}(e, P, P, P)
\end{equation}
and for the conformal case the pressure and the energy density are related by the equation of state $P(e)=e/3$. A vanishing inverse Reynolds number $\text{Re}^{-1}\to0$ corresponds to $\pi^{\mu\nu}\to 0$, which characterises a fully equilibrated system. On the other limit, a free streaming evolution leads to an increase of the inverse Reynolds number.\\
Notice that, up to a numerical constant, the inverse Reynolds number is identical to the $Q_0$ parameter introduced in \cite{Kurkela:2019kip}:
\begin{equation}
	Q_0= \sqrt{ \dfrac{\pi^{\mu\nu}\pi_{\mu\nu}}{T^{\mu\nu}_{id} T_{\mu\nu, id} } } = \sqrt{\dfrac{\pi^{\mu\nu}\pi_{\mu\nu}}{4/3\, e^2}} = \frac{\sqrt{3}}{2} \text{Re}^{-1}.
\end{equation}
In order to observe the evolution of systems with different values of opacity, we show two contour plots of $\text{Re}^{-1}$ with respect to $t/R$ and $x_{\perp}/R$ in Figure\,\ref{fig:Q0_contour}. 
Both plots refer to the same initial distribution, which is chosen to be isotropic in the transverse plane, with $Y=\eta_s$ and $R=2.5$ fm, but with different $\eta/s$, respectively $1/4\pi$ and $4/4\pi$. These values correspond respectively to $\hat \gamma = 7.12$ (left panel) and $\hat \gamma = 1.8$ (right panel). It is straightforward to observe that the larger the opacity, the wider the dark regions extend, thus showing that more interacting systems, i.e. with larger opacity values $\hat \gamma$, remain close to equilibrium in a wider spatial region and for longer times. It is also possible to see that at very small times and for large radii the system is dominated by the free streaming expansion, which causes a larger inverse Reynolds number. Qualitatively, these results are similar to what is shown in \cite{Kurkela:2019kip}.\\
It is interesting to study the $\invRe$ for different initial conditions and to look for universal behaviour in its evolution. In particular, we analyse systems with  different initial longitudinal anisotropy, which correspond to different initial $P_L/P_T$, for the two opacity values $\hat \gamma = [1.8, 7.12]$, corresponding respectively to $R=2.5$ fm and $\eta/s=4/4\pi$ and $R=5.5$ fm and $\eta/s=1.8/4\pi$. In order to do so, we change $\xi_0=[-0.5, 0, 10, +\infty]$; since the system is azimuthally invariant, the inverse Reynolds number is a function of time and transverse radius $x_{\perp}$: $\text{Re}^{-1}=\text{Re}^{-1}(t,x_\perp)$. We consider two anuli $0<x_\perp<1$ fm and 2 fm $<x_\perp<$ 3 fm at midrapidity $|\eta_s|<0.125$. As said before, the inverse Reynolds number quantifies the deviation from equilibrium; as shown in Figure\,\ref{fig:reynolds_attractors}, after $t\approx 1$ fm, the curves approach a universal attractor regardless the initial conditions, therefore suggesting that, at this point, different systems reach the same degree of equilibration. Furthermore, we verified that the same evolution towards universality is reached also if a non-zero initial azimuthal anisotropy in momentum space is given to the distribution function. This is a new hint of the possible presence of universal attractors also in {more differential probes of the phase-space distribution function, such as its momentum moments}.\\

\section{Role of the transverse flow}\label{subsec:transverse_flow}

\begin{figure}[t]
	\centering
	\includegraphics[width=0.5\textwidth]{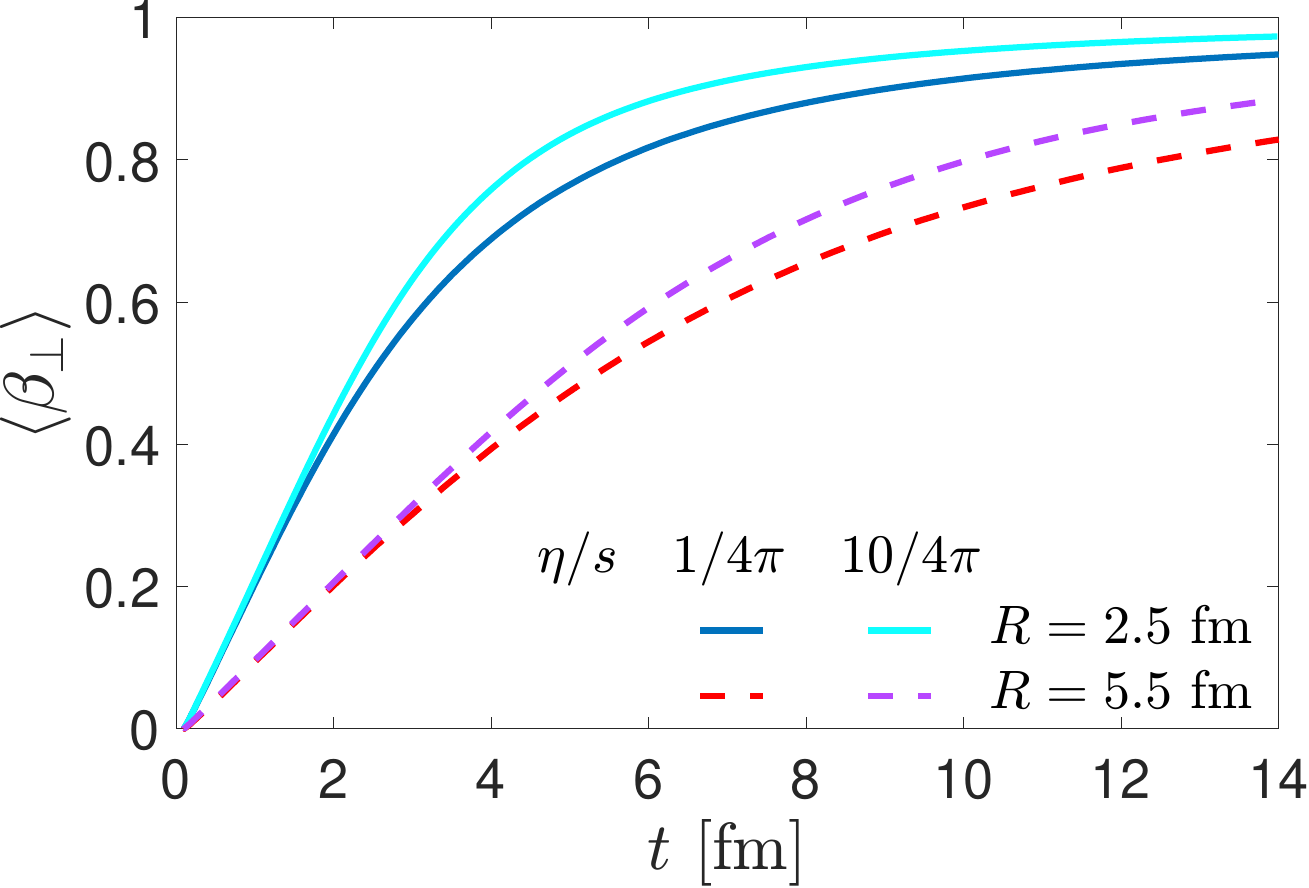}
	\includegraphics[width=0.439\textwidth]{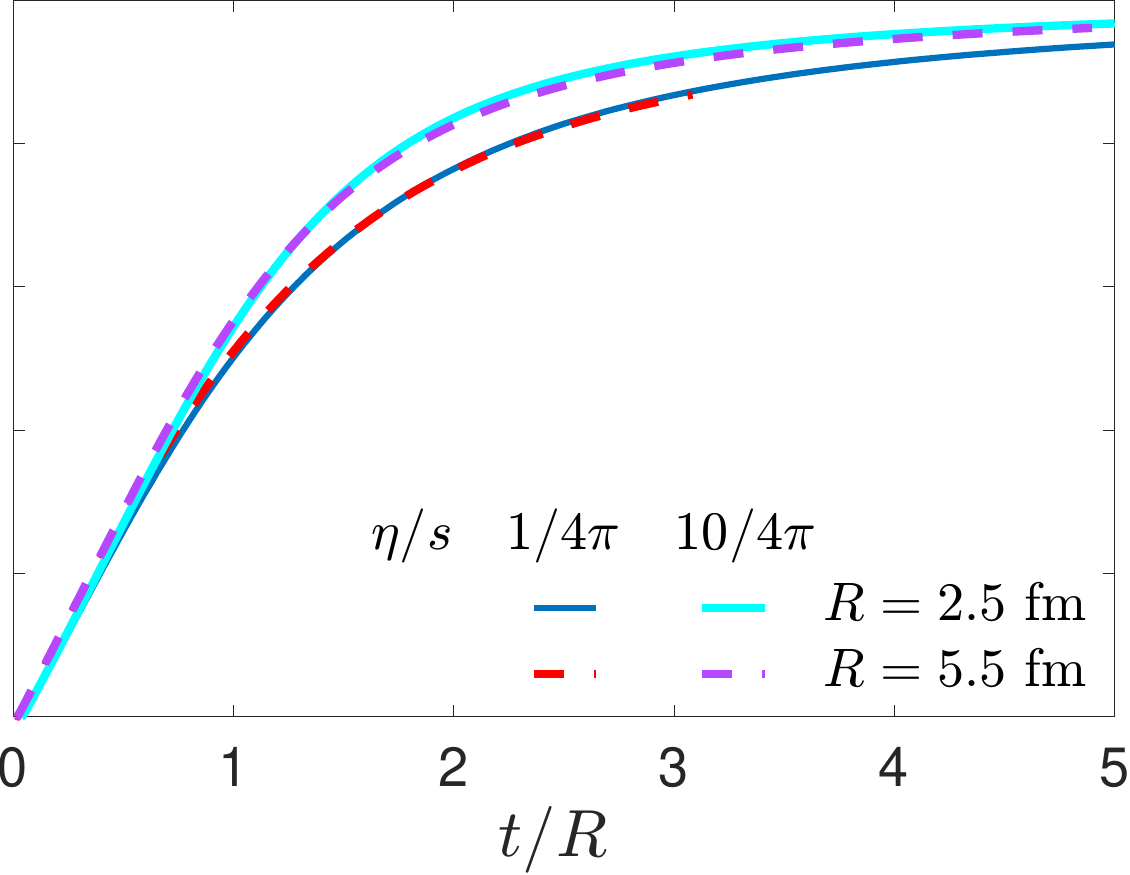}
	\caption{Average transverse flow $\langle \beta_\perp\rangle$  as a function of time $t$ (left panel) and scaled time $t/R$ (right panel). Simulations have same $T_0=0.5$ GeV, but different initial transverse size $R=2.5$ fm (solid lines), $R=5.5$ fm (dashed lines) and different specific viscosity $\eta/s=1/4\pi, 10/4\pi$. }
	\label{fig:beta}
\end{figure}

\begin{figure}[h]
	\centering
	\includegraphics[width=0.6\linewidth]{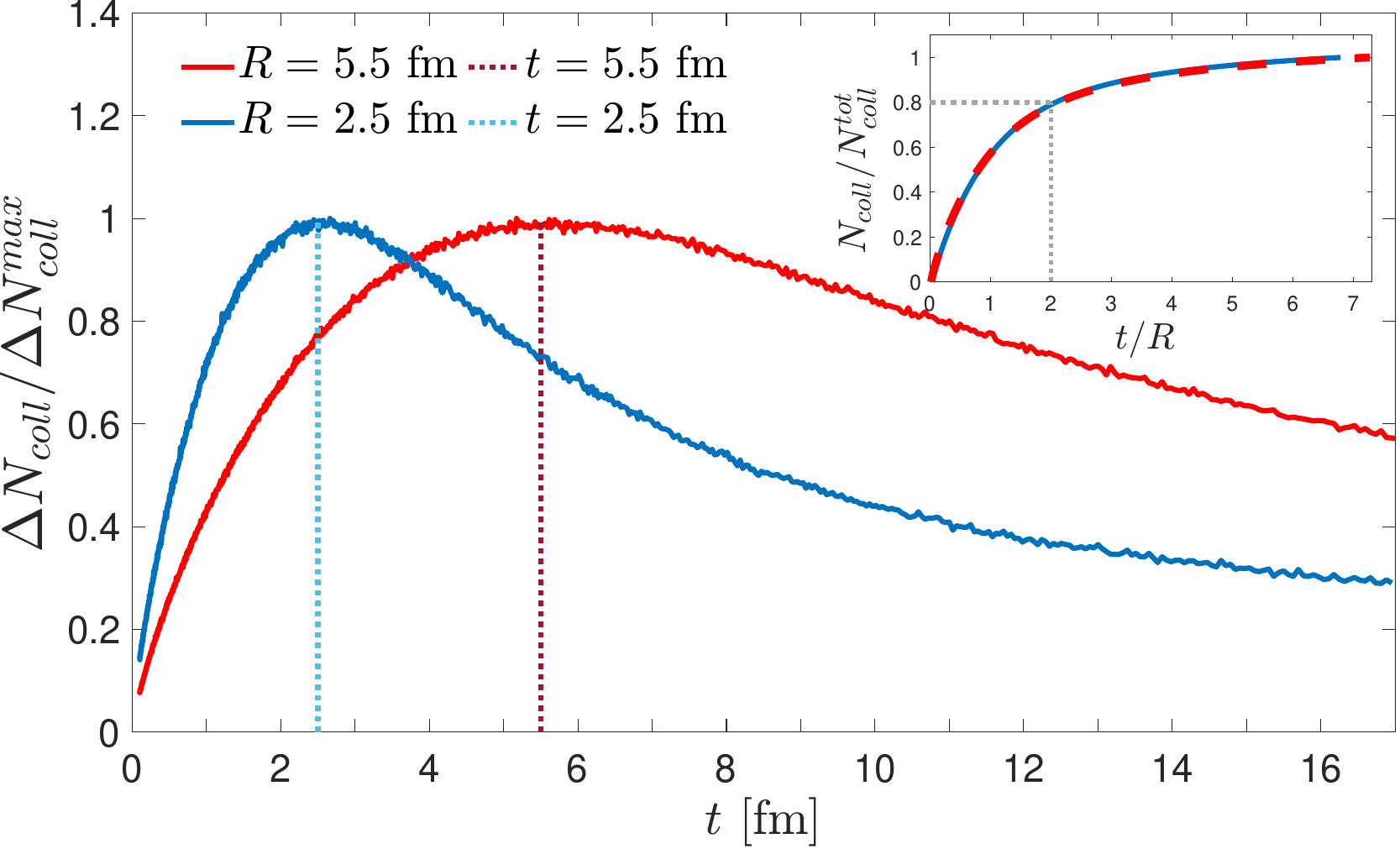}
	\caption{{Number of collisions in each time interval normalised by its maximum value and integrated over the transverse plane at mid-rapidity for two simulation with $\eta/s=1/4\pi$ and radii $5.5$ fm (red line) and $2.5$ fm (blue line). In the inset: the cumulative number of collisions as a function of $t/R$.}}
	\label{fig:collisions}
\end{figure}

Before delving into the detailed analysis of the moments, we must have a look to the onset of the local flow $u^\mu(x,t)$. Since we are going to focus on the midrapidity region, which is symmetric in $\eta_s$ (and therefore in $z$), for symmetry reason it will be $\beta_z(x_\perp,t)=0$, as we checked.  Therefore, in the rapidity region [-0.125, 0.125] the fluid velocity can be written as $u^\mu(x_\perp, t)=\gamma(1, \boldsymbol\beta_\perp(x_\perp, t), 0)$. In Fig.\,\ref{fig:beta} we show the average transverse flow $\langle\beta_\perp(t)\rangle$ in the midrapidity region for four different simulations with $R=[2.5, 5.5]$ fm and $\eta/s=[1/4\pi, 10/4\pi]$ . By looking at the left panel, it is interesting to notice that systems with different initial size $R$ and $\eta/s$ show a different evolution of the $\langle\beta_\perp(t)\rangle$. However, at $t \approx R$, $\langle \beta_\perp\rangle \approx 0.5$ as shown in the right panel of Fig.\,\ref{fig:beta}, which indicates also that, for $t>R$ the development of the flow is significantly non-zero. It is not surprising that curves corresponding to a larger value of $\eta/s$ show a faster development of the flow, since the system is closer to the free-streaming case.  {More generally, if one rescales the time with respect to the initial transverse radius $R$ the curves at fixed $\eta/s$ are almost indistinguishable: it must be emphasised that this scaling property does not depend on $\eta/s$ and is valid in a wide range of specific viscosity. It is interesting to notice that at $t\sim 2 R$, $\langle \beta_\perp\rangle \approx 0.8$, which means that the system is very close to a free streaming regime, in which $\langle \beta_\perp\rangle \lesssim 1$.  This fact could be interpreted along with what seen in Fig.\,\ref{fig:collisions}, where we show the total number of collisions in the transverse midrapidity plane for the two simulations with $\eta/s=1/4\pi$ and initial transverse radius $R=2.5$ fm (blue line) and $R=5.5$ fm (red line). Obviously, the same considerations are valid for the case with $\eta/s=10/4\pi$ as well.}
At $t\sim R$ the system makes the maximum number of collisions and afterwards starts to decouple with fewer and fewer collisions. By looking at the inset, in which the cumulative number of collisions as a function of $t/R$ is shown, one can see that at $t\sim 2R$ both simulations have performed $\sim$80\% of the total number of collisions. At this time scale, indeed, the system is almost completely decoupled: the flat curve at late times indicates that no more collisions occur and thus the system is in a free-streaming regime. This is coherent with what we see in the average transverse flow $\langle \beta_\perp\rangle$. {In the time range $t=R- 2R$, in summary, the system turns from being dominated by the longitudinal expansion to an almost transverse free streaming, as previously observed in \cite{Ambrus:2022koq}}. Because of this, we choose as the final time of our simulations $t=2R-3 R$.\\

\section{Attractors for Moments of  \texorpdfstring{$f(x,p)$}{f(x,p)}}
\label{subsec:3+1d_moments}
Here we extend the analysis performed in 0+1D and 1+1D to a more realistic case of 3+1D simulations studying systems with different transverse sizes, ranging from a typical fireball created in $AA$ collisions to smaller systems like $pA$ and $pp$ at LHC and RHIC energies. For the results shown in this section we consider only azimuthally isotropic initial conditions in coordinate and momentum space, so that $f=f(x_\perp, p_T, p_w, \tau)$.
With the intent to account for the greater complexity of the distribution function, the momentum moments are generalised as follows:
\begin{equation}
	M^{nm}(x_\perp,t) = \int dP (p\cdot u)^n (p\cdot z)^{2m} f(x_\perp,p,t),
\end{equation}
where the moments are given a $x_\perp$ dependence following the $f(x,p)$.
It is useful also to define the integrated moments:
\begin{equation}\label{eq:moments_integrated}
	\mathcal M^{nm}(t) = \int_{A_\perp} d\mathbf{x_\perp} M^{nm}(x_\perp,t)
\end{equation}
where $A_\perp$ is the extension of the fireball in the transverse plane.
Exactly as already done in the previous Chapter, we look for attractors in the normalised moments $\overline M^{nm} = M^{nm}/M^{nm}_{eq}$, where

\begin{equation}\label{eq:moments_eq}
	M^{nm}_{eq} (x_\perp, t) = \Gamma \int dP (p\cdot u)^n (p\cdot z)^{2m} e^{-(p\cdot u)/T},
\end{equation}
and $\Gamma=\Gamma(x_\perp, t)$, $T=T(x_\perp, t)$, $u^\mu=u^\mu(x_\perp, t)$ are respectively the local fugacity, temperature and fluid 4-velocity. In the space-time discretisation of the code, the latter are the primary fluid dynamic variables which are extracted locally in space and time via the resolution of the eigenvalue problem of the energy-momentum tensor (see Paragraph\,\ref{subsec:collision}), by exploiting the Landau matching conditions. Numerically speaking, Eq.\,\eqref{eq:moments_integrated} becomes:
\begin{eqnarray}
	\mathcal M^{nm}(t) = \sum_{\text{cell}}  M^{nm}(\mathbf x_{\perp}^{\text{cell}},t).
\end{eqnarray}
The same is done also for the equilibrium moments in Eq.\,\eqref{eq:moments_eq} where $\Gamma=\Gamma(x_\perp^\text{cell}, t)$, $T=T(x_\perp^\text{cell}, t)$, $u^\mu=u^\mu(x_\perp^\text{cell}, t)$. Once more the integrated moments are given by:
\begin{eqnarray}
	\mathcal M^{nm}_{eq}(t) = \sum_{\text{cell}}  M^{nm}_{eq}(\mathbf x_{\perp}^{\text{cell}},t).
\end{eqnarray}
We recall that, due to the matching conditions, $\overline {\mathcal M}^{10}=1$ and $\overline{\mathcal M}^{20}$=1, while $\overline {\mathcal M}^{01} = P_L/P_{eq} = 3P_L/e$, which is strictly linked with $P_L /P_T = 2 P_L/(e - P_L)$.\\
\begin{figure}[t]
	\centering
	\includegraphics[width=.8\linewidth]{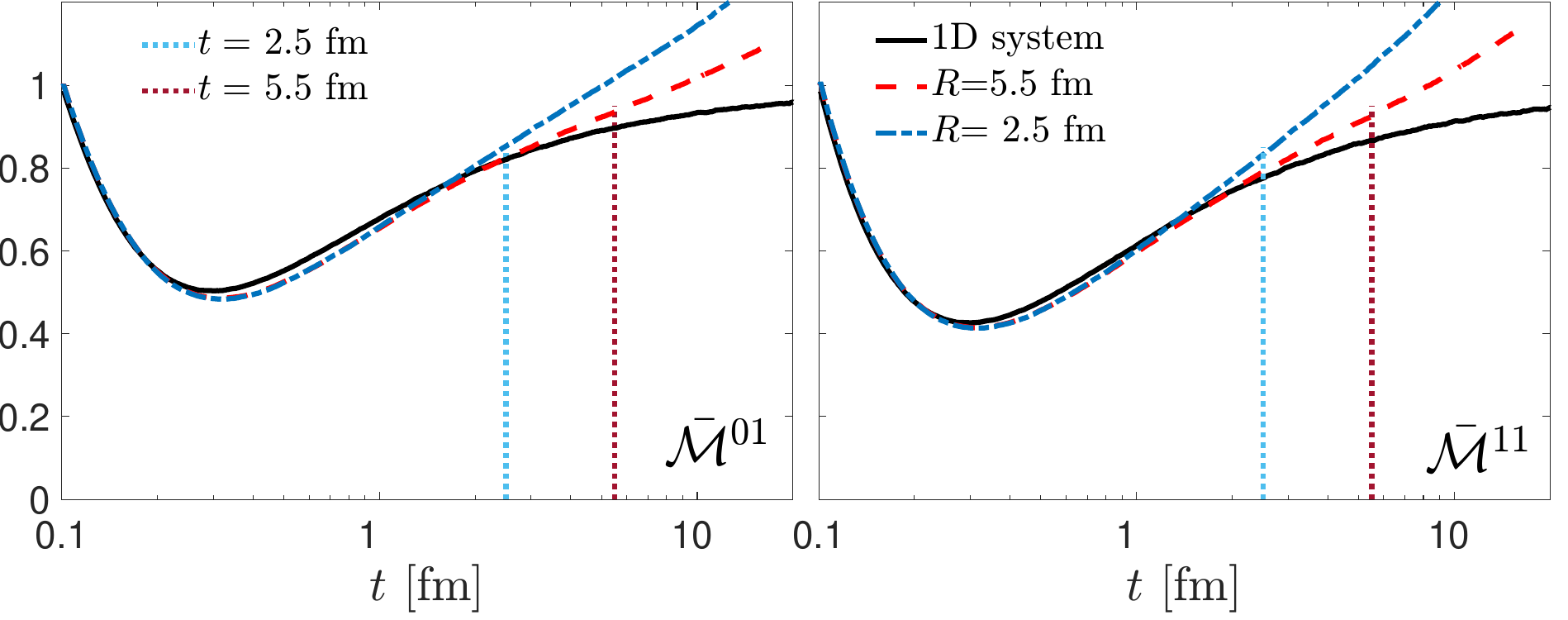}
	\caption{Normalised moments $\overline{ \mathcal M}^{01}$ and $\overline{ \mathcal M}^{11}$ as a function of time $t$. The 3D curves depart from the 1D one at a time scale $t\sim R$. All the simulations are performed with $T_0=0.5$ GeV, $\tau_0 =0.1$ fm, $\eta/s=1/4\pi$. It is possible to notice a slight dependence on the order of moments of the departure time, which is smaller for the larger order ones.}
	\label{fig:1d_vs_3d}
\end{figure}

In Fig.\,\ref{fig:1d_vs_3d} we show the integrated normalised moments $\overline {\mathcal M}^{nm}$ at midrapidity as a function of time. We compare the time evolution of the moments for two different system sizes $R=2.5$ fm and $R=5.5$ fm to the 1D case. The three simulations share the same $\eta/s= 1/4\pi$ and same initial conditions: $T_0=0.5$ GeV, $\xi_0=0$.  At very early times $t\ll R$ the three curves show the same evolution: as studied in 1D, there is a  departure from the initial equilibrium due to the longitudinal quasi-free streaming which dominates in the initial stages, and a later increase of the normalised moments when the collisions start to play a role. At time $t\approx R= 2.5$ fm the small system moments (blue dot-dashed line) depart from the 1D case; the same happens for the large ones at $t\approx R= 5.5$ fm (red dashed line). The deviation from the 1D case can be traced back to the development of the transverse flow $\beta_\perp$ observed in Fig.\,\ref{fig:beta}: as one can expect it is the transverse expansion which discriminate a 3D from a 1D simulation. Larger systems, which tend to develop later a transverse flow, behave for a longer time like a 1D system, which can be seen as a medium undergoing a longitudinal expansion with an infinite extension in the transverse plane.

\subsection{Forward Attractors}
\begin{figure}[t]
	\centering
	\includegraphics[width=.5\linewidth]{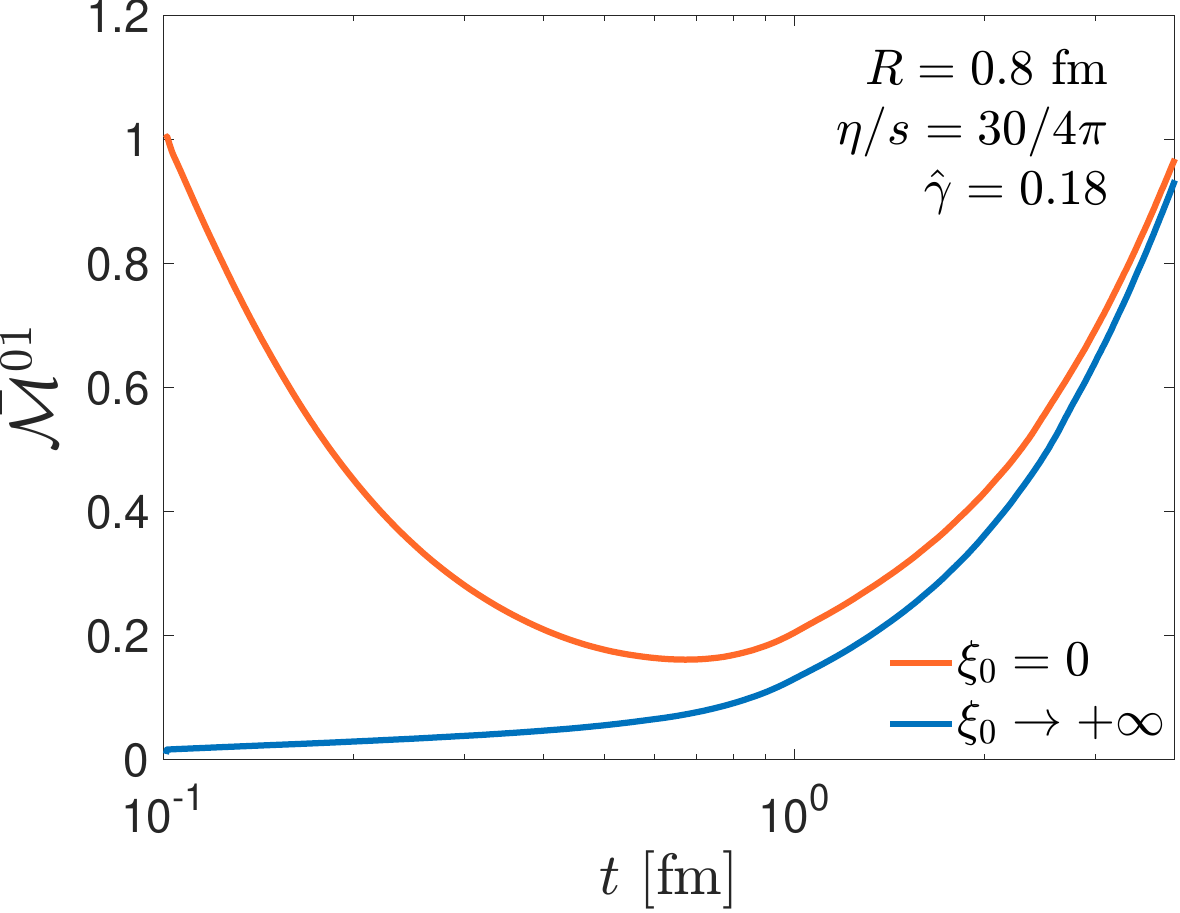}
	\caption{{Normalised integrated moment $\overline {\mathcal M}^{01} = P_L /P_{eq}$ for a simulation with very small opacity $\hat \gamma = 0.18$. Notice that there is no attractor even going up to $t=4$ fm, i.e. $t=5R$.}}
	\label{fig:moments_noattractor}
\end{figure}

Following the nomenclature introduced in Paragraph\,\ref{subsec:forward_attractors_1D} in this Section we study the normalised moments evolution for the same system sizes discussed above with $\eta/s=1/4\pi$ and for different initial anisotropy values $\xi_0=[-0.5, 0, 10, \infty]$ in order to highlight the emergence of universal behaviour.\\
\begin{figure}[t]
	\centering
	\includegraphics[width=.7\linewidth]{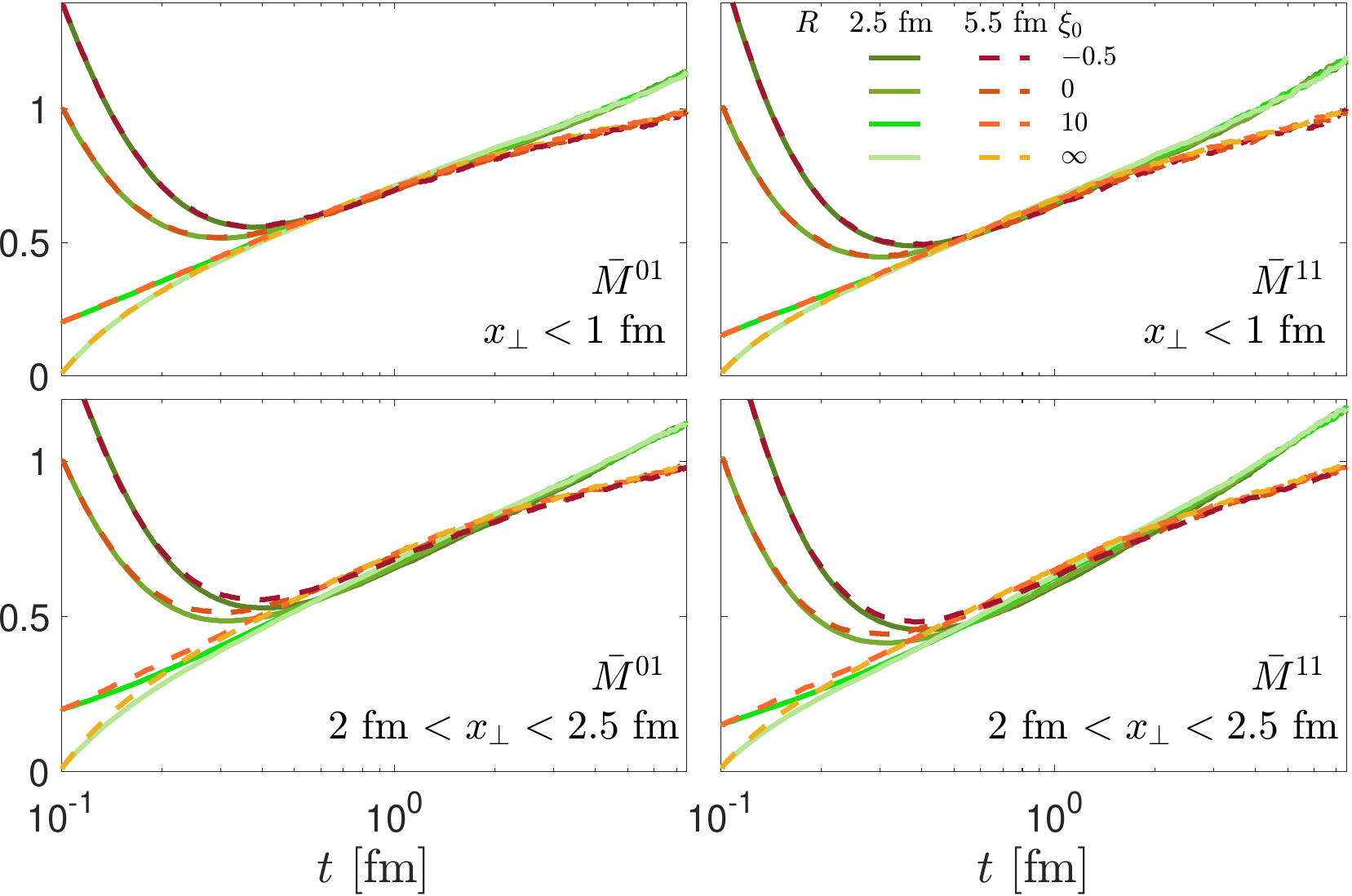}
	\caption{Forward attractors as a function of time in the two moments $\overline M^{01}, \overline M^{11}$ of the distribution function, for two different anuli: $x_\perp<1$ fm (upper panels) and 2.0 fm $<x_\perp<$ 2.5 fm (lower panels) and two different initial transverse sizes $R=2.5$ fm (green solid lines) and $R=5.5$ fm (orange dashed lines). Both computations are performed with $\eta/s=1/4\pi$. }
	\label{fig:forward_3d}
\end{figure}
In Fig.\,\ref{fig:forward_3d}, we show the normalised moments $\overline M^{nm} (x_\perp, t)$ for two different anuli $x_\perp<1$ fm (upper panels) and 2.0 fm $<x_\perp< 2.5$ fm (lower panels). Irrespectively of the initial size $R$, the systems reach the attractor at the same time scale, thus forgetting about the different initial anisotropy values immediately after the minimum in the curves. Since, as seen in Fig.\,\ref{fig:1d_vs_3d}, the initial expansion in these cases is identical to that of a 1D system, the same loss of memory observed in 1D has to occur also in these 3D systems.\footnote{Notice that the universality here shown for different anuli is obviously present also in the integrated normalised moments $\overline {\mathcal M}^{nm}$.}\\
However, at later times the curves depart from the typical 1D trend and therefore the late-time attractor sensitively depends on the system size $R$, despite forgetting about the initial anisotropy.\\
One can imagine a more extreme scenario in which the transverse expansion takes place before the longitudinal one has been completed. Obviously this is a combined effect of the interaction strength of the system, encoded in the specific viscosity $\eta/s$, which determines how long the longitudinal expansion lasts, and of the initial transverse radius $R$, that sets the time scale at which the transverse flow sets off. By accurately choosing such quantities, one can get a system which keeps memory of the initial anisotropy, with the consequent lost of the universal behaviour, {as shown in Fig.\,\ref{fig:moments_noattractor}. It is possible to see that the attractor is not reached even for $t=4-5R$ fm, 10 times larger than $t=0.4$ fm, at which the forward attractor is reached in Fig.\,\ref{fig:forward_3d}. For this simulation we choose a very small radius $R=0.8$ fm and $\eta/s=30/4\pi$ (the system is very weakly interacting), in order to have $\hat \gamma = 0.18$.  Notice that such an extreme case is far from possible systems created in uRHICs: reasonable estimates of opacity in $pp$ collisions are of about 0.7 \cite{Werthmann:2023zms} and even considering an extreme case for a possible $\hat \gamma$ in $pp$ ($R=0.8$ fm, $4\pi\eta/s = 10$, $T_0=0.3$ at $\tau_0=0.1$ fm) one gets an estimate of $\approx$ 0.45, which is more than twice larger than the value we considered for this investigation. Nonetheless this example is quite instructive in order to understand the interplay of the two time scales, i.e. the time needed for the longitudinal expansion and the transverse size $R$, which marks the onset of the transverse expansion.} 

\subsection{Pullback Attractors}

\begin{figure}[t]
	\centering
	\includegraphics[width=.7\linewidth]{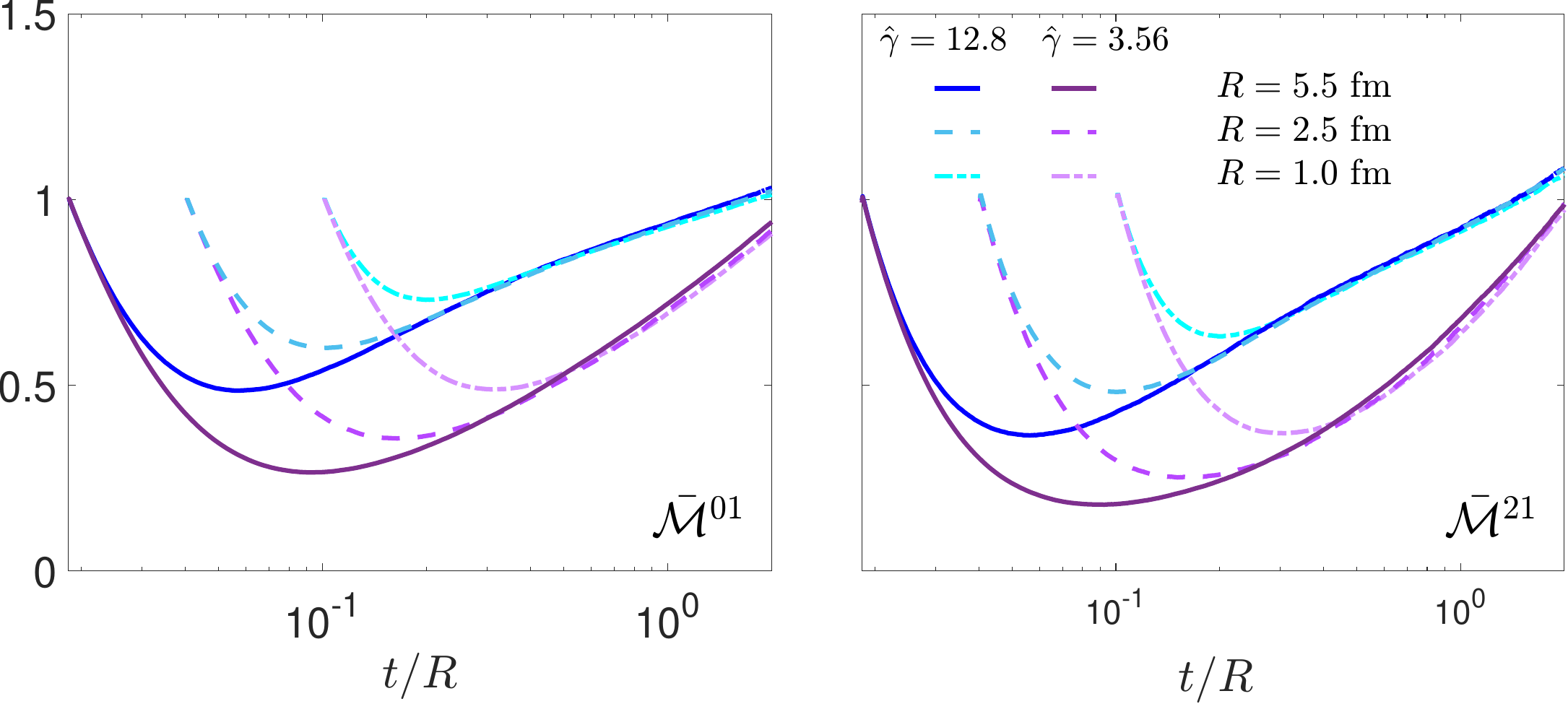}
	\caption{Comparison of two different pullback attractors for different values of the opacity $\gamma$. Notice that the equilibrium line is crossed at very similar scaled times.}
	\label{fig:double_pullback}
\end{figure}

In the 1D framework we show the appearance of an attractor by changing the initial scaled time $(\tau/\tau_{eq})$, which we called pull-back attractor and differentiate from the forward (or late-time) attractor since it is governed by rescaling of the system evolution in terms of the only relevant time scale $\tau_{eq}$. As mentioned above, in the 3+1D case there is a second time-scale strongly characterising the system, which is related to the expansion in the transverse direction and can be specified by the transverse size $R$. As anticipated before, one can make use of the dimensionless parameter $\hat \gamma = T_0 R^{3/4} \tau_0^{1/4} / (5\eta/s)\approx R/\lambda_{\text{mfp}}$ which includes also information about $R$, see Eq.\,\eqref{eq:our_opacity}. It may be of interest to rewrite the opacity as:
\begin{equation}\label{eq:opacity_link_with_1D}
	\hat \gamma = \frac{\tau_0 T_0}{5 \eta/s} \left( \frac{R}{\tau_0} \right)^{3/4} = (\tau/\tau_{eq})_0 \left( \frac{R}{\tau_0} \right)^{3/4}.
\end{equation}
In such a way, the new parameter $\hat \gamma$ is directly related to the ratio $(\tau/\tau_{eq})_0$, which characterises the 1D systems, while the factor $(R/\tau_0)^{3/4}$ accounts for the transverse extension; for this reason in this section we continue talking about opacity even though the conclusions are exactly the same if one talks about Knudsen number. It is easily to argue that, differently from the 1D case, we cannot expect the simulations to exhibit universality by fixing $\tau_0 T_0/(\eta/s)$, as will be clear below. 
\\	
In Fig.\,\ref{fig:double_pullback} we show the integrated normalised moments as a function of $t/R$ for simulations corresponding to several system size values $R$ and viscosity $\eta/s$ ranging from typical $pp$ to $AA$ collisions. It is straightforward to see that the three violet curves converge to a unique behaviour and the same is true for the three blue curves, in both cases for $t/R<1$: these are the equivalent of the pull-back attractors observed in the 1D case. The two groups are characterised by two different values of opacity $\hat \gamma$, respectively $\hat \gamma = 3.56$ and $\hat \gamma = 12.8$, and the same behaviour has been observed for a wider range of $\hat \gamma$, down to $\hat \gamma \lesssim 1$. Therefore the opacity defines a universality class in which systems share the same behaviour in terms of normalised moments after a certain scaled time $t/R$ which always occurs immediately after the minimum, that is when the collisions start to dominate with respect to the initial longitudinal expansion, exactly as happens in 1D.\\
It is meaningful, however, to understand how this pull-back attractor compares to the one seen in 1D. Indeed, one could have expected that some information about the relaxation time should appear in the scaling of the time $t$, as suggested by the 1D systems. Unfortunately, this cannot be the case if $R$ is finite, since, even if initially the moments follow the 1D curves, as observed at $t\approx R$ they depart from them due to the transverse expansion: it happens always at $t/R=1$, but at a different $t/\tau_{eq} \propto t/\mfp$. By considering the expression of $\hat \gamma$ in Eq.\,\eqref{eq:opacity_link_with_1D} one sees that the 1D case ($R\to \infty$) corresponds to a case with infinite opacity. Therefore, one could look at the ratio $\hat \gamma/R = (\tau/\tau_{eq})_0$ which gives a finite quantity. As depicted in the previous Chapter, if one keeps this ratio constant in 1D the full dynamics in terms of $\tau/\tau_{eq}$  is fixed and one single curve is observed. Moreover, in the 1D case this scaled time can be considered by itself an estimation of the Knudsen number, defined as the ratio between the inverse of a gradient and the mean free path. Indeed, $\tau_{eq}\sim\lambda_{\text{mfp}}$ and $\p_\mu u^\mu = \tau^{-1}$, therefore:
\begin{equation}
    \IKn = \frac{(\p_\mu u^\mu)^{-1}}{\lambda_{\text{mfp}}}=\frac{\tau}{\tau_{eq}}.
\end{equation}
Thus, when we fix the $(\tau/\tau_{eq})_0$ we are basically choosing an universality class in terms of $\IKn$. Differently from what happens in the full 3+1D case, however, the very simple 0+1D dynamics and the very few relevant scales allow a rescaling across universality classes as well.

\chapter{Universality in the collective flows}
\label{chap:knudsen}

\subsection*{Collective flows}

Immediately after the collision, an initial distribution with a huge energy density is left in the central region. However, the shape of such a medium is far from being isotropic. In the naive picture of two disks colliding with a certain impact parameter $b$, this region is characterised by a typical almond shape, which could be characterised by the geometrical ellipticity. In a more refined picture which includes event-by-event fluctuations, the initial shape of the fireball has a fully irregular shape, which is expected to reduce to an almond only if an average over the events, namely in the same centrality class, is performed. However, the event-by-event fluctuations do have a measurable impact on the final observables and therefore one has to take into account the irregularity of the initial state to reproduce experimental data. In particular, different models have been developed to generate realistic initial profiles. Here we have already briefly illustrated the Monte Carlo Glauber (Par.\,\ref{subsec:mc-glauber}) and in the following other models such as \trento\,are going to be used. This irregular shape, however, can be described via an (infinite) series of coefficients, called eccentricities:
\begin{equation}
	\varepsilon_n = \frac{\int d^2x\, r^n\, e^{in\phi} f(x,p)}{\int d^2x\, r^n\, f(x,p)}.
\end{equation}
Namely, the ellipticity $\varepsilon_2\ne 0$ indicates an almond shape, $\varepsilon_3\ne 0$ a triangular shape,  $\varepsilon_4\ne 0$ a quadrupolar shape and so on. Obviously, a realistic initial condition has several $\varepsilon_n\ne 0$, even though only a few of them generate a significantly large effect in the final observables. Moreover, this information about the initial state is directly inaccessible, since one has to deal only with the final distribution of the produced hadrons. The initial eccentricities, however, determine late-time collective observables, such as the anisotropic flows $v_n$. The interaction happening within the medium, indeed, convert this spatial anisotropy, which manifests as pressure gradients within the medium, into a corresponding momentum anisotropy, and in the meanwhile tend to spatially isotropise the medium. The efficiency of this conversion is the first indirect measure ever performed of the interaction strength of the hot QCD matter, which was possible thanks to the successful hydrodynamic descriptions at very small $\eta/s$. The observation of such collective flows also in small systems as $pp$ and $pA$ collisions has challenged the community, since these really small and short-living systems are not expected to behave like a fluid or to develop a pseudo-hydro response to the initial deformation. It is interesting, for instance, to disentangle the initial state effects, since a certain momentum anisotropy can be present already in the initial stage of the collision, and the late-time contribution due to the conversion of $\varepsilon_n$ to $v_n$. \\
An expansion of the distribution function can be performed in terms of the anisotropic flows:
\begin{equation}
	\frac{dN}{d\phi\,p_T\,dp_T} \propto 1 + 2 \sum_{n=1} v_n (p_T) \cos[ n(\phi_p - \Psi_n) ].
\end{equation}
where $\Psi_n$ are the reaction plane orientations for the different flows. The differential Fourier coefficients $v_n(p_T)$ at a specific time $t$ and rapidity $\eta_s$ can be computed as:
\begin{equation}\label{eq:differential_vn}
	v_n (p_\perp,t,\eta_s) =  \frac{\displaystyle\int d^2\mathbf x_\perp \displaystyle\int \dfrac{dp_z}{2\pi} \displaystyle\int d\phi_p\,  \cos [ n(\phi_p - \Psi_n) ] \,f(\vet x_\perp,t,\eta_s,p)}{\displaystyle\int d^2\mathbf x_\perp \displaystyle\int \dfrac{dp_z}{2\pi} \displaystyle\int d\phi_p\, f(\vet x_\perp,t,\eta_s,p)}.
\end{equation}
In the following, the dependence on $t$ and $\eta_s$ will be dropped to simplify the notation.
If one wants the integrated flows it is necessary to integrate in the formula above both the numerator and the denominator, not only over $p_z$ and $\phi_p$ but over the three momentum $d^3\mathbf p$:
{
	\begin{equation}\label{eq:intergrated_vn}
		v_n= \frac{\displaystyle\int d^2\mathbf x_\perp \displaystyle\int \dfrac{d^3\mathbf p}{(2\pi)^3}  \cos[ n(\phi_p - \Psi_n) ] f(x,p)}{\displaystyle\int d^2\mathbf x_\perp \displaystyle\int \dfrac{d^3\mathbf p}{(2\pi)^3} f(x,p)}.
	\end{equation}
}

It has been found that, for a broad range of $\varepsilon_n$ and interaction strength, the $v_n$ response  to a given $\varepsilon_n$ is approximately linear.\\
In this Chapter we are going to investigate the development of the collective flows in a broad range of collision systems. In the beginning, in order to start with the simplest scenario and to disentangle different effects, we deal with a conformal medium with smooth initial conditions slightly deformed in coordinate space so to have only one eccentricity sensitively different from 0. In this case we also study the impact of an initial $v_2$ on the final observable. Later on, we move to progressively more realistic situations, including a non-conformal equation of state and event-by-event fluctuations.

\subsection*{Initial conditions}
\label{subsec:eccentricities}
Differently for what has been done in the previous Chapter, the distribution function is now given an asymmetric initial condition also in the transverse plane. In particular the spatial initial symmetric distribution function $f_0(x,p)$ at $\tau=\tau_0$ used in the previous Chapter:
\begin{equation}
	f_0(x,p) \propto\gamma_0 \exp(-x_\perp^2/R^2) \theta(|\eta_s|- 2.5),   
\end{equation}
characterised by a finite extension in $\eta_s$ and a Gaussian profile in the transverse plane, is deformed in order to induce eccentricities $\varepsilon_n$ able to mimic anisotropies due to the initial elliptic shape and initial state fluctuations~\cite{Alver:2010gr, Plumari:2015cfa} as firstly done in \cite{Plumari:2015sia}. In the literature, this is usually achieved by deforming the initial energy density, as in hydrodynamics or RTA kinetic theory works. In our transport simulation, instead, the initial conditions are specified by the test particles' position and thus it is sufficient to shift by a small amount these coordinates. Introducing the complex notation $z=x+iy$, an anisotropy in transverse plane $\varepsilon_n$ is generated by shifting $z$ according to 
\begin{equation}
	\label{deformation}
	z\to z+\delta z\equiv z-\alpha_n \bar z^{n-1},
\end{equation}
where 
$\bar z\equiv x-iy$, and $\alpha_n$ is a small real positive quantity chosen to get the desired correction. More in detail, to the leading order in $\alpha_n$, one obtains
\begin{eqnarray}\label{eq:eccentricity}
	\varepsilon_n &=& \dfrac{ \sqrt{ \langle x_\perp^{n} \cos(n\phi)\rangle^2 + \langle x_\perp^{n} \sin(n\phi)\rangle^2 } }{\langle x_\perp^n \rangle} \simeq \nonumber \\ 
	&\simeq &n\alpha_n\dfrac{\langle x_\perp^{2(n-1)}\rangle}{\langle x_\perp^n\rangle}.
	\label{eq:ecc_first_order}
\end{eqnarray}
{Specifically, starting with a Gaussian distribution in the transverse plane as we do, Equation\,\eqref{eq:ecc_first_order} can be analytically evaluated to find  the $\varepsilon_n$ as a function of the $\alpha_n$:
	\begin{equation}\label{eq:epsilon_vs_alpha}
		\varepsilon_n = \frac{n\,\Gamma(n)}{ \Gamma(1+n/2)} \alpha_n R^{n-2}.
	\end{equation}
	Apart from $\alpha_2$ which is dimensionless, for every other $n$ its value depends on the initial root-mean-square radius $R$:
	\begin{gather}
		\varepsilon_2 \simeq 2\, \alpha_2;\\
		\varepsilon_3 \simeq \frac{8\, \alpha_3\, R}{\sqrt{\pi}};\\
		\varepsilon_4 \simeq 12\, \alpha_4\, R^2.
	\end{gather}
}
This artefact allows to study the response functions $v_n/\varepsilon_n$ by isolating the contribution from each of the eccentricities. In particular, the cases with $n=[2,3,4]$ will be addressed.\\
The details on the momentum dependence of the distribution functions will be given in the different sections.

\section{Universality in conformal systems in terms of \texorpdfstring{$\hat \gamma$}{opacity}}

In this Section, following what has been done in the literature, we concentrate in the study of the anisotropic flows at fixed opacity $\hat \gamma$. As said in the previous Chapter, when looking at the moments $M^{mn}$, actually little or no difference can be seen between the opacity and the Knudsen number. We are going to see, instead, that quantities particularly sensitive to the transverse dynamics can help to identify the scaling parameter in the RBT framework.

\subsection*{Initial conditions in momentum space}

In momentum space, following \cite{Romatschke:2003ms, Nopoush:2014pfa}, we use a more general version of the Romatschke-Strickland distribution function:
\begin{equation}\label{eq:modifiedRS}
	f(p) \propto \gamma_0\exp{ -\left[ \sqrt{p_x^2 (1+\psi_0) + p_y^2 + p_w^2 (1+\xi_0)}/\Lambda_0 \right]}
\end{equation}
Here $p_w$ is the momentum associated with $\eta_s$, defined as $p_w = p_z \cosh(\eta_s) - p_0 \sinh(\eta_s)$.\\
We determine $\gamma_0$ and $\Lambda_0$ in order to have the initial particle  and energy density associated to the given $T_0$ and $\Gamma_0$. These conditions correspond to determine the parameter $\Lambda_0$ and $\gamma_0$ by means of the following equations:
\begin{eqnarray}\label{eq:Lambdacondition}
	\Lambda_0 &=& T_0 \Bigg[ \frac{1}{2\sqrt{1+\xi_0}} + \frac{\sqrt{(1+\psi_0)(1+\xi_0)}}{4\pi} \int_0^{2\pi} d\phi \dfrac{ \arctan \sqrt{\frac{\xi_0 - \psi_0 \cos^2\phi}{1+\psi_0\cos^2\phi}} }{ \sqrt{(\xi_0 - \psi_0 \cos^2\phi)(1+\psi_0\cos^2\phi)^3 }  } \Bigg]^{-1} \nonumber\\
	\gamma_0 &=& \frac{\Gamma_0 T_0^3}{\Lambda_0^3} 
\end{eqnarray}
We fix thorough this section $T_0=0.5$ GeV and $\Gamma_0=1$. The two parameters $\xi_0$ and $\psi_0$ determine respectively the longitudinal and azimuthal anisotropy in momentum space. In particular, a $\psi_0\ne 0$ allows us to mimic an initial azimuthal anisotropy which corresponds to a finite $v_2$, in order to simulate scenarios that in approaches like the Colour Glass Condensate could contribute to the measured finite $v_2$ in $pp$ and $pA$ \cite{Krasnitz:2002ng, Schenke:2015aqa, Lappi:2015vta,Mantysaari:2017cni, Greif:2017bnr, Schenke:2019pmk}.\\

A more refined study can be done, however, if one is able to assign a precise $v_2(p_T)$ to the initial distribution function. In particular, one can adapt the artefact used to mimic an initial $\varepsilon_n$ to deform the momentum distribution function in a similar way. Despite the basic idea is the same, the formula\,\eqref{eq:epsilon_vs_alpha} cannot be trivially used in the momentum case, since the expression of the $v_n$ is qualitatively different from the $\varepsilon_n$ one. By performing the deformation of $z= p_x + ip_y$ into $z - \alpha\bar z$, indeed, one finds simply that
$$ v_2(p_T) = -\alpha. $$
Please notice that this shifting must be performed in a self-consistent way, given that it slightly changes the modulus of the momentum of the particle as well:
\begin{gather}
	p_T ' = p_T \sqrt{ 1- 2\alpha\cos(2\varphi_p) };\\
	v_2 (p_T'(\alpha)) = -\alpha,
\end{gather}
with the latter being the equation to be solved in terms of $\alpha$ given a certain $v_2(p_T)$.

\subsection{Response functions \texorpdfstring{$v_n/\varepsilon_n$}{}}

\label{sec:flows}
In this section we analyse the time evolution of the anisotropic flow coefficients $v_n$ for several systems with different values of transverse size $R$ and specific viscosity $\eta/s$, in order to explore a wide range of opacity values $\hat \gamma$, going from small systems like $pp$ or $pA$ to larger ones ($AA$).  The initial eccentricities are fixed by choosing the $\alpha_n$ values for different $n$ in order to get $\varepsilon_n=[0.1, 0.2]$ which correspond to typical hypothesised initial eccentricities produced in central and mid-peripheral collisions \cite{Plumari:2015cfa}. In this section we will make use of the integrated anisotropic flows as in Eq.\,\eqref{eq:intergrated_vn}; since there is no $n$-dependent event plane, $\Psi_n=0$ by construction. 

\begin{figure*}[t]
	\centering
	\includegraphics[width=.32\linewidth]{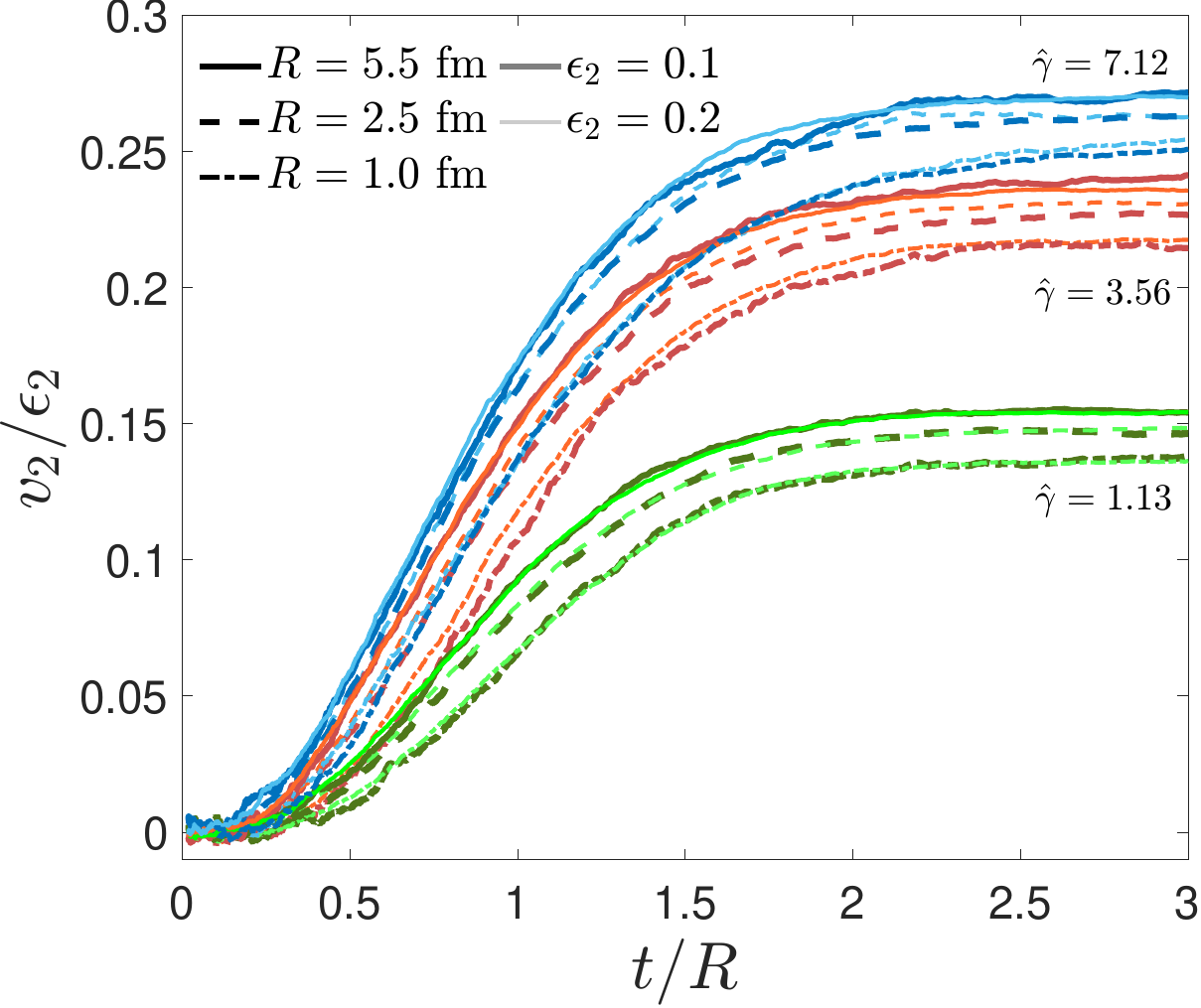}
	\includegraphics[width=.32\linewidth]{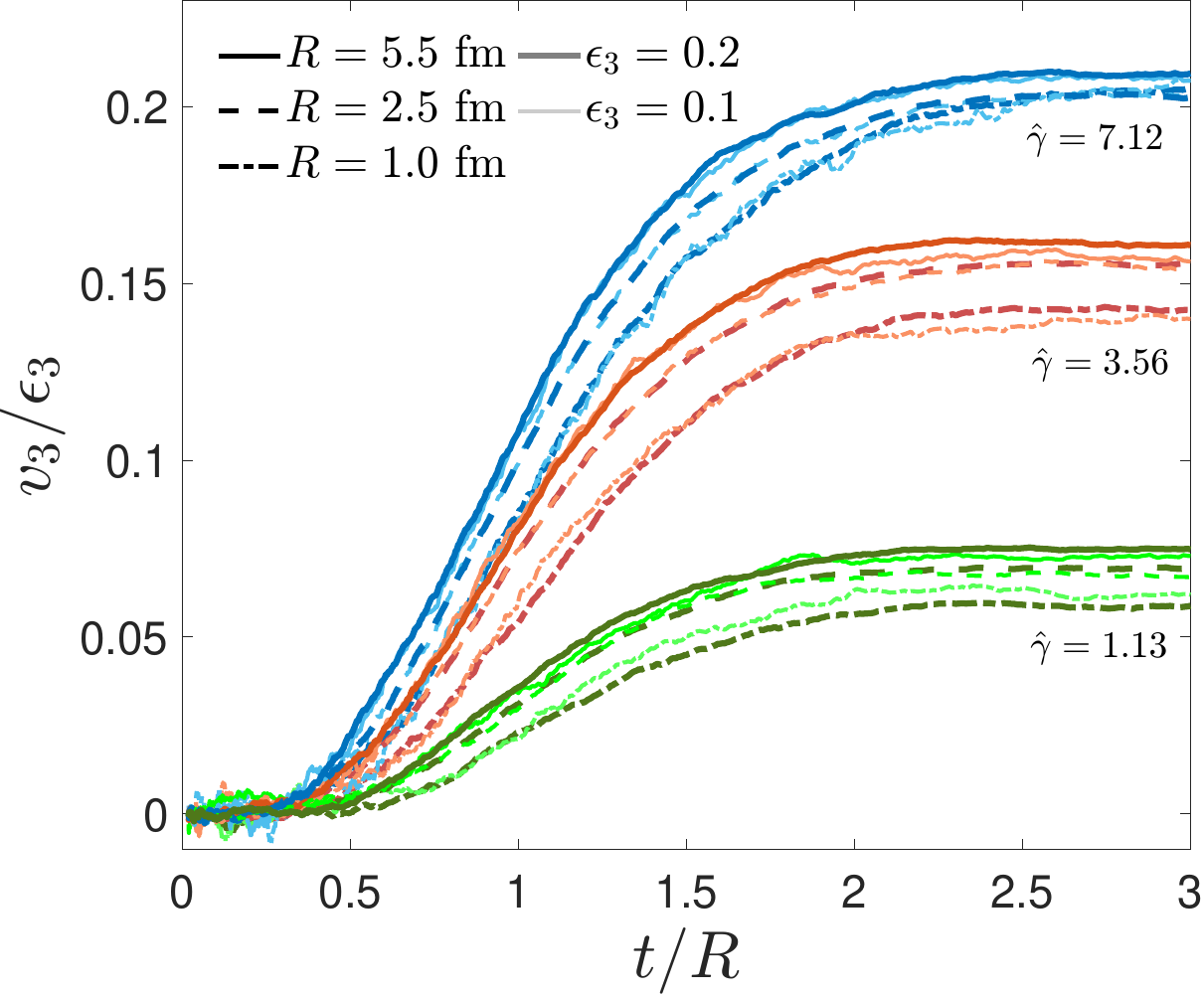}
	\includegraphics[width=.32\linewidth]{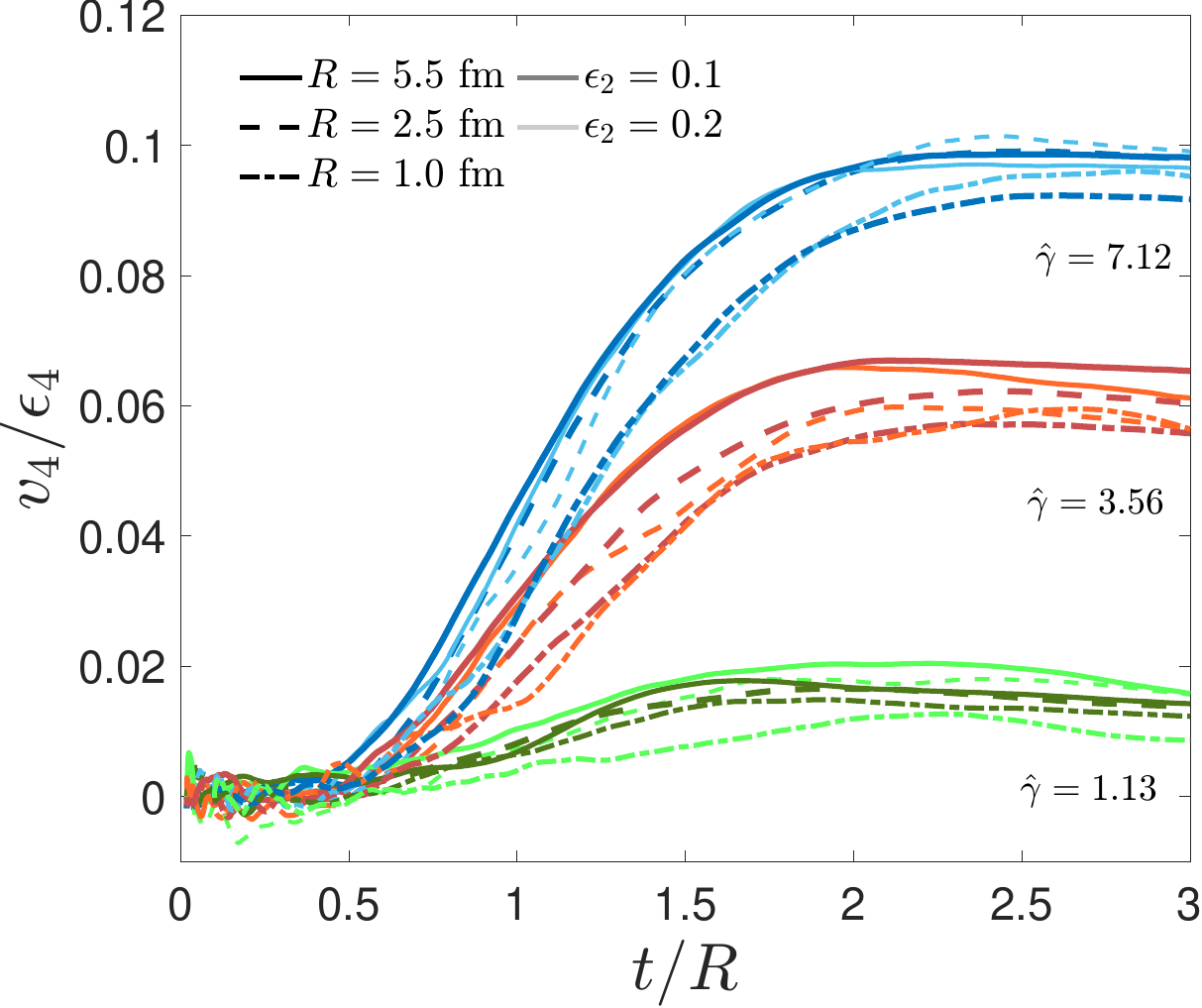}
	\caption{Response function $v_n/\epsilon_n$ for different values of initial eccentricities and different values of opacity $\hat \gamma$. The details on the corresponding $\eta/s$ can be found in Table\,\ref{tb:opacity_values}.}
	\label{fig:response_vn}
\end{figure*}

We study the time evolution of the linear response ratio $v_n/\varepsilon_n$ for $n=[2,3,4]$. In Fig.\,\ref{fig:response_vn} we show the response functions as a function of $t/R$ for three different values of $\hat \gamma$ = [1.13, 3.56, 7.12]. For fixed $\hat \gamma$ we explore three different initial sizes $R=[1.0, \, 2.5, \, 5.5]$ fm so that the value of $\eta/s$ is determined by the definition of $\hat \gamma$ (see Table\,\ref{tb:opacity_values}). Thinner, lighter lines correspond to $\varepsilon_n=0.1$, while bolder, darker lines to $\varepsilon_n=0.2$.\\
We observe that the curves cluster in three branches according to the opacity $\hat \gamma$, thus agreeing, in a first approximation, with the approach of other studies from RTA and ITA \cite{Ambrus:2022oji, Kurkela:2019kip, Kurkela:2020wwb}. As expected, the larger the value of the opacity, the more efficiently the system is able to convert the initial anisotropy in coordinate space to the momentum space.
Notice that the $v_n/\varepsilon_n$ curves all saturate at $t\sim 2R$, that is what we expect since at this time scale the system is about to decouple (see the inset in Fig.\,\ref{fig:collisions}). As expected, we observe that the curves are independent of the initial eccentricity $\varepsilon_n$, at least in the regime of small deformations.
Even though we are mainly interested in the behaviour of the response functions at $t\to \infty$, we can also look at early times $t<1$ fm to analyse how the anisotropic flows are developed. We checked that, in agreement with previous studies \cite{Borghini:2022qha, Borrell:2021cmh}, in the particle-like regime (small $\hat \gamma$) $v_n\propto t^{n+1}$, while in the hydrodynamic regime (large $\hat \gamma$)  $v_n\propto t^{n}$, with the exponent smoothly going from $n$ to $n+1$ with increasing $\hat \gamma$. In addition, we found that, for a fixed $\hat \gamma$ value, a similar dependence is present also on $R$, with larger values of $R$ providing smaller exponents. As far as the $v_2$ is concerned, we found, within the investigated range of $R=1-5.5$ fm, $n=1.9-2.3$ for $\hat \gamma = 7.12$, while $n=2.4-2.9$ for $\hat \gamma=1.13$. A similar trend is observed also for $v_3$ and $v_4$.\\
{One may also wonder whether there is a dependence on $\xi_0$ in the development of the response function, i.e. if a longitudinal anisotropy in the pressure has an impact on the transverse flow dynamics. In fact, we checked that the results are independent on the initial parameter $\xi_0$ once the initial transverse energy density ${dE^0_\perp}/{d\eta_s}$ is fixed, as one can see from Eq.\,(\ref{eq:our_opacity}). In other words, if one fixes $T_0$ and changes $\xi_0$ a different ${dE^0_\perp}/{d\eta_s}$ is assigned initially, which results in a different value for the opacity (or Knudsen number) and therefore to a different response; on the contrary, if  $\xi_0$ is modified but the initial energy at midrapidity is kept fixed by a suitable modification of $T_0$, no impact is seen on the collective flows.}\\
Going  more into detail, however, we must focus on the difference in the response functions, which is up to 10\% within the same group with fixed $\hat \gamma$, with a monotonic behaviour with the initial transverse size $R$. Since $\hat \gamma \propto R^{3/4}/(\eta/s)$, keeping fixed $\hat \gamma$ and moving to higher $R$ means also going to higher $\eta/s$. This suggests that, for a fixed value of opacity, going to larger values of the transverse radius has a major impact on the flows than moving to larger $\eta/s$, even though the dependence on the radius is via the $R^{3/4}$ power.\\
These observations show that the perfect universality emerging in the context of RTA and ITA in terms of opacity is somehow lost in our approach, which is not surprising, since, as said previously, if a natural dimensionless parameter has to be identified in the RBT method, the most likely candidate would be the $\KnR=R/\langle \lambda_{\text{mfp}}\rangle$.
{Indeed, as one can see by looking in detail at Fig.\,\ref{fig:gamma_def}, different curves in the same opacity class show a slight deviation in $\KnR$, which can be held responsible for the discrepancies observed in the $v_n/\varepsilon_n$. Moreover, the deviation in the response functions increase with the increasing of the difference in $\KnR$. Such a discrepancy can be found to be regular in Fig.\,\ref{fig:response_comparison} which shows the RBT results in the limit $t\gg R$, corresponding to the saturation value of the $v_n/\varepsilon_n$ for a much wider range of opacity values. For each value of the opacity, three different points correspond to the different radii used above. All the values are obtained with an initial distribution $Y=\eta_s$ and $\varepsilon_n=0.05$. For the range of opacity values here considered, the spreading for different $R$ values is $<10\%$, except for the quasi-free-streaming cases of very small opacity ($\hat\gamma <1$); in general, the relative difference between the results becomes smaller at larger opacity, i.e. closer to a fully hydrodynamic behaviour. It is also possible to see that the ordering in $R$ previously observed is well respected.\\

\begin{figure}
	\centering
	\includegraphics[width=.6\linewidth]{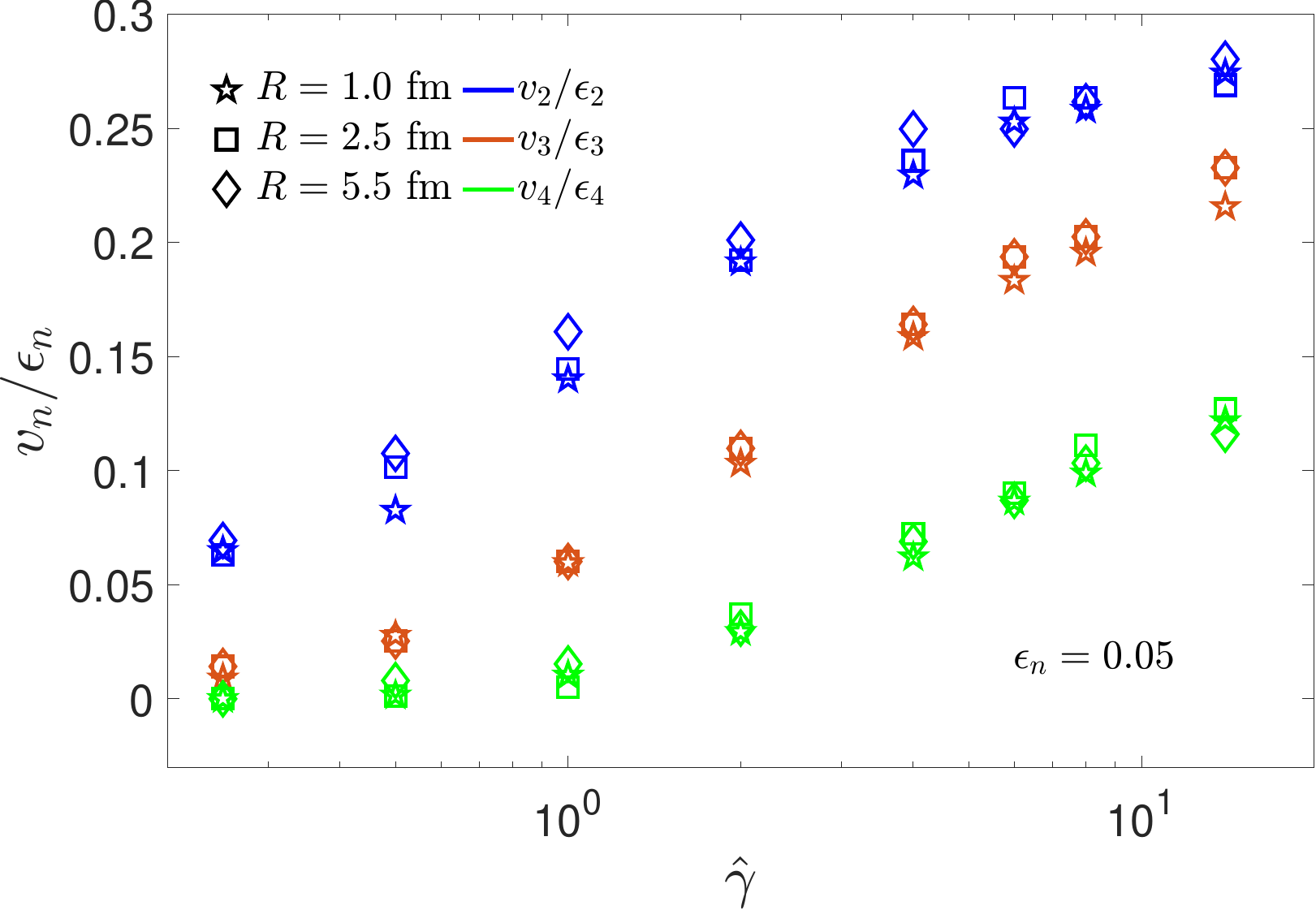}
	\caption{Integrated $v_n/\epsilon_n$ as a function of the opacity $\hat \gamma$ in the limit $t\to \infty$. Different colours correspond to different harmonics, while different point styles to different radii.}
	\label{fig:response_comparison}
\end{figure}

It is therefore legitimate to investigate whether the Knudsen number behaves as the true scaling parameter of the model. This could be done by empirically fixing the $\KnR$ and looking at how different system respond to initial eccentricities: this study is addressed in Section\,\ref{subsec:fixed_kn}, where it is also extended to non-conformal systems. In the remaining part of this section, instead, we show further results obtained at fixed opacity in which an initial non-vanishing $v_2$ is assigned to the system.

\subsection{Dissipation of initial \texorpdfstring{$v_2$}{v2}}

\begin{figure}
	\centering
	\includegraphics[width=.515\linewidth]{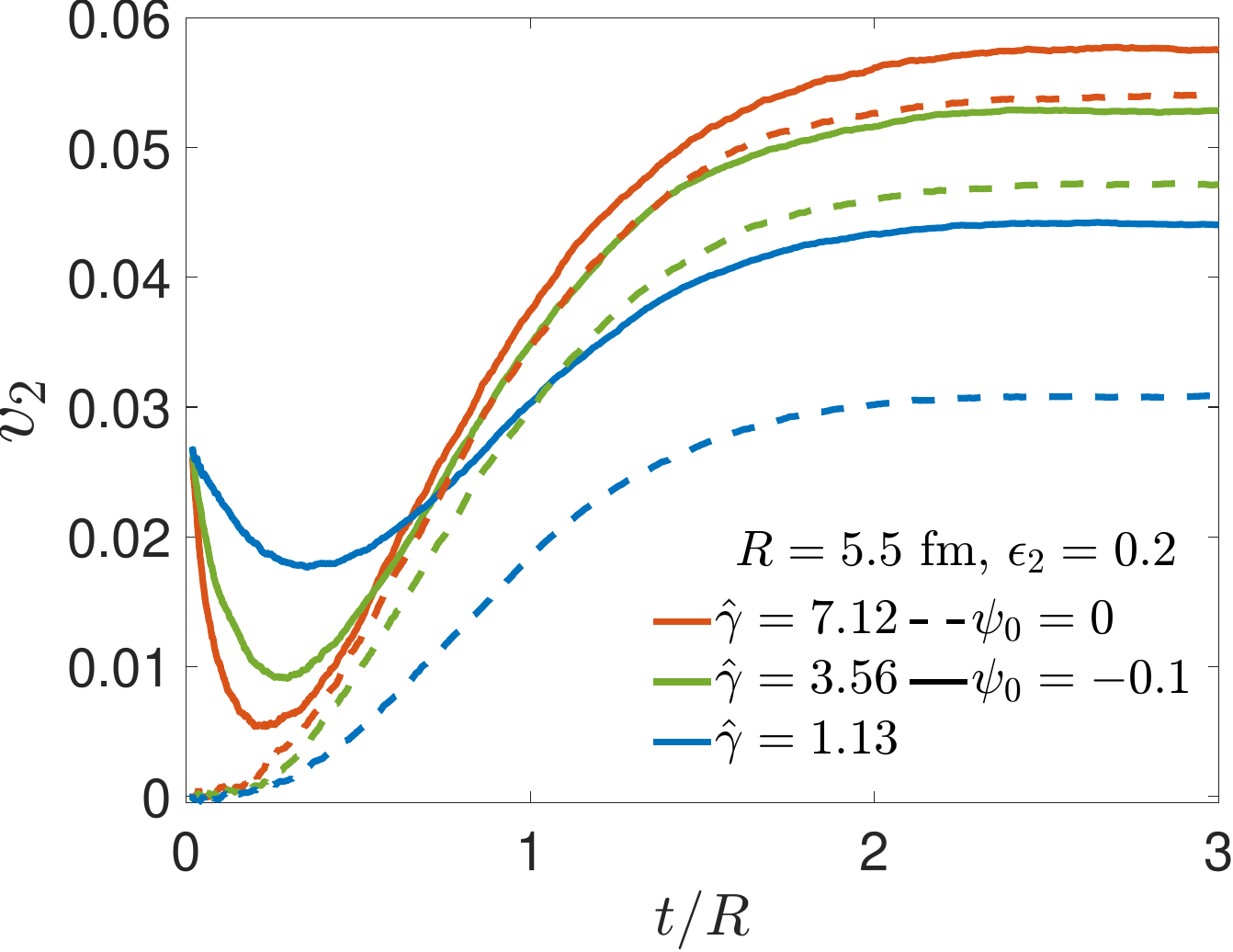}
	\includegraphics[width=.445\linewidth]{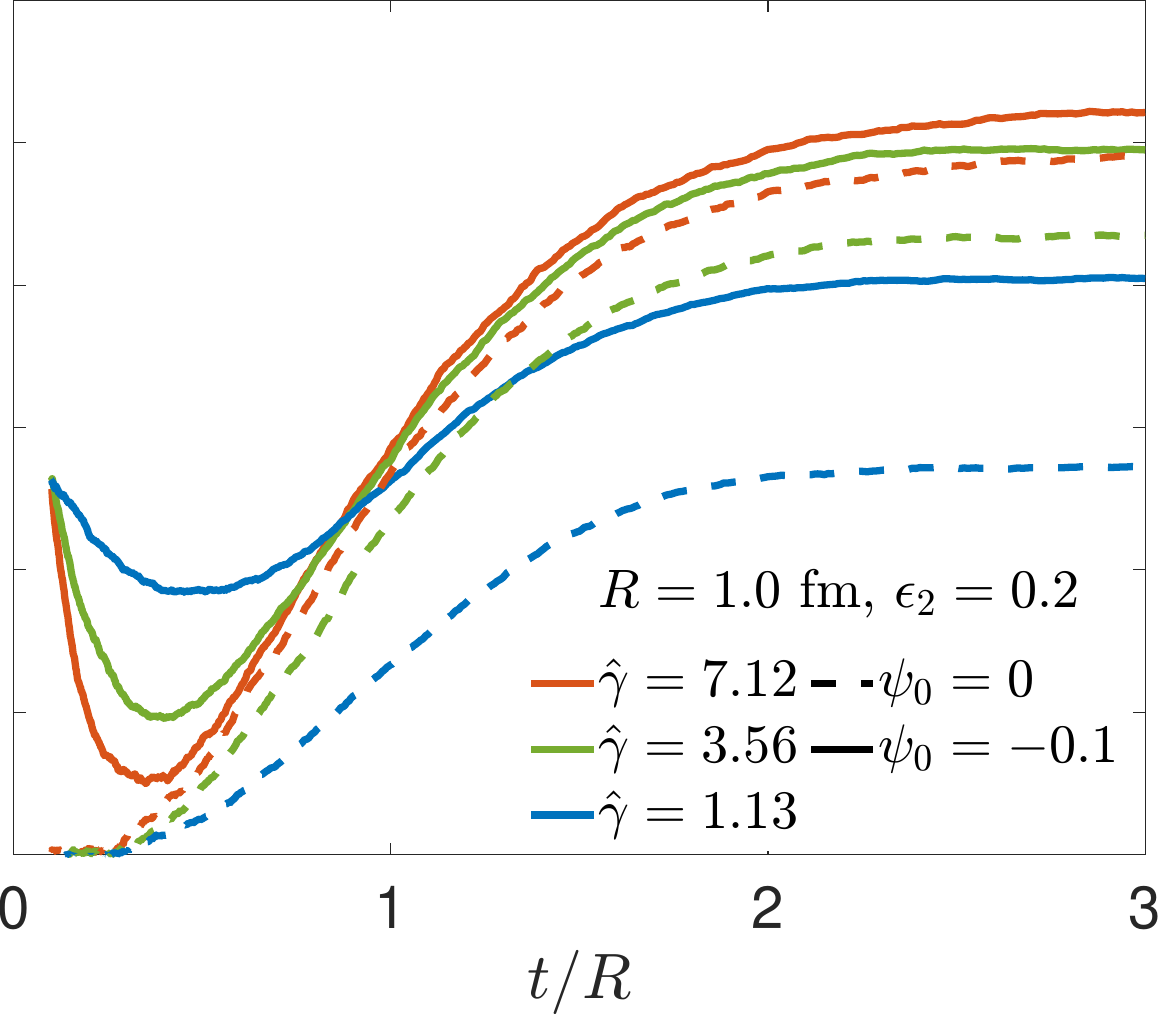}
	
	\caption{ Elliptic flow $v_2$ as a function of scaled time $t/R$ for $R=5.5$ fm (left panel) and $R=1.0$ fm (right panel), with initial eccentricity $\varepsilon_2=0.2$ and two different initial conditions in the momentum space $\psi_0=-0.1$ (solid lines) and $\psi_0=0$ (dashed lines). Different colours refer to different opacity $\hat \gamma$; the corresponding $\eta/s$ values can be found in Table\,\ref{tb:opacity_values}.}
	\label{fig:vn_dissipation}
\end{figure}

We now consider non-azimuthally symmetric initial distributions in the momentum space, in order to mimic an initial correlation in momentum space typical of CGC initial conditions \cite{Krasnitz:2002ng, Schenke:2015aqa, Lappi:2015vta,Mantysaari:2017cni, Schenke:2019pmk}: the system starts its evolution with a non zero elliptic flow $v_2(\tau_0)$. As explained above, this is easily achieved by fixing $\psi_0 \ne 0$ in Eq.\,\eqref{eq:modifiedRS}: choosing $\psi_0=-0.1$ provides an initial $v_2(\tau_0)\approx 0.025$, of the same order of that found by CGC calculations \cite{Greif:2017bnr}.\\
Firstly, we investigate whether and how the system loses memory about these correlations in momentum space. In Fig.\,\ref{fig:vn_dissipation}, we consider the same three values of opacity $\hat \gamma = [7.12, 3.56, 1.13]$ and two different system transverse sizes: $R=5.5$ fm in the left panel, which corresponds to a typical $AA$ collision, and $R=1$ in the right one, resembling small collision systems.The initial eccentricity in coordinate space is $\epsilon_2=0.2$. For fixed $\hat \gamma$, the behaviour of the elliptic flow is quite independent of the transverse size; we checked it also for intermediate values of $R$. The initial elliptic flow is strongly dissipated in the early times of the evolution at about {$t\approx 0.3 - 0.5$ fm}, depending on the opacity class considered. When the $v_2$ production rate, which converts the initial anisotropy in coordinate space $\epsilon_2(\tau_0)$ to the momentum space, exceeds the $v_2$ dissipation rate, the curve starts to rise; at $t/R\approx 2$ the system begins to decouple and the $v_2$ saturates. In the high opacity case ($\hat \gamma=7.12)$ the system lose almost completely memory of the initial anisotropy, as one can see by comparing the final $v_2$ to the one obtained with $\psi_0=0$, that is for isotropic initial conditions in momentum space. This corresponds to $4\pi\eta/s=1-2$ for $R=5.5$ fm, as estimated for QGP, while for $R=1$ fm one needs to go down to $4\pi\eta/s\approx 0.5$ to have the same $\hat \gamma$. Even though the competition between the two processes of $v_2$ dissipation and production is present also for smaller opacity, for $\hat \gamma=3.56$ and even more for $\hat \gamma=1.18$ the system keeps memory of the initial anisotropy, with a final $v_2$ which keeps memory of $v_n(\tau_0)$. For instance, in a system with $R=1$ fm and $4\pi\eta/s = 1$, there is an impact on the final integrated $v_2$ which is $\gtrsim 15\%$, which goes up to $30-40\%$ for $4\pi\eta/s\approx 3$.  Not surprisingly, therefore, we find that, for the range of $\eta/s$ supposed to be explored in QGP, smaller size systems can be sensitive to initial correlations in momentum space. 

\subsection{Attractors in \texorpdfstring{$v_n/v_{n,eq}$}{vn/vn,eq}}

\begin{figure*}[t]
	\centering
	\includegraphics[width=.35\linewidth]{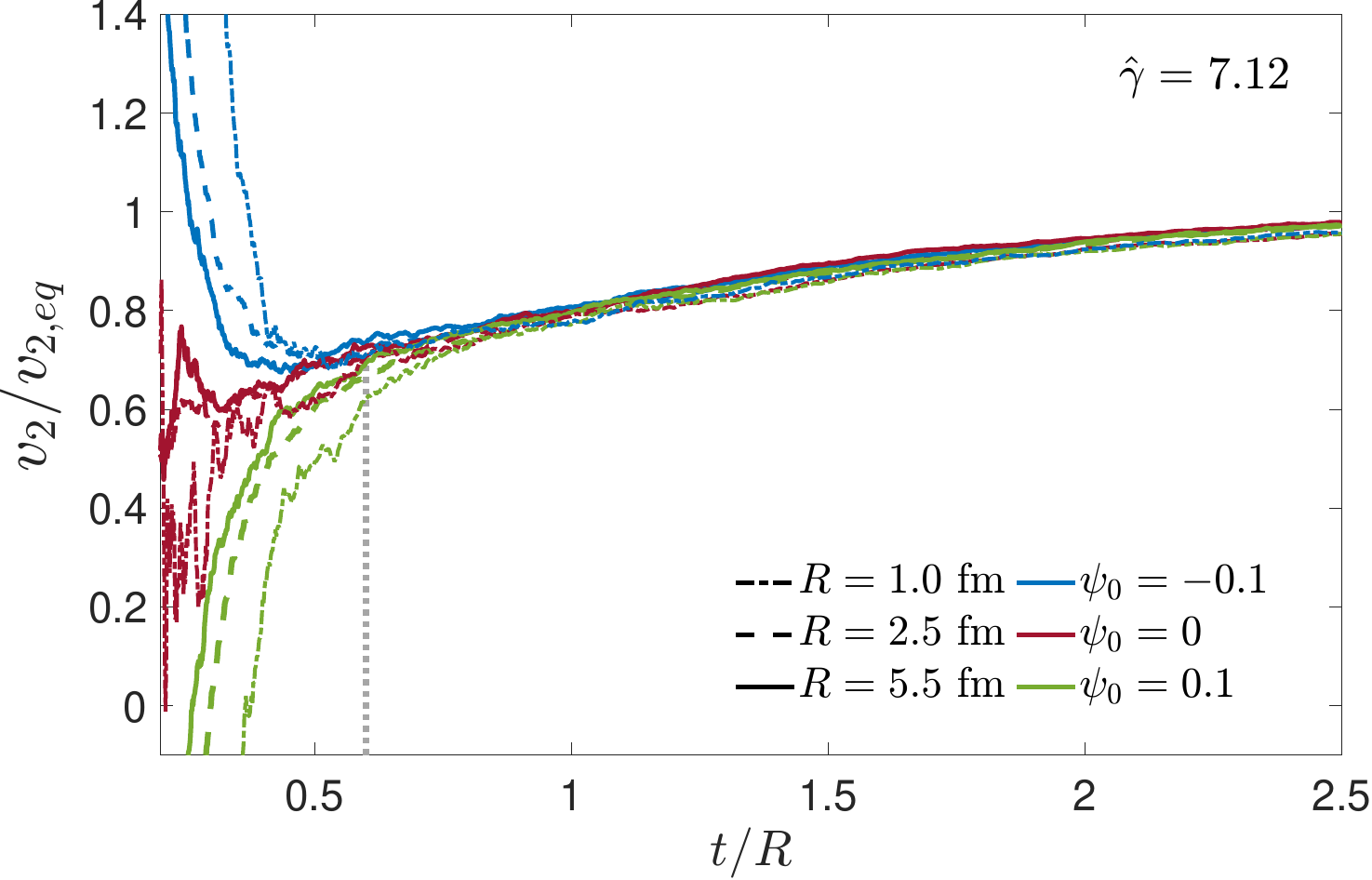}
	\includegraphics[width=.31\linewidth]{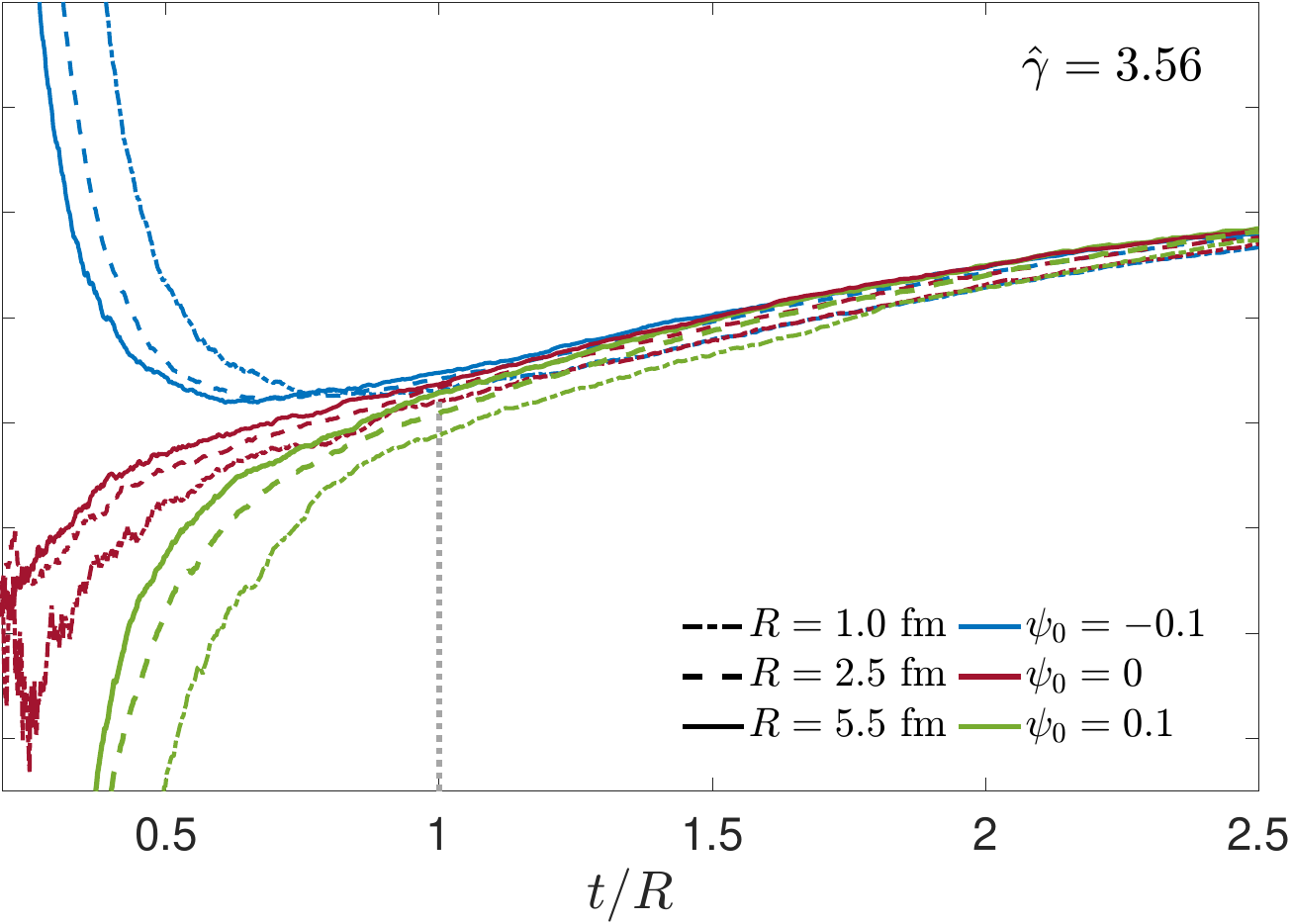}
	\includegraphics[width=.31\linewidth]{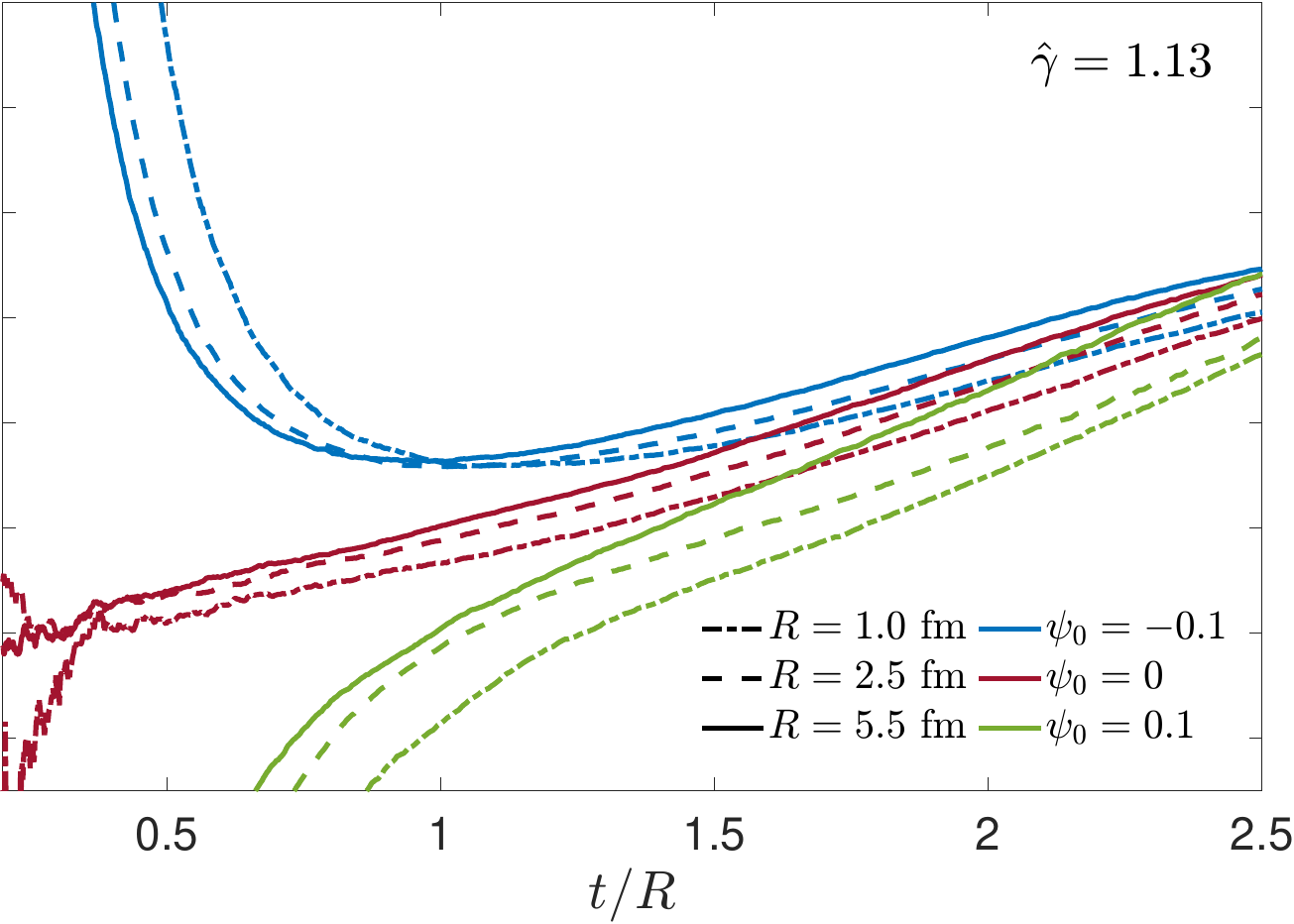}
	\caption{Normalised elliptic flow $v_2/v_{2,eq}$ for three different universality classes (from left to right) $\hat \gamma = [7.12, 3.56, 1.13]$. Different colours correspond to the three initial $\psi_0=[-0.1, 0, 0.1]$, while different line styles to the three radii $R=[1.0, 2.5, 5.5]$ fm. The corresponding $\eta/s$ values can be found in Table\,\ref{tb:opacity_values}.}
	\label{fig:v2veq}
\end{figure*}

Following what has been done for the momentum moments of the distribution function, it would be interesting to quantify how much the computed anisotropic flows deviate from those obtained by assuming a local thermal distribution function $f(x,p)\sim \Gamma(x)\exp( - p_\mu\cdot u^\mu (x) /T(x) )$, where as usual the primary variables $u^\mu(x)$, $\Gamma(x)$ and $T(x)$ are extracted locally in space and time from the simulation. The anisotropic flows computed at equilibrium $v_n^{eq}$ at midrapidity and at every specific time step are given by:
\begin{equation}\label{eq:equilibrium_flows}
	v_n^{eq}(t) = \frac{\displaystyle\int d^2\mathbf x_\perp \displaystyle\int \dfrac{d^3\mathbf p}{(2\pi)^3} \cos(n\phi) \Gamma(x_\perp,t) \exp\left( -\frac{ p_\mu\cdot u^\mu (\mathbf x_\perp,t)}{T(\mathbf x_\perp,t)} \right)}{\displaystyle\int d^2\mathbf x_\perp \displaystyle\int \dfrac{d^3\mathbf p}{(2\pi)^3} \Gamma(x_\perp,t)\exp\left( -\frac{ p_\mu\cdot u^\mu (\mathbf x_\perp,t)}{T(\mathbf x_\perp,t)} \right)}.
\end{equation}

One can thus define the normalised anisotropic flows $\overline v_n$:
\begin{equation}
	\overline v_n (t) = \frac{v_n (t)}{v_n^{eq}(t)}.
\end{equation}
When $t\to t_0$, since we have not introduced an initial flow $u_0^\mu=0$, $v_n^{eq}(t_0)=0$. At the same time, the initial $v_n$ is vanishing when we start with an azimuthally isotropic distribution, therefore the limit $\lim_{t\to t_0} \overline{v}_n$ is not well-defined and we have some numerical instabilities; on the other hand, if $v_n(t_0)\ne 0$ (due to $\psi_0\ne 0$), since still $v_n^{eq}(t_0)=0$, $\lim_{t\to t_0} |\overline v^n|=\infty$.
In the limit $t\to \infty$, if the system reaches thermalisation, we expect $\lim_{t\to \infty} \overline v_n=1$ for all the cases studied.\\
We want to study whether, similarly to what shown for the momentum moments of the distribution function, there exist universality classes in terms of the opacity $\hat \gamma$ in the evolution of the normalised anisotropic flows with attractors also for the $v_2$, seeing also if the anisotropic flows developed by the medium reach those expected assuming local equilibrium. In fact $\overline v_n$ quantifies the deviation of the integrated anisotropic flow with respect to the one assuming a thermal distribution with the same $T(x), \Gamma(x)$ and $u^\mu(x)$. \\
In Fig.\,\ref{fig:v2veq}, we show three different plots of $\overline v_2$ in terms of $t/R$ for three different values of opacity $\hat \gamma=[7.12, 3.56, 1.13]$ (left, middle and right panel respectively). The curves correspond to different systems with three radii $R=[1.0, 2.5, 5.5]$ fm (dot-dashed, dashed and solid line respectively) and different initial momentum anisotropies $\psi_0 = [-0.1, 0, 0.1]$ (blue, dark red and green lines respectively); see Table\,\ref{tb:opacity_values} for the corresponding $\eta/s$. We can see that for the two largest $\hat \gamma$ the plots are qualitatively similar: all the curves, irrespectively of the transverse size, $\eta/s$ and initial $v_2(t_0)$, converge to the same behaviour when $\overline v_2 \sim 0.6 -0.7$ and then saturate to 1. We observe that the scaled time $\bar t= t/R$ at which all the curves approach the attractor depends on the opacity class considered, in particular, the smaller the opacity, the later the system will reach the universal curve: specifically, $\bar t\approx0.6$ for $\hat \gamma=7.12$ and $\bar t \approx 1$ for $\hat \gamma=3.56$. For these cases, at $t=2R$, i.e. when we expect that the system is almost completely decoupled (see Sec.\,\ref{subsec:transverse_flow}), $v_2\approx 0.8 - 0.9\, v_{2,eq}$. Notice that the time scale at which this attractor is reached is larger than the one characterising the moments: this is due to the fact that the development of anisotropic flows is related to the transverse expansion, whose typical time scale is $R$, and not the initial longitudinal expansion.\\
At small-opacity ($\hat \gamma=1.13$), instead, the attractor behaviour seems to be partially broken, since different curves at $t=2R$ converge within a band of width $\sim 15 \%$ to a value $\bar v_2 \approx 0.8\pm 0.1$. This means that these systems thermalise more slowly and with different trends depending on the initial conditions, $\psi_0, R$ and $\eta/s$. According to Ref.1,\cite{Kurkela:2019kip}, the small opacity regime corresponds to the particle-like behaviour: therefore it is not surprising that the final collective flow is sensitively different with respect to that generated by a locally equilibrated system.

\section{Universality in non-conformal systems}

\begin{figure}
	\centering
	\includegraphics[width=0.7\linewidth]{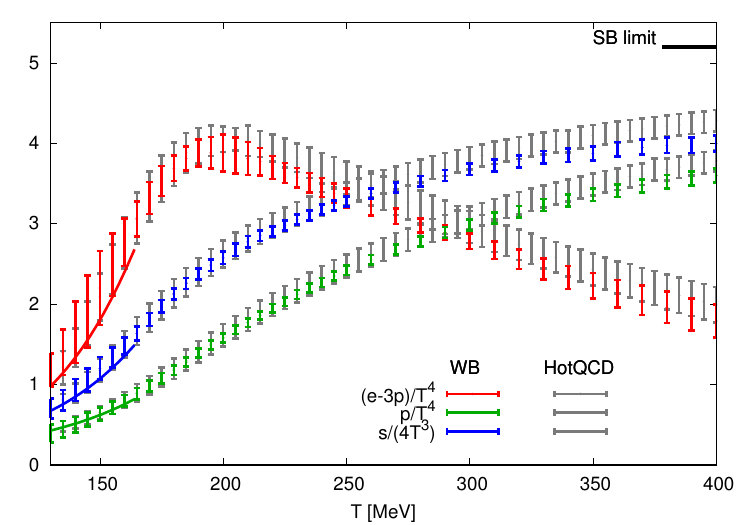}
	\caption{Lattice QCD results for trace anomaly, entropy density and pressure. The gray points are from the HotQCD collaboration \cite{HotQCD:2014kol}, while the coloured ones are from the WB collaboration \cite{Borsanyi:2013bia}. The figure also shows the Stefan-Boltzmann limit for the pressure and the scaled entropy and the Hadron Resonance Gas model predictions at low temperatures. Figure from \cite{Ratti:2018ksb}.}
	\label{fig:interactionmeasure}
\end{figure}

From lattice QCD results we know that the Equation of State (EoS) of the hot QCD matter is sensitively non-conformal, as visible via the interaction measure predictions in Figure\,\ref{fig:interactionmeasure}. Despite also conformal studies can capture many interesting features of this medium, a more realistic study has to include the effect of a finite mass. In the RBT approach, as shown in Chapter\,\ref{chap:RBT}, the implementation of a finite mass is straightforward, and one can easily investigate the impact of a different EOS in the dynamic. The presence of a finite mass strongly affects the relevant scales governing the system evolution: in particular, as discussed diffusively below,  one has to take into account that the speed with which a perturbation propagates within a medium depends on the equation of state, and therefore on the mass.\\
Notice that, in the non-conformal case, the system also has a bulk viscosity $\zeta/s$, that depends locally on mass, temperature and cross section. In Sec.\,\ref{sec:collision}  we explained how the local cross section is fixed to keep $\eta/s$ constant. As a side effect, the bulk viscosity is locally fixed as well, as a function of temperature, mass, cross section and fugacity. However, since the system is number-conserving, the bulk viscosity is expected to be almost vanishing; as one can estimate with analytical approaches by following for instance \cite{Ambrus:2023qcl}, $\zeta/s$ is of the order $10^{-4}$. We estimated it also numerically, by computing the bulk viscous pressure $\Pi$ and comparing it with viscous hydro results obtained with MUSIC \cite{Schenke:2010nt, Schenke:2010rr, Paquet:2015lta}: we observe a very good agreement between the two models if we fix $\zeta/s\sim 10^{-4}$ in MUSIC. A further extension of the code including also inelastic collisions will allow for a finite and realistic bulk viscosity, with the possibility to generate and destroy particles in order to keep the system in chemical equilibrium.

\subsection{Initial conditions in momentum space and code setup}
\label{sec:RBT}

In this section we keep a symmetric transverse distribution function in momentum space with $Y = \eta_s$ and implement the presence of a finite mass:
\begin{equation}
	f_0(x,p)=\gamma_0 \exp(-x_\perp^2/R^2) \theta(|\eta_s|- 2.5)\delta(Y-\eta_s)\exp{ \left(- \sqrt{p_x^2 + p_y^2 + m^2}/\Lambda_0 \right)}        
\end{equation}
In coordinate space, initial eccentricities are generated with the same artefact explained in Par.\,\ref{subsec:eccentricities}.
To get the results shown here, the number of test particles is of order 15M-20M. For each simulation (single curve) we run 30-60 events (until convergence is reached). Each space cell has $\Delta\eta=0.16$ and $\Delta x=\Delta y=$0.2 fm.

\begin{table}
	\caption{Parameters for the smooth initial conditions.}
	\centering
	\begin{tabular}{ccccc}
		\toprule
		$\IKn$ & $R$ [fm] & $4\pi\eta/s$ & $\tau_0$ [fm] & $T_0$ [GeV] \\
		\midrule
		1.30 & 5.0 & 12.57 & 0.5 & 0.50 \\
		1.30 & 3.0 & 5.03 & 0.3 & 0.33 \\
		\midrule
		3.20 & 5.0 & 5.03 & 0.5 & 0.50 \\
		3.20 & 3.0 & 2.012 & 0.3 & 0.33 \\
		\midrule
		6.45 & 5.0 & 2.515 & 0.5 & 0.50 \\
		6.45 & 3.0 & 1.006 & 0.3 & 0.33 \\
		\bottomrule
	\end{tabular}
	\label{tab:parameters}
\end{table}

\subsection{Knudsen Number in the massive case}
In the previous chapter, we presented for the conformal case the universality in the evolution of the Inverse Knudsen number. Here we use the different configurations illustrated in Table\,\ref{tab:parameters} with exactly fixed $\KnR$: in the left panel of Fig.\,\ref{fig:fig_1} the darkest lines demonstrate the universality of the evolution. For the non-conformal simulations the same initial conditions are used with a different mass: what one clearly sees (left panel of Figure\,\ref{fig:fig_1}) is that by changing the $m/T_0$ ratio the functional form of $\IKn(t/R)$ sensitively changes, with a consequent loss of universality. By looking at the Figure, the comparison between red solid and green dashed lines shows that the scaling behaviour persists only if the ratio $m/T_0$ is kept fixed.\\
This observation suggests that in this broader context a key role could be played by the speed of sound $c_s$: in the non-conformal case, indeed, $c_s=c_s(m/T)$, which means that information travels through the medium with a velocity governed by the $m/T$ ratio. Notice that, since $T$ depends on space-time coordinates, $c_s$ does as well.
One can define the characteristic time scale of a system as the time interval necessary for a perturbation to travel across the whole system, that is roughly $R/c_s$.
In the conformal case there is obviously no difference between different systems since $c_s^2=1/3$ without any temperature dependence, which allows to recover the typical scaling variable $t/R$; for a non conformal system, instead, one should replace it with $c_s t/R$. As shown in the right panel of Figure\,\ref{fig:fig_1}, with this choice of the scaled time it is possible to recover a quite good scaling, which is loss only when (see coloured circles) $t>2R$, i.e. when the system is largely decoupled.\\
	
\begin{figure}
	\centering
	\includegraphics[width=0.8\linewidth]{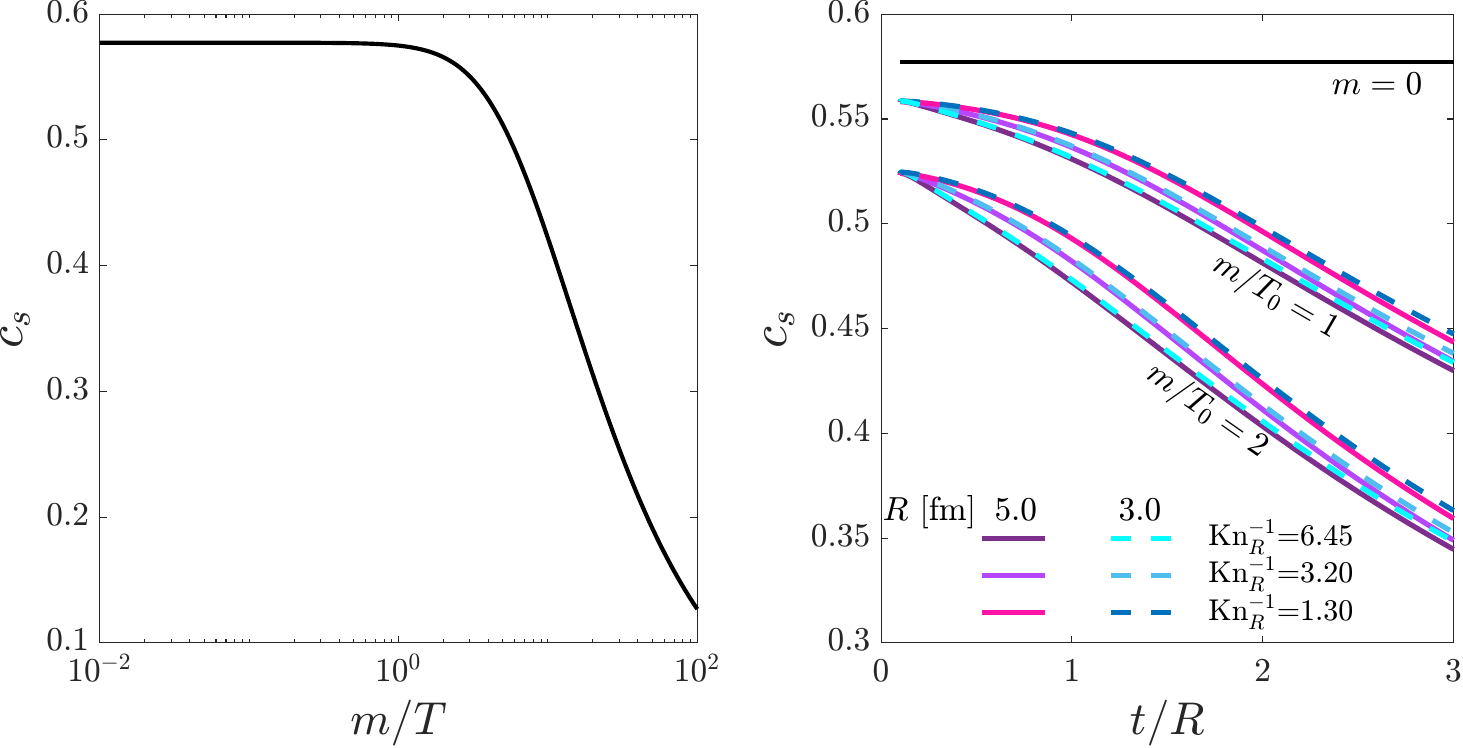}
	\caption{Left panel: Analytical curve of the speed of sound as a function of $m/T$ according to Eq.\,\eqref{eq:sound_velocity}. Right panel: global sound velocity as defined in the text for the different cases investigated. The two clusters are for the two values of $m/T_0=[1,2]$; solid lines are for systems with $R=5.0$ fm, dashed for $R=3.0$ fm (see Table\,\ref{tab:parameters} for more details); different colours account for different Knudsen number.}
	\label{fig:soundvelocity}
\end{figure}

The expression for the speed of sound in a number-conserving system of massive particles \cite{Ambrus:2023qcl} reads:
\begin{equation}\label{eq:sound_velocity}
	c_s^2 =\frac{4+z^2- g^2(z) + 3g(z)}{(3+z^2- g^2(z) + 3g(z))(g(z) +1)},
\end{equation}
where $g(z)$ is the ratio $e/P$:
\begin{equation}
	g(z)=\frac{e}{P} = z\frac{K_3(z)}{K_2(z)} -1,
\end{equation}
and $K_n(z)$ are the modified Bessel functions of order $n$. As one can see from the left panel of Fig.\,\ref{fig:soundvelocity} the $c_s$ is a decreasing function of the $m/T$ ratio, which converges to $1/\sqrt{3}$ in the massless limit.
In the systems under examination the mass is fixed, but the temperature depends on space-time coordinates. By following the assumption of local thermal equilibrium, we can compute the local $c_s(x)=c_s(m/T(x))$. In order to have a `global' sound velocity which only depends on time, we define
$c_s(t)=c_s(m/T_{\text{avg}}(t))$, where $T_{\text{avg}}(t)$ is the time-dependent weighted average of the temperature. We checked that the temperature average can be weighted by energy density or particle density, with a difference $<1\%$ between the two. In the right panel of Fig.\,\ref{fig:soundvelocity} the global $c_s$ for the cases under examination is plotted: as one can see, in terms of the scaled time $t/R$ curves with the same Knudsen number and same $m/T_0$ show a universal pattern also for the sound velocity evolution, while there is a sensitive difference between curves with same Knudsen number and different $m/T_0$, which make it necessary to adopt $c_s t/R$ as the new scaled time. Finally, the discrepancy between curves with same $m/T_0$ and distinct $\KnR$ are due to the difference in the temperature evolution $T(t)$. Notice also that the solid lines refer to systems with initial temperature $T_0=0.5$ GeV and masses $m=0.5, 1.0$ GeV, while dashed lines to $T_0=0.333$ GeV and masses $m=0.333, 0.666$ GeV, which proves once more that the key quantity is the $m/T_0$ ratio.

\subsection{Collective flows at fixed Inverse Knudsen number \texorpdfstring{$\KnR$}{}}
\label{subsec:fixed_kn}

\begin{figure}
	\centering
	\includegraphics[width=.7\linewidth]{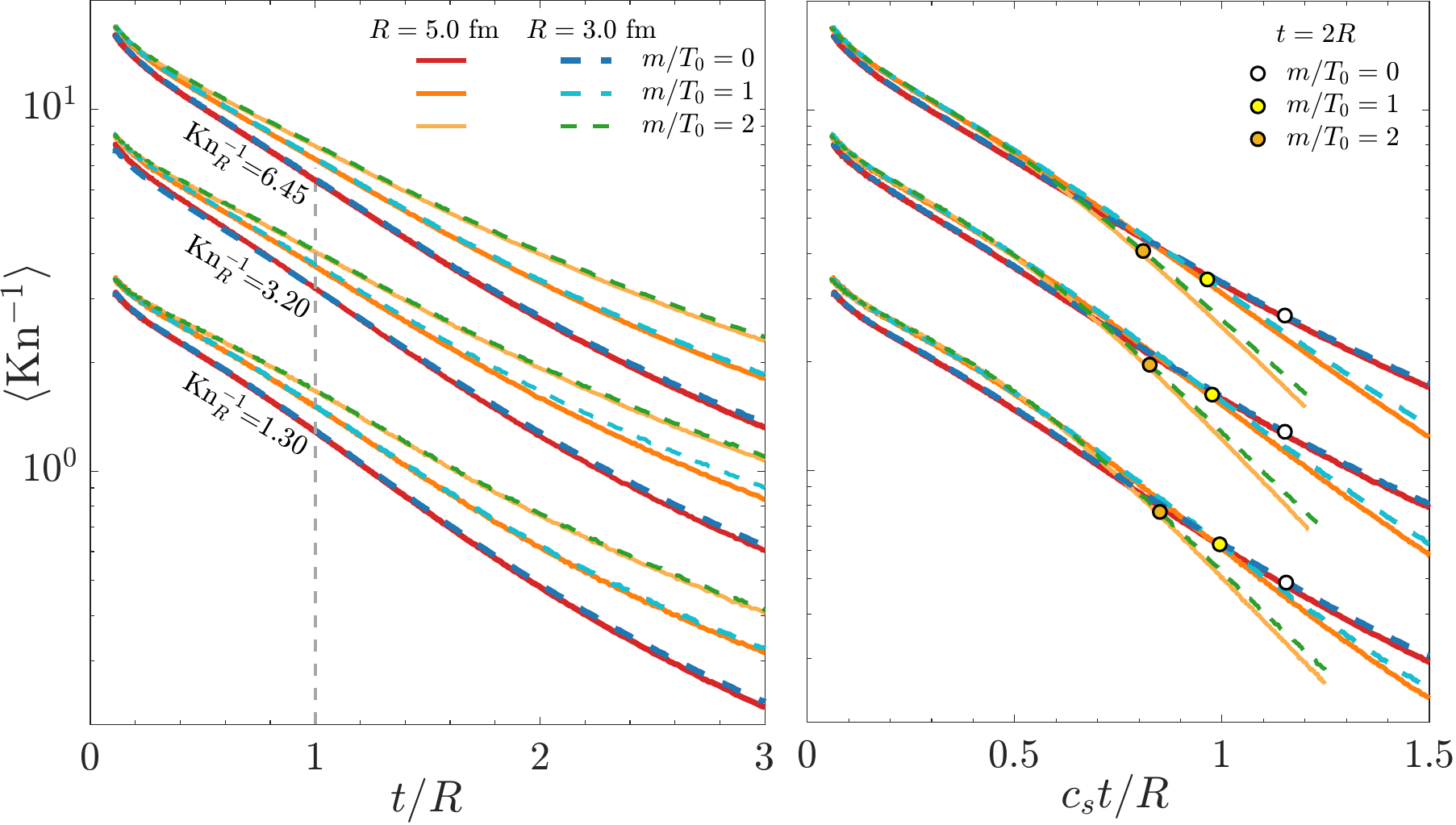}
	\caption{Inverse Knudsen Number for three different universality classes $\IKn=[1.30, 3.20, 6.45]$ and three different $m/T_0=[0,1,2]$ with respect to $t/R$ (left panel) and  $c_st/R$, which grants the partial scaling, broken only for $t>2R$ (right panel).}
	\label{fig:fig_1}
\end{figure}

In this section we report the response curves $v_n/\varepsilon_n$ ($n=2,3,4$) at fixed $\KnR$. We demonstrate how the universality is now perfectly achieved and investigate whether it were the case also in the non-conformal case, which has not been addressed before in the literature in the context of kinetic theory.\\
All the curves shown in Figure\,\ref{fig:fig_2} and\,\ref{fig:vn_pt_etast} have been obtained with $\varepsilon_n=0.2$; following what has been presented in the previous section, we checked that the results are independent of $\varepsilon_n$ in $[0.05-0.4]$, which is the range where both hydrodynamics and transport approaches show a linear dependence of the $v_n$ on the $\varepsilon_n$ \cite{Noronha-Hostler:2015dbi, Roch:2020zdl, Plumari:2015cfa}. Therefore, for clarity, we decide to show only the intermediate case with $\varepsilon_n=0.2$.
In the top panel of Figure\,\ref{fig:fig_2} we plot the scaled-time evolution of elliptic (left panels), triangular (central panels) and quadrupole (right panels) flow for three different $\IKn_R$ (see setups in Table\,\ref{tab:parameters}) and three values of $m/T_0=[0,1,2]$. Perfect universality is observed for $v_2$ and $v_3$ in results for the conformal case: the thick dark red lines ($m=0, R=5.0$ fm) and the thick dashed dark green lines ($m=0, R=3.0$ fm) nicely overlap for the range of explored $\IKn$.\\
In the non-conformal case, one clearly observes a mass ordering in the response functions: larger masses lead to a smaller $v_n/\varepsilon_n$. This can be traced back to the fact that at larger masses the speed of sound is smaller and the conversion from $\varepsilon_n$ to $v_n$ is slower; then the transverse finite size stops the conversion at $t\gtrsim2R$, when the decoupling becomes dominant. This effect is more evident in the range of small $\IKn$, where the system exhibits particle-like behaviour, while in the hydrodynamic limit (large $\IKn$) the splitting between different $m/T_0$ is far less pronounced. Indeed, the difference in the relative splitting is even larger; for instance at $\IKn=6.45$ for $v_2$ the splitting is $\lesssim3\%$, while at $\IKn=1.3$ it is $\approx 15\%$; similar deviations are observed for $v_3$ and $v_4$.\\
To corroborate the interpretation given above, we plot in the bottom panel of Figure\,\ref{fig:fig_2} the response curves with respect to $c_s t/R$. This has been observed to be the scaling variable for the collective flows also in the ideal hydro limit \cite{Bhalerao:2005mm}.  One clearly sees that, by taking into account the role of the speed of sound in the conversion $\varepsilon_n\to v_n$, it is possible to recover the universality in the response functions of $v_2$ and $v_3$. Notice that, in agreement with Figure\,\ref{fig:fig_1}, the departure from universality occurs at $t\approx 2R$, as it is clearly visible by looking at the coloured points.\\

\begin{figure}
	\centering
	\includegraphics[width=\linewidth]{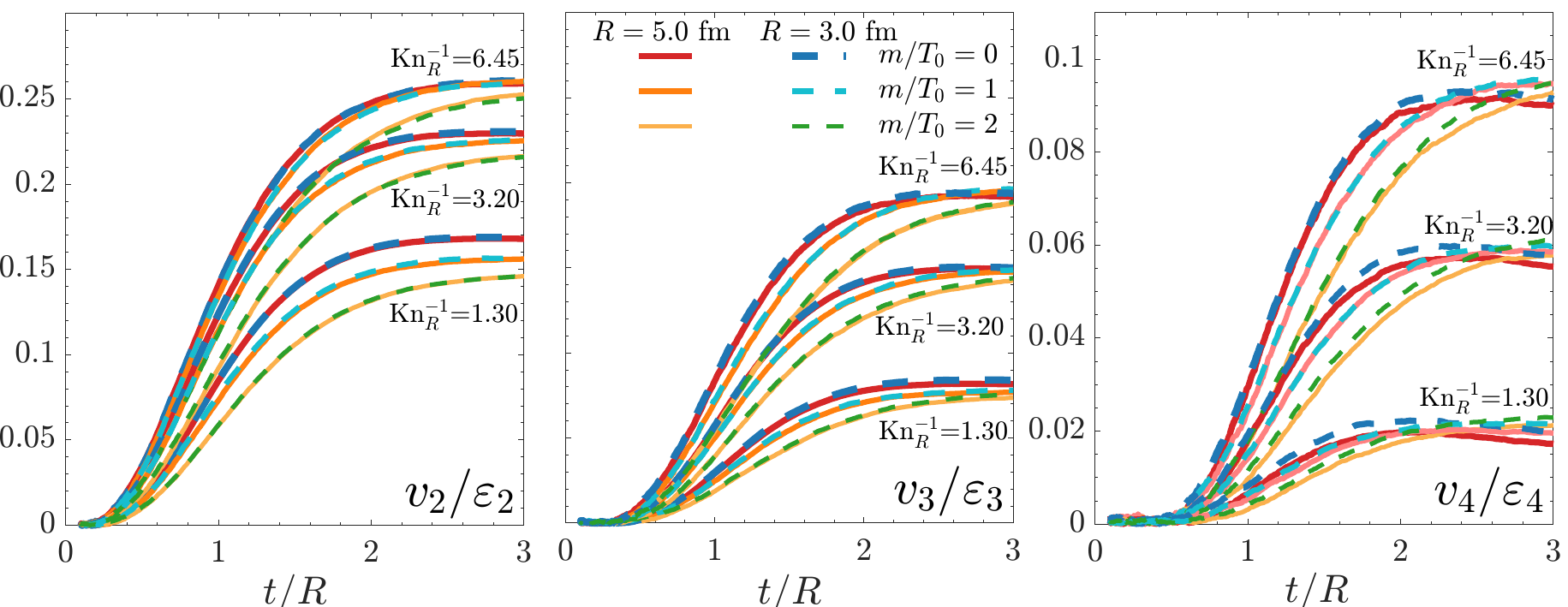}\vspace{8pt}
	\includegraphics[width=\linewidth]{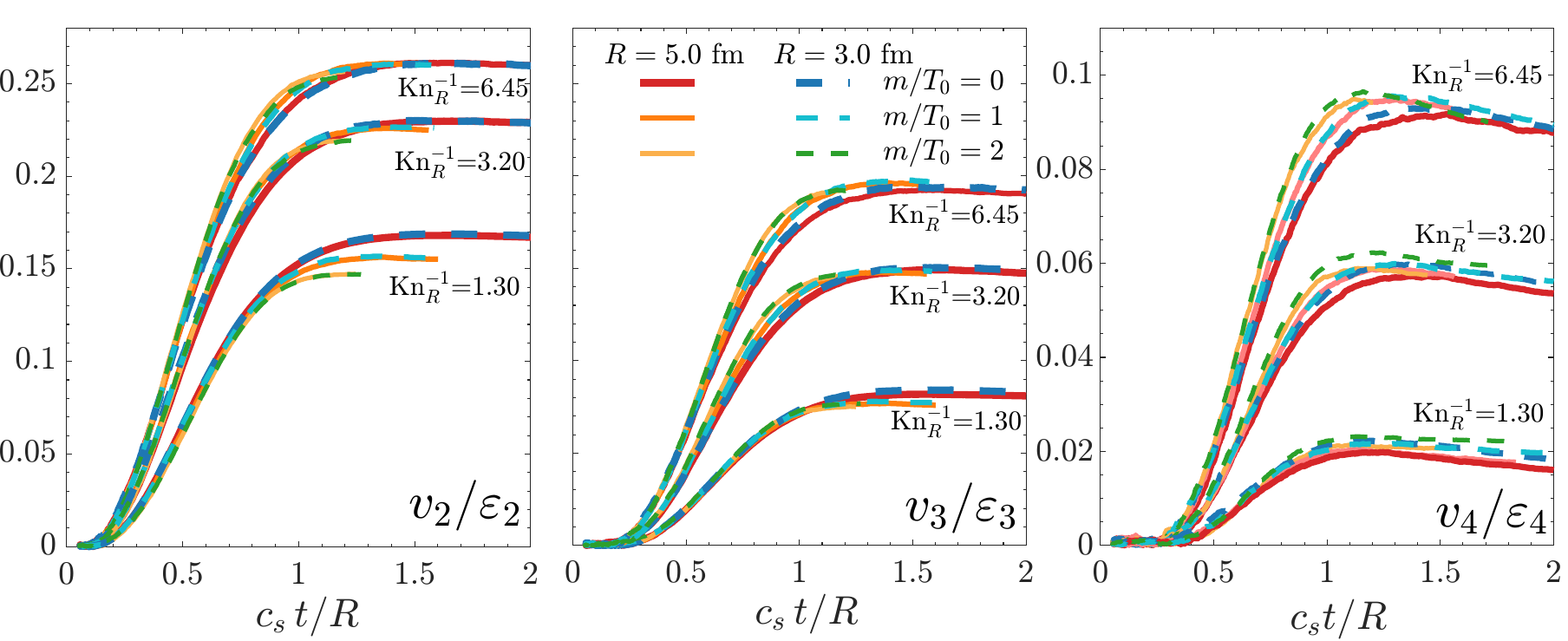}
	\caption{Response curves for elliptic (left panels), triangular (central panels) and quadrupole (right panels) flows  in systems with fixed $\IKn$ and radii $R=5.0$ fm (red and orange curves) and $R=3.0$ fm (blue and green curves) (see Table\,\ref{tab:parameters} for the details) and different values of $m/T_0=[0,1,2]$. The top panels are plotted with respect to $t/R$, the bottom panels with respect to $c_st/R$, to take into account the different speed at which information propagates with dependence on the EOS.}
	\label{fig:fig_2}
\end{figure}

In the right panels of Figure\,\ref{fig:fig_2} we also show the responses for the $v_4$, with respect to both $t/R$ (top panel) and $c_s t/R$ (bottom panel). A very good scaling is seen for large $\KnR$, while it gets progressively worse for smaller values of the parameter. This is not surprising since the response to the quadrupole eccentricity $\varepsilon_4$ is known to be much smaller and slower than for the lower-order flows. Indeed, by comparing this Figure to Fig.\,\ref{fig:fig_2} one sees that the $v_n$ build up takes more time for $n=4$: at $t=R$, for instance, more than 30\% of the $v_2$ and $v_3$ have been already created, while $<20\%$ of the $v_4$ has been developed. This can explain the observed discrepancy, since the $v_4$ develops ($t\lesssim R$) when the system has already sensitively expanded and is almost at saturation at $t=2R$, with a major role played by the system geometry. It would be therefore be of interest to go deeper in the study of the $v_4/\varepsilon_4$ and to investigate the role of the system size. Nevertheless, the build up of the flow in terms of $c_s t/R$ is still really similar also for the smallest value of $\KnR$, which once more confirms the role played by the speed of sound in this context. It would be of interest, moreover, to study how the $v_4$ is affected by the initial $\varepsilon_2$, since we expect a quadratic dependence on the eccentricity. However, the procedure we are using to reproduce an initial $\varepsilon_2\propto \alpha_2$ also creates an $\varepsilon_4 \propto \alpha_2^2$. This means that the initial state of the simulation will have $\varepsilon_2^2\propto\alpha_2^2$ and $\varepsilon_4\propto \alpha_2^2$, making it impossible to identify how much of the observed $v_4$ is a quadratic response to the $\varepsilon_2$ or a linear response to the $\varepsilon_4$.

Moving to more differential observables, the study of the $v_n(p_T)$ turns out to be more sensitive to the microscopic details of the system, despite, fixing $\KnR$, one expects to recover the universality once more. In such observables the only relevant scale is the average transverse energy $\langle E_T\rangle= \langle (p_T^2 + m^2)^{1/2} \rangle$. Similarly to what has been observed by Ref.\,\cite{Muncinelli:2024izj} in the spectra and by Ref.\,\cite{Jia:2025rab} in the radial flow $v_0$, we analyse the $v_n(p_T/\langle E_T\rangle)$ and show in the left panels of Figure\,\ref{fig:vn_pt_etast} the results for $\IKn=3.20$ at $t=3R$. Interestingly, different systems (dashed for $R=3.0$ fm and solid for $5.0$ fm) at fixed $m/T_0$ show a perfect scaling, pointing out that the universality seen in integrated observables (e.g. Figure\,\ref{fig:fig_2}) is still present even for differential quantities. Since in this paper we are interested in the bulk, we show the $v_n(p_T/\langle E_T\rangle)$ in the range $p_T/\langle E_T\rangle<2$, which includes $>90\%$ of the spectrum. We notice that this observable is strongly sensitive to the $m/T_0$ ratio, which translates in a direct dependence on the speed of sound. In particular, as well-known \cite{Danielewicz:1994nb}, at $p_T<\langle E_T\rangle$ we observe the typical linear vs quadratic increase for the massless vs massive bulk: the larger the mass, the slower the small $p_T$ dependence. On the other hand, for $p_T\gtrsim\langle E_T\rangle$ we observe the inversion of the mass-ordering, which can be partially explained by the fact that the values of $\langle E_T\rangle$ are different for the various curves. More in detail, if we consider for instance the $R=5.0$ fm simulations, we get $\langle E_T\rangle=[1.1, 1.3, 1.6]$ respectively for $m/T_0=[0, 1, 2]$. Nevertheless, this cannot  fully explain the inverse mass ordering, which is observed also for the non-scaled $v_n(p_T)$ (the grey curves in right panels of Fig.\,\ref{fig:vn_pt_etast}). This inversion is indeed the consequence of the non-physical fixing of $\eta/s$ also for $t>2R$ and $r>2R$, when the system is mostly decoupled and has already cooled down beyond the critical temperature: at small temperatures the local cross-section has to be much larger in the massive case in order to keep $\eta/s$ constant \cite{Plumari:2012ep} and this results in a larger response. As one can check by looking at the coloured curves in the right panels of Fig.\,\ref{fig:vn_pt_etast} this effect indeed disappears with a more realistic $\eta/s(T)$ which naturally implements a freeze-out for $T<T_c$ \cite{Plumari:2019gwq}:
\begin{equation}
	\eta/s =
	\begin{cases}
		(\eta/s)_0 & T\ge T_c\\
		- \vartheta \dfrac{T - T_c}{T_c} + (\eta/s)_0 & T<T_c
	\end{cases}
\end{equation}
where the coefficient $\vartheta$ is fixed to 25 if $4\pi\eta/s=5.03$ for the simulations shown with $R=5.0$ fm and $\KnR=3.20$; $v_n$ vs $p_T$ are plotted so to remove the effect of $\langle E_T\rangle$ rescaling. These outcomes are obviously independent of the system size and $\KnR$. By comparing the grey and the coloured curves one can understand that the massless results are nearly unaffected, since the $\sigma_{22}$ cross sections involved at large $r$ and $t$ are already quite small without the freeze-out; on the contrary, there is a large impact on the massive curves, whose $v_n(p_T)$ is strongly dumped for $p_T\gtrsim1.5$ GeV. Finally, as one can expect, at large $p_T$ the results are almost independent on the mass, since the larger the energy scale involved the less relevant the mass becomes. These results demonstrate that a non-conformal realistic kinetic simulation needs an $\eta/s(T)$ if more differential observables want to be addressed.

\begin{figure}[t]
	\centering
	\includegraphics[width=.48\linewidth]{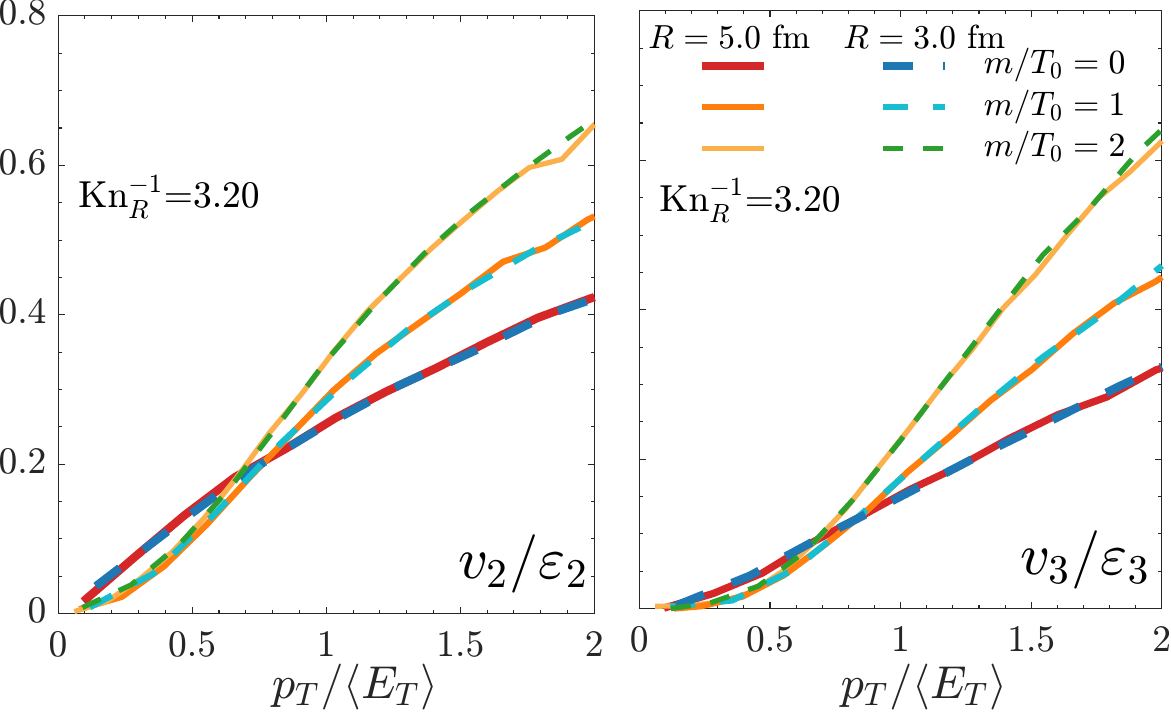}
	\includegraphics[width=.48\linewidth]{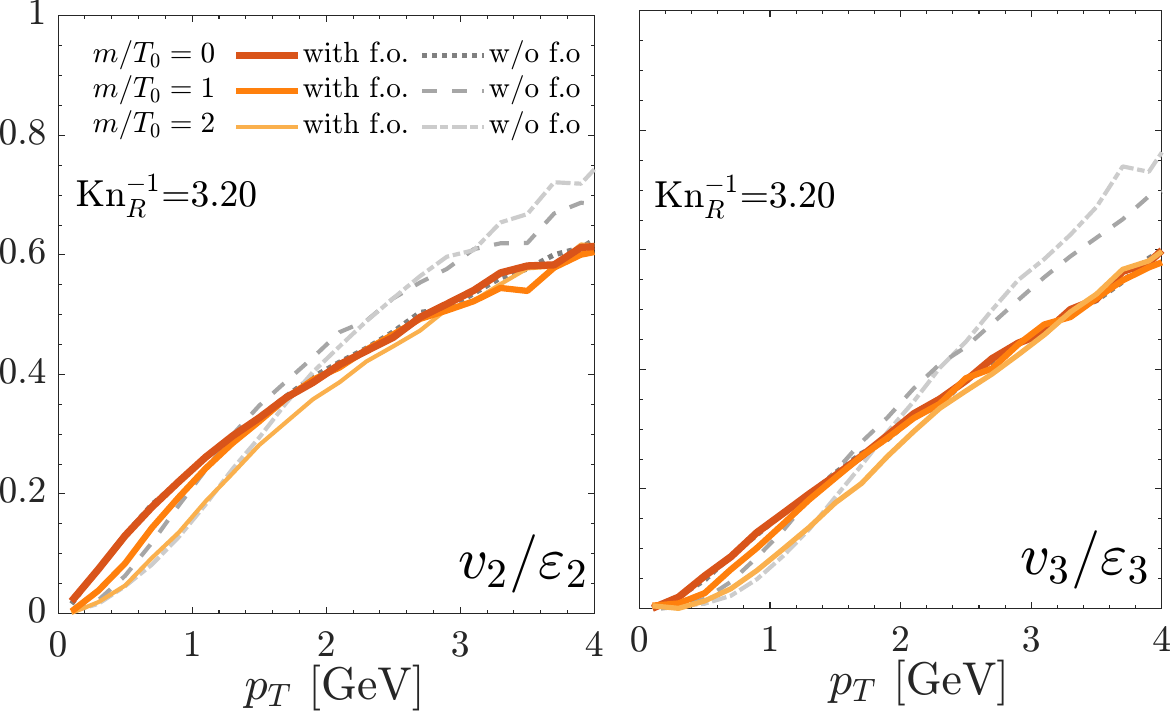}
	\caption{Left panels: Differential response curves for elliptic (left panel) and triangular (right panel) plotted with respect to the scaled momentum $p_T/\langle E_T\rangle$. The colour map is the same as Figure\,\ref{fig:fig_2}. Right panels: Differential responses plotted with respect to $p_T$. By comparing the grey and coloured curves the effect of the increasing of $\eta/s$ for temperature $T<T_c$ can be seen to strongly affect the high-$p_T$ behaviour of the massive simulations, while the massless ones are almost unaffected.}
	\label{fig:vn_pt_etast}
\end{figure}

\section{Universality with event-by-event initial fluctuations}
\label{sec:fluctuations}

\subsection*{Initial conditions}
In order to have a more realistic description of the initial state of the collisions, we employ the \trento\:model \cite{Moreland:2014oya} to generate initial profiles of several collision systems (see Table\,\ref{tab:event_by_event}). Following the Bayesian analysis posterior performed for viscous hydro simulations \cite{Liyanage:2023nds}, we fix the \trento\:parameters $w = 0.985$ and $p = 0.038$, which represent respectively the width of the Gaussian modelling the nucleon density profile and the reduced-thickness parameter which regulates the energy deposition of the participants nucleons. $5\times 10^5$ events are generated and then classified in ten classes of ten percentiles each (0-10\%, 10-20\%...), by making use of the matching between the initial energy density and the final multiplicity introduced in Ref. \cite{Giacalone:2019ldn}, following in particular Ref. \cite{Andronic:2025ylc} :
\begin{equation}
	\frac{dN_{ch}}{d\eta_s}= \frac 43 C_\infty^{3/4} \tonde{4\pi\eta/s \frac{\pi^2}{30} \nu_{eff}}^{1/3} \frac{N_{ch}}{S} \int d^2\vet x\, [e(\vet x)\tau]_0^{2/3} ,
\end{equation}
where $S/N_{ch}=(dS/dY)/(dN_{ch}/dy)\approx 6.7-8.5$ is the entropy per charged particle at freeze-out and $C_\infty$ is a constant of order unit which characterises the Bjorken attractor at small scaled time $\tau/\tau_{eq}$. Since we are going to keep $\eta/s$ constant in the following, we have simply that  $dN_{ch}/d\eta_s \propto \int d^2x \,[e(x)\tau]_0^{2/3}$. The generated outputs are interpreted as energy-density profiles and used to map the initial distribution function. To construct the full $f(x,p)$, an assumption has to be made concerning the equation of state and the initial momentum distribution: as a starting point we consider a conformal system ($m=0$) with an effective number of degrees of freedom $n_{\text{dof}}=49$ and an initial momentum distribution with $Y=\eta_s$ (see Section\,\ref{sec:RBT}). Moreover, since \trento\:provides a 2D profile, the system is assumed to be boost-invariant in the space-time rapidity interval $\eta_s\in[-2.5, 2.5]$. The only parameter to be computed in the matching \trento+RBT is the overall scaling factor. It is  by comparing the final $(dE_T/d\eta_s)_{\tau_\text{freeze}}$ extracted from the code and available experimental data or predictions.
The freeze-out surface, on which the $(dE_T/d\eta_s)_{\tau_\text{freeze}}$ is computed, is defined by a condition on the energy density: when one cell is below the threshold $e<0.182$ GeV/fm$^3$, its energy is added to $(dE_T/d\eta_s)_{\tau_\text{freeze}}$. The freeze-out energy-density corresponds to the value $e(T_c)$ assumed at the critical temperature according to the lQCD EOS \cite{HotQCD:2014kol, Borsanyi:2010bp}. In particular, for the Pb-Pb collisions at $\sqrt {s_{NN}}={2.76}$ TeV we use the experimental data for $dE_T/d\eta$ from ALICE results \cite{ALICE:2016igk}; for Au-Au at $\sqrt {s_{NN}}={200}$ GeV we use PHENIX results \cite{PHENIX:2013ehw}. For O-O and Ne-Ne collisions we use existing predictions and give a rough estimation of $dE_T/d\eta\approx 3/2\, \langle p_T\rangle\, dN_{ch}/d\eta$. Since there is not agreement among the different predictions yet, we label by (1) the simulations gauged on the predictions by \cite{Khan:2024fef, Behera:2021zhi} and by (2) the ones following \cite{Giacalone:2024luz}. Moreover, to take into account the peculiar nucleon configurations of $^{16}$O and $^{20}$Ne, the Nuclear Lattice Effective Field Theory (NLEFT) configuration files with positive weights from \cite{Summerfield:2021oex, Giacalone:2024luz} are employed.\\

\begin{figure}
	\centering
	\subfloat[][{Pb-Pb 10-20\%}]
	{\includegraphics[width=.8\linewidth]{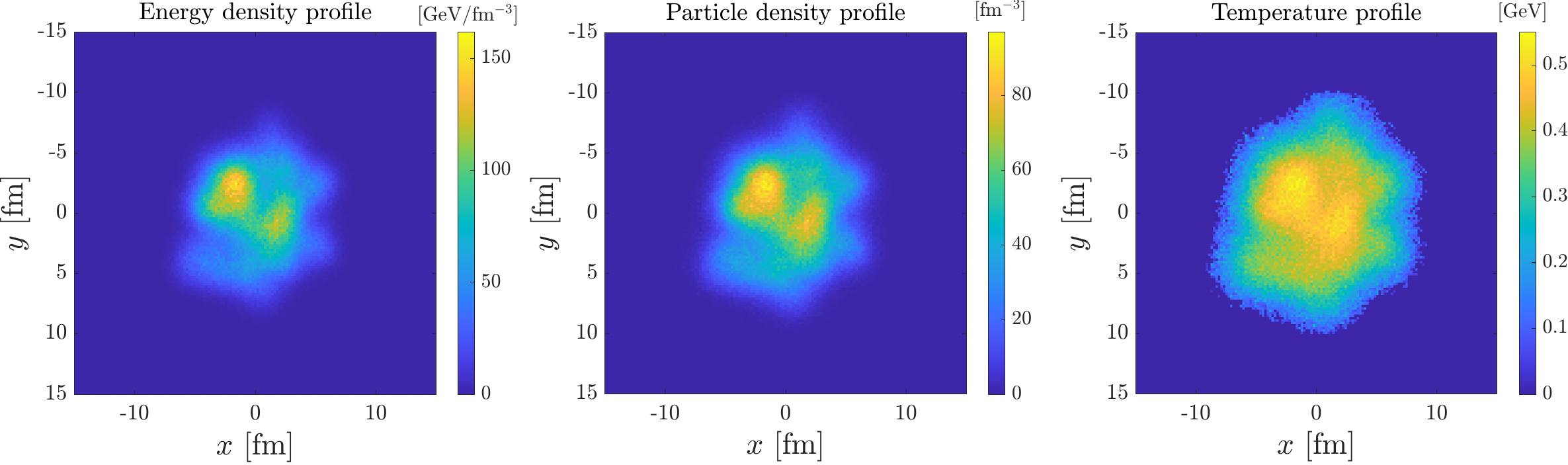}}\\
	\subfloat[][{Pb-Pb 60-70\%}]
	{\includegraphics[width=.8\linewidth]{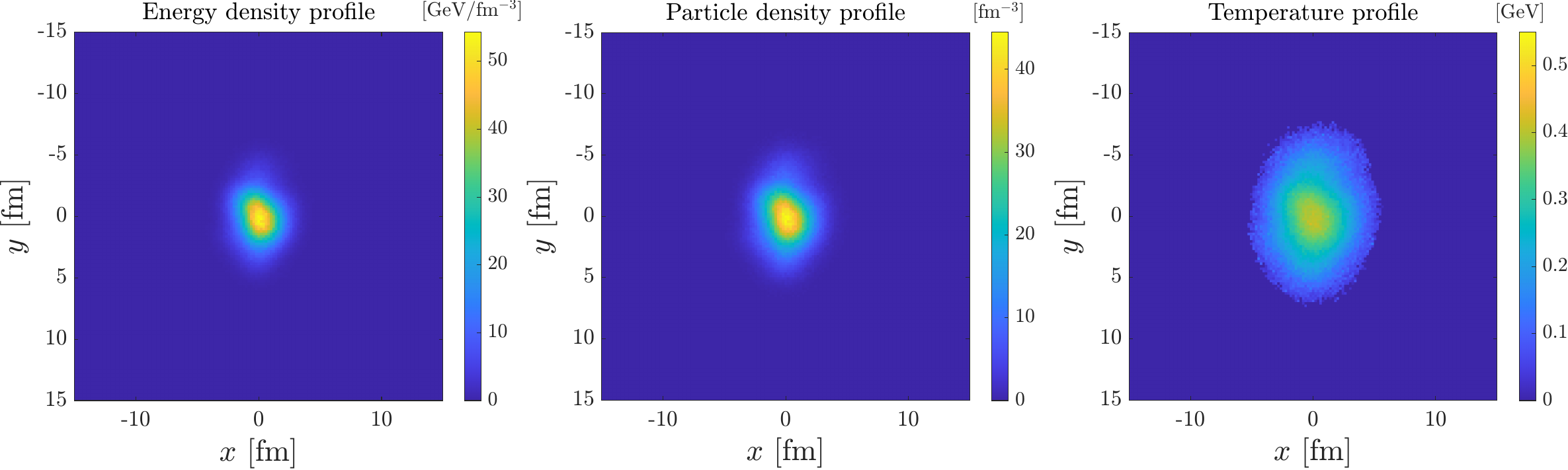}}\\
	\subfloat[][{NeNe 20-30\%}]
	{\includegraphics[width=.8\linewidth]{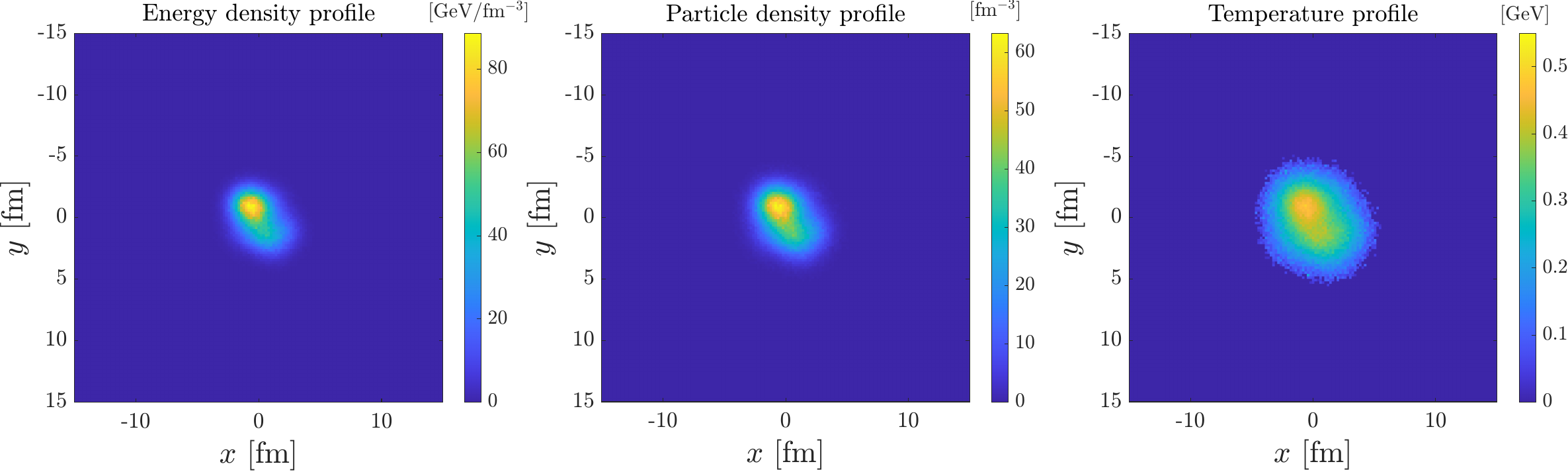}}\\
	\subfloat[][{O-O 60-80\%}]
	{\includegraphics[width=.8\linewidth]{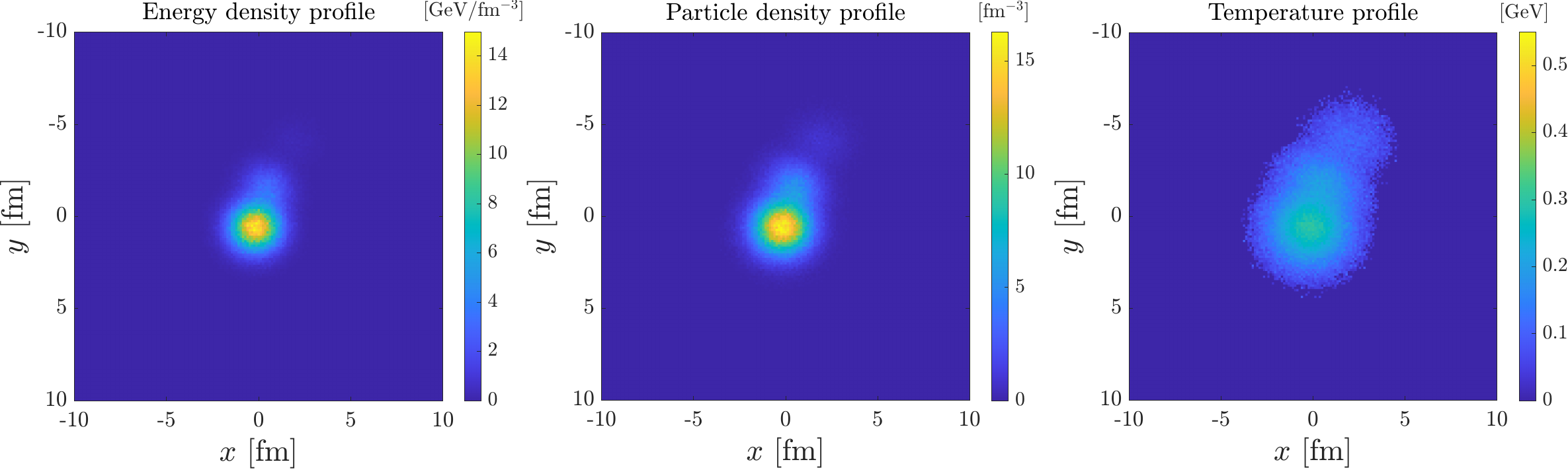}}\\
	\caption{Initial profiles for different collision systems. The three columns correspond to energy density (left), particle density (central) and temperature (right); while the rows to different systems: central Pb-Pb, peripheral Pb-Pb, semi-central Ne-Ne and peripheral O-O. Notice that the plots in the first two columns have each a different colour scale to make them readable, while the colour scale is unified for the temperature. }
	\label{fig:profileoo60-80}
\end{figure}

\subsection{Knudsen number evolution}
In Fig.\,\ref{fig:knudsen_evolution_contour} we show some space-time evolution plots of the Knudsen number in a set of collision systems with different $\KnR$. In particular, these plots are obtained by starting from an initial state generated by the \trento\,code and then evolved with $\eta/s=0.13$. They make clear the dependence of the $\IKn$ on the transverse coordinate $x_\perp$ and on time and therefore highlight the regions were hydrodynamics can be rigorously applied (yellow-orange), the transition regime (violet) and the areas where the behaviour is particle-like (blue). In principle, the value of the (inverse) Knudsen Number is only a necessary condition to model the system as a fluid, since one should also impose a constraint on the nearly-equilibration of the system by looking for instance at the Reynolds number: here we focus only on the Knudsen number, while the Reynolds was analysed in the previous Chapter, by starting from smooth initial conditions and was also shown to exhibit an attractor is rather short times. It is nonetheless true that these constraints on the regime of applicability of hydrodynamics have been proved to be more loose than expected.
By looking at the Figure, it is patent the distinction between a large collision system (such as a central PbPb) and a more peripheral PbPb or one involving lighter nuclei (NeNe or OO), since the applicability of hydrodynamics is very well justified for large time and radius in the former, while is in principle forbidden above $t\sim 1.5 R$ also for the very central region of small systems. It is however of great interest to see that, if one compares the plots corresponding to same $\KnR$ (namely top right - central left and central right -- top left), one sees that, beyond the universality observed in the evolution of the averaged Knudsen number, a really similar pattern can be observed in this much more detailed perspective, in which also the geometry plays a role. The top left and bottom right plots, instead, are for the two extreme cases considered, the former corresponding to a large system which behaves as a fluid for most of its evolution, while the latter exhibits an extremely limited region of applicability for hydro.

\subsection{Collective Flows}

\begin{figure}[t]
	\centering
	\includegraphics[width=.8\linewidth]{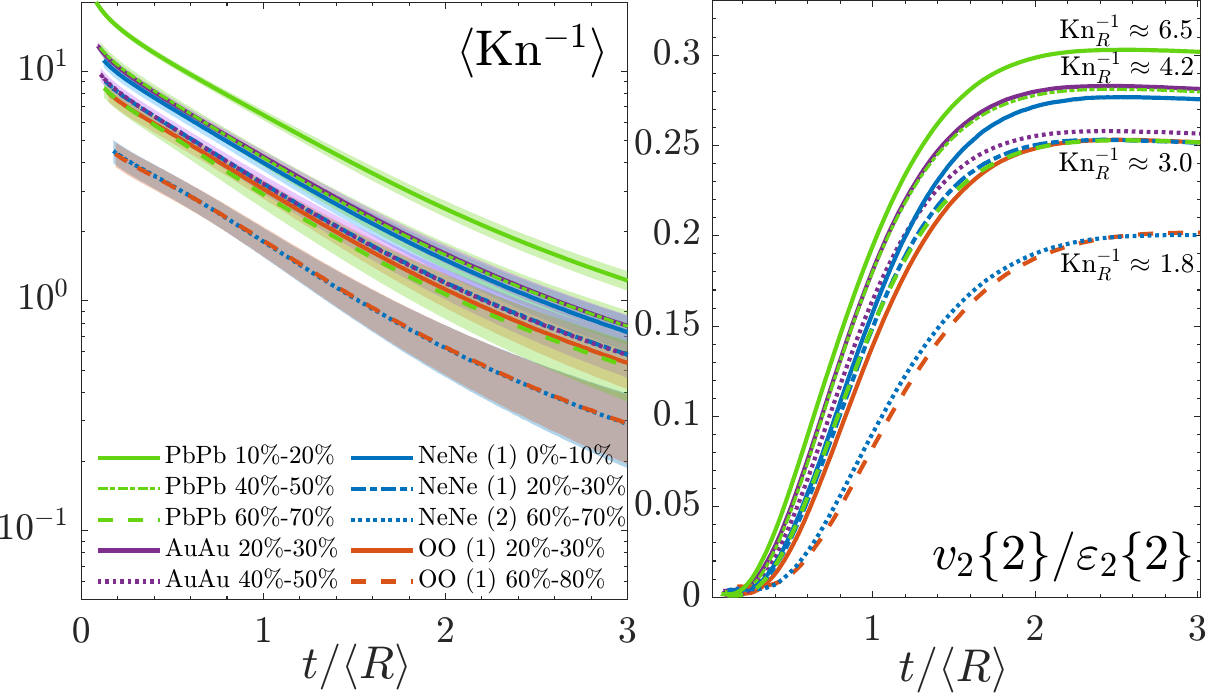}
	\caption{Left panel: Inverse Knudsen Number evolution for different collision systems, energies and centrality classes. From the top to the bottom, they correspond to a $\IKn_R\approx6.5, 4.2, 3.0, 1.6$. More details on the event-by-event configurations in Table\,\ref{tab:event_by_event}. Right panel: Response functions for the same simulations of left panel. It is evident the scaling behaviour irrespectively of energy, system size and centrality.} 
	\label{fig:fig_4}
\end{figure}

Following the discussion of the previous sections, we want to study whether, also in the more realistic case with initial state fluctuations, one can cluster different events in the same universality class on the basis of the $\KnR$ value and, in general, of the $\IKn$ evolution. To have a more direct comparison, the centrality classes are chosen in order to get an average Knudsen number close to the ones discussed above (6.45, 3.20, 1.30); we also added a fourth value (4.2) to probe further configurations. The results shown here are obtained with 200 events for each ensemble. Obviously, each event has a different Knudsen number evolution; in the left panel on Figure\,\ref{fig:fig_4} we plot the $\IKn$ averaged over the different events and a band whose width corresponds to one standard deviation. The $R$ used is the average of the root-mean-square radii of the different events. In Table\,\ref{tab:event_by_event} we list and give details about these simulations, reminding that they are performed in the conformal ($m=0$) case.\\
As far as the collective flows are concerned, we show in the right panel of Figure\,\ref{fig:fig_4} the two-particle correlation elliptic flow \cite{Borghini:2000sa, Borghini:2001vi} $$v_2\{2\} = ( c_{v_2}\{2\})^{1/2} \approx (\langle v_2^2\rangle)^{1/2},$$
scaled by the two-particle correlation  $\varepsilon_2\{2\}$, defined in an equivalent way. Here $\langle\cdot\rangle$ indicates an average over the ensembles. This choice is justified by the fact that, on the contrary with respect to what happens in the academic cases studied before, the reaction plane is not fixed and there are several non-vanishing initial eccentricities in most of the cases. Experimentally, only the relative azimuthal angle between particles can be measured:
\begin{equation}
	\langle e^{in(\phi_1 - \phi_2)} \rangle = \dfrac{ \int e^{in(\phi_1 - \phi_2)} f(\vet p_1, \vet p_2)\, d^3\vet p_1 \, d^3\vet p_2 }{ \int f(\vet p_1, \vet p_2)\, d^3\vet p_1 \, d^3\vet p_2}
\end{equation}
In the initial conditions implemented there are no two-particle correlations, therefore $f(\vet p_1, \vet p_2) = f(\vet p_1) f(\vet p_2)$, which results in:
\begin{equation}
		\langle e^{in(\phi_1 - \phi_2)} \rangle = \langle e^{in(\phi_1)} \rangle \langle e^{in(\phi_2)} \rangle = v_n^2.
\end{equation}
Moreover, even considering two-particle non-flow correlations, the correction is found to be $\sim 1/N$ \cite{Borghini:2000sa, Borghini:2001vi}, which can be safely neglected due to the huge number of test particles involved in the simulations. Finally, notice that in computing the average $\langle e^{in(\phi)} \rangle = \langle \sqrt{ \sin^2(n\phi) + \cos^2(n\phi) } \rangle$, both sin and cos contributions are to be included. It is trivial to check that the $v_n\{2\}$ simply reduce to the computed $v_n$ if no event-by-event fluctuations are implemented and the reaction plane is known to be $\Psi_n=0$ as in the previous paragraphs.\\
By comparing the results in Figure\,\ref{fig:fig_4} with Figure\,\ref{fig:fig_2}, we observe regularly larger responses in the event-by-event simulations with respect to the ones with smooth initial conditions. This is not surprising since the geometry of the energy and particle density distribution has to play a major role in building the anisotropic flows. By looking at the space-time development of the collective flows, we observed that this difference is mainly due to the presence of temperature gradients (which are absent in the smooth initial conditions defined in Section\,\ref{sec:RBT}, in which the profile has a constant $T_0$). Indeed, when particles exchange momenta in a region with higher $T$, they have averagely a larger $p_T$ than particles in the surrounding area, therefore they can easily cross it and carry the generated flow out of the hot spot, producing a larger response. On the contrary, if no such peaks are present, particles continue colliding and exchanging momenta until they reach the edges of the distribution, making less efficient the building of the collective flow.\\
Thus, it is non-trivial that the scaling behaviour observed within the context of events with smooth initial conditions, all sharing the same geometry, is now recovered among event-by-event simulations. The geometry of each initial profile is different, with the likely formation of peaks and strong anisotropies, especially for small systems.  Nonetheless, when one averages on an ensemble of events, universality in the response function is fully recovered, even among systems with different nuclei and collision energies: from $^{16}$O to $^{208}$Pb, from 200 GeV (typical of Au collision at RHIC) to 7 TeV and even involving the strongly deformed $^{20}$Ne.\\
{The only free parameter is $\eta/s$, which however is fixed for all the simulations and of course has an impact in the determination of the overall scaling factor used for the initial conditions. It would be of great interest to study how a more realistic $T$-dependent $\eta/s$ and a non-conformal equation of state affect these results, aiming to a fully realistic simulation, as well as a detailed analysis of higher order anisotropic flows in these more realistic scenarios.}
\begin{table}
	\caption{Different event-by-event simulations parameters.}
	\centering
	\begin{tabular}{ccccc}
		\toprule
		system & energy  & centrality class  & $\IKn_R$ & $\langle R\rangle$ [fm] \\
		\midrule
		PbPb & 2.76 TeV & 10\%-20\% &  6.5 & 4.20 \\
		PbPb & 2.76 TeV & 40\%-50\% &  4.2 & 3.25 \\
		PbPb & 2.76 TeV  & 60\%-70\%  & 3.0 & 2.72 \\
		\midrule
		AuAu& 200 GeV & 20\%-30\% & 4.3  & 3.73 \\
		AuAu& 200 GeV & 40\%-50\% & 3.2  & 3.2 \\
		\midrule
		NeNe (1) & 7 TeV & 0\%-10\% & 4.1 & 2.75 \\
		NeNe (1) & 7 TeV & 20\%-30\% & 3.2 & 2.5 \\
		NeNe (2) & 7 TeV & 60\%-70\% & 1.8  & 1.92 \\
		\midrule         
		OO (1) & 7 TeV & 20\%-30\% & 3.0  & 2.3 \\
		OO (1) & 7 TeV & 60\%-80\% & 1.8 & 1.80 \\
		\bottomrule
	\end{tabular}
	\label{tab:event_by_event}
\end{table}

\begin{figure}[t]
	\centering
	\subfloat[][\emph{}]
	{\includegraphics[width=.45\textwidth]{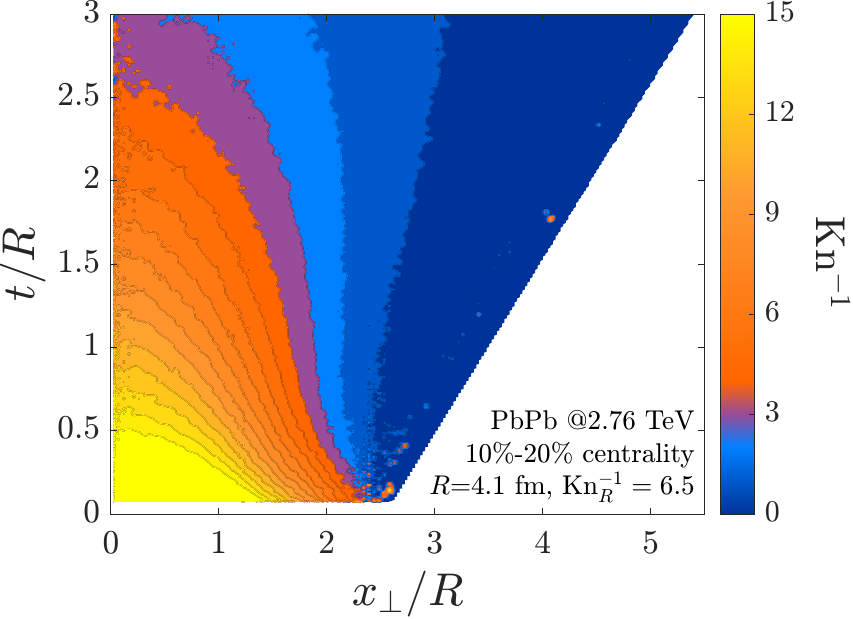}}\quad
	\subfloat[][\emph{}]
	{\includegraphics[width=.45\textwidth]{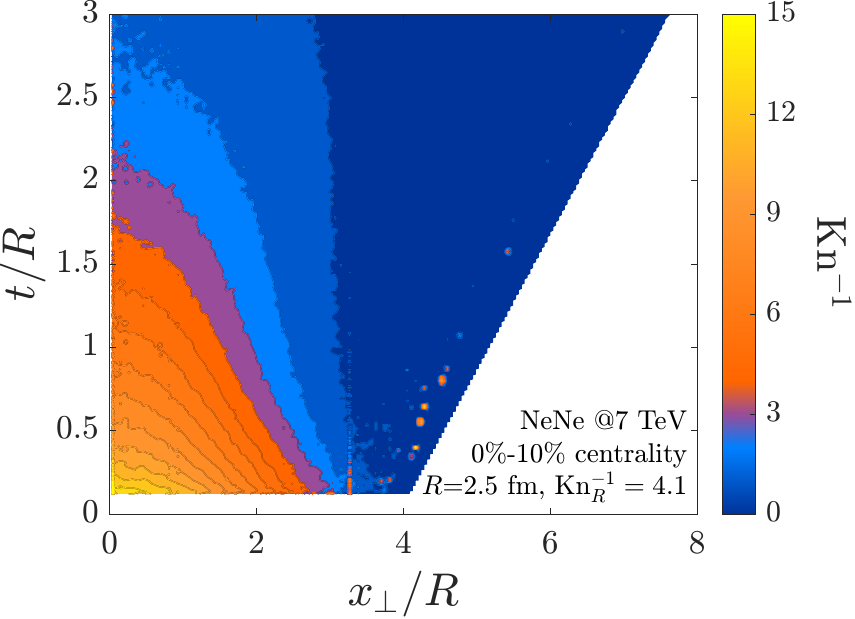}}\\
	\subfloat[][\emph{}]
	{\includegraphics[width=.45\textwidth]{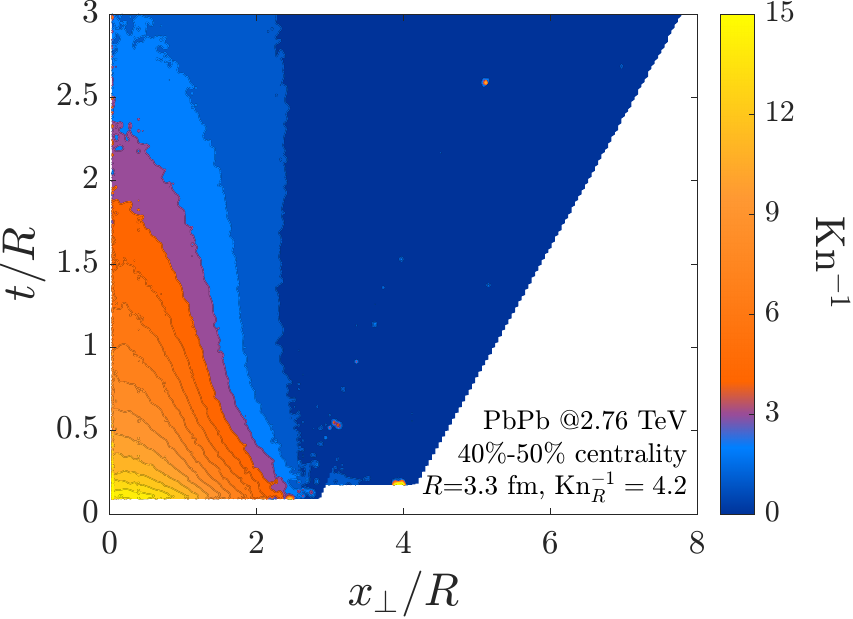}}\quad
	\subfloat[][\emph{}]
	{\includegraphics[width=.45\textwidth]{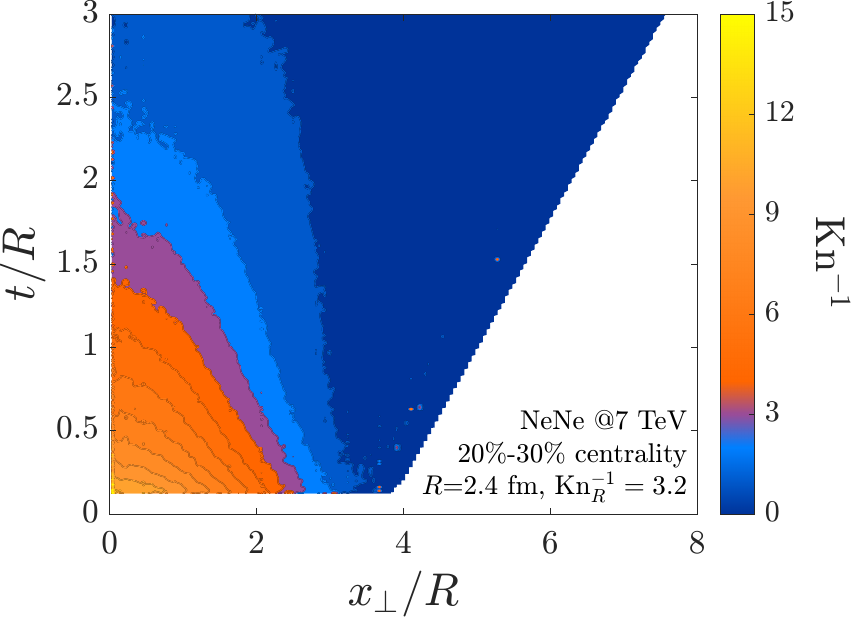}}\\
	\subfloat[][\emph{}]
	{\includegraphics[width=.45\textwidth]{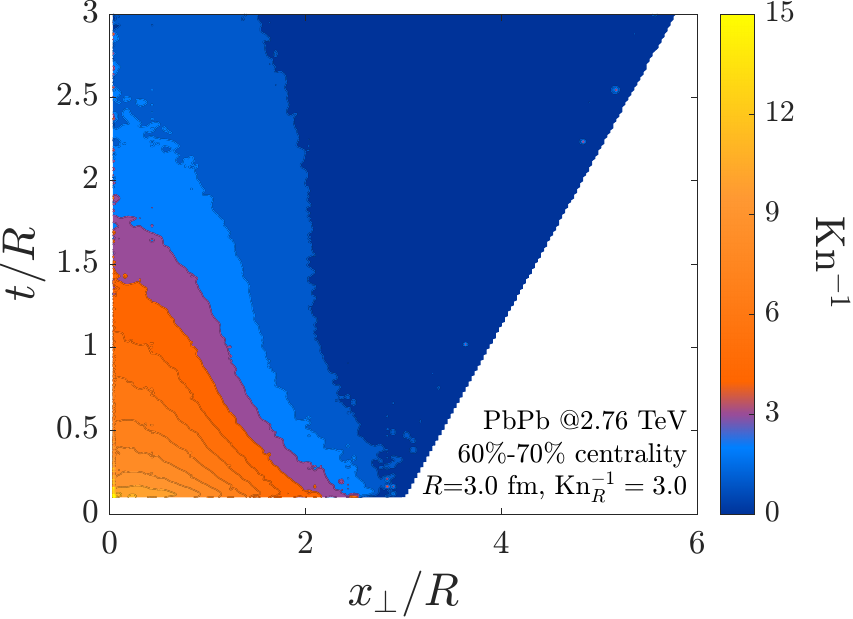}}\quad	
	\subfloat[][\emph{}]
	{\includegraphics[width=.45\textwidth]{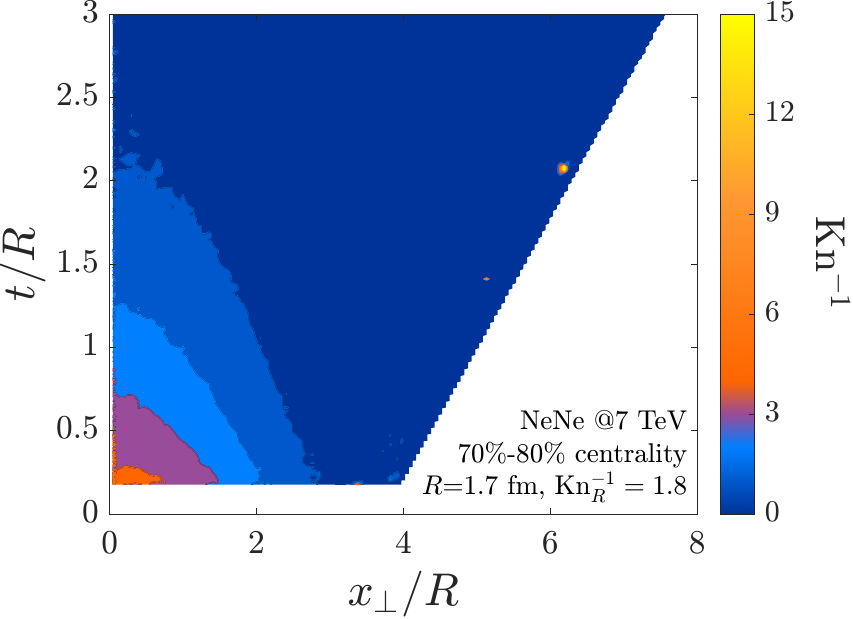}}\\
	
	\caption{Knudsen number evolution for different PbPb (left panels) and NeNe (right panels) initial conditions, corresponding to different regimes from hydro like (top left corner) to particle like  (bottom right corner).}
	\label{fig:knudsen_evolution_contour}
\end{figure} 

\chapter*{Conclusions}
\addcontentsline{toc}{chapter}{Conclusions}
\markboth{CONCLUSIONS}{CONCLUSIONS}

The understanding of the emergence of collectivity in the context of heavy-ion collisions has considerably advanced during the last two decades. The observation of a sizeable elliptic flow and afterwards of higher order anisotropic flows is commonly considered one of the pieces of evidence of the formation of a fluid-like medium in these systems and challenged also our comprehension of this probable new state of matter, namely the Quark Gluon Plasma, who exhibits some of the most extreme features in different quantities, from the vorticity to the energy density, from the magnetisation to the shear viscosity. The quite accepted standard model of the evolution of the hot QCD matter pictures a pre-equilibrium phase in which the medium strongly expands because of the huge pressure gradients and then rapidly enters a collective phase which is successfully described by hydrodynamics. The developments in relativistic hydrodynamics, however, delved deeply into its theoretical foundations and showed how this effective long-range theory can be related to the microscopic description of the medium, thanks to the kinetic theory formalism. This allowed us to bridge the gap between two apparently separated regimes (the dilute one for the kinetic theory, the strong interacting one for fluid dynamics) giving also a broader interpretation to the Relativistic Boltzmann Equation itself. The latter, indeed, can be conceived as a mesoscopic theory which does not need to know about all the microscopic details of the system, but can suitably model the evolution of the distribution function seen as a mesoscopic quantity. In this interpretation, the distribution function is not founded on the actual microscopic degrees of freedom of the system (namely the partons) neither with macroscopic quantities such as energy density or pressure, as done by hydrodynamics. In this context, the proposed Relativistic Boltzmann Transport (RBT) model aims to give a unified description of the medium evolution, from the pre-equilibrium phase ($\approx$ 0.1 fm) to the particlisation, i.e. when the system can no longer be considered a collective medium and hadrons freely stream towards the detectors. This would make it possible to describe the different stages of the evolution (pre-equilibrium, collective phase, hadronisation, free streaming) without switching from one theoretical framework to the other. Moreover, recent investigations and experimental results on small systems challenged the picture that was assumed to be valid for $pp$, $pA$ or light-ion collisions. Signals of collectivity, indeed, have been observed in these systems which may suggest the formation of a medium. This does not necessarily mean, however, that the generated system is a droplet of QGP, since collective behaviour could in principle emerge also for a confined medium; nevertheless, other observables cannot exclude this fascinating hypothesis. What is surprising is that the modelling of these systems can be once more carried out by means of hydrodynamics, which is not expected to be valid in regimes that are surely far away from equilibrium and thermalisation. Once more, the basic foundations of fluid dynamics were challenged and the connection to kinetic theory, which is expected to be valid also in this limit, proves to be extremely useful. This led to the theoretical development of the attractor concept: a universal behaviour shared by systems with different initial conditions that evolve together, i.e. with the same time evolution of observables, towards equilibration \emph{well before} equilibration occurs. A further hypothesis suggests that the reaching of the attractor could explain this surprising success of hydrodynamics, with the so-called hydrodynamisation process, that, however, is still under debate. Apart from this purely theoretical application, the attractors could be useful to identify which observables forget and which keep memory about the initial conditions, since we are interested in reconstructing the initial state of the collision starting from the final hadrons' spectra.\\
In this context, the work presented in this thesis starts with the investigation of a conformal 0+1D system with fixed $\eta/s$, which was the scenario in which studies concerning the attractors have been carried out in several frameworks, from hydrodynamics to RTA kinetic theory, effective kinetic theory and AdS/CFT. The solution of the Relativistic Boltzmann Equation with the implementation of the full collision integral confirms the presence of the attractor for the whole distribution function and for each of its momentum moments, consolidating what had been already observed for instance in RTA kinetic theory. There are mainly two mechanisms which bring the system to the attractor: the nearly-free initial expansion (pull-back or early-time attractor) and the medium interaction (forward or late-time attractor), which could be considered in principle two competing processes but are actually dominated by a unique time scale, the relaxation time $\tau_{eq}$. This quantity, introduced in the context of RTA and hydrodynamics, was found to coincide in the RBT framework with the average collision time per particle, that in the conformal case coincides with the mean free path. Since the system is 0+1D expanding, the only parameter we need to know to characterise its evolution is $\mfp$. Furthermore, for the first time this analysis was extended towards two possible directions: on the one hand, going from 0+1D to 1+1D, i.e. relaxing the hypothesis of boost-invariance, proved not to break the attractor at all, neither the forward nor the pull-back; on the other hand, the impact of a more realistic $\eta/s(T)$ was investigated, finding that only a partial breaking of the attractor can be observed in terms of $\tau/\tau_{eq}$, while the late-time behaviour is always constrained to the attractor itself.\\
The most immediate extension, in this sense, was aimed at the full 3+1D dynamics, since the presence of the Bjorken attractor could be of great interest in modelling the collision. Moving to this more realistic scenario means basically adding a new scale to the system description, i.e. the transverse size $R$, which governs the onset of the transverse expansion and therefore determines how long the Bjorken picture should be valid. Not surprisingly, the analysis of the system dynamics leads to a three-stage picture of the expansion: it is nearly 1D for $t<R$; mostly dominated by transverse dynamics for $R<t<2R$; almost free streaming in the transverse plane for $t>2R$. This suggests that if the dynamical attractor is reached for $t<R$, then it behaves exactly as in the 1D case and information on the initial conditions, e.g. the initial pressure anisotropy $P_L/P_T$ is washed out. This was found to happen in basically every case of physical interest, since one has to go to extreme unphysical scenarios to make the transverse expansion prevent the attractor from being reached. As far as the universality in terms of scaled time is concerned, however, the unique attractor seen in 1D is replaced by a class of pull-back attractors in terms of $t/R$, one for each value of the $R/\langle \mfp \rangle$ ratio computed at a fixed $t/R$. This quantity, which is nothing more than the well-known (inverse) Knudsen number, is therefore a good candidate to be the key parameter of the framework. On the one hand it is possible to see that for conformal systems with roughly the same initial geometry and sharing the value of the Knudsen number for a certain value of $t/R$, the evolution of Kn itself is universal, meaning that completely different systems, with distinct sizes and interaction measures, can be described exactly in the same way with respect to $t/R$, suggesting the formation of universality classes in terms of the $\IKn$. On the other hand, the Knudsen number itself has a strongly theoretical meaning, being the key quantity able to distinguish between a particle-like and a fluid-like behaviour, and thus giving us basic information about the regime in which the system under analysis lies. This concept is put in comparison with the opacity parameter $\hat \gamma$, emerged in the context of RTA and ITA as the only scaling parameter of the theory and having in those frameworks also the physical meaning of distinguishing between different dynamical regimes. By looking at the longitudinal dynamics no large differences are visible between the two parameters, while discrepancies emerge when the transverse flows are taken into account, specifically in the study of the anisotropic flow coefficients. In the context of RBT, indeed, the Knudsen number appears as the key scaling parameter, since perfect universality is recovered in the response function $v_n/\varepsilon_n$ across different systems once they belong to one of the universality classes defined above. Moreover, the study of the Knudsen number allows a \emph{local} discrimination, in space and time, between hydrodynamic and particle-like behaviour, with the possibility to look at the whole evolution of different scenarios and give a quantitative insight into this issue.\\
Two possible extensions of this analysis were carried out: the study of non-conformal systems and the implementation of realistic event-by-event fluctuations. The former allowed a detailed analysis of this Knudsen number universality, which apparently is lost once a finite mass is considered in the system under consideration. However, a nearly complete recovery of this universality is possible once the key role of the speed of sound is taken into account, as it governs the time within which information propagates through the medium. The conformal scaled time $t/R$ is therefore replaced by $c_s t/R$, which, above all, allows to observe a perfect agreement in the build-up of collective flows across systems with different masses and fixed Knudsen number. The investigation of differential $v_n(p_T)$, moreover, underscored that, especially in the case of a non-conformal equation of state, a more realistic $\eta/s(T)$ must be taken into account to avoid unphysical behaviour at high-$p_T$. Eventually, the RBT code was used to evolve realistic event-by-event initial conditions generated by the \trento\, package back in the conformal configuration. This made it possible to see how the very good scaling in the $v_n/\varepsilon_n$ observed for idealised systems with smooth initial conditions is not cancelled by the event-by-event fluctuations; systems sharing the same Knudsen number evolution (i.e. Pb-Pb @2.76 TeV 60-70\%, Au-Au @200 GeV 40-50\%, O-O @7 TeV 20-30\%, Ne-Ne @7 TeV 20-30\%) show the same response curves $v_n/\varepsilon_n$. It is important to stress that this scaling is perfectly valid also in a regime in which the Knudsen number is relatively small (Kn $\approx 3-4$) and is sensitively broken only for very small Kn $\sim$ 2 for $v_4/\varepsilon_4$, while it still works well for $n=2,3$. This suggests that this universality is by no means due to the hydrodynamic behaviour of the systems, but can safely survive also in the transition region and in the particle regime.\\
These studies are particularly interesting in view of the recent developments in the context of quark matter experiments, which are heading towards the investigation of small systems. More in detail, in order to bridge the gap between heavy-ion collisions and smaller systems involving protons such as $pp$ and $pA$, light ion collisions have already been performed and are planned for the near future. As shown by looking at the Knudsen number plots, for the most part of these systems' evolution hydrodynamics could not in principle be applied, not just because the system is far from equilibrium but due to their small size. In this context, the mesoscopic interpretation of the kinetic theory appears as a very good candidate to model the observed collective behaviour, naturally interpolating from the hydrodynamic to the particle-like regime. Moreover, as has already been done in literature, it can also be used as a benchmark to test the effectiveness of the hydrodynamical description in these challenging scenarios. Going back to the attractor context, it would nonetheless be interesting to investigate whether information about the initial conditions, which are highly determined by the detailed nuclear structure, can still be accessible in the final observables. More refined initial state models have been developed in recent years, that are able to take into account, for instance, the deformation and clustered structure of small nuclei; moreover, for these systems the detailed longitudinal structure can be determinant for a realistic description and a 2+1D boost-invariant model could no longer be enough. The RBT model can easily host 3D initial conditions and potentially study which is the impact of these longitudinal details on the physical predictions.\\
The near future offers several opportunities in the field of ion collisions which will require a deeper comprehension of the concepts here analysed, above all the hydrodynamisation, the emergence of attractors and the universal scaling of different collision systems. This more theoretical inquiry could furthermore be paired by a more phenomenological investigation, in which the RBT approach may be used to its full potential (event-by-event fluctuations with non-conformal equation of state, realistic transport coefficients and final hadronisation) to make predictions for incoming experimental data.

\addcontentsline{toc}{chapter}{Bibliography}
\bibliography{biblio}

\end{document}